\documentclass[12pt,epsf]{report}

\usepackage{graphicx}
\newcommand{\be}{\begin{equation}}
\newcommand{\ee}{\end{equation}}
\newcommand{\bea}{\begin{eqnarray}}
\newcommand{\eea}{\end{eqnarray}}
\newcommand{\ba}{\begin{array}}
\newcommand{\ea}{\end{array}}
\newcommand{\p}[1]{(\ref{#1})}

\def\bbox{{\,\lower0.9pt\vbox{\hrule \hbox{\vrule height 0.2 cm
\hskip 0.2 cm \vrule height 0.2 cm}\hrule}\,}}
\newcommand{\dsl}{\pa \kern-0.5em /}

\newcommand{\ep}{\epsilon}

\newcommand{\nn}{\nonumber \\}

\def\tr{{\rm tr}}
\def\Tr{{\rm Tr\,}}

\def\ds{\raise.15ex\hbox{/}\kern-.57em\partial}
\def\Ds{\,\raise.15ex\hbox{/}\mkern-13.5mu D}
\newcommand{\balpha}{{\mbox{\boldmath $\alpha$}}}

\newcommand{\bbeta}{{\mbox{\boldmath $\beta$}}}
\newcommand{\bgamma}{{\mbox{\boldmath $\gamma$}}}

\newcommand{\bphi}{{\mbox{\boldmath $\phi$}}}

\newcommand{\btau}{{\mbox{\boldmath $\tau$}}}
\newcommand{\bsigma}{{\mbox{\boldmath $\sigma$}}}
\def\sbsigma{\mbox{\boldmath $\scriptstyle\sigma$}}
\def\sbbeta{\mbox{\boldmath $\scriptstyle\beta$}}
\def\sbmu{\mbox{\boldmath $\scriptstyle\mu$}}
\def\sbalpha{\mbox{\boldmath $\scriptstyle\alpha$}}

\newcommand{\beq}{\begin{equation}}
\newcommand{\eeq}{\end{equation}}
\newcommand{\beqn}{\begin{eqnarray}}
\newcommand{\eeqn}{\end{eqnarray}}

\newcommand{\betabf}{{\mbox{\boldmath $\beta$}}}
\newcommand{\lambdabf}{{\mbox{\boldmath $\lambda$}}}

\newcommand{\sn}[2]{\,\mbox{sn}_#1\/#2}
\newcommand{\cn}[2]{\,\mbox{cn}_#1\/#2}
\newcommand{\dn}[2]{\,\mbox{dn}_#1\/#2}
\def\high{\vphantom{\Biggl(}\displaystyle}

\newcommand{\bI}{{I}}
\newcommand{\bJ}{{J}}
\newcommand{\bmu}{{m}}
\newcommand{\bnu}{{n}}

\makeatletter

\def\vereq#1#2{\lower3pt\vbox{\baselineskip1.5pt \lineskip1.5pt
\ialign{$\m@th#1\hfill##\hfil$\crcr#2\crcr\sim\crcr}}}
\def\lesssim{\mathrel{\mathpalette\vereq<}}

\def\gtrsim{\mathrel{\mathpalette\vereq>}}

\@addtoreset{equation}{section}
\makeatother
\newcommand{\sech}[1]{\,\mbox{sech}\/#1}
\begin{document}

\begin{titlepage}
\vfill
\begin{flushright}
{\normalsize CU-TP-1155}\\
{\normalsize KIAS-P06041}\\
{\normalsize hep-th/0609055}\\
\end{flushright}

\vfill
\begin{center}
{\Large\bf Magnetic Monopole Dynamics, Supersymmetry, and Duality}

\vskip 1.5cm

{ \bf Erick J. Weinberg$^{a,b}$\footnote{\tt ejw@phys.columbia.edu}
and Piljin Yi$^b$\footnote{\tt piljin@kias.re.kr}}

\vskip 1.5cm
$^a${\it Physics Department, Columbia University, New York, N.Y.
10027, U.S.A.}

\vskip 0.5cm
$^b${\it School of Physics, Korea Institute for Advanced Study,} \\
{\it 207-43, Cheongnyangni-2dong, Dongdaemun-gu, Seoul 130-722,
Korea}

\vfill

\end{center}

\vfill

\begin{abstract}
\normalsize\noindent

We review the properties of BPS, or supersymmetric, magnetic
monopoles, with an emphasis on their low-energy dynamics and their
classical and quantum bound states.

After an overview of magnetic monopoles, we discuss the BPS limit and
its relation to supersymmetry.  We then discuss the properties and
construction of multimonopole solutions with a single nontrivial Higgs
field.  The low-energy dynamics of these monopoles is most easily
understood in terms of the moduli space and its metric.  We describe
in detail several known examples of these.  This is then extended to
cases where the unbroken gauge symmetry include a non-Abelian
factor.

We next turn to the generic supersymmetric Yang-Mills (SYM) case, in
which several adjoint Higgs fields are present.  Working first at the
classical level, we describe the effects of these additional scalar
fields on the monopole dynamics, and then include the contribution of the
fermionic zero modes to the low-energy dynamics.  The resulting low-energy
effective theory is itself supersymmetric.  We discuss the
quantization of this theory and its quantum BPS states, which are
typically composed of several loosely bound compact dyonic cores.

We close with a discussion of the D-brane realization of ${\cal N}=4$
SYM monopoles and dyons and explain the ADHMN construction of
monopoles from the D-brane point of view.

\end{abstract}

\vfill

\end{titlepage}
\setcounter{footnote}{0}

\tableofcontents
\newpage

\chapter{Introduction}

The magnetic monopole may be the most interesting, and perhaps the
most important, particle to be never found.

Once the unity of electricity and magnetism was understood, it was
quite natural to conjecture the existence of isolated magnetic poles
that would be the counterparts of electric charges and that would
complete the electric-magnetic duality of Maxwell's equation.
Interest in the possibility of such objects was increased by Dirac's
observation in 1931 \cite{Dirac:1931kp} that the existence of even a
single magnetic monopole would provide an explanation for the observed
quantization of electric charge.

A new chapter opened in 1974, when 't Hooft \cite{'tHooft:1974qc} and
Polyakov \cite{Polyakov:1974ek} showed that in certain spontaneously
broken gauge theories --- a class that includes all grand unified theories ---
magnetic monopoles are not just a possibility, but a prediction.
These objects are associated with solutions of the corresponding
classical field equations and, in the weak coupling regime, their mass
and other properties are calculable.

These theoretical developments were accompanied by experimental
searches for monopoles in bulk matter (including rocks from the Moon)
and cosmic rays as well as by attempts to produce them in particle
accelerators.  To date all of these have been negative, and
theoretical arguments now suggest, at least for GUT monopoles, that
their abundance in the universe is so low as to make the detection of
even one to be extraordinarily unlikely.  Already in 1981, Dirac wrote
\cite{diracquote}, in response to an invitation to a conference on the
fiftieth anniversary of his paper,

\begin{quote}
``I am inclined now to believe that monopoles do not exist.  So
   many years have gone by without any encouragement from the
   experimental side.''
\end{quote}

Yet, in the half-decade preceeding Dirac's statement magnetic
monopoles had inspired two important lines of theoretical inquiry.
First, the attempt to understand why there is not an overabundance of
monopoles surviving from the early universe \cite{Preskill:1979zi} led
Guth to the inflationary universe scenario \cite{Guth:1980zm},
which has revolutionized
our understanding of cosmology.  The second direction, initiated by
the approximation introduced by Prasad and Sommerfield
\cite{Prasad:1975kr} and by Bogomolny \cite{Bogomolny:1975de}, has led
to considerable insights into the properties of supersymmetric field
theories and string theory.  It is this latter line of research that
is the subject of this review.

The Bogomolny-Prasad-Sommerfield (BPS) limit of vanishing scalar
potential was first proposed simply as a means of obtaining an
analytic expression for the classical monopole solution.  However,
over the next few years several remarkable properties of the theory in
this limit emerged.  First, it was shown that solutions of the full
second-order field equations could be obtained by solving the
Bogomolny equation, a kind of self-duality equation that is
first order in the fields.  Solutions of this equation are guaranteed
to have an energy that is exactly proportional to the magnetic charge.
This suggests that there might be static solutions composed of two or
more separated monopoles, with their mutual magnetic repulsion exactly
cancelled by the attractive force mediated by the Higgs field, which
becomes massless in the BPS limit.  This possibility is actually
realized, with there being continuous families of multimonopole
solutions.

In the context of the classical SU(2) theory, the BPS limit seems to
be rather ad hoc, and the properties that follow from it appear to be
curiosities with no deep meaning.  Their relevance for the quantum
theory seems uncertain.  Indeed, it is not even clear that the BPS
limit can be maintained when quantum corrections are included.
However, matters are clarified by the realization that this theory can
be naturally expanded in a way that makes it supersymmetric.  The
resulting supersymmetric Yang-Mills (SYM) theory has a nonvanishing
scalar field potential --- whose form is preserved by quantum
corrections --- but yet yields the same classical field equations.
The classical energy-charge relation is seen to correspond to an
operator relation between the Hamiltonian and the central charges,
with states obeying this relation lying in a special class of
supermultiplets and leaving unbroken some of the generators of the
supersymmetry.  This relation implies the Bogomolny equation in the
weak coupling limit, but is still meaningful when the coupling is so
large that the semiclassical approximation can no longer be trusted.

There is another motivation for making the theory supersymmetric.
Montonen and Olive \cite{Montonen:1977sn} had noted that the classical
mass spectrum was invariant under an electric-magnetic duality
symmetry, and suggested that this might be a symmetry of the full
theory.  However, when particle spins are taken into account, the
spectrum is seen to only be fully invariant if the theory is maximally
expanded, to ${\cal N}=4$ SYM.  If the field theory is viewed as a
low-energy approximation to string theory, this duality symmetry is a
reflection of the S-duality of the string theory.

The supersymmetry brings in other new features.  If the gauge group is
larger than SU(2), the additional scalar fields of the SYM theory give
rise to new classical solutions that can be viewed as loosely bound
collections of two or more dyonic cores.  These can be studied within
the quantum theory with the aid of a low-energy effective theory.
This is a truncation of the full quantum field theory that retains
only the bosonic collective coordinates of the individual monopoles
and their fermionic counterparts associated with fermion zero modes.
The correspondence between the classical and quantum theories turns
out to be rather subtle, revealing some unexpected aspects 
of the spectra of supersymmetric, or BPS, states in SYM theories.

The spectrum of BPS states becomes of particular importance in the
context of duality, both in field theory and in string theory.  In the
1990's, as the notion of duality took on a more definite formulation,
the study of the BPS spectra became a primary tool for checking
whether a duality existed between a pair of theories. This was true
not only for the self-duality of ${\cal N}=4$ SYM field theory, but
for the various dualities between the five string theories in ten
dimensions and M theory in eleven dimensions, where supersymmetric
states involving D-branes are often counted and matched with their
dual BPS states.

Although the initial investigations in this direction were fairly
successful, the counting of more general BPS states turned out to be
anything but straightforward. This was particularly true for those BPS
states that preserve four or fewer supercharges.  The existence of
such states is often sensitive to the choice of vacuum and to the
choice of coupling constants, which are then interwoven with the
duality in a subtle manner. So far, no general theory of BPS spectra
is known, although much effort has been devoted to attacking this
problem in the many guises in which it presents itself, including
D-branes wrapping cycles in Calabi-Yau manifolds, boundary states in
conformal field theories, open membranes ending on appropriately
curved M5 branes, and BPS dyons in the Seiberg-Witten description of
${\cal N}=2$ SYM.

We believe the material presented in this review will shed much light
on this general unsolved problem.  While the methodology used here is
itself somewhat limited, in that it deals with weakly coupled SYM,
some of the qualitative features should prove to be common to all
these related problems.  These include both the marginal stability
domain wall and the large degeneracy, unrelated to any known symmetry,
that is often found in BPS states with large charges.

We begin our review, in Chap.~2, with an overview of the SU(2)
magnetic mono\-pole solution of 't Hooft and Polyakov.  We describe
how it arises as a topological soliton.  We also discuss the zero
modes about the solution, and explain how one of these, related to the
unbroken global gauge symmetry, leads to the existence of dyonic
solutions carrying both electric and magnetic charges.  We introduce
the moduli space of solutions and its metric; these concepts play an
important role in our later discussions.

In Chap.~3, we specialize to the case of BPS solutions.  We discuss in
detail their relation to supersymmetry, and describe how magnetically
charged states that preserve part of the
supersymmetry can be obtained.  The Montonen-Olive duality conjecture
is also introduced here.

Chapter~4 is devoted to the discussion of classical multimonopole
solutions.  After developing the formalism for describing monopoles in
theories with gauge groups larger than SU(2), we show how index theory
methods can be used to determine the dimension of the space of
solutions.   We describe a powerful method, introduced by Nahm,
for constructing multimonopole solutions and illustrate its
use with several examples.

For a given magnetic charge, the multimonopole solutions obtained in
Chap.~4 form a manifold, the moduli space, with the coordinates on
this manifold naturally taken to be the collective coordinates of
the component monopoles.  In the low-energy limit, a good
approximation to the full field theory dynamics is obtained by
truncating to these collective coordinates, whose behavior is governed
by a purely kinetic Lagrangian that is specified by a naturally
defined metric on the moduli space.  The classical motions of the
monopoles correspond to geodesics on the moduli space.   We discuss
the moduli space and its metric in Chap.~5, and describe, with examples,
some methods by which the metric can be determined.

Most of the discussion in this review assumes that the gauge group is
maximally broken, to a product of U(1)'s.   In Chap.~6 we discuss some
of the consequences of the alternate possibility, where there is
an unbroken non-Abelian subgroup.  Among these are the presence of
``massless monopoles'' that are the dual counterparts of the
massless gauge bosons and their superpartners.  These massless monopoles
cannot be realized as isolated classical solitons, but are instead
found as clouds of non-Abelian field surrounding one or more massive
monopoles.

Although SYM theories with extended supersymmetry contain either two
(for ${\cal N}=2$) or six (for ${\cal N}=4$) Higgs fields, the
discussion of classical solutions up to this point assumes that only
one of these is nontrivial.  For an SU(2) gauge theory this can always
be arranged by a redefinition of fields.  However, for larger gauge
groups it need not be the case, a point that was fully appreciated
only relatively recently.  The generic solution then has two
nontrivial Higgs fields, and is typically a dyonic bound state with
components carrying both magnetic and electric charges.  The BPS
solutions preserve only one-fourth, rather than one-half, of the
supersymmetry.  The low-energy dynamics can still be described in
terms of the collective coordinates and the moduli space that they
span, but the moduli space Lagrangian now includes a potential energy
term.  In Chap.~7 we discuss the effects of these additional scalar
fields, and show how to derive the potential energy that they
generate.

Although we are considering supersymmetric theories, the effects of
the fermions have been omitted so far.  This is remedied in Chap.~8,
where we introduce fermionic counterparts to the bosonic collective
coordinates, and derive the effects of the fermion fields on the
low-energy dynamics.  The resulting moduli space Lagrangian possesses
a supersymmetry that is inherited from that of the underlying field
theory.  The discussions thus far, although being set in the context
of a quantum field theory, have been essentially classical.  In this
chapter we also show how to quantize this low-energy Lagrangian.

In Chap.~9, we discuss the quantum BPS dyons that arise as bound
states of this Lagrangian.  In a sense, this is the quantum
counterpart of the classical discussion of Chap.~7.  We present the
exact wavefunctions for several states comprising only two dyonic
cores.  Although it appears to be much more difficult to obtain
explicit wavefunctions for states with many cores, we show how index
theorems can be used to count such states.  An important point that
emerges here is the striking difference between the spectra of the
${\cal N}=2$ and ${\cal N}=4$ SYM theories.  In particular, only the
latter includes certain zero-energy bound states that are required to
satisfy the duality conjecture for larger gauge groups.

Although the monopoles that we discuss arise originally in the context
of field theory, they find a very natural setting when the quantum
field theory is viewed as the low-energy limit of string theory.  In
Chap.~10, we describe how the monopoles and dyons of ${\cal N}=4$ SYM
can be realized, in a rather elegant fashion, in terms of D-branes.
In particular, we describe how this picture provides a very natural
motivation for the multimonopole construction of Nahm.

There are two appendices.  Appendix~A provides some background
material on complex geometry and zero modes.  Appendix B describes the
extension of the discussions of Chap.~8 to ${\cal N}=2$ SYM theories
containing matter hypermultiplets.

\chapter{The SU(2) magnetic monopole}
\label{su2chap}

The first example of a magnetic monopole solution was discovered by 't
Hooft \cite{'tHooft:1974qc} and Polyakov \cite{Polyakov:1974ek},
working in the context of an SU(2) gauge theory.  Although many
examples with larger gauge groups have subsequently been found, the
SU(2) solution remains the simplest, and is perhaps the best suited
for introducing some concepts that will be important for our
subsequent discussions.  Furthermore, this solution will play an
especially fundamental role for us, because in the
Bogomolny-Prasad-Sommerfield (BPS)
\cite{Prasad:1975kr,Bogomolny:1975de} limit the monopole solutions for
larger groups are all built up, in a sense that will later become
clear, from components that are essentially SU(2) in nature.

We start our discussion in Sec.~\ref{sec2-1}, where we describe how
nonsingular magnetic monopoles can arise as topological solitons, and
then focus on the 't Hooft-Polyakov solution in Sec.~\ref{sec2-2}.
These static solutions actually belong to continuous families of
solutions, all with the same energy.  In the one-monopole case
considered in this chapter, these solutions are specified by four
parameters, all related to the symmetries of the theory; in later
chapters we will explore the much richer multimonopole structure that
arises in the BPS limit.  As we explain in Sec.~\ref{zeromodes},
infinitesimal variations of these parameters are associated with zero
modes about a given solution.  Excitation of these modes gives rise to
time-dependent monopole solutions.  Translational zero modes thus give
rise in a straightforward manner to solutions with nonzero linear
momentum.  The case of global gauge modes, which lead to dyonic
solutions with nonzero electric charge \cite{Julia:1975ff}, is
somewhat more subtle, as we describe in Sec.~\ref{dyons}.  In
Sec.~\ref{chaptwoModspace} we describe how a family of degenerate
static solutions can be viewed as forming a manifold, known as the
moduli space, with a naturally defined metric.  Although this concept
is relatively trivial for the one-monopole solutions considered in
this chapter, it proves to be a powerful tool for understanding the
multimonopole solutions we will study in later chapters.  Finally, in
Sec.~\ref{sec2-quant}, we discuss the relevance of these classical
soliton solutions for the quantum theory.

Our main focus in this chapter is on providing the background for the
discussion for monopoles in the BPS limit.  Of necessity, there are
many other aspects of magnetic monopoles that we must omit.  For
further discussion of these, we refer the reader to two classic
reviews \cite{coleman-fifty, Preskill:1986kp}.

\section{Magnetic monopoles as topological solitons}
\label{sec2-1}

We consider an SU(2) gauge theory whose symmetry is spontaneously
broken to U(1) by a triplet Higgs field $\Phi$.  With the
generalization to other gauge groups in mind, we will usually write
the fields as Hermitian matrices in the fundamental representation of
the group.  However, for this SU(2) example it will sometimes be more
convenient to work in terms of component fields defined by
\begin{equation}
         A_\mu = {1\over 2} \tau^a A_\mu^a \, , \qquad \qquad
       \Phi = {1\over 2} \tau^a \Phi^a \, .
\end{equation}
Our conventions will be such that $\Phi^a\Phi^a = 2 \Tr \Phi^2 \equiv
|\Phi|^2$.

The Lagrangian is
\begin{equation}
    {\cal L} =- {1\over 2} \Tr F_{\mu\nu} F^{\mu\nu}
      + \Tr D_\mu \Phi D^\mu \Phi -V(\Phi)
\label{lagrangian}
\end{equation}
where
\begin{equation}
   V(\Phi) = - \mu^2 \Tr \Phi^2  + \lambda (\Tr \Phi^2)^2  \, ,
\end{equation}
\begin{equation}
    D_\mu \Phi = \partial_\mu \Phi +ie [A_\mu, \Phi]  \, ,
\end{equation}
and
\begin{equation}
    F_{\mu\nu} = \partial_\mu A_\nu -\partial_\nu A_\mu
      +ie [A_\mu , A_\nu]  \, .
\end{equation}
It is often convenient to separate the field strength into magnetic
and electric parts
\begin{equation}
    B_i = {1 \over 2} \epsilon_{ijk} F_{jk}
\end{equation}
and
\begin{equation}
    E_i =  F_{0i} \, .
\end{equation}

In order that there be a lower bound on the energy, $\lambda$ must be
positive.  If $\mu^2>0$, as we will assume, there is a degenerate
family of asymmetric classical minima with
\begin{equation}
    |\Phi|= v \equiv \sqrt{\mu^2\over \lambda} \, .
\label{su2vev}
\end{equation}
These preserve only a U(1) subgroup, which we will
describe with the language of electromagnetism.

For definiteness, let us choose the vacuum solution
\begin{eqnarray}
   \Phi(x) &=& v {\tau^3\over 2} \equiv \Phi_0  \cr\cr
   A_\mu(x) &=& 0
\label{vacsoln}
\end{eqnarray}
so that the unbroken U(1) corresponds to the ``$a=3$'' direction of
the SU(2). The physical fields may then be taken to be
\begin{eqnarray}
    {\cal A}_\mu &=& A_\mu^3    \cr \cr
    W_\mu &=& {A_\mu^1 + i A_\mu^2 \over \sqrt{2}} \cr \cr
    \varphi &=&  \Phi^3 \, .
\label{physfields}
\end{eqnarray}
The corresponding elementary quanta are a massless ``photon'', a pair
of vector mesons with mass $m_W= ev$ and electric charges $\pm e$, and an
electrically neutral scalar boson with mass $m_H= \sqrt{2}\,\mu$.

This theory also has nontrivial classical solutions.  The
existence of these can be demonstrated without having to examine the
field equations in detail.  The key fact is that
any static configuration that is a local minimum of the energy is
necessarily a solution of the classical field equations.  Our strategy
will be to identify a special class of finite energy configurations
and then show that the configuration of minimum energy among these
cannot be the vacuum.

It is fairly clear that in a finite energy solution the fields must
approach a vacuum solution as $r \rightarrow \infty$ in any fixed
direction.  However, this need not be the same vacuum solution in
every direction.  Thus, we could allow
\begin{equation}
    \lim_{r\rightarrow \infty} \Phi(r,\theta,\phi)
       \equiv \Phi_\infty(\Omega) = U(\Omega) \Phi_0 U^{-1}(\Omega)
\end{equation}
to vary with direction, provided that it is accompanied by a suitable
asymptotic gauge potential.  (Note that the smoothness of $\Phi$ does not
imply that $U$ must be smooth.  In the cases of most interest to us,
$U$ has a singularity.)  The function $\Phi_\infty(\Omega)$ is a
continuous mapping of the two-sphere at spatial infinity onto the
space of Higgs fields obeying Eq.~(\ref{su2vev}), which happens to
also be a two-sphere.  Such maps can be classified into topologically
distinct classes corresponding to the elements of the homotopy group
$\Pi_2(S^2)$.  Any two maps within the same class can be continuously
deformed one into the other, while two maps in different classes
cannot.

One can show that $\Pi_2(S^2) = Z$, the additive group of the
integers, so that configurations can be labelled by an integer
winding number
\begin{equation}
    n = {1\over 8\pi} \epsilon_{ijk}\, \epsilon_{abc} \int d^2S_i \,
       \hat \phi^a \, \partial_j \hat\phi^b \, \partial_k \hat\phi^c
\label{windingnumber}
\end{equation}
where $\hat \phi^a$ is the unit vector $\Phi^a/|\Phi|$ and the
integration is taken over a sphere at spatial infinity.  In fact,
since the value of the integral is quantized, it must be invariant
under smooth deformations of the surface of integration that do not
cross any of the zeroes of $\Phi$, where $\hat \phi^a$ is undefined.  It
follows that any configuration with nonzero winding number must have
at least $|n|$ zeroes of the Higgs field (precisely $n$ if one
distinguishes between zeroes and ``antizeroes'' and counts the latter
with a factor of $-1$).\footnote{This relation between
topological charge and zeroes of the Higgs field does not have a
simple extension to the case of larger groups.}

The vacuum solution of Eq.~(\ref{vacsoln}) clearly has $n=0$ and falls
within the trivial, identity, element of the homotopy group.

Now consider the set of field configurations such that the Higgs field
at spatial infinity has unit winding number, $n=1$.  Among these
configurations, there must be one with minimum energy.  This cannot be
a vacuum solution (because it has a different winding number) and
cannot be smoothly deformed into the vacuum (because winding number is
quantized).  Hence, this configuration must be a local minimum of the
energy and thus a static classical solution.\footnote{There is
actually a loophole in this argument.  Because the space of field
configurations is not compact, there might not be a configuration of
minimum energy.  For example, there is in general no static solution
with winding number $n=2$, because the minimum energy for a pair of
monopoles is achieved only when the monopoles are infinitely far
apart.  (An exception occurs in the BPS limit.)  Even for $n=1$, the
existence of singular configurations causes the extension of this
argument to curved spacetime to fail if $v$ is too large
\cite{VanNieuwenhuizen:1975tc,Lee:1991vy}.}

Some of the asymptotic properties of this solution can be obtained
from general arguments.  In order that the energy be finite, the
asymptotic Higgs field must be of the form
\begin{equation}
    \Phi^a = v {\hat \phi}^a \, .
\end{equation}
Further, the covariant
derivative $D_i \Phi$ must fall faster than $r^{-3/2}$, which
implies that
\begin{equation}
    \partial_i {\hat \phi}^a -  e \,\epsilon_{abc} \, A_i^b\,\hat\phi^c
          < O(r^{-3/2}) \, .
\end{equation}
This in turn requires that the gauge potential be of the form
\begin{equation}
    A_i^a =  {1 \over e} \epsilon_{abc}\, \hat\phi^b \, \partial_i
          \hat\phi^c
        + f_i({\bf r}) \hat\phi^a  + \cdots
\end{equation}
where the ellipsis
represents terms that fall faster than $r^{-3/2}$.

The corresponding magnetic field is
\begin{equation}
    B_i^d = {1\over 2} \epsilon_{ijk}
          \left[ {1\over e} \epsilon_{abc}\, \hat\phi^a \,
     \partial_j\hat\phi^b  \,   \partial_k\hat\phi^c
   +(\partial_j f_k - \partial_k f_j) \right]\hat\phi^d
     + \cdots \, .
\label{asymfield}
\end{equation}
Its leading terms are proportional to ${\hat \phi}^a$ and thus lie in
the ``electromagnetic'' U(1) defined by the Higgs field.  We define the
magnetic charge by
\begin{equation}
  Q_M =  \int d^2S_i \, \hat\phi^a \, B_i^a \, .
\label{su2magchargedef}
\end{equation}
When Eq.~(\ref{asymfield}) is inserted into this expression, the first
term gives a contribution proportional to the winding number
(\ref{windingnumber}), which we are assuming to be unity, while the
contribution from the second term vanishes as a consequence of Gauss's
theorem; hence, the solution corresponds to a magnetic monopole with
charge $4\pi/e$.  More generally \cite{Arafune:1974uy}, a solution
with Higgs field winding number $n$ has a magnetic charge\footnote{The
Dirac quantization condition would have allowed magnetic charges $2\pi
n/e$.  The more restrictive condition obtained here can be understood
by noting that it is possible to add SU(2) doublet fields to the
theory in such a way that the classical solution is unaffected.  After
symmetry breaking these doublets would have electric charges $\pm e/2$
and the Dirac condition would become the same as the topological
condition obtained here.}
\begin{equation}
  Q_M =  {4 \pi n \over e} \, .
\end{equation}

The classical energy of the solution, which gives the leading
approximation to the monopole mass, is
\begin{equation}
    E = \int d^3x \left[\Tr E_i^2 +  \Tr (D_0\Phi)^2  +\Tr B_i^2
         + \Tr (D_i\Phi)^2 + V(\Phi)  \right] \, .
\label{classenergy}
\end{equation}
For a static solution with no electric charge, one would expect the
first two terms to vanish.  The contribution of the remaining terms
can be estimated by rewriting Eq.~(\ref{classenergy}) in terms of the
dimensionless quantities ${\bf s} = ev{\bf x}$, $\psi= \Phi/v$, and
$a_i = A_i/v$.  This isolates the dependence on $e$ and $v$, and shows
that the mass must be of the form
\begin{equation}
    M =  {4\pi v \over e} f(\lambda/e^2)
\label{monomass}
\end{equation}
where $f(\lambda/e^2)$ is expected to be of order unity.

\section{The 't Hooft-Polyakov solution}
\label{sec2-2}

In trying to proceed beyond this point, considerable simplification is
achieved by restricting to the case of spherically symmetric
solutions.  In a gauge theory this means that the fields must be
invariant under the combination of a naive rotation and a compensating
gauge transformation, which may be position-dependent.  With unit
winding number, $n=1$, this position-dependence can be eliminated by
adopting a special gauge choice that correlates the orientation of the
Higgs field in internal space with the direction in physical space.
In this ``hedgehog'' gauge, rotational invariance requires that the
fields be invariant under combined rotation and global internal SU(2)
transformation.  This gives the ansatz\footnote{Rotational symmetry
would also allow contributions to $A_i^a$ proportional to
$\delta_{ia}$ or $\hat r^i \hat r^a$, both of which have opposite
parity from the terms included in this ansatz.  Including these terms
does not lead to any new solutions.}
\begin{eqnarray}
      A_i^a &=&  \epsilon_{iam} \hat r^m \left[1-u(r)\over er \right]  \cr
      \Phi^a &=&  \hat r^a  h(r)
\label{hedghogansatz}
\end{eqnarray}
for the Higgs field and the spatial components of the gauge potential
\cite{'tHooft:1974qc,Polyakov:1974ek}.  It is easy to verify that for
time-independent fields one can consistently set $A_0=0$ in the field
equations, and we do so now.

The equations obeyed by $u$ and $h$ can be obtained either by
substituting the ansatz (\ref{hedghogansatz}) directly into the field
equations, or by substituting it into the Lagrangian in Eq.~(\ref{lagrangian})
and then varying the resulting expression with respect to these
coefficient functions.  (The latter procedure is allowed because the
ansatz is the most general one consistent with a symmetry of the
Lagrangian.) Either way, one obtains
\begin{eqnarray}
   0 &=&  h'' +{2\over r} h' - {2u^2h \over r^2} + \lambda (v^2-h^2)h
          \cr
   0 &=& u'' - {u(u^2-1)\over r^2 } - e^2 u h^2
\label{su2monoeqs}
\end{eqnarray}
with primes denoting derivatives with respect to $r$.  Finiteness of
the energy requires that $u(\infty) = 0$ and $h(\infty) = v$, while
requiring that the fields be nonsingular at the origin implies that
$u(0)=1$ and $h(0)=0$.

In general, these equations can only be solved numerically. There is a
central core region, of radius $R_{\rm mon} \sim 1/ev$, outside of
which $u$ and $|h-v|$ decrease exponentially with distance. The
function appearing in Eq.~(\ref{monomass}) for the monopole mass is a
monotonic function of $\lambda/e^2$ with limiting values $f(0)=1$ and
$f(\infty) = 1.787$ \cite{Kirkman:1981ck}

It is instructive to gauge transform this solution from the hedgehog
gauge into a ``string gauge'' where the Higgs field direction is uniform.
This can be done, for example, by the gauge transformation
\begin{equation}
     U = e^{-i\phi \tau_3/2} e^{i\theta \tau_2/2}
           e^{i\phi \tau_3/2} \, .
\label{hhtostringgauge}
\end{equation}
(This gauge transformation is singular along the negative
$z$-axis.  Such a singularity is an inevitable consequence of any
transformation that changes the homotopy class of the Higgs field at
infinity.)  In terms of the physical fields defined in
Eq.~(\ref{physfields}), this leads to
\begin{eqnarray}
    {\cal A}_i &=&- \epsilon_{ij3}\, {\hat r_j\over er}
      {1\over (1+\cos\theta)} \cr
    W_i&=&  {u(r)\over er} v_i \cr \cr
    \varphi &=&  h(r)
\label{stringansatz}
\end{eqnarray}
where the complex vectors
\begin{eqnarray}
    v_1 &=&- {i\over \sqrt{2}} \left[1- e^{i\phi}\cos\phi
      (1-\cos\theta) \right]   \cr
    v_2 &=& {1\over \sqrt{2}} \left[1 + ie^{i\phi}\sin\phi
      (1-\cos\theta) \right]      \cr
    v_3 &=&  {i\over \sqrt{2}} e^{i\phi}\sin\theta
\label{stringvectors}
\end{eqnarray}
obey $v_j^*v_j = 1$.

The gauge transformation that connects the string and hedgehog gauges
is not uniquely determined.  If we had multiplied the gauge
transformation of Eq.~(\ref{hhtostringgauge}) on the left by
$e^{i\alpha \tau_3/2}$, the only effect would have been to multiply
$W_j$ by a phase factor $e^{-i \alpha}$.  This freedom to rotate by an
arbitrary phase while staying within the string gauge is a reflection
of the unbroken U(1) symmetry.

The ${\cal A}_j$ that appears in Eq.~(\ref{stringansatz}) is just the
Dirac magnetic monopole potential and yields a Coulomb magnetic field
corresponding to a point magnetic monopole.  Usually such a field
would imply a Coulomb energy that diverged near the location of the
monopole.  This divergence is avoided because the charged massive
vector field gives rise to a magnetic moment density
\begin{equation}
     \mu_{ij} = -ie (W^*_i W_j - W^*_j W_i)
\end{equation}
that orients itself relative to the magnetic field in such a way as
to cancel the divergence in the Coulomb energy.

\section{Zero modes and time-dependent solutions}
\label{zeromodes}

The unit monopole described in the previous section should be viewed
as just one member of a four-parameter family of solutions.  Three of
these parameters correspond to spatial translation of the monopole and
are most naturally chosen to be the coordinates $\bf z$ of the
monopole center.  The fourth parameter is the U(1) phase noted in the
discussion below Eq.~(\ref{stringvectors}).  Infinitesimal variation
of these parameters gives field variations $\delta A_i$ and $\delta
\Phi$ that leave the energy unchanged and preserve the field
equations.  Hence, they correspond to zero-frequency modes of fluctuation
(or simply ``zero modes'') about the monopole.

In addition to these four modes, there are an infinite number of zero
modes corresponding to localized gauge transformations of the
monopole.  However, these are less interesting because the new
solutions obtained from them are physically equivalent to the original
solution.  These zero modes are eliminated once a gauge condition is
imposed.

One might wonder why we should want to retain the mode associated with
the U(1) phase, since it is also a gauge mode.  If we were only
concerned with static solutions, then we could indeed ignore this
mode.  The distinction between this global gauge mode and the local
gauge modes only becomes important when we consider time-dependent
excitations of these modes.  As we will describe below, excitation of
the global gauge mode leads to solutions with nonzero electric charge.
By contrast, the solutions obtained by time-dependent excitations of
the local gauge modes are still physically equivalent to the original
solution.\footnote{The distinction between the two types of gauge
modes is also seen in the corresponding Noether charges.  The
conserved quantity corresponding to the global phase invariance is the
electric charge.  The conservation laws associated with the local
gauge symmetries are simply equivalent to the Gauss's law constraint at
each point in space, and yield no additional conserved quantities.}

We begin by considering time-dependent excitations of the
translational zero modes.  These should yield solutions of the field
equations with nonvanishing linear momentum.  At first thought, one
might expect to obtain a time-dependent solution by simply making the
substitution ${\bf r} \rightarrow {\bf r} - {\bf v}t$ in the static
solution.  This is almost, but not quite, right.  First, the Lorentz
contraction of the monopole modifies the field profile.  However,
since this is an effect of order ${\bf v}^2$, we can ignore it for
sufficiently low velocities.  Second, we must ensure that the Gauss's
law constraint
\begin{equation}
    0 = D_j F^{j0} - i e [\Phi, D^0\Phi]
\label{gauss}
\end{equation}
is obeyed.  For most choices of gauge, this implies a nonzero
$A_0$.\footnote{Neither the hedgehog gauge nor the string gauge is
well-suited for dealing with moving monopoles.  In the former the
configurations develop gauge singularities once the zero of the Higgs
field moves away from the origin, while in the latter the position of
the Dirac string must be allowed to move.  There are other gauges
(e.g., axial gauge) that avoid these difficulties
\cite{Christ:1976cg}; since this issue is peripheral to our main
focus, we will not pursue it further.}

Even without solving for $A_0$, we can still calculate the kinetic
energy associated with this linear motion of the monopole.  The
time-dependence of $A_k$ and of $\Phi$ comes solely from the factors
of ${\bf v}t$ in the solution.  Hence,
\begin{eqnarray}
    F_{0i} &=& \partial_0 A_i  - D_i A_0  \cr
           &=& -v^k\partial_k A_i - D_i A_0  \cr
           &=& -v^k F_{ki} - D_i(v^k A_k + A_0)
\end{eqnarray}
\begin{eqnarray}
     D_0 \Phi &=& \partial_0 \Phi + ie[A_0, \Phi] \cr
           &=&  -v^k\partial_k \Phi + ie[A_0, \Phi] \cr
           &=&  -v^k D_k\Phi + ie[(A_0 + v^kA_k), \Phi] \, .
\end{eqnarray}
The kinetic energy can then be written as
\begin{eqnarray}
    \Delta E &=& \int d^3x \Tr \left\{ F_{0i}^2 + (D_0 \Phi)^2
                \right\} \cr
      &=& \int d^3x \Tr \left\{  (v^k F_{ki})^2 + (v^k D_k\Phi)^2
          \right\}  \cr
     &+&  \int d^3x \Tr \left\{ (A_0 + v^jA_j) \left[ D^i (F_{0i}
         - v^k F_{ki}) + ie[\Phi, (D_0\Phi - v^k D_k\Phi)]\right]
\right\}  \, .  \cr
   &&
\end{eqnarray}
The second factor in the last integral vanishes as a result of Gauss's
law and the field equations obeyed by the static solution.  Using the
rotational invariance of the static solution, we can rewrite the
remaining terms to obtain
\begin{equation}
    \Delta E = {\bf v}^2 \int d^3x \Tr \left[ {2\over 3} B_i^2
          + {1\over 3} (D_i\Phi)^2 \right] \, .
\end{equation}
The fact that the static solution must be a stationary point of the
energy under rescalings of the form ${\bf x} \rightarrow \rho {\bf x}$,
$A_i \rightarrow \rho^{-1} A_i$ implies a virial theorem
\begin{equation}
    0 = \int d^3x \left[ \Tr B_i^2  - \Tr  (D_i\Phi)^2 - 3 V(\Phi)
            \right] \, .
\end{equation}
Multiplying this by ${\bf v}^2/6$, subtracting the result from the previous
equation, and then recalling Eq.~(\ref{classenergy}) for the mass, we
find that the translational kinetic energy is
\begin{equation}
   \Delta E = {1\over 2} M {\bf v}^2 \, ,
\end{equation}
in perfect accord with non-relativistic expectations.

\section{Dyons}
\label{dyons}

Let us now consider time-dependent excitations of the zero mode that
rotates the phase of the charged vector meson fields.  Just as
excitation of the translational zero modes leads to solutions with
nonzero values of the linear momentum, the conserved quantity
corresponding to translational symmetry, excitation of this U(1)-phase
zero mode produces a nonzero value for the corresponding Noether
charge.  The physical significance of this lies in the fact that the
Noether charge of a gauged symmetry is (up to a factor of the gauge
coupling) also the source of the gauge field.  In the case at hand,
the Noether charge
is the electric charge of the unbroken U(1), and the solutions
produced by excitation of this zero mode are dyons, objects carrying
both electric and magnetic charge.

At least to start, it is easiest to work in the string
gauge.  In terms of the field variables defined in
Eq.~(\ref{physfields}), the electric charge is
\begin{equation}
     Q_E =   -i e \int d^3 x  \left[ W^j{}^* ({\cal D}_0 W_j -{\cal D}_j
     W_0) - W^j ({\cal D}_0 W^*_j -{\cal D}_j W^*_0)\right]
\label{fieldcharge}
\end{equation}
where the U(1) covariant derivative ${\cal D}_\mu W_\nu = (\partial_\mu
-ie{\cal A}_\mu) W_\nu$.

To construct dyon solutions, we begin with a static
solution in the string gauge and multiply the $W$-field by
a uniformly varying U(1) phase factor $e^{i\omega t}$.  The U(1) Gauss's
law  requires a nonzero ${\cal A}_0$ [which itself contributes to $Q_E$
through the covariant derivative in Eq.~(\ref{fieldcharge})]. It also
guarantees that the asymptotic U(1) electric field ${\cal F}_{0j}=
\partial_0 {\cal A}_j -\partial_j {\cal A}_0$ satisfies
\begin{equation}
     Q_E = \int d^2S_i {\cal F}_{0i} = \int d^2S_i \hat\phi^a
     E_i^a
\label{gausscharge}
\end{equation}
where the integrals are over a sphere at spatial infinity.  Plugging
the resulting ${\cal A}_0$ back into the other field equations yields
corrections to the field profiles that are proportional to $\omega^2$.
These are analogous to the $O(v^2)$ corrections due to Lorentz
contraction that were noted in the previous section, and can be
neglected for sufficiently small $Q_E$.

To be more explicit, we start with the string-gauge form of the
spherically symmetric solution, given in Eq.~(\ref{stringansatz}), and
assume a spherically symmetric ${\cal A}_0(r)$.  Gauss's law,
Eq.~(\ref{gauss}), reduces to
\begin{equation}
     0= {\cal A}_0'' + {2 \over r}{\cal A}_0' -{2 u^2\over r^2}
     \left({\cal A}_0 - {\omega\over e} \right) \, .
\label{calAeq}
\end{equation}
Recalling that $u(0)=1$, we see that we must require ${\cal A}_0(0) =
\omega/e$ in order to avoid a singularity at the origin.  We also
require\footnote{It is not necessary to require that ${\cal
A}_0(\infty)$ vanish.  However, after following through steps
analogous to those shown below, one finds that starting with a nonzero
${\cal A}_0(\infty)$ is equivalent to starting with ${\cal
A}_0(\infty)=0$ and a different value for $\omega$.}  ${\cal
A}_0(\infty)=0$.  With these boundary conditions imposed, ${\cal
A}_0(r)$, and hence $Q_E$, are proportional to $\omega$.

This time-dependent solution can be transformed into a static solution
by a U(1) gauge transformation of the form
\begin{eqnarray}
     W_i &\rightarrow& \tilde W_i = e^{i\Lambda} W_i     \cr
     {\cal A_\mu} &\rightarrow& {\tilde {\cal A}_\mu}
               = {\cal A_\mu} + {1\over e}\partial_\mu\Lambda
\end{eqnarray}
with $\Lambda = -\omega t$.  This shifts the scalar potential by a
constant, so that $\tilde {\cal A}_0(0) = 0$ and $\tilde {\cal A}_0
(\infty) = -\omega/e$.  In this static form it is easy to transform
the solution into the manifestly nonsingular hedgehog gauge, with
$A_i^a$ and $\Phi^a$ as in Eq.~(\ref{hedghogansatz}) and
\begin{equation}
    A_0^a = \hat r^a j(r)
\end{equation}
with $j(r) = \tilde {\cal A}_0(r)$.  After this modification to the
spherically symmetric ansatz, the static field equations become
\cite{Julia:1975ff}
\begin{eqnarray}
   0 &=&  h'' +{2\over r} h' - {2u^2h \over r^2} + \lambda (v^2 -h^2)h
       \label{dyonHeq}   \\
   0 &=&  u'' - {u(u^2-1)\over r^2 } - e^2 u (h^2  - j^2)
       \label{dyonUeq}  \\
   0 &=&  j''  + {2\over r} j' - {2u^2j \over r^2} \, .
       \label{dyonJeq}
\end{eqnarray}
The first of these equations is the same as in the purely magnetic
case, while the second differs only by the addition of the $O(Q_E^2)$
term $2e^2 u j^2$, in accord with the remarks above.  The last is
equivalent to Eq.~(\ref{calAeq})

In this static form, which was the approach used by Julia and Zee
\cite{Julia:1975ff} in their original discussion of the dyon solution,
the electric charge does not directly appear as a consequence of a
rotating phase.  Instead, the spectrum of electric charges corresponds
to the existence of a one-parameter family of solutions to
Eqs.~(\ref{dyonHeq}) - (\ref{dyonJeq}) characterized by $j(\infty) = -
\omega/e$.  From the large distance behavior of the second of these
equations, we see that the existence of a solution requires that
$|j(\infty)| < h(\infty)$, and hence that $|\omega| < ev$.

To obtain a relation between $Q_E$ and $\omega$, we return to
Eq.~(\ref{fieldcharge}).  Substituting our ansatz into the right-hand
side of that equation, and recalling that ${\cal W}_0=0$, we obtain
\begin{equation}
     Q_E = {8\pi \omega\over e} \int dr\, u(r)^2 \left[{j(r)
                    \over j(\infty)} \right]
     \equiv I \omega \, .
\label{QandIandOmega}
\end{equation}
The integral can be estimated by noting that
$u(r)$ falls exponentially outside a region of
radius $\sim 1/ev$, implying that
\begin{equation}
       I = {4\pi k \over e^2 v }
\end{equation}
with $k$ of order unity.  For $|Q_E| \ll Q_M $ the 
the field profiles are, apart from an overall rescaling, only weakly dependent
on the charge, and so $I$ is essentially independent of $\omega$.  However, 
as $|\omega|$ approaches its limiting value $ev$, the profiles are deformed so
that $I$ grows without bound.   As a result, the upper bound on $|\omega|$ 
does not imply an upper bound on $|Q_E|$.

As a consistency check, let us verify that the electric charge
given in Eq.~(\ref{fieldcharge}) agrees with that obtained from
the asymptotic behavior of the electric field, whose radial component
is equal to $-j'(r)$.  Integrating Eq.~(\ref{dyonJeq})
leads to
\begin{equation}
   -j'(r) = - {2\over r^2} \int_0^r ds \, u(s)^2 j(s) \, .
\end{equation}
For large $r$, where the integrand is
exponentially small, we introduce a negligible
error by replacing the upper limit of the integral by infinity.
Together with Eq.~(\ref{QandIandOmega}), this gives
\begin{equation}
   -j'(r) =  {Q_E\over 4\pi r^2}  \, ,
\end{equation}
as required.

Finally, let us calculate the correction to the mass associated with
the electric charge.  To lowest order,
\begin{eqnarray}
     \Delta E &=& \int d^3 x \Tr F_{0i}^2 \cr
      &=& \int d^3 x \left[ -\partial_i(\Tr A_0F_{0i})
        + \Tr A_0 D_iF_{0i} \right] \cr
       &=& -{1\over 2}j_0(\infty) Q_E = {\omega \over 2e} Q_E \, .
\label{dyonDeltaE}
\end{eqnarray}
(We have used the equations of motion and the fact that $D_0\Phi =0$
to eliminate the final term in the integrand on the second line.)
Recalling Eqs.~(\ref{classenergy}) and (\ref{dyonDeltaE}), we then
obtain
\begin{equation}
     \Delta E = {Q_E^2 \over 2e I}
         = \left({e \over 4\pi}\right)^2 {Q_E^2 M \over 2kf}
         \sim {Q_E^2 \over 2 Q_M^2} M \, .
\end{equation}

\section{The moduli space and its metric}
\label{chaptwoModspace}

The results of the previous two sections can be reformulated
by introducing the concept of a moduli space.  While the advantage
for these relatively simple examples may seem slight, this formalism
will be of considerable utility when we turn to less trivial cases.

To motivate this, let us first consider a purely bosonic non-gauge
theory whose fields, which we assume to all be massive, are
combined into a single multicomponent field $\psi({\bf x},t)$.  Let us
suppose that there is a family of degenerate static solutions,
parameterized by $n$ collective coordinates $z_r$, that we denote by
$\psi^{\rm cl}({\bf x};z)$.  These static solutions may be viewed as
forming a manifold, known as the moduli space, with the $z_r$ being
coordinates on the manifold.  This manifold is itself a subspace of
the full space of field configurations.

An arbitrary field configuration can be decomposed as
\begin{equation}
    \psi({\bf x},t) = \psi^{\rm cl}({\bf x};z(t))
              + \delta \psi({\bf x};z(t),t)
\label{fielddecomp}
\end{equation}
with $\delta \psi$ required to be orthogonal to motion on the
moduli space, in the sense that at any time $t$
\begin{equation}
    0 = \int d^3x \, {\partial \psi^{\rm cl}
            \over \partial z_r}\, \delta\psi
\end{equation}
for all $r$.  Thus, $\delta \psi$ measures how far the configuration
is from the moduli space.  Stated differently, if $\psi$ is expanded
in terms of normal modes of oscillation about $\psi^{\rm cl}({\bf
x};z(t))$, only the modes with nonzero frequency contribute to $\delta
\psi$; the zero-mode contribution is included by allowing the $z_r$ to
be time-dependent.

If the kinetic energy is of the standard form, with ${\cal L} = (1/2)
{\dot \psi}^2 + \cdots$, then substitution of Eq.~(\ref{fielddecomp})
into the Lagrangian leads to
\begin{equation}
    L = -E_{\rm static} + {1 \over 2} g_{rs}(z) \dot z_r \dot z_s
      + L_{\rm quad} + \cdots
\label{solitonquantLag}
\end{equation}
where $E_{\rm static}$ is the energy of the static solutions,
$L_{\rm quad}$ is quadratic in $\delta \psi$ and the ellipsis denotes
terms that are cubic or higher in $\delta \psi$.  The coefficients
$g_{rs}(z)$ are given by
\begin{equation}
    g_{rs}(z) = \int d^3x {\partial \psi^{\rm cl} \over \partial z_r}
       \, {\partial \psi^{\rm cl} \over \partial z_s}
\end{equation}
and may be viewed as defining a metric on the moduli space.

Now assume that the collective coordinates are slowly varying and that
the energy is small compared to the lowest nonzero normal frequency.
The deformations of the solution corresponding to excitation of the
modes with nonzero frequency are then negligible, and the field
configuration will never wander far from the moduli space.  A good
approximation to the dynamics is then given by the moduli space
Lagrangian \cite{Manton:1981mp}
\begin{equation}
   L_{\rm MS} =  {1\over 2} g_{rs}(z)\, \dot z_r \dot z_s \, .
\end{equation}
In this approximation, the time dependence of the field comes only
through the collective coordinates; i.e.,
\begin{equation}
    \psi({\bf x},t) = \psi^{\rm cl}({\bf x};z(t))
\end{equation}
with $z(t)$ being a solution of the Euler-Lagrange equations that
follow from $L_{\rm MS}$.  If $g_{rs}(z)$ is viewed as a metric, these
equations require that $z(t)$ be a geodesic motion on the moduli
space.

We now turn to the SU(2) gauge theory in which we are actually
interested.  It will be convenient to adopt a Euclidean
four-dimensional notation in which $A_j$ and $\Phi$ are combined into
a single field $A_a$, with $a$ running from 1 to 4.  In this notation
$D_a$ and $F_{ab}$ have their usual meanings if $a$ and $b$ are 1, 2,
or 3, while
\begin{eqnarray}
    D_4 A_a &=& - ie [\Phi, A_a]  \cr \cr
    F_{a4} &=& -F_{4a} = D_a \Phi \, .
\end{eqnarray}
Note that $A_a$ does not include $A_0$; in this notation zero
subscripts on fields and derivatives will always be explicitly
displayed.

This theory differs from the above example in two significant aspects.
First, because there are massless fields in the theory, the spectrum
of normal frequencies extends down to zero.  Hence, there is no range
of energies that is small compared to all of these frequencies, and so
one might wonder whether this invalidates the moduli space
approximation.  We will postpone a detailed discussion of this point
until Sec.~\ref{trajectories}.  We will see there that the presence of
massless fields does not have a significant effect on the
approximation for the monopoles in the SU(2) theory, although it does
have consequences in some other theories.

The second difference is that the fact that we are dealing with a
gauge theory means that there is an infinite-dimensional family of
static solutions.  From these, we pick out a finite-dimensional set of
gauge-inequivalent configurations $A_a^{\rm cl}({\bf x};z)$.  The
specific choice that is made here is essentially a specification of
gauge, and therefore cannot affect any physical results.  Now let us
introduce a time-dependence by allowing the $z_r$ to be slowly
varying.  As we saw in Secs.~\ref{zeromodes} and \ref{dyons}, Gauss's
law,
\begin{equation}
    0 = D_a F^{a0} = D_a \left[ D^a A^0 -
        \dot z_r {\partial (A^{\rm cl})^a \over
         \partial z_r}\right] \, ,
\label{zeromodeGauss}
\end{equation}
then requires a nonzero $A_0$.

From the form of this equation, it is clear that $A_0$ is proportional
to the collective coordinate velocities, and so can be written in the
form
\begin{equation}
    A_0 = \dot z_r \epsilon_r \, .
\end{equation}
Hence,
\begin{equation}
    F^{a0} = -\dot z_r \,\delta_r A^a
\label{MSAfa0}
\end{equation}
where
\begin{equation}
    \delta_r A^a =  {\partial (A^{\rm cl})^a \over\partial z_r}
           - D^a \epsilon_r \, .
\label{formofzeromode}
\end{equation}
The second term in $\delta_r A^a$ has the same form as an
infinitesimal gauge transformation.  This suggests a second approach
to the motion on the moduli space, in which we work in the temporal
gauge, $A_0=0$.  Because of the Gauss's law constraint, the time
evolution of the fields can no longer be restricted to the family of
configurations $A^{\rm cl}_a({\bf x};z)$ with which we started.
Instead, the fields must also move ``vertically'' along some purely
gauge directions, with the specific choice of gauge function being
dictated by Eq.~(\ref{zeromodeGauss}).

Whichever approach one takes, $\delta_r A^a$ has two important
properties:

1) It is a zero mode of the linearized static field equations.  This
   follows immediately from the fact that $z_r$ is a collective
   coordinate if $\epsilon_r$ vanishes.  Since a time-independent
   gauge transformation preserves the static equations, $\delta_r A^a$
   must still be a zero mode even if $\epsilon_r \ne 0$.

2) It obeys the ``background gauge'' condition
\begin{equation}
    D_a \delta_r A^a = 0
\label{chap2backgroundgauge}
\end{equation}
   as a result of Eq.~(\ref{zeromodeGauss}).

The moduli space Lagrangian can be obtained by substituting
Eq.~(\ref{MSAfa0}) into the Lagrangian of Eq.~(\ref{lagrangian}).  The
resulting metric is
\begin{equation}
   g_{rs}  = 2 \int d^3x  \Tr \delta_r A^a  \delta_s A^a
\end{equation}

Let us now specialize to the case of a single SU(2) monopole.  Not
counting local gauge modes, there are four zero modes
\cite{Mottola:1978pz}, and thus a four-dimensional moduli space whose
coordinates can be chosen to be the location $\bf R$ of the monopole
center and a U(1) phase $\alpha$.  From the discussion in
Secs.~\ref{zeromodes} and \ref{dyons} we find that
\begin{equation}
    g_{rs}(z)\, \dot z_r \dot z_s = M \dot{\bf R}^2
                + {I\over e}\, \dot \alpha^2 \, .
\label{singledyonmetric}
\end{equation}
where $M$ is the mass of the monopole and $I$ is defined by
Eq.~(\ref{QandIandOmega}).  The factor of $e$ enters the second
term on the right hand side because $I$ was defined with reference
to the electric charge $Q_E$, whereas the canonical momentum conjugate
to $\alpha$ (i.e., the Noether charge),
\begin{equation}
      P_\alpha = {I \dot \alpha \over e} = {Q_E \over e}  \, ,
\end{equation}
differs from $Q_E$ by a factor of the gauge coupling.

The metric in Eq.~(\ref{singledyonmetric}) is manifestly flat.
Because $\alpha$ is a periodic variable, the moduli space is a
cylinder, $R^3 \times S^1$.  The geodesic motions are straight lines
with constant ${\bf v} = \dot{\bf R}$ and $\omega = \dot\alpha$, and
correspond to dyons moving with constant velocity.  The special cases
$\omega=0$ and ${\bf v}=0$ give the moving monopole of
Sec.~\ref{zeromodes} and the stationary dyon of Sec.~\ref{dyons},
respectively.

\section{Quantization}
\label{sec2-quant}

The relevance of these classical solutions for the quantum theory is
most easily understood in the weak coupling limit.  For small $e$ the
radius of the monopole core, $R_{\rm mon} \sim 1/ev$, is much greater
than the monopole Compton wavelength, $1/M_{\rm mon} \sim e/v$.
Consequently, the quantum fluctuations in the monopole position can be
small enough relative to the size of the monopole for the classical
field profile to be physically meaningful.

In this weak coupling limit the quantum corrections to the monopole
mass can be calculated perturbatively.  The calculation follows the
standard method for quantizing fields in the presence of a soliton
\cite{Goldstone:1974gf,Dashen:1974ci,Christ:1975wt,Tomboulis:1975gf, 
Klein:1975yh,Creutz:1975qt,Callan:1975yy,Gervais:1975pa, 
Gervais:1974dc,Tomboulis:1975qt}.
For a theory with only bosonic fields, one decomposes the fields as in
Eq.~(\ref{fielddecomp}) and takes the $z_r$ and $\delta\psi({\bf x})$
as the dynamical variables to be quantized, thus leading to the
expression for the Lagrangian given in Eq.~(\ref{solitonquantLag}).
The first term on the right-hand side of Eq.~(\ref{solitonquantLag})
is a number, the classical energy of the soliton.  The next two terms
can be taken as the unperturbed Lagrangian; note that to lowest order
the $z_r$ and $\delta\psi$ do not mix, and so these two terms can be
treated separately.  Finally, the terms represented by the ellipsis can be
treated as perturbations.

If $\delta\psi$ is expanded in terms of normal modes about the
soliton, the quad\-ratic term $L_{\rm quad}$ is diagonalized and
becomes a sum (or, more precisely, an integral) of simple harmonic
oscillator Lagrangians.  The contribution to the soliton mass from the
zero-point energies of these oscillators might seem to be divergent.
However, one must subtract from this the zero-point oscillator
contributions to the vacuum energy.  The difference between the two
--- i.e., the shift in the zero-point energies induced by the presence
of the soliton --- is finite and, for weak coupling, suppressed
relative to the classical energy.

For the specific case of the monopole, the classical energy is, as we
have seen, of order $v/e= m_W/e^2$.  The contribution from the shift
of the zero point energies is of order $m_W$.  Because the metric is
flat, the quantization of the collective coordinates is particularly
simple.  The position variables range over all of space, and so their
conjugate momenta $\bf P$ take on all real values.  The phase angle
$\alpha$ has period $2\pi$, implying that $P_\alpha$ is quantized in
integer units and that the electric charge is of the form $Q_E = ne$.

The monopole energy can thus be written as
\begin{eqnarray}
     E &=& M_{\rm cl} + (\Delta M)_{\rm zero-point}
       + {{\bf P}^2 \over 2M_{\rm cl}}  + {e Q_E^2 \over 2I}
                    + \cdots \cr\cr
    &=& m_W \left[ O(1/e^2) + O(1) + O({\bf v}^2/e^2)
            + O(n^2 e^2) + \cdots \right] \, .
\label{Eforsolitonquant} \,
\end{eqnarray}
The terms represented here by the ellipsis are due to the
perturbations, and contain additional powers of $e^2$.  They include
terms that are quartic in the momenta, and so cannot be neglected if
either ${\bf v}$ or $Q_E$ is too large.  This last condition can be
made more precise by requiring that the terms quadratic in the momenta
be at most of order unity, which implies ${\bf v}^2 \lesssim e^2$ and
$n \lesssim 1/e$.

Now imagine that the theory is extended to include additional fields,
with the couplings of these being such that the previous monopole
solution, with all of the new fields vanishing identically, remains a
solution of the classical field equations.  If the new fields are
bosonic, the analysis is unchanged except for the addition of new
eigenmodes.  The same is true for the nonzero-frequency modes of any
fermion fields, apart from the usual restriction that the occupation
numbers must be 0 or 1.  However, the spectrum of a fermion field in
the presence of a monopole typically also contains a number of
discrete zero modes \cite{Jackiw:1975fn}.  Because the energy of the
system is independent of whether these modes are occupied or not, it
is not useful to interpret an occupation number of 0 or 1 as
corresponding to the absence or presence of a particle.  Instead, a
set of $N$ fermion zero modes should be viewed as giving rise to
multiplets containing $2^N$ degenerate states that all have equivalent
status.  In particular, the monopole ground state becomes a degenerate
set of states with varying values for the spin angular momentum.

\chapter{BPS Monopoles and Dyons}
\label{BPSchap}

For the remainder of this review we will concentrate on monopoles and
dyons in the BPS limit \cite{Prasad:1975kr,Bogomolny:1975de}, which we
introduce in this chapter.  As we describe in Sec.~\ref{BPSasLimit},
this limit was originally invented as a trick for obtaining an
analytic expression for the one-monopole solution to
Eqs.~(\ref{su2monoeqs}).  It was soon realized
\cite{Bogomolny:1975de,Coleman:1976uk} that the solutions thus
obtained saturate an energy bound and satisfy a generalized
self-duality equation, as we explain in Sec.~\ref{BPSandEbounds}.
These insights led to the discovery that the BPS limit gives rise to a
rich array of classical multimonopole and multidyon solutions with
very interesting properties.  Further, it turns out that the special
features of this limit can be naturally explained in terms of
supersymmetry.  This connection, which is described in
Sec.~\ref{susyconnectionSec}, allows these features to be seen as
properties, not simply of the classical field equations, but also of
the underlying quantum field theory.  In particular, one is naturally
led to conjecture a duality symmetry of the theory, as was first done
by Montonen and Olive \cite{Montonen:1977sn}; we discuss this in
Sec.~\ref{OM}.

\section{BPS as a limit of couplings}
\label{BPSasLimit}

We begin by recalling Eqs.~(\ref{su2monoeqs}) for the coefficient
functions entering the spherically symmetric monopole ansatz of
Eq.~(\ref{hedghogansatz}).  These equations depend on the three
parameters $e$, $\lambda$, and $v$.  Two of these parameters can be
eliminated by rescaling $h$ and $r$, but the combination $\lambda/e^2$
still remains.

In general, these equations cannot be solved analytically.  However,
one might hope to be able to proceed further for special values of
$\lambda/e^2$.  In particular, Prasad and Sommerfield
\cite{Prasad:1975kr} proposed considering the limit $\lambda/e^2
\rightarrow 0$.  More precisely, they took the limit $\mu^2
\rightarrow 0$, $\lambda \rightarrow 0$, but with $v^2 = \mu^2 /
\lambda$ held fixed so as to maintain the boundary condition on
$h(\infty)$.  The last term in the first of Eqs.~(\ref{su2monoeqs})
then disappears, and by trial and error one can find the solution
\begin{eqnarray}
     u(r) &=& {evr \over \sinh(evr) } \cr
     h(r) &=& v\coth(evr) - {1 \over er} \, .
\label{monoPSsoln}
\end{eqnarray}
Notice that $h(r)$ only falls as $1/r$ at large distance, in contrast
with its usual exponential decrease.  This is a consequence of the fact
that $m_H = \sqrt{2}\, \mu$ vanishes in this ``BPS limit''.  Because the
Higgs field is now massless, it mediates a long-range force, a fact
that turns out to be of considerable significance.

These results can be easily extended to the case of nonzero electric
charge.   The dyon Eqs.~(\ref{dyonHeq}) - (\ref{dyonJeq}) are solved by
\begin{eqnarray}
     u(r) &=& {e\tilde vr \over \sinh(e\tilde vr)  } \cr\cr
     h(r) &=& {\sqrt{Q_M^2 + Q_E^2}\over Q_M}
               \left[\tilde v\coth(e\tilde vr) - {1 \over er}\right] \cr\cr
     j(r) &=& -{Q_E \over Q_M}
               \left[\tilde v\coth(e\tilde vr) - {1 \over er}\right] 
\label{dyonPSsoln}
\end{eqnarray}
where
\begin{equation}
     \tilde v = v {Q_M \over \sqrt{Q_M^2 + Q_E^2} }
\end{equation}

\section{Energy bounds and the BPS limit}
\label{BPSandEbounds}

Further special properties associated with this limit were pointed out
by Bogomolny and by Coleman et
al.~\cite{Bogomolny:1975de,Coleman:1976uk}.  Although the argument was
first formulated in terms of the SU(2) theory, it immediately
generalizes to any gauge group, provided that the Higgs field is in
the adjoint representation.  As in the SU(2) case, we take all
parameters in the Higgs potential to zero, but keep appropriate ratios
fixed so that the Higgs vacuum expectation value is unchanged.

With the Higgs potential omitted, and $A_\mu$ and $\Phi$ written as
elements of the Lie algebra, the energy is
\begin{eqnarray}
    E &=& \int d^3x \left[\Tr E_i^2 +  \Tr (D_0\Phi)^2  +\Tr B_i^2
         + \Tr (D_i\Phi)^2  \right] \cr
      &=& \int d^3x \left[\Tr(B_i \mp \cos \alpha D_i \Phi)^2
          + \Tr(E_i \mp \sin \alpha D_i \Phi)^2 + \Tr (D_0\Phi)^2
         \right] \cr  &&\qquad
       \pm 2 \int d^3x \left[ \cos\alpha \Tr (B_i D_i\Phi)
                 + \sin\alpha \Tr (E_i D_i\Phi) \right]
\end{eqnarray}
where $\alpha$ is arbitrary.  If we integrate by parts in the last
integral and use the Bianchi identity $D_iB_i=0$ and
Gauss's law, Eq.~(\ref{gauss}), we obtain
\begin{eqnarray}
      E &=& \int d^3x \left[\Tr(B_i \mp \cos \alpha D_i \Phi)^2
          + \Tr(E_i \mp \sin \alpha D_i \Phi)^2 + \Tr (D_0\Phi)^2
         \right] \cr  &&\qquad
     \pm \cos\alpha \, {\cal Q}_M \pm \sin\alpha \, {\cal Q}_E  \cr
    &\ge& \pm \cos\alpha \, {\cal Q}_M \pm \sin\alpha \, {\cal Q}_E
\label{su2Ebound}
\end{eqnarray}
where
\begin{eqnarray}
     {\cal Q}_M &=& 2 \int d^2S_i \Tr(\Phi B_i)    \cr
     {\cal Q}_E &=& 2 \int d^2S_i \Tr(\Phi E_i) \, .
\label{higgscharge}
\end{eqnarray}
with the integrations being over the sphere at spatial infinity.
For the case of SU(2), these quantities are related to the magnetic
and electric charges defined in Eqs.~(\ref{su2magchargedef}) and
(\ref{gausscharge}) by ${\cal Q}_M = v Q_M$ and ${\cal Q}_E = v Q_E$.

The inequality (\ref{su2Ebound}) holds for any choice of signs and of
$\alpha$.  The most stringent inequality,
\begin{equation}
      E \ge \sqrt{{\cal Q}_M^2 + {\cal Q}_E^2}
\label{bpsEbound}
\end{equation}
is obtained by setting $\alpha =\tan^{-1}({\cal Q}_E/{\cal Q}_M)$ and
choosing the upper or lower signs according to whether ${\cal Q}_M$
is positive or negative; without loss of generality, we can take
${\cal Q}_M >0$.  This lower bound is achieved by configurations
obeying the first-order equations
\begin{eqnarray}
      B_i &=& \cos \alpha \, D_i\Phi    \cr
      E_i &=& \sin \alpha \, D_i\Phi    \cr
      D_0\Phi&=& 0 \, .
\label{dyonBPSeq}
\end{eqnarray}
Configurations that minimize the energy for fixed values of ${\cal
Q}_M$ and ${\cal Q}_E$ are solutions of the full set of second-order
field equations, provided that they also obey the Gauss's law
constraint.  Using the Bianchi identity, together with the fact that
$E_i$ is proportional to $B_i$, one readily verifies that this latter
condition is satisfied here.  Hence, solutions of the first-order
Eqs.~(\ref{dyonBPSeq}) are indeed classical solutions of the
theory.\footnote{There are also solutions of the second-order field
equations that are not solutions of these first-order equations
\cite{Taubes:1982ie}.  However, these correspond to saddle points of
the energy functional, and are therefore not stable.}  They are
referred to as BPS solutions, and their energy is given by the BPS
mass formula
\begin{equation}
      M = \sqrt{{\cal Q}_M^2 + {\cal Q}_E^2}  \, \, .
\label{bpsEformula}
\end{equation}

We will be particularly concerned with the case ${\cal Q}_E=0$, and
hence with static configurations with $A_0=0$ that satisfy the
Bogomolny equation
\begin{equation}
     B_i = D_i\Phi \, .
\label{bogomolny}
\end{equation}
This equation is closely related to the self-duality equations
satisfied by the instanton solutions \cite{Belavin:1975fg} of
four-dimensional Euclidean Yang-Mills theory.  The latter equations
can be reduced to Eq.~(\ref{bogomolny}) by taking the fields to be
independent of $x_4$ and writing $A_4 = \Phi$.  Because of this
analogy, solutions of Eq.~(\ref{bogomolny}) are often referred to as
being self-dual.

We have thus found that going to the BPS limit leads to two striking
results.  First, we have found a set of first-order field equations
whose solutions actually satisfy the full set of second-order field
equations of the theory.  Second, the energy of these classical
solutions is simply related to their electric and magnetic charges.
Analogous properties are actually found in a number of other settings,
including Yang-Mills instantons \cite{Belavin:1975fg}, Ginzburg-Landau
vortices at the Type I-Type II boundary \cite{Weinberg:1979er}, and
certain Chern-Simons vortices \cite{Hong:1990yh,Jackiw:1990aw}.  These
examples all have in common the fact that they can be simply extended
to incorporate supersymmetry, an aspect that we turn to next.

\section{The supersymmetry connection}
\label{susyconnectionSec}

The approach to the BPS limit described above is somewhat
artificial and unsatisfactory. One introduces a potential to induce a
nonzero Higgs field vacuum expectation value, but then works in a
delicately tuned limit in which the potential vanishes.  Aside from the
conceptual difficulties at the classical level, it is hard to see how
this limit would survive quantum corrections.

These difficulties can be overcome by enlarging the theory.  To start,
consider the bosonic Lagrangian
\begin{equation}
    {\cal L} =- {1\over 2} \Tr F_{\mu\nu}^2 + \sum_{P=1}^k \Tr (D_\mu
      \Phi_P)^2  + {e^2\over 2} \sum_{P,Q =1}^k \Tr [ \Phi_P, \Phi_Q]^2
\label{susyLag}
\end{equation}
where $\Phi_P$ ($P= 1, \dots, k$) are a set of Hermitian adjoint
representation scalar fields.  The scalar potential vanishes whenever
the $\Phi_P$ all commute, leading to a large number of degenerate
vacua.  In particular, let us choose a vacuum with $\Phi_1 \ne 0$ and
$\Phi_P=0$ for all $P \ge 2$ and seek soliton solutions with
corresponding boundary conditions.\footnote{In the SU(2) theory, any
symmetry-breaking vacuum can be brought into this form by an SO($k$)
transformation of the scalar fields.  For gauge groups of higher rank
there are more possibilities, to which we will return in Chap.~7.}
If we impose the constraint that $\Phi_P({\bf x})$ vanish identically
for $P \ge 2$, then the field equations reduce to those of the BPS
limit described in the previous subsections.

The form of the potential in Eq.~(\ref{susyLag}) is not in general
preserved by quantum corrections.  However, for $k=2$ ($k=6$),
Eq.~(\ref{susyLag}) is precisely the bosonic part of the Lagrangian
for a SYM theory with ${\cal N}=2$ (${\cal N}=4$) extended
supersymmetry \cite{Brink:1976bc}.  Adding the fermionic terms
required to complete the supersymmetric Lagrangian will not affect the
field equations determining the classical solutions, but will ensure,
via the nonrenormalization theorems, that quantum corrections do not
change the form of the potential.

The BPS self-duality equations take on a deeper meaning in this
context of extended supersymmetry.  We illustrate this for the case of
${\cal N}=4$ supersymmetry.\footnote{Our conventions in this section
generally follow those of Sohnius~\cite{Sohnius:1985qm}, although our
$\gamma^5$ differs by a factor of $i$.}  It is convenient to write the
six Hermitian spinless fields as three self-dual scalar and three
anti-self-dual pseudoscalar fields obeying
\begin{equation}
    G_{rs} = -G_{sr} ={1\over 2} \epsilon_{rstu}G_{tu}  \qquad\qquad
    H_{rs} =  -H_{sr}= -{1\over 2} \epsilon_{rstu}H_{tu}
\end{equation}
(with $r, s = 1, \dots, 4$),
while the fermion fields are written as four Majorana fields $\chi^r$.
The Lagrangian
\begin{eqnarray}
    {\cal L} &=& \Tr\left\{ -{1\over 2} F_{\mu\nu}^2
     + {1\over 4}D_\mu G_{rs}^2
    + {1\over 4}D_\mu H_{rs}^2    \right. \cr\cr &&\quad
     +{e^2 \over 32}[G_{rs},G_{tu}]^2 +{e^2 \over 32}[H_{rs},H_{tu}]^2
     +{e^2 \over 16}[G_{rs},H_{tu}]^2  \cr\cr &&\quad \left.
     +{i}\bar \chi_r \gamma^\mu D_\mu \chi_r
     +{ie}\bar \chi_r [\chi_s, G_{rs}]
      +{e}\bar \chi_r \gamma^5[\chi_s, H_{rs}] \right\}
\label{Nequal4Lag}
\end{eqnarray}
is invariant under the supersymmetry transformations
\begin{eqnarray}
     \delta A_\mu &=& i \bar \zeta_r \gamma_\mu \chi_r  \cr\cr
     \delta G_{rs} &=& \bar\zeta_r\chi_s - \bar\zeta_s\chi_r
           +\epsilon_{rstu} \bar\zeta_t\chi_u   \cr\cr
     \delta H_{rs} &=&-i \bar\zeta_r \gamma^5\chi_s
          +i \bar\zeta_s\gamma^5\chi_r
           +i\epsilon_{rstu} \bar\zeta_t\gamma^5\chi_u   \cr\cr
     \delta \chi_r &=& -{i\over 2}\sigma^{\mu\nu} \zeta_r F_{\mu\nu}
         + i \gamma^\mu D_\mu(G_{rs} -i \gamma^5 H_{rs})\zeta_s  \cr
           &&\quad
         +{ie\over 2}\, [G_{rt} +i \gamma^5 H_{rt}, \,
             G_{ts} -i \gamma^5 H_{ts}]  \zeta_s
\label{susytransformation}
\end{eqnarray}
where the $\zeta_r$ are four Majorana spinor parameters and
$\sigma^{\mu\nu} = (i/2)[\gamma^\mu,\gamma^\nu]$.

Now consider the effect of such a transformation on an arbitrary classical
(and hence purely bosonic) configuration.  The variations of the bosonic
fields are proportional to the fermionic fields and so automatically vanish.
The variations of the $\chi_r$, on the other hand, are in
general nonzero.  However, for certain choices of the $\zeta_r$ there are
special configurations for which the fermionic fields are also invariant,
so that part of the supersymmetry remains unbroken.

To illustrate this, let us suppose that $G_{12} = G_{34} \equiv b$ and
$H_{12} = -H_{34} \equiv a$ are the only nonzero spin-0 fields.
Requiring that the $\delta \chi_r$ all vanish gives two pairs of
equations, one involving $\zeta_1$ and $\zeta_2$ and one involving
$\zeta_3$ and $\zeta_4$.   Using the identity $\gamma^5 \sigma^{ij}=
i\epsilon^{ijk}\sigma^{0k} \equiv 2i \epsilon^{ijk} S^k$, we can write
these as
\begin{eqnarray}
    0&=&2 \left\{{\bf S}\cdot \left[({\bf B}\gamma^5 - i {\bf E}) \delta_{rs}
      + {\bf D}(b +i \gamma^5 a) \gamma^0 \epsilon_{rs}\right] \right.
    \cr & & \quad  \left.
      + i D_0(b +i \gamma^5 a) \gamma^0 \epsilon_{rs}
      -e \, [b,a]\gamma^5 \delta_{rs} \right\}\zeta_s
                \, , \quad r,s = 1,2 \cr \cr
    0 &=& 2 \left\{{\bf S}\cdot \left[({\bf B}\gamma^5 - i {\bf E})
      \delta_{rs}
      + {\bf D}(b -i \gamma^5 a) \gamma^0 \epsilon_{rs}\right] \right.
     x\cr & & \quad  \left.
      + i D_0(b -i \gamma^5 a) \gamma^0 \epsilon_{rs}
      +e \, [b,a]\gamma^5 \delta_{rs} \right\}\zeta_s
                \, ,  \quad r,s = 3,4
\label{susyinvar}
\end{eqnarray}
where $\epsilon_{12}=-\epsilon_{21}=\epsilon_{34}=-\epsilon_{43}=1$.
We will list three special solutions to these equations:

1)  Suppose that the four $\zeta_r$ are related by
\begin{eqnarray}
    \zeta_1 &=& - e^{i\alpha \gamma^5} \gamma^5 \gamma^0
        \zeta_2 \cr
    \zeta_3 &=& -  e^{i\alpha \gamma^5} \gamma^5 \gamma^0
        \zeta_4 \, .
\label{halfBPSzetas}
\end{eqnarray}
Equation~(\ref{susyinvar}) then requires
\begin{eqnarray}
    B_i &=& \cos \alpha \, D_i b     \cr
    E_i &=& \sin \alpha \, D_i b     \cr
    D_0 b &=& 0  \cr
    D_\mu a &=& [b,a] = 0 \, .
\label{halfBPSeqs}
\end{eqnarray}
Thus, the BPS solutions of Eq.~(\ref{dyonBPSeq}), possibly
supplemented by a constant field $a$ that commutes with all the other
fields, are invariant under a two-parameter set of transformations,
and thus preserve half of the supersymmetry.

2)  If we further restrict the $\zeta_r$ by requiring
\begin{eqnarray}
    \zeta_1 &=& - \gamma^5 \gamma^0  \zeta_2 \cr
    \zeta_3 &=&\zeta_4 =0 \, ,
\label{quarteronezetas}
\end{eqnarray}
then Eq.~(\ref{susyinvar}) requires
\begin{eqnarray}
    B_i &=& D_i b     \cr
    E_i &=& -D_i a     \cr
    D_0 b &=& ie\,  [b,a]  \cr
    D_0 a &=& 0 \, .
\label{quarterone}
\end{eqnarray}
In contrast with case 1, these equations do not guarantee that the
fields satisfy Gauss's law,
\begin{equation}
    D_i E_i = e^2\, [b, [b,a]] \, ,
\end{equation}
which must be imposed separately.  However, any configuration that
satisfies both Gauss's law and Eq.~(\ref{quarterone}) is also a
solution of the full set of field equation.

3)  Alternatively, one can require that
\begin{eqnarray}
    \zeta_1  &=& \zeta_2 =0  \cr
    \zeta_3 &=& - \gamma^5 \gamma^0  \zeta_4 \, .
\label{quartertwozetas}
\end{eqnarray}
This leads to
\begin{eqnarray}
    B_i &=& D_i b     \cr
    E_i &=&  D_i a     \cr
    D_0 b &=& - ie \, [b,a]  \cr
    D_0 a &=& 0 \, .
\label{quartertwo}
\end{eqnarray}
As with case 2, Eq.~(\ref{quartertwo}) must be supplemented by the
Gauss's law constraint in order to guarantee a solution of the field
equations.

In both case 2 and case 3, there is only one independent $\zeta_r$,
and so only one fourth of the ${\cal N}=4$ supersymmetry is preserved by the
solution.  We will return to these 1/4-BPS solutions in Chap.~7.  The
case of ${\cal N}=2$ supersymmetry is obtained by restricting the values of
the indices $r$ and $s$ to 1 and 2.  In this case, both solutions 1
and 2 preserve half of the supersymmetry, while solution 3 breaks all
of the supersymmetry.

The significance of a configuration's preserving a portion of the
supersymmetry can be illuminated by considering the supersymmetry
algebra.  Recall that the the most general form of the algebra of the
supercharges can be written as
\begin{equation}
     \{Q_{r\alpha}, \bar Q_{s\beta}\} = 2
     \delta_{rs}(\gamma^\mu)_{\alpha\beta} P_\mu
      +  2i \delta_{\alpha\beta} X_{rs}
           - 2 (\gamma^5)_{\alpha\beta} Y_{rs}
\label{superalgebra}
\end{equation}
where $X_{rs}=-X_{sr}$ and $Y_{rs}= -Y_{sr}$ are central charges that
commute with all of the supercharges and with all of the generators of
the Poincar\'e algebra.  These central charges can be calculated by
writing the supercharge as the spatial integral of the time component of
the supercurrent $S_r^\mu$.
Performing a supersymmetry transformation on $S^0_s(x)$ gives
$\{Q_r,S^0_s(x)\}$.  A spatial integral then gives $\{Q_r,Q_s\}$. The
central charges arise as surface terms that are nonvanishing in the
presence of electric or magnetic charges.  Explicitly,
\begin{eqnarray}
     X_{rs} &=& 2\int d^2S_i \, \Tr[ G_{rs} E_i + H_{rs} B_i]  \cr
     Y_{rs} &=& 2\int d^2S_i \, \Tr[ G_{rs} B_i + H_{rs} E_i].
\label{explicitcentral}
\end{eqnarray}

Multiplying Eq.~(\ref{superalgebra}) on the right by
$\gamma^0_{\beta\gamma}$ we obtain $\{Q_r, \bar Q_s \gamma^0\}$.
Because this is a positive definite matrix, its eigenvalues must all
be positive, thus implying a lower bound on the mass.  This bound is
most easily derived by multiplying this matrix by its adjoint and
then taking the trace to obtain
\begin{equation}
      M^2 \ge  {1\over 4} [ X_{rs} X_{rs} + Y_{rs} Y_{rs} ] \, .
\label{susymassbound}
\end{equation}
For the case of a single nonzero scalar field, this is equivalent to
the BPS bound, Eq.~(\ref{bpsEbound}), that we obtained previously.

For a state to actually achieve this lower bound, it must be
annihilated by a subset of the supersymmetry generators.  To see how
this works, let $\eta_r$ and $\eta'_r$ be a set of supersymmetry
parameters that satisfy relations of the form of Eq.~(\ref
{halfBPSzetas}).  Within the subspace of states annihilated by the
corresponding combinations of supersymmetry transformations, the
matrix elements of
\begin{equation}
     F = \bar \eta_r \gamma^\mu P_\mu \eta'_r
         + X_{rs} \bar \eta_r \zeta'_s
         +i Y_{rs} \bar \eta_r \gamma^5 \eta'_s
\end{equation}
must vanish for all choices of $\eta_r$ and $\eta'_r$.  By considering
in turn the cases $\eta_4=\eta'_4=0$ and $\eta_2=\eta'_2=0$, we find
that within this subspace
\begin{equation}
    0 = F = 2 \bar \eta_2 \left( \gamma^\mu P_\mu
           -\sin\alpha \gamma^0 X_{12}
           - \cos\alpha \gamma^0 Y_{12} \right)\eta'_2
\end{equation}
and
\begin{equation}
    0 = F = 2 \bar \eta_4 \left( \gamma^\mu P_\mu
           -\sin\alpha \gamma^0 X_{34}
           - \cos\alpha \gamma^0 Y_{34} \right)\eta'_4 \, .
\end{equation}
In order that these hold for all allowed choices of $\eta_r$ and
$\eta'_r$, the spatial momentum $\bf P$ must vanish and
\begin{equation}
     P_0 = M = \sin\alpha X_{12} + \cos\alpha Y_{12}
         = \sin\alpha X_{34} + \cos\alpha Y_{34} \, .
\label{energyforhalfbps}
\end{equation}
Further, by considering the cases $\eta_2 = \eta'_4 = 0$ and $\eta_4 =
\eta'_2 = 0$, one can show that all of the other independent
components of $X_{rs}$ and $Y_{rs}$ must vanish.  Combining the two
parts of Eq.~(\ref{energyforhalfbps}), and recalling that $G_{rs}$ and
$H_{rs}$ are self-dual and anti-self-dual, respectively, we obtain
\begin{equation}
     M = 2\sin \alpha \int d^2S_i \, \Tr (G_{12} E_i)
       + 2\cos \alpha \int d^2S_i \, \Tr (G_{12} B_i) \, .
\end{equation}
The integrals in this equation are, in fact, just the quantities
${\cal Q}_E$ and ${\cal Q}_M$ that were defined in
Eq.~(\ref{higgscharge}), with $G_{12}$ playing the role of $\Phi$.
Recalling now the relation between the electric and magnetic charges
that follows from Eq.~(\ref{halfBPSeqs}), we see that the energy bound
is indeed achieved by these BPS states.

Although this relation between the mass and the charges is the same as
we found in Sec.~\ref{BPSandEbounds}, the crucial difference is that
we have now obtained it as an operator expression, rather than by
relying on the classical solutions.  Indeed, the connection between
the BPS conditions and the central charges guarantees that there are
no corrections to Eq.~(\ref{bpsEformula}).  In the absence of central
charges, massless supermultiplets are smaller than massive ones.  The
analogous result in the presence of central charges is that states
preserving half of the supersymmetry form supermultiplets that are
smaller than usual; with $\cal N$-extended supersymmetry, a minimal
supermultiplet obeying Eq.~(\ref{susymassbound}) has $2^{\cal N}$ states,
compared to $2^{2\cal N}$ states otherwise.  In the weak coupling regime,
where one would expect perturbation theory to be reliable, it would
not seem surprising if one-loop effects gave a small correction to
Eq.~(\ref{bpsEformula}).  However, this would imply an increase in the
size of the supermultiplet, which would be quite surprising.  Hence,
we conclude \cite{Witten:1978mh} that the BPS mass formula must be
preserved by perturbative quantum corrections.\footnote{It was first
pointed out in Ref.~\cite{D'Adda:1978mu} that the bosonic and fermionic
corrections to the supersymmetric monopole mass should cancel.
However, there turn out to be a number of subtleties involved in
actually verifying that the BPS mass formula is preserved by quantum
corrections.  For recent discussions of these, see
Refs.~\cite{Rebhan:2004vn,Rebhan:2005yi,Rebhan:2006fg}.}

\section{Montonen-Olive duality}
\label{OM}

Montonen and Olive \cite{Montonen:1977sn} pointed out that the
particle spectrum of the SU(2) theory defined by
Eq.~(\ref{lagrangian}) has an intriguing symmetry in the BPS limit.
Table~1 shows the masses and charges for the elementary bosons of the
theory, together with those of the monopole and antimonopole.  If one
simultaneously interchanges magnetic and electric charge ($Q_M
\leftrightarrow Q_E$) and weak and strong coupling ($e \leftrightarrow
4\pi/e$), the entries for the $W$-boson are exchanged with those for
the monopole, but the overall spectrum of masses and charges is
unchanged.  [This reflects the fact that the elementary particles of
the theory obey the BPS mass relation of Eq.~(\ref{bpsEformula}).]

It is tempting to conjecture that this symmetry of the spectrum
reflects a real symmetry of the theory, one that generalizes the
electric-magnetic duality symmetry of Maxwell's equations.  Such a
symmetry would interchange the $W$ states corresponding to quanta of
an elementary field with the monopole states arising from a classical
soliton.  This may seem strange, but it may well be that the apparent
distinction between these two types of states is merely an artifact of
weak coupling.  In other words, there could be a second formulation of
the theory in which the monopole, rather than the $W$, corresponds to
an elementary field.  For large $e$ (and hence small $4\pi/e$), this
second formulation would be the more natural one, and the $W$ would be
seen as a soliton state.  This would be analogous to the equivalence
between the sine-Gordon and massive Thirring models
\cite{Coleman:1974bu}, except that the two dual formulations of the
theory would both take the same form; i.e., the theory would be
self-dual.

\begin{center}
\begin{tabular}{lcccccc}
&& \underline{Mass\vphantom{$Q_{E}$}} && \underline{$Q_E$}
&&\underline{$Q_M$} \\\\
photon && 0 && 0 && 0 \\\\
$\phi$ && 0 && 0 && 0 \\\\
$W^\pm$ && $ev$ && $\pm e$ && 0 \\\\
Monopole &&$ \high{4\pi v \over e}$ && 0 && $\pm \high {4\pi v
\over e}$
\end{tabular}
\begin{quote}
{\bf Table 1:} {\small The particle masses and charges in the BPS
limit of the SU(2) theory.}
\end{quote}
\end{center}

In addition to the self-duality of the particle spectrum, further
evidence for this conjecture can be obtained by considering low-energy
scattering.  As we will see in the next chapter, there is no net force
between two static monopoles, because the magnetic repulsion is
exactly cancelled by an attractive force mediated by the massless
Higgs scalar.  The counterpart of this in the elementary particle
sector can be investigated by calculating the zero-velocity limit of
the amplitude for $W$-$W$ scattering.  Two tree-level
Feynman diagrams contribute in this limit --- one with a single photon
exchanged, and one with a Higgs boson exchanged.  Their contributions
cancel, and so there is no net force \cite{Montonen:1977sn}.

There is, however, one very obvious difficulty.  The $W$-bosons have spin
1, whereas the quantum state built upon the spherically symmetric
monopole solution must have spin 0.  (Had the solution not been
spherically symmetric, there would have been rotational zero modes
whose excitation would have led to monopoles with spin.)  The
resolution is found by recalling that the BPS limit is most naturally
understood in the context of extended supersymmetry.  On the
elementary particle side, the additional fields of the supersymmetric
Lagrangian clearly add new states.  For ${\cal N}=2$ supersymmetry, the
massive $W$ becomes part of a supermultiplet that also contains a
scalar and the four states of a Dirac spinor, all with the same mass
and charge; for ${\cal N}=4$, there are five massive scalars and eight
fermionic states, corresponding to two Dirac spinors.

New states also arise in the soliton sector, although by a more subtle
mechanism.  Recall that the existence of a fermionic zero-mode about a
soliton leads to two degenerate states, one with the mode occupied and
one with it unoccupied; with $k$ such modes there are $2^k$ degenerate
states.  In the presence of a unit monopole (BPS or not) an adjoint
representation Dirac fermion has two zero modes.  (We will prove this
statement for the BPS case in the next chapter, but note that these
modes can be obtained by acting on the bosonic BPS solution with the
supersymmetry generators that do not leave it invariant.)  The ${\cal
N}=2$ SYM theory has a single adjoint Dirac field, and thus two zero
modes giving rise to four degenerate states.  These have helicities 0,
0, and $\pm 1/2$, and so the magnetically charged supermultiplet does
not match the electrically charged one.  With ${\cal N}=4$
supersymmetry, on the other hand, there are 16 states, and one can
check that their spins exactly match those of the electrically charged
elementary particle supermultiplet \cite{Osborn:1979tq}.  Thus, the
${\cal N}=4$ theory is a prime candidate for a self-dual theory.

\chapter{Static multimonopole solutions}
\label{multimonChap}

We now want to discuss BPS solutions with more structure than
the unit SU(2) monopole, including both solutions with higher magnetic
charge in the SU(2) theory and solutions in theories with
larger gauge groups.

Within the context of SU(2), one might envision two classes of
multiply-charged solutions.  The first would be multimonopole
solutions comprising a number of component unit monopoles.  At first
thought, one might expect that the mutual magnetic repulsion would
rule out any such solutions.  However, this is not obviously the case
in the BPS limit, because the massless Higgs scalar carries a
long-range attractive force that can counterbalance the magnetic
repulsion \cite{Manton:1977er,Goldberg:1978ee,O'Raifeartaigh:1979ca}.
In fact, it turns out that there are static solutions for any choice
of monopole positions.

One might also envision localized higher charged solutions that were
not multimonopole configurations and that would give rise, after
quantization, to new species of magnetically charged particles.  This
possibility is not realized, at least for BPS solutions.  While there
are localized higher charge solutions, the parameter counting
arguments that we give in Sec.~\ref{indexsection} show that these are
all multimonopole solutions in which the component unit monopoles
happen to be coincident.

For larger gauge groups, there turn out to be not one, but several,
distinct topological charges.  Associated with each is a ``fundamental
monopole'' \cite{Weinberg:1979zt} carrying a single unit of that
charge.  These fundamental monopoles can be explicitly displayed as
embeddings of the unit SU(2) monopole.  As with the SU(2) case, we
find that there are static multimonopole solutions, which may contain
several different species of fundamental monopoles.  Also as before,
there are no intrinsically new solutions beyond these multimonopole
configurations.

We begin our discussion in Sec.~\ref{generalgroupformalism} by
reviewing some properties of Lie algebras and establishing our
conventions for describing monopoles in larger gauge groups.  The
fundamental monopoles are described in this section.  Next, in
Sec.~\ref{indexsection}, we use index theory methods to count the
number of zero modes about an arbitrary solution with given
topological charges.  We find that an SU(2) solution with $n$ units of
magnetic charge, or a solution in a larger group with topological
charges corresponding to a set of $n$ fundamental monopoles, has
exactly $4n$ zero modes.\footnote{There are some complications if the
unbroken gauge group contains a non-Abelian factor, as we will explain
in Chap.~\ref{unbrokenNonAb}.}  It is therefore described by $4n$
collective coordinates and corresponds to a point on a
$4n$-dimensional moduli space.  These collective coordinates have a
natural interpretation as the positions and U(1) phases of the
component monopoles.

While these methods determine the number of parameters that must enter
a solution with arbitrary charge, they do not actually show that any
such solutions exist.  In Sec.~\ref{generalsolutionremarks} we present
some general discussion of the problem of finding explicit solutions.
Then, in Sec.~\ref{nahmsection}, we describe a method, due to
Nahm \cite{Nahm:1982jb,Nahm:1981xg,Nahm:1981nb,Nahm:1983sv},
that establishes a correspondence between multimonopole solutions and
solutions of a nonlinear differential equation in one variable.
Not only does this method yield some multimonopole solutions more
readily than a direct approach, but it also provides insights in
some cases where an explicit solution cannot be obtained.  Some
examples of the use of this construction are described in
Sec.~\ref{nahmexamplesection}.

\section{Larger gauge groups}
\label{generalgroupformalism}

The topological considerations that give rise to monopole solutions in
the SU(2) theory can be generalized to the case of an arbitrary of
gauge group $G$, with the Higgs field $\Phi$ being in an arbitrary
(and possibly reducible) representation.\footnote{For a detailed discussion,
see Ref.~\cite{Coleman:1975qj}.}  If the vacuum expectation
value of $\Phi$ breaks the gauge symmetry down to a subgroup $H$, then
the vacuum manifold of values of $\Phi$ that minimize the scalar field
potential is isomorphic to the quotient space $G/H$.  Topologically
nontrivial monopole configurations exist if the second homotopy group
of this space, $\Pi_2(G/H)$, is nonzero.  This homotopy group is most
easily calculated by making use of the identity $\Pi_2(G/H) =
\Pi_1(H)$, which holds if $\Pi_2(G)=0$ (as is the case for any
semisimple $G$) and $\Pi_1(G)=0$ (which can be ensured by taking $G$
to be the covering group of the Lie algebra).

Because we are interested in BPS monopoles, our discussion in this
review will be restricted to the case where $\Phi$ transforms under
the adjoint representation of the gauge group.

\subsection{Lie algebras}

Let us first recall some results concerning Lie groups and algebras.
Let $G$ be a simple Lie group of rank $r$.  A maximal set of mutually
commuting generators is given by the $r$ generators $H_i$ that span
the Cartan subalgebra; it is often convenient to choose these to be
orthogonal in the sense that
\begin{equation}
   \Tr (H_i H_j) = {1\over 2} \, \delta_{ij}   \, .
\label{CartanStdNorm}
\end{equation}
(This normalization agrees with the conventions we have used in the
preceding chapters.)
The remaining generators can be taken to a set of ladder operators
$E_{\sbalpha}$ that are generalizations of the raising and lowering
operators of SU(2).  These are associated with roots $\balpha$ that
are $r$-component objects defined by the commutation relations
\begin{equation}
  [{\bf H} , E_{\sbalpha}] = \balpha E_{\sbalpha}
\end{equation}
of the ladder operators with the $H_i$.
These roots may be viewed as vectors forming a lattice in an
$r$-dimensional Euclidean space.  We will also make use of the dual
roots, defined by
\begin{equation}
  \balpha^*  \equiv \balpha/\balpha^2  \, .
\end{equation}

Any root can be used to define an SU(2) subgroup with
generators
\begin{eqnarray}
   t^1(\balpha) &=& {1 \over \sqrt{2 \balpha^2}}
        \left(E_{\sbalpha} + E_{-\sbalpha} \right)  \cr
   t^2(\balpha) &=& -{i \over \sqrt{2 \balpha^2}}
        \left(E_{\sbalpha} - E_{-\sbalpha} \right)  \cr
   t^3(\balpha) &=&{1 \over 2 \balpha^2}  \balpha \cdot H \, .
\label{embeddingGenerators}
\end{eqnarray}
The remaining generators of $G$ fall into irreducible representations
of this SU(2).  The requirement that they correspond to integer or
half-integer values of $t^3$ implies that any pair of roots $\balpha$
and $\bbeta$ must satisfy
\begin{equation}
   {\balpha \cdot \bbeta \over \balpha^2}
      =\balpha^* \cdot \bbeta
        = {n \over 2}
\label{embeddingsubgp}
\end{equation}
for some integer $n$.  Further, if $\bbeta^2\ge\balpha^2$, then
$\bbeta^2/ \balpha^2 \equiv k$ must equal 1, 2, or 3, and $|n| \le k$.
It follows that at most two different root lengths can occur for a
given Lie algebra.

\begin{figure}[h,t]
\begin{center}
\scalebox{0.78}[0.78]{\includegraphics{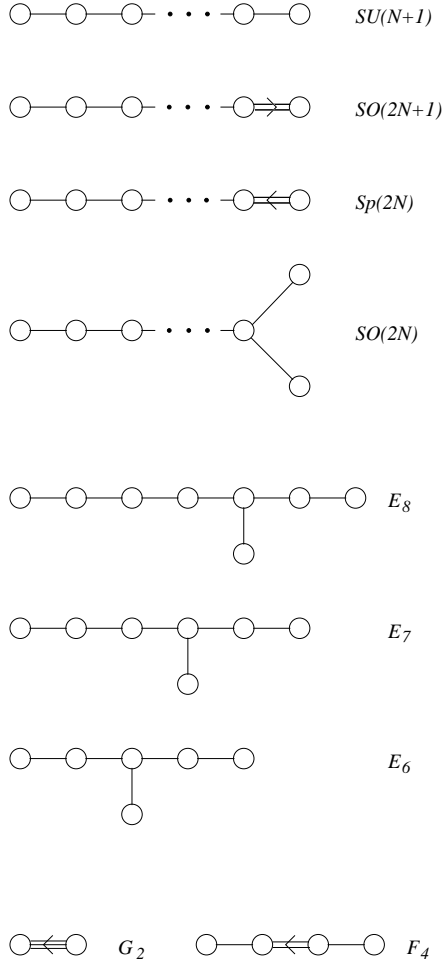}}
\par
\vskip-2.0cm{}
\end{center}
\begin{quote}
\caption{\small Dynkin diagrams for all simple Lie algebras. SU($N$),
SO($2N$), and $E_k$  are simply laced, meaning that
all roots have the same length. For the remaining groups, the arrow
points toward the long root(s).
\label{dynkin1}}
\end{quote}
\end{figure}

One can choose a basis for the root lattice that consists of $r$ roots
$\bbeta_a$, known as simple roots, that have the property that all
other roots are linear combinations of these with integer coefficients
all of the same sign; roots are termed positive or negative according
to this sign.
The inner products between the simple roots
characterize the Lie algebra and are encoded in the Dynkin diagram.
This diagram consists of $r$ vertices, one for each simple root, with
vertices $a$ and $b$ joined by
\begin{equation}
   m_{ab} = {4 (\bbeta_a \cdot \bbeta_b)^2
             \over \bbeta_a^2 \bbeta_b^2}
\end{equation}
lines.  The Dynkin diagrams for the simple Lie algebras are shown
in Fig.~\ref{dynkin1}.

The choice of the simple roots is not unique (although the $m_{ab}$
are).  However, it is always possible to require that the $\bbeta_a$
all have positive inner products with any given vector.  If these
inner products are all nonzero, then this condition picks out a unique
set of simple roots.

\subsection{Symmetry breaking and magnetic charges}

By an appropriate choice of basis, any element of the Lie algebra ---
in particular the Higgs vacuum expectation value $\Phi_0$ --- can be
taken to lie in the Cartan subalgebra.  We can use this fact to
characterize the Higgs vacuum by a vector $\bf h$ defined by
\begin{equation}
       \Phi_0 =  {\bf h}\cdot  {\bf H} \, .
\label{defOFh}
\end{equation}

The generators of the unbroken subgroup are those generators of
$G$ that commute with $\Phi_0$.  These are all the generators of
the Cartan subalgebra, together with the ladder operators corresponding
to roots orthogonal to $\bf h$.  There are two cases
to be distinguished.  If none of the $\balpha$ are orthogonal to $\bf
h$, the unbroken subgroup is the U(1)$^r$ generated by the Cartan
subalgebra.  If instead there are some roots $\bgamma$ with $\bgamma
\cdot {\bf h}=0$, then these form the root diagram for some
semisimple group $K$ of rank $r'$, and the unbroken subgroup is
$K\times {\rm U(1)}^{r-r'}$.

For the time being we will concentrate on the former case, which we
will term maximal symmetry breaking (MSB), leaving consideration of
the case with a non-Abelian unbroken symmetry to
Chap.~\ref{unbrokenNonAb}.  Because $\Pi_2(G/H) = \Pi_1[{\rm U(1)}^r]=
Z^r$, the single integer topological charge of the SU(2) case is
replaced by an $r$-tuple of integer charges.

To define these charges, we must examine the asymptotic form of the
magnetic field.  At large distances $B_i$ must commute with the Higgs
field.  Hence, if in some direction $\Phi$ is asymptotically of the
form of Eq.~(\ref{defOFh}), we can choose $B_i$ to also lie in the
Cartan subalgebra, and can characterize the magnetic charges by a
vector $\bf g$ defined by
\begin{equation}
   B_k =  \frac{{\hat r}_ k}{ 4\pi r^2 } {\bf g}\cdot {\bf H } \, .
\label{defOFg}
\end{equation}
The generalization of the SU(2) topological charge quantization is the
requirement that
\begin{equation}
      e^{ i e{\bf g}\cdot {\bf H } } =I
\end{equation}
for all representations of $G$
\cite{Goddard:1976qe,Englert:1976ng}. This is equivalent to requiring
that $\bf g$ be a linear combination
\begin{equation}
    {\bf g} = \frac{4\pi}{e} \sum_{a=1}^k  n_a \bbeta_a^*
\label{bfGdecomp}
\end{equation}
of the duals of the simple roots.
The integers $n_a$ are the desired topological charges.

We noted above that there are many possible ways to choose the simple
roots.  Each leads to a different set of $n_a$, with the various
choices being linear combinations of each other.  A particularly
natural set is specified by requiring that the simple roots all
satisfy
\begin{equation}
    \bbeta_a \cdot {\bf h} > 0 \, .
\label{hdefinessimple}
\end{equation}
Associated with this set are $r$ fundamental monopole
solutions, each of which is a self-dual BPS solution carrying one unit of
a single topological charge.  Thus, the $a$th fundamental monopole has
topological charges
\begin{equation}
    n_b = \delta_{ab}
\end{equation}
and, by the BPS mass formula of Eq.~(\ref{bpsEformula}), has mass
\begin{equation}
    m_a = {4 \pi \over e} {\bf h}\cdot\bbeta^*_a \, .
\label{fundmonomass}
\end{equation}

This fundamental monopole can be obtained explicitly by embedding
the unit SU(2) solution in the subgroup defined by $\bbeta_a$
via Eq.~(\ref{embeddingsubgp}).  If $A^{(s)}_i({\bf r}; v)$ and
$\Phi^{(s)}_i({\bf r}; v)$ ($s = 1,2,3$) are the gauge and scalar fields
of the SU(2) monopole with Higgs expectation value $v$, then the
$a$th fundamental monopole solution is
\begin{eqnarray}
    A_i &=& \sum_{s=1}^3 A^{(s)}_i({\bf r}; \bbeta_a \cdot {\bf h})
        \, t^s(\bbeta_a)     \cr
    \Phi &= & \sum_{s=1}^3 \Phi^{(s)}_i({\bf r}; \bbeta_a \cdot {\bf h})
         \, t^s(\bbeta_a)
   +\left[{\bf h} - ({\bf h}\cdot \bbeta^*_a) \, \bbeta_a \right]
           \cdot {\bf H} \, \, .
\label{fundmonosolution}
\end{eqnarray}
(The second term in $\Phi$ is needed to give the proper asymptotic
value for the scalar field.)

With the aid of Eqs.~(\ref{bfGdecomp}) and (\ref{fundmonomass}), the
energy of a self-dual BPS solution with topological charges $n_a$ can
be written as a sum of fundamental monopole masses,
\begin{equation}
    M = n_a m_a \, .
\label{massAsSum}
\end{equation}
While it may not be obvious that such solutions actually exist for
all choices of the $n_a$ (an issue that we will address later in this
chapter), some higher charge solutions can be written down immediately.
Since every root, simple or not, defines an SU(2) subgroup, the embedding
construction used to obtain the fundamental monopoles can be carried out
for any composite root $\balpha$.  The topological charges of the
corresponding solution are the coefficients in the expansion
\begin{equation}
    \balpha^* = n_a \bbeta_a^* \, .
\end{equation}

At first sight, the embedded solutions based on composite roots seem
little different than the fundamental monopole solutions.  However,
there is an essential, although quite surprising, difference.  Whereas
the fundamental monopoles are unit solitons corresponding to
one-particle states, the index theory results that we will obtain in
the next section show that the solutions obtained from composite roots
are actually multimonopole solutions.  They correspond to several
fundamental monopoles that happed to be superimposed at the same
point, but that can be freely separated.

The ideas of this section can be made a bit more explicit by focussing
on the case of SU($N$), which has rank $N-1$.  Its Lie algebra can be
represented by the set of traceless Hermitian $N\times N$ matrices.
The $H_i$ can be taken to be the $N-1$ diagonal generators.  The
$E_{\sbalpha}$ are then the $N(N-1)$ matrices that have a single
nonzero element in an off-diagonal position.  The Higgs expectation
value $\Phi_0$ can be taken to be diagonal, with matrix
elements\footnote{While this ordering of the eigenvalues is the most
convenient one for our purposes, it should be noted that it
corresponds to an ordering of the rows and columns of the matrices in
the Cartan subalgebra that is the opposite of the usual one; e.g., for
SU(2), $H_1 = {\rm diag}(-1,1)$.}
\begin{equation}
    s_1 \le s_2 \le \cdots  \le s_{N} \, .
\label{SUNphielements}
\end{equation}
If any $k\ge 2$ of the $s_j$ are equal, there is an unbroken SU($k$)
subgroup.  Otherwise, the symmetry breaking is maximal, and the simple
roots defined by Eq.~(\ref{hdefinessimple}) correspond to the matrix
elements lying just above the main diagonal.  The fundamental
monopoles are embedded in $2 \times 2$ blocks lying along the
diagonal, with the $a$th fundamental monopole lying at the
intersections of the $a$th and $(a+1)$th rows and columns and having
a mass proportional to $s_{a+1} - s_a$.  In a direction where the
asymptotic Higgs field is diagonal with matrix elements obeying
Eq.~(\ref{SUNphielements}), the asymptotic magnetic field is
\begin{equation}
   B_k =  {1\over 2e}\, \frac{{\hat r}_ k}{r^2 } \,
     {\rm diag}\left(-n_1, n_1-n_2, \dots,
         n_{N-2} -n_{N-1}, n_{N-1} \right) \, .
\end{equation}

Finally, we conclude this section with a brief note about
normalizations.  It is sometimes convenient to modify the
normalization given by Eq.~(\ref{CartanStdNorm}), so that the
$H_i$ obey
\begin{equation}
   \Tr (H_i H_j) = {c^2\over 2} \, \delta_{ij}
\end{equation}
with $c \ne 1$.
Under a rescaling of the normalization constant $c$, the roots
$\balpha \sim c$, while their duals $\balpha^* \sim
c^{-1}$.  To maintain the correct quantization of the
topological charge, $\bf g\sim c^{-1}$.

\subsection{Generalizing Montonen-Olive duality}

At the end of Chap.~\ref{BPSchap} we discussed the duality 
conjecture of Montonen and Olive.  This conjecture was motivated by
the invariance of the BPS spectrum if the transformation $e
\leftrightarrow 4\pi/e$ is accompanied by a simultaneous interchange
of electric and magnetic charges.  We are now in a position to ask how
this duality conjecture might be generalized to the case of larger
gauge groups.

The first step is to identify the particles to be exchanged by the
duality.  In the SU(2) theory these are the massive electrically
charged gauge boson and its superpartners, on the one hand, and the
magnetically charged supermultiplet obtained from the unit monopole
and the various possible excitations of the fermion zero modes in its
presence.  With a larger gauge group, the particle spectrum of the
electrically-charged sector is again composed of the gauge bosons that
acquire masses through the Higgs mechanism, and their superpartners.
On the magnetically charged side, matters are more subtle.  The
simpest guess is that the dual states should be built upon the
classical solutions obtained by using the various roots of the Lie
algebra to embed the SU(2) unit monopole.  The problem with this is that, as
we noted in the previous section, the embedded solution is
actually a multimonopole solution if the root $\balpha$ used for the
embedding is composite; only embeddings via the simple roots yield
one-monopole solutions.

As we will explain in Chap.~\ref{Bound}, this difficulty is resolved
by the existence of threshold bound states of the appropriate
fundamental monopoles.  These arise by a rather subtle mechanism
involving the fermion fields, and are only possible if the theory has
${\cal N}=4$ supersymmetry.  The existence of these bound states has
been explicitly demonstrated for the case where the embedding root
$\balpha$ is the sum of two simple roots
\cite{Lee:1996if,Gauntlett:1996cw}.  Because the construction used for
this case becomes much more tedious when more than two simple roots
are involved, the existence of the bound states for these cases has
not been verified, although there seems little doubt that they are
present.  For the remainder of this discussion, we will simply assume
their existence.

The next step is to look more closely at the masses of these particles.
The gauge boson associated with the root $\balpha$ has a mass
\begin{equation}
     M_{\balpha} = e {\bf h} \cdot \balpha   \, .
\end{equation}
This should be compared with the mass
\begin{equation}
     m_{\balpha} = {4\pi \over e} {\bf h} \cdot \balpha^*
      = {4\pi \over e} \, {{\bf h} \cdot \balpha \over\balpha^2}
\end{equation}
of the magnetically charged state (whether a fundamental monopole or a
threshhold bound state) associated with the same root.  The crucial
point to note here is the appearance of the root in the former case, but of
its dual in the latter.

For gauge groups whose root vectors all have the same length $\mu$
(the so-called ``simply laced'' groups), the roots and their duals
differ by a trivial factor of $\mu^2$.  All that is necessary to
generalize the duality conjecture is to replace the transformation $e
\leftrightarrow 4\pi/e$ by $e \leftrightarrow 4\pi/e \mu^2$; indeed, the
necessity of the additional factor becomes clear as soon as one recalls
that the normalization of the gauge coupling depends on the convention
that determines the root length.

However, the situation is not so simple if the gauge group has roots
of two different lengths, since in this case the roots and their duals
are not related by a common rescaling factor.  Instead, replacing all
of the roots by their duals is equivalent, up to an overall rescaling,
to simply interchanging the short and the long roots.  It is a
remarkable fact \cite{Goddard:1976qe,Englert:1976ng} that the new set
of roots obtained in this fashion is again the root system of a Lie
algebra, although not necessarily the original one.

This is easily demonstrated for the rank $N$ algebras SO($2N+1$) and
Sp($2N$).  Let $e_i$, $i=1,2,\dots,N$ be a set of unit vectors in
$N$-dimensional Euclidean space.  The roots of SO($2N+1$) can then be
written as $\pm \sqrt{2}\,e_i$ and $(\pm e_i \pm e_j)/\sqrt{2}$ ($i\ne
j$), while those of Sp($2N$) can be chosen to be $\pm e_i/\sqrt{2}$
and $(\pm e_i \pm e_j)/\sqrt{2}$ ($i\ne j$).  Replacing each root of
SO($2N+1$) by its dual then simply yields the roots of Sp($2N$), and
conversely.  The other two non-simply laced Lie algebras, $F_4$ and
$G_2$, are self-dual, up to a rotation; i.e., replacing the roots by
their duals yields the initial root system, but rotated.  [The same is
actually true of the algebras SO(5) and Sp(4), which are identical.]

The generalized Montonen-Olive conjecture can now be stated as follows
\cite{Bais:1978fw}.  Theories with simply laced gauge groups are
self-dual under the interchange of electric and magnetic charges and
weak and strong coupling.  If the gauge group has a Lie algebra that
is not simply laced, but is still self-dual, the theory is again
self-dual, but with appropriate relabeling of states.  In the
remaining cases [SO($2N+1$) and Sp($2N$) with $N \ge 3$], the duality
maps the gauge theory onto the theory with the dual gauge group.

\section{Index calculations}
\label{indexsection}

In Chap.~\ref{su2chap} we noted the existence of four zero modes (in
addition to those due to local gauge transformations) about the unit
monopole, and related that fact to the existence of a four-dimensional
moduli space of solutions.  We will now consider the zero modes about
BPS solutions of arbitrary charge.  In contrast with the previous
case, we do not know the form of the unperturbed solution.  Also
unlike the case of unit charge, the zero modes do not all arise from
the action of symmetries on the monopole solution.  It is,
nevertheless, possible to determine the number of these zero modes
\cite{Weinberg:1979zt,Weinberg:1979ma}.

The first step, which we describe in Sec.~\ref{indexperteqs}, is to
formulate the problem in terms of a matrix differential operator $\cal
D$, and to define a quantity $\cal I$ that counts the normalizable
zero modes of $\cal D$.  Next, in Sec.~\ref{indexDiracmodes}, we
rewrite the problem in terms of a Dirac equation.  This translation
from bosonic to fermionic language both simplifies the calculation and
illuminates some important properties of the moduli space.  The actual
evaluation of $\cal I$ is described in Sec.~\ref{indexevaluation}.

\subsection{Perturbation equations}
\label{indexperteqs}

For the calculations in this section, it
will be convenient to adopt a notation where the fields $A_i$ and
$\Phi$ of the unperturbed solution are written as anti-Hermitian
matrices in the adjoint representation of the group, while the
perturbations $\delta A_i$ and $\delta\Phi$ are written as column
vectors.  Using this notation, we expand Eq.~(\ref{bogomolny}).
Keeping terms linear in the perturbation gives
\begin{equation}
     0 = D_i \delta\Phi - e\Phi \delta A_i - \epsilon_{ijk} D_j
        \delta A_k
\label{selfdualpert}
\end{equation}
where
\begin{equation}
   D_i = \partial_i + eA_i
\end{equation}
is the covariant derivative with respect to the unperturbed
solution.

The solutions of Eq.~(\ref{selfdualpert}) include perturbations
that are local gauge transformations of the form
\begin{equation}
      \delta A_i = D_i\Lambda \, , \qquad \qquad
        \delta\Phi = e\Phi \Lambda \, .
\end{equation}
We are not interested in these, and so require that our
perturbations be orthogonal to such gauge transformations, in
the sense that
\begin{eqnarray}
   0 &=& \int d^3 x \left[ (D_i \Lambda)^\dagger \delta A_i  +
             e (\Phi \Lambda) ^\dagger  \delta \Phi\right]   \cr
      &=&  -\int d^3 x \, \Lambda^\dagger \left[D_i \delta A_i
                   + e\Phi \delta \Phi \right]
          +  \int d^2 S_i \, \Lambda^\dagger \delta A_i \, .
\end{eqnarray}
(The last integral is to be taken over a
surface at spatial infinity.)  For gauge functions $\Lambda(x)$
that fall off sufficiently rapidly that the surface term
vanishes, orthogonality is ensured by imposing the background
gauge condition
\begin{equation}
     0 =D_i \, \delta A_i +e \Phi \, \delta \Phi \, .
\label{backgroundgauge}
\end{equation}
This does not eliminate all of the gauge modes for which
$\Lambda(\infty)$ is nonzero.  The surviving modes correspond to
global gauge transformations in the unbroken subgroup.  These
modes are physically significant, as we saw in the analysis of
the U(1) phase mode and the related dyons in Sec.~\ref{dyons}.

Our goal is to count the number of linearly independent
solutions of Eqs.~(\ref{selfdualpert}) and (\ref{backgroundgauge}).
These equations can be
combined into a single matrix equation
\begin{equation}
     0 = {\cal D} \Psi
\label{bosonperteq}
\end{equation}
where $\Psi = (\delta A_1, \delta A_2, \delta A_3,
\delta \Phi)^t$ and
\begin{equation}
    {\cal D} = \left( \matrix{ - e\Phi & D_3 & -D_2 & D_1  \cr
                               - D_3  & -e\Phi & D_1 & D_2  \cr
                                 D_2 & -D_1 & -e\Phi & D_3  \cr
                               - D_1 & -D_2 & -D_3 & e\Phi \cr}
            \right) \,\, .
\end{equation}
The quantity that we want is the number of normalizable zero
modes of $\cal D$.  These are the same as the
normalizable zero modes of ${\cal D}^\dagger {\cal D}$, where
${\cal D}^\dagger$ is the adjoint of $\cal D$.  (Note that
${\cal D}^\dagger$ differs from $\cal D$ only in the signs of the
diagonal elements.)

Let us define
\begin{equation}
    {\cal I}  = \lim_{M^2 \to 0} {\cal I}(M^2)
\label{indexdef}
\end{equation}
where
\begin{equation}
     {\cal I}(M^2) =
     \Tr \left({ M^2 \over {\cal D}^\dagger {\cal D} +M^2}\right)
     -  \Tr \left({ M^2 \over {\cal D}{\cal D}^\dagger+M^2} \right)
\label{IofMdef}
\end{equation}
and $\Tr$ indicates a combined matrix and functional
trace.\footnote{On a compact space, where ${\cal D}^\dagger {\cal D}$
and ${\cal D}{\cal D}^\dagger$ would have discrete spectra with
identical nonzero eigenvalues, ${\cal I}(M^2)$ would be independent of
$M^2$.  The fact that it is not is a consequence of the continuum
spectra.}  Each normalizable zero mode of ${\cal D}^\dagger {\cal D}$
contributes 1 to the right-hand side of Eq.~(\ref{IofMdef}), while
each normalizable zero mode of ${\cal D}{\cal D}^\dagger$ contributes
$-1$. However, by making use of the fact that $A_i$ and $\Phi$ obey the
Bogomolny equation, it is easy to show that
\begin{equation}
   {\cal D}{\cal D}^\dagger  = - D_j^2 - e^2\Phi^2  \, .
\end{equation}
This is a manifestly positive definite operator (remember that $\Phi$
is an anti-Hermitian matrix) and therefore has no normalizable zero
modes.  It would thus seem that $\cal I$ is precisely the quantity
that we want.

There is one potential complication.  Because we are dealing with
operators that have continuum spectra extending down to zero, we must
worry about a possible contribution to $\cal I$ from the continuum.
Such a contribution would be of the form
\begin{equation}
     {\cal I}_{\rm cont} = \lim_{M^2 \to 0} M^2
       \int  {d^3k  \over (2\pi)^3 }
       {1 \over k^2  +M^2}
      \left[ \rho_{{\cal D}^\dagger {\cal D}}(k^2)
       - \rho_{{\cal D}{\cal D}^\dagger}(k^2) \right]
\label{Icont}
\end{equation}
where $\rho_{\cal O}(k^2)$ is the density of continuum eigenvalues of
the operator $\cal O$.  For this to be nonvanishing, the $\rho_{\cal
O}(k^2)$ must be rather singular at $k^2 =0$.

Singularities of this sort are absent when there is maximal symmetry
breaking.  This is most easily understood by viewing the theory in a
string gauge, where the correspondence between particles and field
components is clearest.  First, note
that the small-$k$ behavior of the densities of states is determined
by the large-distance structure of the differential operators.  Hence,
terms in $\cal D$ and ${\cal D}^\dagger$ that fall exponentially with
distance can be ignored.  The potentially dangerous terms that fall as
inverse powers of $r$ can only arise from field components
corresponding to the massless gauge and Higgs bosons.  Further, the
only modes that can have eigenvalues near zero are those with
components corresponding to perturbations of these massless fields.
Because the unbroken theory is Abelian, the massless fields do not
interact with themselves.  The long-range terms in $\cal D$ and ${\cal
D}^\dagger$ therefore have negligible effect on the small-$k$ behavior
and cannot give rise to any singularities.  Hence, ${\cal I}_{\rm
cont} = 0$.

If, instead, the unbroken gauge group is non-Abelian, the long-range
fields in the unperturbed solution can act on the massless
perturbations, and the above arguments no longer apply
\cite{Weinberg:1982ev}.  We will return to this issue in
Sec.~\ref{masslessIntroSection}, where we will see that a nonzero
continuum contribution to $\cal I$ actually does arise in certain
situations.

\subsection{Connection to Dirac zero modes and supersymmetry}
\label{indexDiracmodes}

Let us define a $2 \times 2$ matrix $\psi$ by
\begin{equation}
    \psi = I \delta\Phi + i \sigma_j \delta A_j \, .
\label{bosetofermi}
\end{equation}
Equations~(\ref{selfdualpert}) and (\ref{backgroundgauge})
can then be rewritten as the Dirac-type equation \cite{Brown:1977bj}
\begin{equation}
   0= (-i\sigma_j D_j + e\Phi) \psi \equiv  {\cal D}_f \psi \, .
\label{Diraczeromodeeq}
\end{equation}
Note, however, that when Eq.~(\ref{bosetofermi}) is inverted to give
$\delta A_j$ and $\delta \Phi$ in terms of $\psi$, the bosonic
perturbations obtained from $i \psi$ are linearly independent of those
obtained from $\psi$.  The number of normalizable zero modes of $\cal
D$ is thus twice the number of normalizable zero modes of the Dirac
operator ${\cal D}_f$.  Hence, if ${\cal I}_f$ is defined in the same
manner as $\cal I$, but with ${\cal D}$ replaced by ${\cal D}_f$, the
two quantities will be related by
\begin{equation}
     {\cal I} = 2 {\cal I}_f \, .
\end{equation}

This shows that the number of bosonic zero modes must be even.  In
fact, an even stronger result holds.  If $\psi$ is a solution to
Eq.~(\ref{Diraczeromodeeq}), then so is $\psi U$, where $U$ is any
unitary $2 \times 2$ matrix.  By this means a second linearly
independent $\psi$ can be constructed from the first.  Together, this
complex doublet of Dirac modes implies a set of four linearly
independent bosonic zero modes.  To make this explicit, let us use a
four-dimensional Euclidean notation, similar to that introduced in
Sec.~\ref{chaptwoModspace}, where $\delta\Phi = \delta A_4$.  If
$\delta A_a$ is the bosonic zero mode corresponding to the Dirac
solution $\psi$, then the zero mode corresponding to $\psi' =  i\psi
\sigma_r$ has components
\begin{equation}
     (\delta A)'_a = - \bar \eta^r_{ab} \delta A_b
\label{transformingzeromodes}
\end{equation}
where the anti-self-dual tensor $\bar\eta^r_{ab}$ and its self-dual
counterpart $\eta^r_{ab}$ (with $r=1,\,2,\,3$) are defined by
$\eta^r_{ij} = \bar\eta^r_{ij}=\epsilon_{rij}$,
$\eta^r_{a4} = -\bar\eta^r_{a4} = \delta_{ra}$.
Because of the
antisymmetry of the $\bar\eta^r_{ab}$, $\delta A'$ is orthogonal to
$\delta A$ at each point in space.

The zero modes form a basis for the tangent space at a given
point on the moduli space.  We thus have three maps $J^{(r)}$ of
this tangent space onto itself, with
\begin{equation}
    {J^{(r)}}_m{}^n \, \delta_n A_a
           = - \bar\eta^r_{ab} \delta_m A_b \, .
\label{quaternionicAction}
\end{equation}
These obey the quaternionic algebra
\begin{equation}
    J^{(r)} J^{(s)}  = - \delta^{rs} + \epsilon^{rst} J^{(t)}
\end{equation}
and thus define a local quaternionic structure on the moduli space.
In Sec.~\ref{moduliSpaceProperties} we will obtain an even stronger
result, that the moduli space is hyper-K\"ahler.\footnote{A discussion
of quaternionic and hyper-K\"ahler manifolds is given in Appendix~A.}

The existence of these multiplets of zero modes, and of the
hyper-K\"ahler structure that follows from them, can be understood in
terms of supersymmetry.  We have seen that by the addition of
appropriate fermion and scalar fields the Lagrangian can be extended
to that of ${\cal N}=4$ SYM theory, and that this is
the most natural setting for the Bogomolny equation.  Because the BPS
solution breaks only half of the supersymmetry, the zero modes about
any solution must fall into complete multiplets under the unbroken
${\cal N}=2$ supersymmetry.  The smallest possible multiplet has four real
bosonic and four real fermionic components, with the fermionic
components transforming as a complex doublet under the SU(2)
R-symmetry.

In fact, there are always four bosonic zero modes that can be
obtained directly from a supersymmetry transformation.  We saw in
Sec.~\ref{susyconnectionSec} that half of the supersymmetry in the
${\cal N}=2$ or ${\cal N}=4$ SYM theories is preserved by the BPS
solutions.  Acting on these solutions with the generators of the
broken supersymmetry produces Dirac zero modes.  In particular, by
examining Eqs.~(\ref{susytransformation}) -- (\ref{halfBPSzetas}), with
$\alpha=0$ and $\zeta_1= i \gamma^5 \gamma^0 \zeta_2$, we see that
\begin{equation}
     \psi = i \bsigma \cdot {\bf B}
\end{equation}
should be a solution of Eq.~(\ref{Diraczeromodeeq}), as can be easily
verified.  Guided by Eq.~(\ref{bosetofermi}), we immediately read off
the bosonic zero mode
\begin{equation}
     \delta A_i = B_i  \, , \qquad \delta \Phi = 0
\end{equation}
that corresponds to a global U(1) phase rotation.  Acting on this
mode with the three $J^{(p)}$ yields the three translation zero modes
\begin{equation}
     \delta_p A_i = F_{pi}  \, , \qquad \delta \Phi = D_p \Phi \, .
\end{equation}
[These differ from the naive form of the translation zero mode,
$\delta_p A_i = \partial_p A_i$ and $\delta_p \Phi =\partial_p\Phi$,
by a local gauge transformation with gauge function $\Lambda = - A_p$,
and thereby satisfy the background gauge condition,
Eq.~(\ref{backgroundgauge}).]

\subsection{Evaluation of ${\cal I}$}
\label{indexevaluation}

Returning to our calculation, let us define a set of Euclidean
Dirac matrices
\begin{equation}
    \Gamma_k= \left(\matrix{ 0 & -i \sigma_k \cr  i\sigma_k & 0}
             \right) \, , \qquad
    \Gamma_4= \left(\matrix{ 0 & I\cr  I & 0}
             \right)
\label{chap4eucdirac}
\end{equation}
obeying
\begin{equation}
    \{ \Gamma_a, \Gamma_b\} = 2 \delta_{ab} \, ,
\end{equation}
as well as
\begin{equation}
    \Gamma_5 = \Gamma_1 \Gamma_2 \Gamma_3 \Gamma_4 =
        \left(\matrix{ I & 0 \cr  0 & -I}   \right) \, .
\end{equation}
As in the preceding section, it will sometimes be convenient to use
the four-dimen\-sional notation in which $A_4= \Phi$, $F_{a4} = D_a
\Phi$, and (because we are using adjoint representation matrices for
the unperturbed fields in this section) $D_4= e\Phi$.  However, we
will at times need to switch back to the three--dimensional notation.
To distinguish between the two, we will let indices $a, b, \dots$
range from 1 to 4, while $i, j, \dots$ will range from 1 to 3.

With this notation,
\begin{equation}
    \Gamma \cdot D = \Gamma_a D_a
     = \left(\matrix{ 0 &{\cal D}_f \cr  -{\cal D}_f^\dagger & 0}
     \right)
\label{chap4eucdiracoperator}
\end{equation}
and
\begin{eqnarray}
    {\cal I}_f(M^2) &=& -\Tr \Gamma_5
    { M^2 \over - (\Gamma\cdot  D)^2  + M^2}  \cr \cr \cr
     &=& - \int d^3 x\, \tr  \left\langle x \left| \Gamma_5
          { M^2 \over - (\Gamma\cdot  D)^2  + M^2}
          \right| x\right\rangle  \cr  \cr \cr
     &=& - \int d^3 x\, \tr  \left\langle x \left| \Gamma_5
          { M\left[- (\Gamma\cdot  D)  + M\right] \over
        - (\Gamma\cdot  D)^2  + M^2}
          \right| x\right\rangle  \cr  \cr \cr
    &=& - \int d^3 x\, \tr  \left\langle x \left| \Gamma_5
           { M \over  (\Gamma\cdot  D)  + M}
           \right| x \right\rangle
\label{Ifastrace}
\end{eqnarray}
where $\tr$ indicates a trace only over Dirac and group indices.  (To
obtain the third equality, one must use the cyclic property of the
trace and the fact that $\Gamma_5$ anticommutes with an
odd number of $\Gamma$-matrices.)

The trick to the evaluation of $\cal I$ is to show that the integrand
on the right-hand side is a total divergence.  This allows ${\cal
I}_f(M^2)$ to be written as a surface integral at spatial infinity
that depends only on the asymptotic behavior of the fields.  To this
end, we define a nonlocal current
\begin{equation}
     J_i(x,y) =  \tr  \left\langle x \left| \Gamma_5 \Gamma_i
           { 1 \over  (\Gamma\cdot  D)  + M}
           \right| y \right\rangle \, .
\end{equation}
Using the identities
\begin{eqnarray}
    \delta(x-y) &=& \left[ \Gamma_i {\partial \over \partial x_i} +
        e \Gamma \cdot A +M \right]
        \left\langle x \left| { 1 \over  (\Gamma\cdot  D)  + M}
           \right| y \right\rangle     \cr \cr
        &=& \left\langle x \left| { 1 \over  (\Gamma\cdot  D)  + M}
           \right| y \right\rangle
         \left[ -\Gamma_i {\overleftarrow\partial \over \partial x_i} +
        e \Gamma \cdot A +M \right]
\end{eqnarray}
and the cyclic property of the trace, we find that
\begin{eqnarray}
   \left({\partial \over \partial x_i}
    + {\partial \over \partial y_i} \right) J_i(x,y)&=&
   - \tr \left[2M - e\Gamma\cdot A(x) + e\Gamma\cdot A(y) \right]
   \left\langle x \left| \Gamma_5 { 1 \over  (\Gamma\cdot  D)  + M}
           \right| y \right\rangle . \cr\cr  &&
\label{splitcurrentdiv}
\end{eqnarray}

The manipulations here are analogous to those used in the calculation
of the divergence of the four-dimensional axial current.  In four
dimensions, there are short-distance singularities as $y$ approaches
$x$ that produce an anomaly.  In three dimensions these singularities
are weaker and there is no anomaly.  We can therefore set $x=y$ in
Eq.~(\ref{splitcurrentdiv}) and, by comparing with
Eq.~(\ref{Ifastrace}), obtain
\begin{eqnarray}
     {\cal I}_f(M^2) &=& {1 \over 2} \int d^3x \, \partial_i J_i(x,x)
          \cr  \cr
     &=& {1 \over 2} \lim_{R \to \infty} \int_R dS_i \,J_i (x,x)
\label{Ifasdivergence}
\end{eqnarray}
where the surface of integration in the last line is a sphere of
radius $R$.

We now rewrite $J_i(x,x)$ as
\begin{equation}
     J_i(x,x) = - \tr  \left\langle x \left| \Gamma_5 \Gamma_i
          \,(\Gamma\cdot  D) \, { 1 \over  -(\Gamma\cdot  D)^2  + M^2}
           \right| x \right\rangle \, .
\label{Jxx}
\end{equation}
Because
\begin{equation}
  -(\Gamma\cdot  D)^2  + M^2 = - D_j^2 - e^2\Phi^2 + M^2
      -{e \over 4} [\Gamma_a, \Gamma_b] F_{ab}
\end{equation}
the last factor in Eq.~(\ref{Jxx}) can be expanded as
\begin{eqnarray}
  { 1 \over  -(\Gamma\cdot  D)^2  + M^2} &=&
   { 1 \over  - D_j^2 - e^2\Phi^2 + M^2 }  \cr \cr \cr
   &+& \,
  { 1 \over  - D_j^2 - e^2\Phi^2 + M^2 }
   \left({e \over 4} [\Gamma_a, \Gamma_b] F_{ab} \right)
  { 1 \over  - D_j^2 - e^2\Phi^2 + M^2 }   \cr \cr \cr
     &+& \, \cdots \, .
\end{eqnarray}
When this expansion is inserted into Eq.~(\ref{Jxx}), the
contribution from the first term vanishes after the trace over
Dirac indices is taken, while the $1/x^2$ falloff of $F_{ab}$
implies that the terms represented by the ellipsis do not contribute to
the surface integral.  To evaluate the remaining term, we write
\begin{equation}
      F_{ij} = \epsilon_{ijk} F_{k4} =  \epsilon_{ijk} B_k
       = \epsilon_{ijk} {\hat x_k \over 4\pi \, x^2} Q  + O(1/x^3)
\end{equation}
where $Q$ is an element of the Lie algebra specifying the magnetic
charge.  Inserting this
into Eq.~(\ref{Jxx}) and performing the Dirac trace leads to
\begin{eqnarray}
    \hat x_i J_i(x,x) &=& - {e^2 \over \pi \, x^2}
         \,\tr  \left\langle x \left| \Phi
     { 1 \over  - \nabla_j^2 - e^2\Phi^2 + M^2 }  Q
     { 1 \over  - \nabla_j^2 - e^2\Phi^2 + M^2 }
     \right| x \right\rangle  \cr\cr &&   + O(1/x^3)
\label{groupindepresult}
\end{eqnarray}
where now $\tr$ indicates a trace only over group indices.

By an appropriate gauge transformation, we can put the asymptotic
Higgs and magnetic fields into forms corresponding to those in
Eqs.~(\ref{defOFh}) and (\ref{defOFg}).  Keeping track
of the sign changes that arise because the $H_i$ are
Hermitian, we obtain
\begin{equation}
    \hat x_i J_i(x,x) =  {e^2 \over \pi \, x^2}
         \,\tr  \left\langle x \left|
     { ({\bf h} \cdot {\bf H}) ({\bf g} \cdot {\bf H}) \over
       \left[ - \nabla_j^2 + e^2({\bf h} \cdot {\bf H})^2 + M^2
              \right]^2 }
      \right| x \right\rangle    + O(1/x^3) \, .
\end{equation}
In the adjoint representation the matrix elements
of the generators are given by the structure constants.  In
particular, the matrix elements of the $H_i$ are determined by
the roots.  The matrix trace in the above equation thus
leads to a sum over roots, and
\begin{equation}
    \hat x_i J_i(x,x) =  {e^2 \over \pi \, x^2}
         \, \sum_{\sbalpha}  \left\langle x \left|
     { ({\bf h} \cdot \balpha) ({\bf g} \cdot \balpha) \over
       \left[ - \nabla_j^2 + e^2({\bf h} \cdot \balpha)^2 + M^2
              \right]^2 }
      \right| x \right\rangle    + O(1/x^3) \, .
\end{equation}
Using the identity
\begin{equation}
   \left\langle x \left| {1 \over
       \left(- \nabla_j^2 + \mu^2 \right)^2 } \right| x \right\rangle
    = \int {d^3 k \over (2\pi)^3} \, {1 \over (k^2 + \mu^2)^2}
    = {1 \over 8 \pi \mu}
\end{equation}
then gives
\begin{equation}
  \hat x_i J_i(x,x)  =  {e^2 \over 8\pi^2 \, x^2}
         \, \sum_{\sbalpha}
     { ({\bf h} \cdot \balpha) ({\bf g} \cdot \balpha) \over
       \left[ e^2({\bf h} \cdot \balpha)^2 + M^2
              \right]^{1/2} } + O(1/x^3) \, .
\end{equation}

The contributions to the sum from the positive and negative roots
are clearly equal, so we can insert a factor of two and
restrict the sum to the positive roots.  Then substituting this last
equation into Eq.~(\ref{Ifasdivergence}), we obtain
\begin{equation}
    {\cal I}_f(M^2) = {e^2 \over 2 \pi} \,
          {\sum_{\sbalpha}}'  { ({\bf h} \cdot \balpha)
            ({\bf g} \cdot \balpha) \over
       \left[ e^2({\bf h} \cdot \balpha)^2 + M^2
              \right]^{1/2} } \, .
\label{intermediateindexcalc}
\end{equation}
Taking the limit $M^2 \rightarrow 0$, using the fact that all positive
roots satisfy $\balpha \cdot {\bf h} > 0$, and inserting
Eq.~(\ref{bfGdecomp}), we then find
\begin{eqnarray}
    {\cal I}_f  &=&  {e \over 2 \pi} \,{\sum_{\sbalpha}}'
           {\bf g} \cdot \balpha   \cr
          &=&  2 \, \sum_a n_a  \,{\sum_{\sbalpha}}'
            {\bbeta}_a^* \cdot \balpha
\end{eqnarray}
where the prime indicates that the sum is only over the positive roots.
We now make use of the fact that if ${\bbeta}_a$ is any simple root,
and $\balpha \ne {\bbeta}_a$ is a positive root, then reflection in a
hyperplane orthogonal to ${\bbeta}_a$ gives another positive root
${\balpha}'$ with ${\balpha}' \cdot {\bbeta}_a = - \balpha \cdot
{\bbeta}_a$.  Hence, all the terms in the sum over $\balpha$
cancel pairwise, except for the one with $\balpha = {\bbeta}_a$,
and we have
\begin{equation}
        {\cal I}_f = 2 \sum_a n_a \, .
\end{equation}
Finally, the quantity that we want is
\begin{equation}
        {\cal I} = 2 {\cal I}_f = 4 \sum_a n_a \, .
\label{finalIndex}
\end{equation}

Equation~(\ref{finalIndex}) tells us that the SU(2) solutions with $n$
units of magnetic charge lie on a $4n$-dimensional moduli space.  The
natural interpretation is that there are four moduli for each of $n$
independent monopoles.  One would expect three of these to specify the
position of the monopole.
The fourth modulus should be analogous to the fourth zero mode of the
unit monopole, which is associated with a global U(1) phase.  It is
perhaps most easily understood by considering its conjugate momentum.
For the unit monopole, this is the electric charge divided by $e$; for
solutions with higher charge, excitation of these ``U(1)-phase modes''
leads to independent electric charges on each of the $n$ component
monopoles.\footnote{Note, however, that while a dyon with unit
magnetic charge satisfies the BPS mass formula of
Eq.~(\ref{bpsEformula}) for any value of $Q_E$, a multidyon solution
is only BPS if each of the component dyons has the same electric
charge.}

The story is very much the same with larger gauge groups.  The
fundamental monopoles are obtained by embeddings of the SU(2), and
they have only the four position and U(1) modes inherited from that
solution.  All solutions with higher charge live on higher-dimensional
moduli spaces, and are thus naturally understood as multimonopole
solutions.

\section{General remarks on higher charge solutions}
\label{generalsolutionremarks}

The BPS mass formula suggests the possibility of static multimonopole
solutions, with the Higgs scalar field mediating an attractive force
that exactly cancels the magnetic repulsion between the monopoles.
Further, the index theory calculations of the previous section show
that the number of parameters entering solutions with higher magnetic
charge --- if any such solutions exist --- is just what might be
expected for a collection of noninteracting static monopoles.

While suggestive, neither of these considerations actually establishes
that multimonopole solutions exist.  However, an existence proof has
been given by Taubes \cite{Jaffe:1980mj}.  In the context of the SU(2)
theory, he showed that there is a finite distance $R$ such that, given
arbitrary points ${\bf r}_1, {\bf r}_2 \dots, {\bf r}_N$ with all
$|{\bf r}_i -{\bf r}_j| > R$, there is a magnetic charge $N$ solution
with zeroes of the Higgs field at the given locations.\footnote{The
existence of the minimum distance $R$ is not simply a technical
restriction that might be eliminated from the proof by further
analysis.  We will see below that when the monopole cores overlap, the
simple connection between zeroes of the Higgs field and monopole
positions can be lost.}

In the next section we will describe a construction for obtaining
these higher charge solutions.  Before doing so, we present here some
general remarks concerning the nature of these solutions.  While
we will focus on SU(2) solutions, similar considerations apply with
larger gauge groups.  We start by considering solutions with $N$
monopoles whose mutual separations are all large compared to the
monopole core radius.  There are $N$ zeroes of the Higgs field, with a
monopole core surrounding each zero and the massive fields falling
exponentially outside these cores.  These solutions are in a sense
both rather complex and yet quite simple.

The complexity becomes evident as soon as one considers the twisting
of the Higgs field. In any nonsingular gauge the Higgs field
orientation in the neighborhood of each individual monopole must look
like that for a singly-charged monopole.  However, these Higgs fields
must join up at large distances to give a configuration with winding
number $N$.  The analytic expression for such a configuration cannot
be simple.

At the same time, there is an underlying simplicity arising from the
fact that, apart from the exponentially small massive fields, the
physical fields outside the cores are purely Abelian.  These obey
linear field equations, and so it should be possible to obtain
approximate solutions by superposition.  This is most easily done by
working in a gauge with uniform Higgs field direction, $\Phi =
(0,0,\varphi)$, and defining electromagnetic and massive vector fields
${\cal A}_\mu$ and $W_\mu$ as in Eq.~(\ref{stringansatz}).  In this
gauge there is a Dirac string originating at each of the zeros of the
Higgs field and running off to spatial infinity.  The specific paths
of the strings are gauge-dependent; let us assume that they are
chosen to avoid all monopole cores except the one in which they
originate.

For a single monopole centered at the origin, the electromagnetic field
in this gauge is ${\cal A}_j({\bf r}) = A_j^{\rm Dirac}({\bf r})$.
The Higgs field can be written as
\begin{equation}
    \varphi({\bf r} - {\bf x}^{(a)})
         = v  + \tilde\varphi({\bf r}; v) \, .
\end{equation}
For $|{\bf r}| \gg M_W^{-1}$,
\begin{equation}
    \tilde\varphi({\bf r}; v) = - {1 \over er} + O(e^{-evr})
\end {equation}
and the massive vector field $W_j({\bf r}; v)$ is exponentially small.

Now consider a solution for which the Higgs field has zeros at
${\bf x}^{(a)}$, with $a=1,2, \dots, N$.  The linearity of the Abelian
theory implies that outside the $n$ core regions
\begin{eqnarray}
   {\cal A}_j({\bf r})
         &=& \sum_{a=1}^N A_j^{\rm Dirac}({\bf r} -{\bf x}^{(a)})
    + \cdots  \cr\cr
   \varphi({\bf r}) &=& v -
       \sum_{a=1}^N {1 \over e |{\bf r} -{\bf  x}^{(a)}|}
        +\cdots
\end{eqnarray}
where the ellipsis represents terms that, like the $W_j$ field itself,
fall exponentially with distance from the cores.

The fields inside the core regions are similar to those for a
single monopole, but with a few notable differences.  First, the
scalar field tails of the other monopoles reduce
the Higgs expectation value seen in the $a$th core from $v$ to
\begin{equation}
     v_{\rm eff}^a  = v
    - \sum_{b \ne a} {1 \over e |{\bf x}^{(a)} -{\bf  x}^{(b)}|}
     +  O(e^{-evr_{\rm min}}) \, .
\end{equation}
This produces an increase in the core radius, and implies that the
Higgs and W fields inside this core are approximately
\begin{eqnarray}
     \varphi({\bf r})  &\approx&
       v_{\rm eff}
          + \tilde\varphi({\bf r}- {\bf x}^{(a)}; v_{\rm eff}^a)  \cr\cr
      W_j({\bf r})  &\approx&
        W_j({\bf r}- {\bf x}^{(a)}; v_{\rm eff}^a) \, .
\end{eqnarray}

Because the massive fields fall exponentially with distance, they have
negligible effects on the interactions between the monopoles.
However, they can have a curious effect on the symmetry of the
solutions, with interesting physical consequences.  This can be seen
most clearly by considering a solution containing two monopoles, one
centered at $(0,0,-R)$ and one at $(0,0,R)$, with the Dirac string of
the first (second) chosen to run along the $z$-axis from the monopole
to $z=-\infty$ ($z=\infty$).  Since a single monopole is spherically
symmetric, it would be natural to expect that this solution would be
axially symmetric under rotations about the $z$-axis.

Let us examine this in more detail.  If only the first monopole were
present, its Higgs and electromagnetic fields would be invariant under
rotation by an angle $\alpha$ about the $z$-axis.  Its $W$ field
would also be invariant if this rotation were accompanied by a global
U(1) gauge transformation with gauge function $\Lambda =
e^{-i\alpha}$.  Similarly, if only the second monopole were present,
it would be invariant under the same rotation, except that the change
in direction of the Dirac string would require that $\Lambda =
e^{i\alpha}$.

The mismatch between the two gauge transformations means that, despite
naive expectations, the solution cannot be axially symmetric.  A
gauge invariant measure of this is given by the scalar product ${\bf
W_{(1)}^*}\cdot {\bf W}_{(2)}$, where ${\bf W}_{(a)}$ denotes the
field due to the $a$th monopole; this is $O(e^{-evR})$.

One consequence of this breaking of axial symmetry is that the spectrum
of fluctuations about the solution must include a zero mode
corresponding to spatial rotation about the $z$-axis.  A
time-dependent excitation of this mode gives a solution with nonzero
angular momentum oriented along the axis joining the two monopoles.
This can be understood by noting that, because of the mismatch in
$\Lambda$'s noted above, this rotation also corresponds to a shift in
the relative U(1) phase between the monopoles.  When done in a
time-dependent fashion, this turns the monopoles into a pair of dyons
with equal and opposite electric charges.  The angular momentum is
just the usual charge-monopole angular momentum, which for a pair of
dyons with electric and magnetic charges $q_j$ and $g_j$ points toward
dyon 1 and has a magnitude $g_1 q_2 - g_2 q_1$.

In contrast to the case of widely separated monopoles, where the
general properties of the solutions could have been anticipated,
some surprising features arise when several monopoles are brought
close together.  We will just note a few examples:

1) When two monopoles are brought together
\cite{Ward:1981jb,Forgacs:1980ym,Prasad:1980yw,Rebbi:1980yi}, the axial symmetry,
whose curious absence we have noted, actually emerges when the two
zeros of the Higgs field coincide.  The profiles of the energy density
and of the Higgs field have a toroidal shape.\footnote{There are also
  axially symmetric solutions with more than two units of magnetic
  charge \cite{Forgacs:1981jt,Forgacs:1981ve,Prasad:1980hg,Prasad:1980hi}.  One can
  show that in all such cases the zeros of the Higgs field must all
  coincide \cite{Houston:1980fy}.}

2) There is a solution with tetrahedral symmetry
\cite{Hitchin:1995qw,Houghton:1995bs}, with the energy density
contours looking like tetrahedra with holes in the centers of each
face.  Although the Higgs field has a zero at each vertex of the
tetrahedron, there is also an antizero (i.e., a zero with opposite
winding) at the center.  Thus, this is actually a three-monopole
solution, illustrating quite dramatically that the zeroes of the Higgs
field are not always the same as the monopole positions
\cite{Houghton:1995uj}.

3) Solutions corresponding to the other Platonic solids, but again with
nonintuitive charges, have been found.  There is a cubic $N=4$
solution \cite{Hitchin:1995qw,Houghton:1995bs} for which the Higgs
field has a four-fold zero at the center and no other zeros
\cite{Sutcliffe:1996qz}, an $N=5$ octahedral solution
\cite{octa-dodec} with zeros at the vertices and an antizero at the
center \cite{Sutcliffe:1996qz}, an $N=7$ dodecahedral solution
\cite{octa-dodec} with a seven-fold zero at the center
\cite{Sutcliffe:1996qz}, and an $N=11$ icosahedral solution
\cite{Houghton:1997kg}.

One feature that cannot emerge is spherical symmetry.  Not only are
there no spherically symmetric SU(2) solutions with $N\ge 2$, there
are not even any finite energy configurations with spherical symmetry.
This result was first obtained by a detailed analysis of the behavior
of gauge fields under rotations, including the effects of the possible
compensating gauge transformations \cite{Weinberg:1976eq}.  However, a
much simpler proof can be obtained by considering the properties of
generalized spherical harmonics.  This analysis is best done in the
string gauge used above, with a uniform SU(2) orientation for the
Higgs field, and electromagnetic and massive gauge fields ${\cal
A}_\mu$ and $W_\mu$.  Each of these fields can be expanded in
spherical harmonics, with the coefficients being functions only of
$r$.  A spherically symmetric configuration is one that contains only
harmonics with total angular momentum quantum number $J=0$.

The spin and charge of a field determine what type of harmonics are
appropriate for its expansion.  A neutral scalar field can be expanded
in terms of the $Y_{LM}$, the eigenfunctions of the orbital angular
momentum ${\bf L} = {\bf r} \times {\bf p}$.  For a charged scalar
field one must use the monopole spherical harmonics
\cite{Tamm,Wu:1976ge} that take into account the extra charge-monopole
angular momentum; because the latter contribution is orthogonal to the
usual orbital angular momentum, it places a lower bound on $J$ and
implies that the harmonics for a monopole with $q$ units of charge
have $J \ge q$.  The additional spin angular momentum of a charged
vector field, such as $W_j$, leads to vector monopole spherical
harmonics \cite{Olsen:1990jm,Weinberg:1993sg}.  These all have $J \ge
q-1$, so $W_j$ would vanish identically in any spherically symmetric
configuration with multiple magnetic charge.  This would leave only
the the Higgs and electromagnetic fields, giving an essentially U(1)
configuration that has infinite energy because of the singularity of
the Coulomb field at the origin.\footnote{Spherically symmetric
solutions with higher magnetic charges are possible, however, if the
gauge group is larger than SU(2).  See the discussions of these in
Refs.~\cite{Bais:1978yh,Wilkinson:1978zh,Leznov:1979td,Leznov:1980tz,Ganoulis:1981sx}
and their construction by the ADHMN method in \cite{Bowman:ss}.}

\section{The Atiyah-Drinfeld-Hitchin-Manin-Nahm construction}
\label{nahmsection}

Several methods for constructing multimonopole solutions have been
developed, including ones using twistor methods
\cite{Ward:1981jb,Corrigan:1981fs,Prasad:1980at}, B\"acklund
transformations
\cite{Forgacs:1980su,Forgacs:1981vf,Forgacs:1981bd,Forgacs:1983gr},
and rational maps
\cite{Hitchin:1982gh,Donaldson:1985id,Hurtubise:1985vq,jarvis}.
However, the method due to Nahm
\cite{Nahm:1982jb,Nahm:1981xg,Nahm:1981nb,Nahm:1983sv} has proven to
be the most fruitful.\footnote{For further discussion of the other
construction methods, see Refs.~\cite{Sutcliffe:1997ec} and \cite{Manton:2004tk}.}
  It is readily extended from SU(2) to the other
classical groups \cite{Hurtubise:1989qy}, and also has a natural
string theoretic interpretation in terms of D-branes
\cite{Diaconescu:1996rk}.

Nahm's approach is based on the observation that the
monopole solutions of the Bogomolny equation can be viewed as
dimensionally reduced analogues of the instanton solutions of the
self-dual Yang-Mills equation.  For the latter, the
Atiyah-Drinfeld-Hitchin-Manin (ADHM) construction \cite{Atiyah:1978ri}
gives an equivalence between the solutions of the nonlinear
self-duality differential equations in four variables and a set of
algebraic matrix equations.  From a solution of these matrix
equations, an instanton solution can be obtained by solving linear
equations.

Nahm generalized this construction to the monopole problem.  Instead
of an equivalence between differential equations in four variables and
a set of purely algebraic equations, this
Atiyah-Drinfeld-Hitchin-Manin-Nahm (ADHMN) construction gives an
equivalence between the Bogomolny equation in three variables and the
Nahm equation, which is a nonlinear differential equation in one
variable \cite{Corrigan:1983sv}.  The counterparts of the ADHM
matrices are matrix functions $T_\mu(s)$ ($\mu=0,1,2,3$), known as the
Nahm data.  [We will see that it is always possible to eliminate
$T_0(s)$.  This is usually done, yielding the more familiar form of
the construction in terms of the three $T_j(s)$.]

We begin our discussion in Sec.~\ref{NahmConstSec} by presenting,
without proof, the prescription for constructing a $k$-monopole
solution in the SU(2) theory.  Then, in Sec.~\ref{gaugeactionsection},
we describe some gauge freedoms associated with this construction, and
also describe how the symmetries of spacetime are reflected in the
Nahm data.  In Sec.~\ref{NahmSelfSec} we show that the fields obtained
by the construction are indeed self-dual; the argument is quite
parallel to that for the ADHM construction.  We also verify here that
the solutions have the correct magnetic charge and that they lie in
SU(2).  Next, in Sec.~\ref{NahmCompleteSec}, we demonstrate the
completeness of the construction by showing that any solution of the
Bogomolny equation yields a solution of the Nahm equation.  Finally,
the extension to other classical groups is described in
Sec.~\ref{NahmBiggroups}.

In order to simplify the equations, we will set the gauge coupling $e$
to unity throughout this section.  The factors of $e$ can be easily
restored by simple rescalings.

\subsection{The construction of SU(2) multimonopole solutions}
\label{NahmConstSec}

The ADHMN construction of a BPS SU(2) solution with $k$ units of
magnetic charge can be viewed as a three-step process:

\medskip
\noindent 1) {\bf Solving for the Nahm data}
\smallskip

The first step is to find a quartet of Hermitian $k \times k$ matrices
$T_\mu(s)$ that satisfy the Nahm equation,
\begin{equation}
   0 =   {dT_i\over ds} + i[T_0 , T_i]
        + {i \over 2} \epsilon_{ijk} [T_j , T_k] \, ,
\label{fullNahmequation}
\end{equation}
where the indices $i$, $j$, and $k$ run from 1 to 3, and the auxiliary
variable $s$ lies in the range $-v/2 \le s \le v/2$, with $v$ being
the Higgs vacuum expectation value.  For $k=1$, the $T_i(s)$ are
clearly constants.
For $k > 1$, we impose the condition
that the $T_i(s)$ have poles at the
boundaries of the form
\begin{equation}
     T_i(s) = - \, {L_i^\pm \over s \mp v/2}  + O(1) \, .
\label{boundarypoles}
\end{equation}
The Nahm equation implies that the $L_i^\pm$ form $k$-dimensional
representations of the SU(2) Lie algebra,
\begin{equation}
     [L^\pm_i, L^\pm_j] = i \epsilon_{ijk} L^\pm_k \, .
\end{equation}
The final boundary condition is that these representations be
irreducible; i.e., they must be equivalent to the angular momentum
$(k-1)/2$ representation of SU(2).  There are no boundary conditions
on $T_0$.

\medskip
\noindent 2) {\bf The construction equation}
\smallskip

The next step is to define a linear operator
\begin{equation}
    \Delta(s) = {d \over ds} +iT_0(s)\otimes {\rm I_2}
        - T_i(s) \otimes \sigma_i  + r_i {\rm I_k}\otimes \sigma_i
\end{equation}
and to solve the construction equation
\begin{equation}
     0 = \Delta^\dagger(s) \, w(s, {\bf r})
      = \left[-{d \over ds} -iT_0\otimes {\rm I_2} - T_i \otimes \sigma_i
          + r_i {\rm I_k}\otimes \sigma_i \right] w(s, {\bf r})
\label{SU2construction}
\end{equation}
where $w(s, {\bf r})$ is a $2k$-component vector.  (We will usually
suppress the indices denoting the components of $w$.)

We are only interested in the normalizable solutions of this equation.
We denote by $w_a$ a complete linearly independent set of such
solutions, and require that they obey the orthonormality condition
\begin{equation}
    \delta_{ab} = \int_{-v/2}^{v/2} ds
        \,w_a^\dagger(s,{\bf r})  w_b(s,{\bf r}) \, .
\label{SU2Normalization}
\end{equation}

\medskip
\noindent 3) {\bf Obtaining the spacetime fields}
\smallskip

We assert now, and prove below in Sec.~\ref{NahmSelfSec}, that there
are only two normalizable $w_a(s, {\bf r})$.  The spacetime fields are
obtained as $2 \times 2$ matrices from these by the equations
\begin{equation}
    \Phi^{ab}({\bf r}) = \int_{-v/2}^{v/2} ds \, s
           \, w_a^\dagger(s,{\bf r}) w_b(s,{\bf r})
\label{SU2Phi}
\end{equation}
and
\begin{equation}
     A_j^{ab}({\bf r}) = -i\int_{-v/2}^{v/2} ds  \,w_a^\dagger(s,{\bf r})
           \partial_j w_b(s,{\bf r}) \, .
\label{SU2Afield}
\end{equation}

\medskip

\subsection{Gauge invariances and symmetries}
\label{gaugeactionsection}

Every set of Nahm data satisfying the Nahm equation and boundary
conditions yields a self-dual spacetime solution.  Furthermore, every
$\Phi({\bf r})$ and $A_i({\bf r})$ obeying the Bogomolny equation can
be obtained in this fashion.  However, because of the existence of
gauge freedom on both sides, the correspondence between Nahm
solutions and Bogomolny solutions is not one-to-one.

The usual spacetime gauge transformations correspond to changes in the
basis of the solutions of the construction equation,
\begin{equation}
   w_a(s,{\bf r}) \rightarrow w_a'(s,{\bf r}) = w_b(s,{\bf r})
                     \, U_{ba}({\bf r})
\end{equation}
with the SU(2) matrix $U({\bf r})$ being the usual spacetime gauge
function.

The corresponding gauge symmetry on the Nahm side is an SU($k$) gauge
action that preserves the Nahm equation.  It takes the form
\begin{eqnarray}
   T_j(s) &\rightarrow& T'_j(s) = g(s) \, T_j(s) \, g^{-1}(s)   \cr
   T_0(s) &\rightarrow& T'_0(s) = g(s) \, T_0(s) \, g^{-1}(s)
          + i \,{dg(s) \over ds} \, g^{-1}(s)
\label{gaugeaction}
\end{eqnarray}
where $g(s)$ is an element of SU($k$).  If $w(s,{\bf r})$ is a
solution of the construction equation defined by the $T_\mu$, then
\begin{equation}
    w'(s,{\bf r}) = g(s)\otimes {\rm I}_2 \, w(s,{\bf r})
\label{gaugeactiononW}
\end{equation}
is a solution of the construction equation defined by the $T'_\mu$.
Referring to Eqs.~(\ref{SU2Phi}) and (\ref{SU2Afield}), one then sees
that, just as the spacetime gauge transformation has no effect on the
$T_\mu(s)$, this SU($k$) action leaves $\Phi({\bf r})$ and
$A_j({\bf r})$ unchanged.

By exploiting this SU($k$) gauge action, it is always possible to transform
away any nonzero $T_0(s)$.  We will usually assume that this has been
done, and will write the Nahm equation as
\begin{equation}
     {dT_i\over ds}  = {i \over 2} \epsilon_{ijk} [T_j , T_k]
\label{usualNahmEq}
\end{equation}
and the construction equation as
\begin{equation}
     0  = \left[-{d \over ds}  - T_i \otimes \sigma_i
          + r_i {\rm I}_k\otimes \sigma_i \right] w(s, {\bf r}) \, .
\label{usualSU2construction}
\end{equation}

In addition to these gauge actions, there are also
transformations on the Nahm data that reflect the symmetries of
spacetime.   If $T_i(s)$ is a solution of the Nahm equation,
then so is
\begin{equation}
    T'_i(s) =  T_i(s) + D_i {\rm I}_k \, .
\end{equation}
Referring to the construction equation, we see that the $T'_i(s)$
generate a solution that is translated
in physical space by a displacement $\bf D$.

Similarly, if $R_{ij}$ is an $s$-independent SO(3) matrix,
the replacement of $T_i(s)$ by
\begin{equation}
    T'_i(s) =  R_{ij} \, T_j(s)
\end{equation}
corresponds to a rotation of the spatial coordinates in the
construction equation, and thus to a rotation of the
solution in physical space.

\subsection{Verification of the construction}
\label{NahmSelfSec}

We now show that the fields obtained by the ADHMN construction have the
desired properties.  Thus, we must show that they are self-dual, i.e.,
that they satisfy the Bogomolny equation~(\ref{bogomolny}); that they
lie in SU(2); and that they have $k$ units of magnetic charge.

\medskip
\noindent 1) {\bf Proof of self-duality}
\smallskip

To verify the self-duality of the fields, we separately calculate
$B_i$ and $D_i\Phi$ and show that the two are equal.  The approach
described here \cite{Corrigan:1983sv} closely follows that used in
Ref.~\cite{Christ:1978jy} to demonstrate the self-duality of the
instanton solutions obtained by the ADHM construction.  For the sake
of clarity, we will not explicitly show the dependence on the spatial
position $\bf r$, although we will have to make use of the spatial
derivative $\partial_i$.

We begin with
\begin{eqnarray}
    B_i^{ac} \!\! &=& \!\! {1\over 2} \epsilon_{ijk} F_{ij}^{ac} \cr
      \!\!  &=& \!\!  - i \, \epsilon_{ijk}
     \left[ \int ds \, \partial_j w^\dagger_a(s) \, \partial_k w_c(s)
    + \int ds \, ds' \, w^\dagger_a(s) \, \partial_j w_b(s) \,
          w^\dagger_b(s') \, \partial_k w_c(s')  \right] \cr
 \!\! &=& \!\!  -i \,\epsilon_{ijk} \int ds \, ds' \, \partial_j w^\dagger_a(s)
        \, {\cal F}(s,s') \, \partial_k w_c(s')
\label{NahmForB}
\end{eqnarray}
where
\begin{equation}
   {\cal F}(s,s') = \delta(s-s') -  w_b(s)
         \,  w^\dagger_b(s')
\end{equation}
obeys
\begin{equation}
    \int ds' {\cal F}(s,s') \,  {\cal F}(s',s'') = {\cal F}(s,s'') \, .
\end{equation}
These last two equations show that $\cal F$ is the projection operator
onto the space orthogonal to the kernel of
$\Delta^\dagger$.  It can therefore be written as
\begin{equation}
  {\cal F}(s,s') = \Delta(s) \, G(s,s') \, \Delta^\dagger(s')
\label{NahmFexpression}
\end{equation}
where the Green's function $G= (\Delta^\dagger \Delta)^{-1}$ obeys
\begin{equation}
       \Delta^\dagger(s) \Delta(s) \, G(s,s') = \delta(s,s') \, .
\end{equation}

To show that $G$ actually exists; i.e., that $\Delta^\dagger \Delta$ is
indeed invertible, note that
\begin{equation}
    \Delta^\dagger \Delta
    = -\left({d \over ds} + i T_0 \right)^2 {\rm I}_{2k}
    + (T_i - r_i {\rm I}_k)^2 \otimes {\rm I}_2
     + \left\{ {dT_i\over ds}  + i[T_0,T_i]
       + i \epsilon_{ijk} T_j T_k \right\}
        \otimes \sigma_i \, .
\end{equation}
The last term vanishes because the $T_\mu$ obey the Nahm equation, and
so $\Delta^\dagger \Delta$ is a positive operator.  The vanishing of
this term also means that $\Delta^\dagger \Delta$, and hence $G$,
commute with all of the $\sigma_j$.

Returning to Eq.~(\ref{NahmForB}), we substitute the expression in
Eq.~(\ref{NahmFexpression}) for ${\cal F}$, and then use the
definition of the adjoint to obtain
\begin{equation}
    B_i^{ac} =  -i \,\epsilon_{ijk} \int ds \, ds' \,
       \left[\Delta^\dagger(s) \,  \partial_j w_a(s) \right]^\dagger
         G(s,s') \, \Delta^\dagger(s') \, \partial_k w_c(s') \, .
\label{NahmBtwo}
\end{equation}
Next, by differentiating the construction equation,
Eq.~(\ref{SU2construction}), we obtain the identity
\begin{equation}
    \Delta^\dagger(s) \, \partial_i w(s)
    = - \left[\partial_i \Delta^\dagger (s) \right] w(s)
    = - {\rm I}_k\otimes \sigma_i \, w(s) \, .
\label{NahmDeltaIdentity}
\end{equation}
Substituting this identity into Eq.~(\ref{NahmBtwo}) and using the
facts that the $\sigma_i$ commute with $G$ and obey
$\epsilon_{ijk} \, \sigma_j \sigma_k = 2i \sigma_i$, we obtain
\begin{equation}
    B_i^{ac} = 2 \int ds \, ds' \, w_a(s) \, G(s,s') \,
          \sigma_i \, w_c(s')
\label{NahmBFinal} \, .
\end{equation}

This must be compared with
\begin{eqnarray}
   (D_i \Phi)^{ac} \!\! &=& \!\!
     \int ds\, s \,\partial_i[w^\dagger_a(s) \, w_c(s)]
    - \int ds\,ds' \left[ s'\,  \partial_i w^\dagger_a(s) \,
           w_b(s) \, w^\dagger_b(s') \, w_c(s')  \right.  \cr
   && \qquad   \left.
   + s \, w^\dagger_a(s) \, w_b(s) \, w^\dagger_b(s') \, \partial_i w_c(s')
     \right] \cr\cr
  &=& \!\! \int ds\,ds' \left[
      \partial_i w^\dagger_a(s) \, {\cal F}(s,s') \, s' \, w_c(s')
     + s \, w^\dagger_a(s) \, {\cal F}(s,s') \, \partial_i w_c(s') \right]
      \, . \cr\cr && \qquad
\end{eqnarray}
Proceeding as before, we can rewrite this as
\begin{eqnarray}
(D_i \Phi)^{ac}\!\! &=& \!\! \int ds\,ds' \left\{
      \left[\Delta^\dagger(s) \, \partial_i w_a(s) \right]^\dagger
     G(s,s') \, \Delta^\dagger(s') \, s' \, w_c(s') \right.
      \cr\cr && \qquad  \left.
      + \left[\Delta^\dagger(s) \, s \, w_a(s) \right]^\dagger
    G(s,s') \, \Delta^\dagger(s') \, \partial_i w_c(s')  \right\} \, .
\label{NahmDPhiTwo}
\end{eqnarray}

By making use of Eq.~(\ref{NahmDeltaIdentity}) and the identity
\begin{equation}
    \Delta^\dagger(s) \, s \, w_a(s) = - w_a(s) \, ,
\end{equation}
Eq.~(\ref{NahmDPhiTwo}) can be rewritten as
\begin{equation}
   (D_i \Phi)^{ac} =  2 \int ds\,ds' \, w^\dagger_a(s)
    \, G(s,s') \, \sigma_i \, w_c(s') \, .
\end{equation}
Comparing this with Eq.~(\ref{NahmBFinal}), we see that the Bogomolny
equation is indeed satisfied.

\medskip
\noindent 2) {\bf Proof that the solutions lie in SU(2)}
\smallskip

In general, Eq.~(\ref{SU2construction}) will have $2k$ linearly
independent solutions.  However, in order for $\Phi$ and $A_j$ to be SU(2)
fields, all but two of these solutions must be eliminated as being
non-normalizable.
To do this, we must examine the behavior of
the Nahm data near the endpoints $s = \pm v/2$.

Substituting Eq.~(\ref{boundarypoles}) into the construction
equation~(\ref{SU2construction}), we see that near the endpoints the
latter can be approximated by
\begin{equation}
    0 = \left[{d \over ds}
       - {L^\pm_i \otimes \sigma_i \over s \mp v/2} \right] w \, .
\label{SU2construction_bdy}
\end{equation}
Either by explicit calculation, or by noting that the tensor product
is essentially equivalent to the addition of two angular momenta
[$L=(k-1)/2$ and $S=1/2$], one finds that $L^\pm_i \otimes \sigma_i$
has only two distinct eigenvalues: $(k-1)/2$ with degeneracy $k+1$,
and $-(k+1)/2$ with degeneracy $k-1$.

In a subspace where $L^\pm_i \otimes \sigma_i$ has eigenvalue
$\lambda$, the solutions of Eq.~(\ref{SU2construction_bdy}) behave as
$(s -v/2)^\lambda$.  Hence, a normalizable solution must lie in the
subspace with positive $\lambda$.  Requiring that $w(-v/2, {\bf r})$
be orthogonal to the subspace with eigenvalue $-(k+1)/2$ gives $k-1$
conditions, and the analogous requirement at the other boundary,
$s=v/2$, gives another $k-1$ conditions.  Since $w$ has $2k$
components in all, this leave two independent normalizable solutions,
just as we wanted.

It is at this point that the necessity for the $L^\pm_i$ to be
irreducible arises. Had either of them been reducible, the
construction equation would have had more than two normalizable
solutions.\footnote{One can show that Nahm data with reducible
$L^\pm_i$ correspond to monopole solutions for a larger group, with
the unbroken gauge group having, in general, a non-Abelian component.}

In order to be SU(2) fields, $\Phi$ and $A_i$ must not only have the
correct dimension, but must also be Hermitian and traceless.  The
Hermiticity follows immediately from Eqs.~(\ref{SU2Phi}) and
(\ref{SU2Afield}), but the tracelessness requires a bit more work.  To
show this, we first note that if $\Phi$ and $A_i$ obey the Bogomolny
Eq.~(\ref{bogomolny}), as we have shown, then their traces obey the
Abelian form of this equation,
\begin{equation}
    \partial_i(\Tr \Phi) = \epsilon_{ijk} \, \partial_j (\Tr A_k) \, .
\label{abelianBog}
\end{equation}
It immediately follows that $\Tr \Phi$ is harmonic. Further, since it
is finite at spatial infinity, $\Tr \Phi$ must be constant.
(Evaluation of the asymptotic fields, which we will do in the next
subsection, shows that this constant is actually zero.)
Equation~(\ref{abelianBog}) then implies that $\Tr B_i=0$, and that
$\Tr A_i$ is therefore a pure gradient that can be gauged away by a
U(1) gauge transformation; i.e., by an ${\bf r}$-dependent rotation
of the phases of the $w_a$.

\medskip
\noindent 3) {\bf Evaluation of the magnetic charge}
\smallskip

To verify that the fields obtained by this construction actually have
$k$ units of magnetic charge, all we need to do is to examine the
long-distance behavior of the fields.  We will follow the
approach of Ref.~\cite{Hitchin:1983ay}.  For sufficiently large $r$,
the $T_\mu(s)$ terms in Eq.~(\ref{SU2construction}) are significant
only in the pole regions.  Hence, after using an SU($k$) gauge action
of the form of Eq.~(\ref{gaugeaction}) to set $L^+_i=L^-_i=L_i$, we
can approximate $\Delta^\dagger$ by
\begin{equation}
   \tilde \Delta^\dagger  = -{d \over ds}
      +\left({1 \over s - v/2}
      + {1 \over s + v/2} \right)  L_i \otimes \sigma_i
    + r_i {\rm I}\otimes \sigma_i
\end{equation}
and try to solve the approximate construction equation
\begin{equation}
    \tilde \Delta^\dagger  w = 0 \, .
\label{approxConstruct}
\end{equation}

Because of the spherical symmetry of the asymptotic fields, we can,
without any loss of generality, work on the positive $z$-axis and take
${\bf r} = (0,0,r)$.  Now note that there is a unique vector $\eta_+$
that is both an eigenvector of $L_3$ with the maximum eigenvalue,
$(k-1)/2$, and an eigenvector of $\sigma_3$ with eigenvalue 1.  This
is also an eigenvector of $L_i \otimes \sigma_i$ with eigenvalue
$(k-1)/2$.  Similarly, there is a unique vector $\eta_-$ that is an
eigenvector of $L_3$ with eigenvalue $-(k-1)/2$, of $\sigma_3$ with
eigenvalue $-1$, and of $L_i \otimes \sigma_i$ with eigenvalue
$(k-1)/2$.   Hence, two solutions of Eq.~(\ref{approxConstruct})
are given by
\begin{equation}
    w_\pm(s, r)  = g_\pm(s,r) \eta_\pm
\end{equation}
with
\begin{equation}
     \left[- {d\over ds} + \left({k-1 \over 2}\right)
  \left({1 \over s - v/2}  + {1 \over s + v/2} \right)
     \pm r \right] w = 0 \, .
\end{equation}
This equation is solved by
\begin{equation}
     g_\pm = N\left[ \left(s-{v\over 2}\right)\left(s+{v\over 2}\right)
     \right]^{(k-1)/2}  e^{\pm rs}
\end{equation}
with the constant $N$ being fixed by the normalization condition.
These two solutions are clearly normalizable.  By the arguments we
gave above, the remaining solutions of Eq.~(\ref{approxConstruct})
must be non-normalizable, and so can be ignored for our purposes.

Since $\eta_+$ and $\eta_-$ correspond to different
eigenvalues of $L_3$ and $\sigma_3$, the functions $w_+(s)$ and $w_-(s)$
are pointwise orthogonal.  It follows that $\Phi$ is diagonal,
with eigenvalues
\begin{equation}
    \Phi_\pm({\bf r}) =  {\high \int_{-v/2}^{v/2} ds
            \, s \, g_\pm^2(s,r)  \over
         \high  \int_{-v/2}^{v/2} ds  \, g_\pm^2(s,r) } \,\, .
\end{equation}
The exponential behavior of $g_\pm(s)$ allows us to make some
simplifying approximations in the limit of large $r$.  Because
$g_+(s)$ is concentrated near $s=v/2$, there is little error in
replacing $-v/2$ by $-\infty$ in the lower limits of the integrals
for $\Phi_+$.  Writing $y = (v/2) -s$ and cancelling some common factors,
we then have
\begin{eqnarray}
    \Phi_+({\bf r})
      &=& { \high\int_0^{\infty} dy \, \left({v\over 2} -y\right)
                   y^{k-1} (v-y)^{k-1}  \, e^{-2ry}
               \over \high
      \int_0^{\infty} dy  \,  y^{k-1} (v-y)^{k-1}  \, e^{-2ry} } \cr \cr
     &=& {v \over 2} - { \high \int_0^{\infty} dy \, y^k (v-y)^{k-1}
                       \, e^{-2ry}
       \over \high
      \int_0^{\infty} dy  \,  y^{k-1} (v-y)^{k-1}  \, e^{-2ry} } \,\, .
\end{eqnarray}
To leading order in $1/r$ we can replace the factors of $(v-y)^{k-1}$
by $v^{k-1}$.  The integrals are then easily evaluated to give
\begin{equation}
    \Phi_+({\bf r}) = {v \over 2} - {k \over 2r}
              + O\left({1 \over r^2}\right) \, .
\label{phiplusapprox}
\end{equation}
An analogous argument gives
\begin{equation}
      \Phi_-({\bf r}) =  -{v \over 2} +{k \over 2r}
              + O\left({1 \over r^2}\right) \, .
\label{phiminusapprox}
\end{equation}
This is precisely the behavior expected for the Higgs field in an
SU(2) BPS solution with $k$ units of magnetic charge.  As a bonus,
we see how the eigenvalues of the asymptotic Higgs field are determined
by the location of the boundaries.

\subsection{Completeness of the construction}
\label{NahmCompleteSec}

In the previous subsection we showed that a solution of the Nahm
equation leads, via the ADHMN construction, to spacetime fields that
satisfy the Bogomolny equation.  We now prove the converse; i.e., that
given a solution of the Bogomolny equation, one can obtain a set of
matrices $T_\mu$ that obey the Nahm equation \cite{Corrigan:1983sv}.

Thus, let us assume that $A_j({\bf r})$ and $\Phi({\bf r})$ are a
magnetic charge $k$ solution of the Bogomolny equation, with $v$ being the
vacuum expectation value of $\Phi$.  We define
\begin{eqnarray}
    {\cal D}&=& i\left[\bsigma \cdot {\bf D} - \Phi + z \right]  \cr
    {\cal D}^\dagger &=& i\left[\bsigma \cdot {\bf D} + \Phi -z\right]
\end{eqnarray}
where $D_j$ is the gauge covariant derivative with respect to the
$A_j({\bf r})$ and $z$ is a real number.  Because $A_j$ and $\Phi$ are
self-dual,
\begin{equation}
    {\cal D}^\dagger{\cal D} = - {\bf D}^2 + (\Phi -z)^2 \, .
\end{equation}
It follows from this that ${\cal D}$ has no normalizable zero modes.
However, one can show by an index theorem \cite{Callias:1977kg} that
\begin{equation}
     {\cal D}^\dagger \psi = 0
\label{definitionOfPsi}
\end{equation}
has precisely $k$ linearly independent normalizable solutions if $-v/2
< z < v/2$, and none otherwise.  It is convenient to assemble these
solutions into a $2 \times k$ matrix\footnote{The construction of $\psi$
assumes a particular basis for the solutions of Eq.~(\ref{definitionOfPsi}).
The freedom to make a $z$-dependent change of basis (i.e., to multiply
$\psi$ on the right by a unitary matrix) gives rise to the SU($k$) gauge
action described in Sec.~\ref{gaugeactionsection}.}
normalized so that
\begin{equation}
     \int d^3x \,\psi^\dagger({\bf x},z) \,\psi({\bf x},z)
                 = {\rm I}_k \, .
\end{equation}
Note that Eq.~(\ref{definitionOfPsi}) implies that
\begin{equation}
    \psi^\dagger \overleftarrow{\cal D} = 0 \, .
\end{equation}

We also define Green's function $G_z({\bf x},
{\bf y})$, $S_z({\bf x}, {\bf y})$, and $\bar S_z({\bf x}, {\bf y})$
by
\begin{eqnarray}
    {\cal D}^\dagger{\cal D} \, G_z({\bf x}, {\bf y})
         &=& \delta^{(3)}({\bf x}, {\bf y})  \cr \cr
    {\cal D}^\dagger S_z({\bf x}, {\bf y})
         &=& \delta^{(3)}({\bf x}, {\bf y})  \cr \cr
    {\cal D} \bar S_z({\bf x}, {\bf y})
     &=& \delta^{(3)}({\bf x}, {\bf y})
     - \psi({\bf x},z) \,\psi^\dagger({\bf x},z) \, .
\end{eqnarray}
These Green's functions are related by
\begin{eqnarray}
     {\cal D} G_z({\bf x}, {\bf y}) &=& S_z({\bf x}, {\bf y})  \cr\cr
     G_z({\bf x}, {\bf y}) \overleftarrow{\cal D}
         &=& \bar S_z({\bf x}, {\bf y}) \, .
\end{eqnarray}

Next, we need an expression for $d\psi/dz$.  Differentiating
Eq.~(\ref{definitionOfPsi}) with respect to $z$ gives
\begin{equation}
     {\cal D}^\dagger \, {d \psi \over dz}=  i \psi \, ,
\end{equation}
which implies that
\begin{equation}
     {d \psi({\bf x}) \over dz}
         = i \int d^3y\, S_z({\bf x}, {\bf y}) \psi({\bf y})
         + C \psi({\bf x})
\end{equation}
for some constant $C$.  (For the sake of clarity we have suppressed
the $z$-dependence of $\psi$, here and below.)
To determine $C$, we multiply this equation on the
left by $\psi^\dagger({\bf x})$ and integrate over $\bf x$.  After
noting that the resulting double integral involving $S_z({\bf x}, {\bf
y})$ vanishes, we find that
\begin{equation}
     {d \psi({\bf x}) \over dz}
         = i \int d^3y\, S_z({\bf x}, {\bf y}) \psi({\bf y})
         + \psi({\bf x}) \int d^3 y \,\psi^\dagger({\bf y })
              {d \psi({\bf y}) \over dz} \, .
\label{dpsidz}
\end{equation}

Having completed these preliminaries, we assert that the Nahm data
are given by
\begin{eqnarray}
    T_j(z) &=& -\int d^3x \, x_j \,
              \psi^\dagger({\bf x}) \psi({\bf x}) \cr
    T_0(z) &=& i \int d^3x \, \psi^\dagger({\bf x})
              {d \psi({\bf y}) \over dz} \, .
\end{eqnarray}
These matrices are manifestly Hermitian, and are defined on the
interval $v/2 < z < v/2$.  To verify that they satisfy the Nahm
equation, we first calculate
\begin{eqnarray}
     T_i T_j &=& \int d^3x d^3y \, x_i y_j \,
           \psi^\dagger({\bf x}) \psi({\bf x})
           \psi^\dagger({\bf y}) \psi({\bf y})    \cr \cr
   &=& \int d^3x \, x_i x_j \,\psi^\dagger({\bf x}) \psi({\bf x})
      +\int d^3x d^3y \, x_i y_j \, \psi^\dagger({\bf x})
        {\cal D} \bar S_z({\bf x}, {\bf y}) \psi({\bf y})   \cr\cr
   &=& \int d^3x \, x_i x_j \,\psi^\dagger({\bf x}) \psi({\bf x})
      +\int d^3x d^3y \, x_i y_j \, \psi^\dagger({\bf x})
        {\cal D} G_z({\bf x}, {\bf y}) \overleftarrow{\cal D}\,\psi({\bf y})
            \cr\cr
   &=& \int d^3x \, x_i x_j \,\psi^\dagger({\bf x}) \psi({\bf x})
      -\int d^3x d^3y \,  \psi^\dagger({\bf x}) \sigma_i \sigma_j
        G_z({\bf x}, {\bf y}) \psi({\bf y})
\end{eqnarray}
where the last equality is obtained by integrating by parts twice.
It follows that
\begin{equation}
    [T_i , T_j] = -2i \, \epsilon_{ijk} \int d^3x d^3y \,
      \psi^\dagger({\bf x}) \sigma_k
              G_z({\bf x}, {\bf y}) \psi({\bf y}) \, .
\label{TTcommutator}
\end{equation}

Next,
\begin{equation}
    {d T_k \over dz} = -\int d^3x \, x_k \,{d \psi^\dagger \over dz} \, \psi
       -  \int d^3x \, x_k \, \psi^\dagger \, {d \psi \over dz} \, .
\end{equation}
Using Eq.~(\ref{dpsidz}) and integrating by parts in the last
step, we find that
\begin{eqnarray}
     \int d^3x \, x_k \, \psi^\dagger \, {d \psi \over dz} &=&
    i\int d^3x d^3y \, x_k\,
     \psi^\dagger({\bf x})  S_z({\bf x}, {\bf y}) \psi({\bf y})
        +i T_k T_0   \cr\cr   &=&
    i\int d^3x d^3y \, x_k\,
     \psi^\dagger({\bf x}) {\cal D} G_z({\bf x}, {\bf y}) \psi({\bf y})
        +i T_k T_0  \cr\cr  &=&
    i\int d^3x d^3y \,
     \psi^\dagger({\bf x}) \sigma_k G_z({\bf x}, {\bf y}) \psi({\bf y})
        +i T_k T_0 \, .
\end{eqnarray}
This equation and its adjoint then give
\begin{equation}
    {d T_k \over dz} = -2 \int d^3x d^3y \,
      \psi^\dagger({\bf x}) \sigma_k G_z({\bf x}, {\bf y}) \psi({\bf y})
        +i [T_k,T_0] \, .
\end{equation}
Together with Eq.~(\ref{TTcommutator}), this verifies that the $T_\mu$
satisfy the Nahm equation.

\subsection{Larger gauge groups}
\label{NahmBiggroups}
\subsubsection{SU($N$)}

\begin{figure}[t]
\begin{center}
\scalebox{1.0}[1.0]{\includegraphics{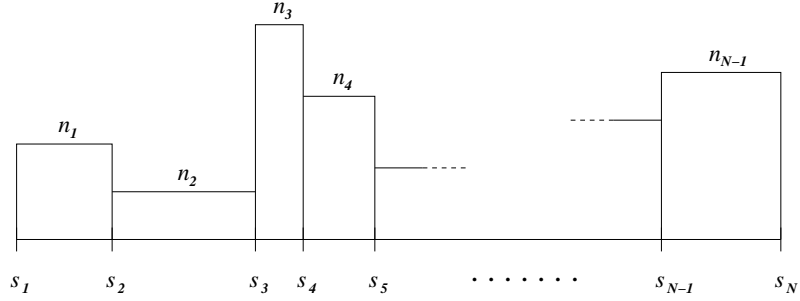}}
\par
\vskip-2.0cm{}
\end{center}
\begin{quote}
\caption{\small An example of a skyline diagram that illustrates the
  magnetic charges and Higgs expectation value for an SU($N$)
  solution.
 \label{skyline}}
\end{quote}
\end{figure}

The next step is to generalize the ADHMN construction for SU(2) to the
case of an arbitrary classical group \cite{Hurtubise:1989qy}. (The
construction does not readily extend to the exceptional groups.) The
natural starting point is SU($N$).  We seek a construction for
solutions with asymptotic Higgs field
\begin{equation}
    \Phi = {\rm diag}\left(s_1, s_2, \dots, s_N \right)
\label{suNasymphi}
\end{equation}
[with the $s_p$ ordered as in Eq.~(\ref{SUNphielements})] and
asymptotic magnetic field\footnote{The factor of $1/e$ is absent from
the magnetic field because we are setting $e=1$ throughout this
section.}
\begin{equation}
   B_k = \, \frac{{\hat r}_ k}{2r^2 } \,
     {\rm diag}\left(-n_1, n_1-n_2, \dots,
         n_{N-2} -n_{N-1}, n_{N-1} \right) \, .
\label{suNasymBfield}
\end{equation}
We will often refer to such SU($N$) solutions as ($n_1$, $n_2$, \dots,
$n_{N-1}$) solutions.  Their asymptotic properties can be captured
graphically in a ``skyline'' diagram \cite{Houghton:1997ei}, such as that shown in
Fig.~\ref{skyline}.

Clearly, the SU(2) construction somehow must be modified so that there
are $N$, rather than just two, normalizable $w_a$.  In addition, the
construction must somehow encode $N$ eigenvalues
for the asymptotic Higgs field and for the magnetic charge obtained
from the asymptotic magnetic field.

Clues as to how to proceed can be found in the last part of
Sec.~\ref{NahmSelfSec}.  We saw there that the positions of the
boundaries corresponded to the two eigenvalues of the asymptotic Higgs
field, and that the dimensions of the SU(2) representations in the
pole terms gave the eigenvalues of the magnetic charge, with a plus or
minus sign depending on whether the pole term was to the right or left
of the boundary.

We start by dividing the interval $s_1 \le s \le s_N$ into $N-1$
subintervals separated by the $s_p$.  On the $p$th subinterval, $s_p
\le s \le s_{p+1}$, the Nahm data are $n_p \times n_p$ matrices
$T^{(p)}_\mu$.  Within a given subinterval, these obey the same Nahm
equation as the SU(2) Nahm data, Eq.~(\ref{fullNahmequation}).  The
behavior at the outer boundaries, $s=s_1$ and $s=s_N$, is just as for
SU(2).  The behavior at the boundaries between the subintervals
depends on the size of the adjacent Nahm data.  Let us first suppose
that $n_p >n_{p+1}$.  The Nahm data just to the left of the boundary
at $s=s_{p+1}$ (i.e., corresponding to $s<s_{p+1}$) are $n_p \times
n_p$ matrices that can be divided into submatrices
\begin{equation}
     T^{(p)}_\mu= \left(\begin{array}{ccc}
         P^{(p)}_\mu & Q^{(p)}_\mu \\ \\
         R^{(p)}_\mu & S^{(p)}_\mu
          \end{array} \right)
\end{equation}
with $S^{(p)}_\mu$ being $n_{p+1} \times n_{p+1}$; i.e., the same size
as $T^{(p+1)}_\mu$.  There are no restrictions on $T^{(p)}_0$. For the
$T^{(p)}_j$, we require that the lower right
corners be continuous across the boundary, and that the upper left
corners have poles with residues forming an irreducible
$(n_p-n_{p+1})$-dimensional representation of SU(2).  This implies
that the off-diagonal blocks must vanish at the boundary, and that
\begin{equation}
     T^{(p)}_j = \left(\begin{array}{ccc}
          -\,\high{L_j^{(p)}\over s-s_{p+1}}+O(1)  &\quad &
      O\left[(s-s_{p+1})^{(m_p-1)/2}\right]
         \\ \\ \high  O\left[(s-s_{p+1})^{(m_p-1)/2}\right] &&
      T_j^{(p+1)}+O(s-s_{p+1}) \end{array} \right) \,\, .
\end{equation}
The prescription is the same if $n_p < n_{p+1}$, except that the
$T^{(p+1)}_j$ are divided into blocks and the pole lies to the right
of the boundary.  The case $n_p = n_{p+1}$ is more complex, and will
be addressed shortly.

The modifications to the construction equation are similar.  Within a
given interval, the construction equation is as before, with
$w_a^{(p)}$ having $2n_p$ components on the $p$th subinterval.  If
$n_p >n_{p+1}$, then the lower $2n_{p+1}$ components of $w_a^{(p)}$
match continuously onto the components of $w_a^{(p+1)}$, while the
upper $2(n_p - n_{p+1})$ components must be such that $w_a$ is
normalizable.  The case $n_p < n_{p+1}$ is similar but, again, the
case $n_p = n_{p+1}$ is more complex.  As long as $n_{p+1} \ne n_p$
for every $p$, the normalization conditions are the obvious
generalization of Eq.~(\ref{SU2Normalization}), with the single
integral being replaced by a sum of integrals over the various
subintervals.

Likewise, the prescription for obtaining the spacetime fields is the
obvious generalization of the SU(2) case, with Eqs.~(\ref{SU2Phi}) and
(\ref{SU2Afield}) now involving a sum of integrals.  Again, this
prescription must be modified, as we describe below, if any two
consecutive $n_p$ are equal.

It is easy to see why this construction works.  To count the $w_a$,
let us divide the $s_p$ (including $s_1$ and $s_N$) into ``rising''
and ``falling'' boundaries according to whether $n_{p+1}$ is greater
or less than $n_p$; at each such boundary we define $\Delta_p =
n_{p} - n_{p-1}$, with the outer boundaries giving
$\Delta_1=n_1$ and $\Delta_N=-n_{N-1}$.  We can
schematically imagine solving the construction equation by starting
with initial data on the left and then integrating to larger values of
$s$.  From this point of view, each rising boundary gives $2 \Delta_p$
initial degrees of freedom.  However, by an obvious generalization of
the arguments given below Eq.~(\ref{SU2construction_bdy}),
normalizability imposes $|\Delta_p| -1 $ conditions at each boundary.
Subtracting these off, each rising boundary gives $\Delta_p +1$
degrees of freedom that must be adjusted to satisfy the constraints
arising at the falling boundaries.  Noting that each of the latter
gives $-\Delta_p -1$ conditions, and that the sum of the $\Delta_p$
vanishes, we find that there are
\begin{equation}
     \sum_{\Delta_p>0} (\Delta_p +1)
         - \sum_{\Delta_p<0} (-\Delta_p -1)
          = \sum_{p=1}^N (\Delta_p +1)
         = N
\end{equation}
independent normalizable $w_a$, as required for an SU($N$) solution.
As in the SU(2) case, the $w_a$ can be chosen so that for large $r$
one $w_a$ is concentrated near each of the $s_p$, giving a diagonal
element of $\Phi$ of the form
\begin{equation}
          \Phi_{pp} = s_p + {n_p-n_{p-1} \over 2r}
               + O\left({1 \over r^2}\right)  \, .
\end{equation}
This is just what is required to satisfy Eqs.~(\ref{suNasymphi}) and
(\ref{suNasymBfield}).

This counting of solutions to the construction equation goes astray if
any of the $\Delta_p$ vanish, since there clearly can't be $\Delta_p -
1= -1$ conditions at the corresponding boundary.  To remedy this, a
new degree of freedom must somehow be introduced.  This is done as
follows.  At every boundary where $\Delta_p$ vanishes we introduce new
``jumping data'' in the form of a $(2n_p)$-component vector
$a_{r\alpha}^{(p)}$, with $r = 1,2,\dots n_p$ and the spinor index
$\alpha=1,2$.  (The subscripts on $a^{(p)}$ correspond exactly to
those on the $w_b$, so it can be thought of as an $n_p$-vector of
two-component spinors.)  Instead of requiring that the Nahm data be
continuous across the boundary, we require that the discontinuity

\begin{equation}
     (\delta T_j)^{(p)} \equiv  T_j^{(p-1)}(s_{p}) -T_j^{(p)}(s_{p})
\end{equation}
be an $n_p \times n_p$ matrix of the form
\begin{equation}
     (\delta T_j)^{(p)}_{rs} = {1 \over 2}\,  a_{r\alpha}^{(p)*}
       \,  (\sigma_j)_{\alpha\beta}\,  a_{s\beta}^{(p)} \, .
\label{Nahmdiscontinuity}
\end{equation}

Correspondingly, the solutions of the construction equation
are allowed to be discontinuous, with the discontinuity
required to be proportional to $a^{(p)}$:
\begin{equation}
     (\delta w_b)^{(p)} \equiv w_b^{(p-1)}(s_{p}, {\bf r})
       - w_b^{(p)}(s_{p}, {\bf r})
       = S_b^{(p)}({\bf r}) a^{(p)} \, .
\end{equation}
The freedom to adjust $S_b^{(p)}({\bf r})$ provides the additional
degree of freedom that was needed.   The orthonormality conditions
on the solutions of the construction equation now take the
form
\begin{equation}
    \delta_{ab} = \sum_{p=1}^{N-1}\int_{s_p}^{s_{p+1}}
         w^{(p)\dagger}_a(s,{\bf r}) w^{(p)}_b(s,{\bf r}) ds
        + \sum_p S^{(p)*}_a({\bf r}) S^{(p)}_b({\bf r}) \, .
\label{jumpingOrtho}
\end{equation}
The second sum, of
course, only receives contributions from the boundaries at
which $\Delta_p=0$.

There are similar modifications in the prescriptions for the
spacetime fields, which are now given by
\begin{equation}
    \Phi^{ab} = \sum_{p=1}^{N-1}\int_{s_p}^{s_{p+1}}
        ds \, s\,  w^{(p)\dagger}_a(s,{\bf r}) w^{(p)}_b(s,{\bf r})
         + \sum_p s_p \,S^{(p)*}_a({\bf r}) S^{(p)}_b({\bf r})
\label{jumpingPhi}
\end{equation}
and
\begin{equation}
    A_j^{ab} = -i\sum_{p=1}^{N-1}\int_{s_p}^{s_{p+1}}
       ds \, w^{(p)\dagger}_a(s,{\bf r})\partial_j w^{(p)}_b(s,{\bf r})
                -i\sum_p S^{(p)*}_a \partial_j S^{(p)}_b  \, .
\label{jumpingA}
\end{equation}

At first glance, the introduction of the jumping data and the
modifications to the ADHMN construction seem quite strange.  We can
gain some physical insight \cite{Chen:2002vb} into what is going on by
considering the case of two adjacent subintervals, with $n+1$ and $n$
monopoles, respectively, and then removing one of the $n+1$ monopoles
by a distance $R \gg \ell$, where $\ell$ is the spatial separation of
the remaining $n$ monopoles.  Intuitively, one would expect that as
$R$ tended toward infinity, the last monopole should in some sense
decouple from the rest.  This is reflected as follows in the Nahm
data.  Outside a narrow region of width $\Delta s_R \sim R^{-1} \ln
(R/\ell)$ near the boundary, the Nahm matrices on the $(n+1)$-monopole
side are essentially block diagonal, with nontrivial $1 \times 1$ and
$n \times n$ blocks and exponentially small off-diagonal $1 \times n$
and $n \times 1$ blocks.  The Nahm equation then naturally separates
into two independent parts, as does the construction equation.

Inside the narrow boundary region, however, the off-diagonal blocks
are nontrivial, and the $n \times n$ parts of the Nahm data (which
must be continuous across the subinterval boundary) are rapidly
varying with a net change $\Delta T_j$.  As $R \rightarrow \infty$,
the width $\Delta s_R \rightarrow 0$, but the $\Delta T_j$ have a
finite nonzero limit.  This limiting value is precisely of the form of
Eq.~(\ref{Nahmdiscontinuity}), with the $a_{r\alpha}$ being naturally
obtained from the off-diagonal blocks of the $T_j$ corresponding to
the two spatial directions orthogonal to that along which the
$(n+1)$th monopole was removed.

\subsubsection{SO($N$) and Sp($N$)}

The strategy for obtaining multimonopole solutions for orthogonal and
symplectic gauge groups is based on the fact that SO($N$), for all
$N$, and Sp($N$), for even $N$, are subgroups of SU($N$).
Specifically, if $K$ is an $N\times N$ matrix with $KK^* = I$, then
SO($N$) is the subgroup of ${\rm SU}(N)$ whose elements obey $G^t K G = K$.
Its generators satisfy
\begin{equation}
    T^t K + K T =0 \, .
\end{equation}
If $J$ is an $N\times N$ matrix with $JJ^* = -I$, then Sp($N$) is the
subgroup of ${\rm SU}(N)$ whose elements obey $G^t J G = J$.  Its
generators obey
\begin{equation}
    T^t J + J T =0  \, .
\end{equation}

Hence, the solutions for these groups are also SU($N$) solutions, but
must satisfy certain additional restrictions.  More specifically, note
that the rank of Sp($N$) is $N/2$, while that of SO($N$) is $N/2$ for
even $N$ and $(N-1)/2$ for odd $N$.  By comparison, the rank of
SU($N$) is $N-1$.  This reduction in rank is accompanied by a
reduction in the number of species of fundamental monopoles.  This is
accomplished by the identification of certain pairs of SU($N$)
monopoles, which is manifested by a type of reflection symmetry of the
Nahm data.

Thus, we first require that the eigenvalues of the SU($N$) Higgs field
obey
\begin{equation}
    s_p = s_{N+1 -p}
\end{equation}
and that the topological charges satisfy
\begin{equation}
    n_p = n_{N -p}
\end{equation}
These conditions imply that the skyline diagram will be symmetric
under the reflection $s \rightarrow -s$.
Next, the Nahm data must satisfy
\begin{equation}
     T_j(-s) = C(s) T^t(s) C(s)^{-1}
\end{equation}
where the matrix $C$ obeys
\begin{equation}
     C(-s) = \cases{ C(s)^t \, &\qquad Sp($N$)  \cr \cr
                    -C(s)^t \, &\qquad SO($N$)  }
\end{equation}
and it is understood that dimension of $C(s)$, like that of the $T_j(s)$,
varies in a stepwise fashion according to the value of $s$.

The spacetime solutions corresponding to these Nahm data have two
possible interpretations.  They can be viewed as SU($N$) solutions
with topological charges $n_a$ ($a=1,2,\dots, N-1$) that
have the special property that the locations of certain pairs of
fundamental monopoles happen to coincide.
Alternatively, they can be viewed as Sp($N$) or SO($N$) solutions
with topological charges $\tilde n_a$.  The two sets of topological
charges are related (in an obvious notation) as follows:
\begin{eqnarray}
{\rm Sp}(N):  && n_a = \tilde n_a\, , \qquad a=1,2,\dots, N/2 \cr\cr
{\rm SO}(2k): && n_a = \tilde n_a\, , \qquad a=1,2,\dots, k-2\cr
          && n_{k-1} = \tilde n_+ + \tilde n_- \cr
          && n_k = 2 \tilde n_+ \cr\cr
{\rm SO}(2k+1): && n_a = \tilde n_a\, , \qquad a=1,2,\dots, k-1\cr
                && n_k = 2 \tilde n_k
\end{eqnarray}

\section{Applications of the ADHMN construction}
\label{nahmexamplesection}

In this section we will illustrate the ADHMN construction by applying
it to several examples.  We begin with the SU(2) $k=1$ and $k=2$
solutions, the only SU(2) cases for which a general closed form
solution of the Nahm equation is known.  Once one knows the Nahm data
for these cases, it turns out to be a fairly straightforward matter to
obtain the Nahm data for two cases with larger gauge groups: the (2,
1) solutions for SU(3) broken to U(1)$\times$U(1), and the (1, 1,
\dots, 1) solution for SU($N$) broken to U(1)$^{N-1}$.  Throughout, we
set $T_0 = 0$.

\subsection{The unit SU(2) monopole}

It is natural to begin by showing how the unit BPS solution is
recovered in the ADHMN construction \cite{Nahm:1979yw}.  Because
$k=1$, the $T_i$ are simply numerical functions of $s$.  The
commutator terms in the Nahm equation then vanish, implying that the
$T_i$ must be constants.  From the manner in which the $T_i$ enter the
construction equation, it is evident that these constants are simply
the spatial coordinates of the center of the monopole.  By
translational invariance, we can set $T_i =0$, so that the monopole is
centered at the origin.

The construction equation~(\ref{usualSU2construction}) then reduces to
\begin{equation}
     {dw\over ds}  =  {\bf r}\cdot \bsigma \, .
\end{equation}
A pair of orthonormal solutions are
\begin{equation}
     w_a(s,{\bf r}) = N({\bf r}) e^{s\,{\bf r}\cdot \sbsigma} \eta_a
\end{equation}
where $N$ is a normalization factor and the $\eta_a$ are
orthonormal two-component constant vectors.

Making use of the integrals
\begin{equation}
    \int_{-v/2}^{v/2} e^{2s\, {\bf r}\cdot \sbsigma } \, ds
         = {\sinh vr \over r} {\rm I}_2
\end{equation}
and
\begin{equation}
    \int_{-v/2}^{v/2} s\, e^{2s\, {\bf r}\cdot \sbsigma} \, ds
         = {{\bf r}\cdot \bsigma \over r^3}
            [ vr \cosh vr - \sinh vr] \, ,
\end{equation}
we find that
\begin{equation}
    N = \sqrt{r \over \sinh vr}
\end{equation}
and
\begin{equation}
   \Phi_{ab} = {1\over 2} \left(v \coth vr - {1 \over r} \right)
           \eta_a^\dagger\, \hat {\bf r}\cdot \bsigma \,\eta_b \, .
\end{equation}
A slightly lengthier calculation gives
\begin{equation}
    (A_i)_{ab} = -i \eta_a^\dagger\, \partial_i \eta_b
      - i \epsilon_{ijk}\, \hat r_j\, \eta_a^\dagger \sigma_k \eta_b
          \left( -{1\over 2r} + {v \over 2\sinh vr} \right) \, .
\end{equation}

The hedgehog gauge solution of Eqs.~(\ref{hedghogansatz})
and (\ref{monoPSsoln}) is obtained by choosing $\eta_1 = (1,0)^t$ and
$\eta_2 = (0,1)^t$.  Alternatively, we can take the $\eta_a$ to be
eigenvectors of $\hat {\bf r}\cdot \bsigma$, with
\begin{equation}
    \eta_1 = \sqrt{ r-z \over 2r} \left(
       \matrix{\high {x-iy \over r-z} \cr\cr  1} \right) \, ,
   \qquad \qquad
     \eta_2 =   \sqrt{ r-z \over 2r} \left(
       \matrix{1 \cr\cr \high {x+iy \over r-z}} \right) \, \, .
\end{equation}
This gives the string gauge form of the solution in which the
Higgs field is everywhere proportional to $\sigma_3$.

\subsection{SU(2) two-monopole solutions}
\label{NahmEx2Sec}

The Nahm construction for the $k=2$ case \cite{Brown:1982gz} is much
less trivial.\footnote{For an alternative approach to the general
$k=2$ solution, see Ref.~\cite{Ward:1981ju}.}  From our counting of zero
modes, we know that these solutions should form an eight-dimensional
moduli space.  One of the eight parameters corresponds to an overall
U(1) phase and, as in the $k=1$ case, does not enter the Nahm data.
Of the remaining parameters, three correspond to spatial translations
of the solution and three more to spatial rotations, leaving only a
single nontrivial parameter.

The $T_i$ are Hermitian $2 \times 2$ matrices, and so can be expanded
as
\begin{equation}
    T_i(s) = {1 \over 2} {\bf C}_i(s)
              \cdot \btau + R_i(s) {\rm I}_2 \, .
\end{equation}
Substituting this expression into the Nahm equation shows that the
$R_i$ are constants, while the ${\bf C}_i$ obey
\begin{equation}
    {d {\bf C}_i \over ds}  = {1 \over 2} \epsilon_{ijk} {\bf C}_j
          \times    {\bf C}_k \, .
\label{NahmEqforC}
\end{equation}
It follows that the elements of the real symmetric matrix
\begin{equation}
   M_{ij} = {\bf C}_i \cdot {\bf C}_j
       - {1 \over 3} \delta_{ij} {\bf C}_k \cdot {\bf C}_k
\end{equation}
are independent of $s$.  If $A$ is the orthogonal matrix that
diagonalizes $M$, the three vectors
\begin{equation}
     {\bf B}_i(s) = A_{ji} {\bf C}_j(s)
\end{equation}
are mutually orthogonal and so can be written as
\begin{equation}
    {\bf B}_i(s) = g_i(s) \hat {\bf e}_i(s)
\end{equation}
where the $\hat {\bf e}_i$ are a triplet of orthogonal unit vectors
obeying $\hat {\bf e}_1 \times \hat {\bf e}_2 = \hat {\bf e}_3$, and
no sum over $i$ is implied.  Substituting these expressions into
Eq.~(\ref{NahmEqforC}), we find that the $\hat {\bf e}_i$ are
independent of $s$, while the $g_i(s)$ obey the Euler-Poinsot
equations
\begin{eqnarray}
    {dg_1 \over ds} &=& g_2 \, g_3  \cr \cr
    {dg_2 \over ds} &=& g_3 \, g_1  \cr  \cr
    {dg_3 \over ds} &=& g_1 \, g_2 \,\, .
\end{eqnarray}
These imply that the differences $\Delta_{ij}= g_i^2 - g_j^2$ are
constants.  It follows that
once two of the $g_i(s)$ are known, the third is determined.  Hence,
there are only two independent constants of integration, which must
be chosen so that the $T_i$ have poles at $s = \pm v/2$.  Adopting the
convention that $|g_1(s)|\le |g_2(s)|\le |g_3(s)|$, we can write
\begin{equation}
    g_i(s) = f_i(s + v/2; \kappa, D)
\end{equation}
where the $f_i$ are the Euler top functions
\begin{eqnarray}
    f_1(u; \kappa, D) &=&  -D { \cn\kappa (Du) \over \sn\kappa (Du)
      } \cr \cr
    f_2(u; \kappa, D) &=&  -D { \dn \kappa(Du) \over \sn \kappa(Du)
      } \cr \cr
    f_3(u; \kappa, D) &=&  -  { D \over \sn\kappa (Du) } \, \,.
\label{topfunctions}
\end{eqnarray}

The $f_i$ have poles at $u=0$ and at $Du= 2 K(\kappa)$, where
$K(\kappa)$ is the complete elliptic integral of the first kind.  The
$u=0$ pole gives the required behavior at $s = -v/2$.  Requiring that
the second pole be at $s=v/2$ gives the relation
\begin{equation}
     2 K(\kappa) = Dv
\label{EllipticKCondition}
\end{equation}
between $\kappa$ and $D$.  Because $K(\kappa)$
increases monotonically over the allowed range $0 \le \kappa < 1$, with
$K(0) =  \pi/2 $ and $K(1) = \infty$, we see that $D$ must lie in in
the range
\begin{equation}
       {\pi \over v} \le D < \infty \, .
\end{equation}

Putting all of this together gives
\begin{equation}
    T_i(s) = {1 \over 2}
           \sum_{j} A_{ij} f_j(s + v/2; \kappa, D) \tau'_j
           + R_i {\rm I}_2
\label{generalK2solution}
\end{equation}
where $\kappa$ and $D$ are related by Eq.~(\ref{EllipticKCondition})
and $\tau'_j \equiv (\hat e_j)_k \tau_k$.  Because $E_{jk} = (\hat
e_j)_k$ is an orthogonal matrix, there exists an SU(2) matrix $g$ such
that an $s$-independent gauge action of the form of
Eq.~(\ref{gaugeaction}) will rotate the $\tau'_k$ back into the
standard Pauli matrices.  Hence, the most general $k=2$ Nahm data can
be written as
\begin{equation}
    T_i(s) = {1 \over 2}  \sum_{j} A_{ij} f_j(s+v/2;\kappa , D) \tau_j
           + R_i {\rm I}_2 \, .
\end{equation}

The next step in the construction would be to solve the construction
equation obtained by inserting these $T_i(s)$ into
Eq.~(\ref{usualSU2construction}).  Unfortunately, it has not proven
possible to obtain a closed-form analytic solution of this equation.
(For some progress in this direction, see
Refs.~\cite{Panagopoulos:1983yx,Bruckmann:2004nu,Bruckmann:2004ib}.)

Nevertheless, we can obtain a physical understanding of the parameters
that enter these Nahm data.  As expected, there are seven of these,
with the components of $\bf R$ giving the center-of-mass position and
the three Euler angles in the orthogonal matrix $A$ determining the
spatial orientation of the solution.  All three of these angles enter
the solution (thus verifying the lack of axial symmetry) as long as
the three $f_j$ are all different, as is true for all\footnote{The
exceptional case occurs when $D=\pi/v$, its minimum allowed value.
Recalling that $K(0)= \pi/2$, we see that this implies that $\kappa=0$.
The Jacobi elliptic functions simplify when $\kappa=0$, giving $f_1(u, 0,
\pi/v) = -\cot(\pi u/v)$ and $f_2(u, 0, \pi/v) = f_3(u, 0, \pi/v) = -
csc(\pi u/v)$.} $D> \pi/v$.

The physical meaning of the last parameter, $D$, is most easily seen
by studying the limit $Dv \gg 1$, where the construction equation
simplifies somewhat.  For simplicity, let us set ${\bf R}=0$ and
$A_{ij} =\delta_{ij}$.

Equation~~(\ref{EllipticKCondition}) implies that $\kappa$ must approach unity
for large $D$; specifically,
\begin{equation}
      \kappa \approx \left( 1 - 16 e^{-Dv} \right) \, .
\end{equation}
For $\kappa$ close to unity, and $s$ not too close to 0 or $2K(\kappa)$,
\begin{eqnarray}
     \cn \kappa(s) &\approx& \dn \kappa(s) \approx \sech s  \cr\cr
     \sn \kappa (s) &\approx& \tanh s \, .
\label{constantF}
\end{eqnarray}
Hence, for $v/2 - |s| \gtrsim D^{-1}$ (i.e., away from the poles of the
$T_i$)
\begin{eqnarray}
   T_1 &\approx & T_2 \approx 0 \cr \cr
   T_3 &\approx & {D \over 2} \, \tau_3 \, .
\end{eqnarray}
In the pole regions, we have
\begin{equation}
     T_i \approx {1 \over s +v/2 } \, {\tau_i \over 2}
\label{leftF}
\end{equation}
near $s=-v/2$, and
\begin{equation}
     T_i \approx  - {1 \over s -v/2} \, {\tau_1 \tau_i \tau_1 \over 2}
\label{rightF}
\end{equation}
near $s=v/2$.

In the interval $-v/2 + D^{-1} < s < v/2 - D^{-1}$, where
Eq.~(\ref{constantF}) applies, the construction equation can be
approximated by
\begin{equation}
    {dw \over ds} = {\cal M} w
\label{chargetwoconstNahm}
\end{equation}
with $\cal M$ being the block diagonal matrix
\begin{equation}
    {\cal M} = \left( \matrix{  { \bf r}_+\cdot \bsigma & 0 \cr
                             0 & {\bf r}_-\cdot \bsigma  } \right)
\end{equation}
and ${\bf r}_\pm = {\bf r} - {\bf x}_\pm$ with ${\bf x}_\pm
=(0, 0, \pm D/2)$.  Hence, in this region four independent
solutions of the construction equation are
\begin{eqnarray}
     v_1(s) &=& e^{-r_+(s+v/2)} \,\eta_1   \cr\cr
     v_2(s) &=& e^{-r_+(v/2-s)} \,\eta_2   \cr\cr
     v_3(s) &=& e^{-r_-(s+v/2)} \,\eta_3   \cr\cr
     v_4(s) &=& e^{-r_-(v/2-s)} \,\eta_4
\label{approxConstSol}
\end{eqnarray}
where $\eta_1$ and $\eta_2$ are eigenvectors of $\hat {\bf r}_+\cdot
\bsigma$ with eigenvalues 1 and $-1$, and $\eta_3$ and $\eta_4$ are
eigenvectors of $\hat{\bf r}_-\cdot \bsigma$ with eigenvalues 1 and
$-1$.  Of these solutions, $v_1$ and $v_3$ are of order unity at the
left end of the interval and then decrease monotonically with $s$,
while $v_2$ and $v_4$ are of order unity at the right end and 
monotonically decreasing as one moves back toward the lower limit of
$s$.

These solutions all develop singularities if they are integrated
all the way out to the poles of the Nahm data at $s = \pm v/2$.
However, we know that there must be two linearly independent
combinations of these solutions that remain finite even in the pole
region.  These can be chosen to be of the form
\begin{eqnarray}
     w_1(s) &=& N_1\left[ v_1(s) + b_3 \, v_3(s) + b_4 \, v_4(s)
            \right]  \cr
     w_2(s) &=& N_2\left[ v_2(s) + c_3 \, v_3(s) + c_4 \, v_4(s)
            \right]
\end{eqnarray}
with appropriately chosen constants $b_j$ and $c_j$.

Now consider a point in space that is much closer to ${\bf x}_+$ than
to ${\bf x}_-$, so that $r_+ \ll r_-$.  Here the exponential falloffs
of $v_3$ and $v_4$ are much faster than those of $v_1$ and $v_2$.  As
a result, over most of the central interval (which is itself most of
the total range of $s$) $w_1 \approx v_1$ and $w_2 \approx v_2$.  In
fact, both the normalization integrals and the integrals that give
$\Phi$ and $A_i$ are, to first approximation, the same as they would
be if $v_1$ and $v_2$ were everywhere given by
Eq.~(\ref{approxConstSol}). The result is that the spacetime fields at
this point are approximately those due to an isolated monopole
centered at ${\bf x}_+$.  By a similar analysis, the fields in the
region where $r_- \ll r_+$ are approximately those of a monopole
centered at ${\bf x}_-$.  Hence, for widely separated monopoles $D$ is
simply the intermonopole distance.

\subsection{(2, 1) solutions in SU(3) broken to U(1)$\times$U(1)}
\label{twoOneInSU3}

These solutions contain three fundamental monopoles, two associated
with $\bbeta_1$ and one with $\bbeta_2$, and thus form a
twelve-dimensional moduli space \cite{Houghton:1999qu}.  The two
global U(1) phases do not enter the Nahm data, which therefore depend
on ten parameters: six corresponding to overall spatial translations
and rotations, and four specifying intrinsic properties of the
solutions.

We denote the eigenvalues of the asymptotic Higgs field as $s_1 < s_2
< s_3$.  On the ``left'' interval $s_1 < s < s_2$ the Nahm data are $2
\times 2$ matrices $T_i^L$, while for $s_2 < s < s_3$ the data are a
triplet of numbers $t_i^R$.  The $T_i^L$ obey the same equations as
the Nahm data for the $k=2$ SU(2) solutions, except that they have 
poles only at $s_1$, but not at $s_2$.  The $t_i^R$ are constants, just
like the $k=1$ SU(2) data, with the matching conditions at $s_2$
requiring that $t_i^R$ be equal to the 22 component of $T_i^L(s_2)$.

Thus, by recalling the steps that led to
Eq.~(\ref{generalK2solution}), we find that
\begin{equation}
    T_i^L(s) = {1 \over 2}
           \sum_{j} A_{ij} f_j(s -s_1; \kappa, D) \tau'_j
           + R_i {\rm I}_2 \, .
\label{twoPlusOneT}
\end{equation}
Previously, $\kappa$ was determined by $D$.  Now, the requirement that
the $T_i^L$ be finite at $s=s_2$ (and not have any poles for $s_1 < s
< s_2$) gives the inequality
\begin{equation}
     2 K(\kappa) > D(s_2 - s_1) \, .
\end{equation}
A second difference from the SU(2) case concerns the gauge action.
Before, the full SU(2) gauge action was available to rotate the
$\tau_j'$ into the standard Pauli matrices.  Because the matching
condition at $s_2$ picks out the 22 components of the $\tau_j'$, the
only available gauge action is the U(1) subgroup that leaves these
components invariant.

Thus, the four intrinsic parameters of the solutions can be taken to
be $\kappa$, $D$, and two of the three Euler angles in the matrix
$E_{jk}$ that defines the $\tau_j'$.  We expect the physical
interpretation of these to be clearest when the three monopoles are
well-separated.  Let us see what this means.

It it clear that in this regime the $t_i^R$ specify the position
of the $\bbeta_2$-monopole relative to the two $\bbeta_1$-monopoles.
For the $\bbeta_2$-monopole to be far from the other two, the
$t_i^R$, and hence the $T_i^L(s_2)$, must be large, which
means that $s_2$ must be near the pole in the $f_j$.  The behavior of
the $f_j$ in the pole region then gives, to leading order,
\begin{equation}
   f_1(s_2 - s_1) = - f_2(s_2- s_1  )= - f_3(s_2 - s_1) = 2r
\end{equation}
where $r$ is defined by
\begin{equation}
    2r(s_2 - s_1) = {D(s_2 - s_1) \over 2 K(\kappa) - D (s_2- s_1)}
      \gg 1 \, .
\end{equation}
In order that the two $\bbeta_1$-monopoles be well separated from each
other, we must also require that $D$ be large and hence that $\kappa$
be close to unity.

Assuming these conditions to hold, let us choose our spatial axes so
that $A_{ij}=\delta_{ij}$ and $R_i=0$.  The Nahm data on the left
interval are then
\begin{equation}
    T_i^L(s) = {1 \over 2}
            f_i(s -s_1; \kappa, D) \tau'_i
\end{equation}
with no sum on $i$.  These correspond to two $\bbeta_1$-monopoles
centered at the points $(0,0,\pm D/2)$.  The data on the right
interval are
\begin{equation}
    t_i^R = \left[T_i^L(s_2)\right]_{22}
\end{equation}
and correspond to a $\bbeta_2$-monopole whose center is a distance $r$
from the origin, at $(-r E_{13}, \, r E_{23}, \, r E_{33})$.

\subsection{$\!\!\!$(1, 1, \dots, 1) solutions in maximally broken SU($N$)}
\label{oneoneonecase}

These solutions \cite{Weinberg:1998hn} contain $N-1$ distinct
fundamental monopoles, one of each type.\footnote{For earlier results,
by a different method, on the (1, 1) solutions in SU(2), see
Refs.~\cite{Athorne:1982ke} and \cite{Ward:1982ft}.}  They form a
$4(N-1)$-dimensional moduli space, with $3(N-1)$ parameters entering
the Nahm data.  As with the $k=1$ solution for SU(2), the commutator
terms in the Nahm equation vanish, and so the Nahm data are constant
within each interval.  Thus, on the $p$th interval we write
\begin{equation}
    T^{(p)}_j(s) = x_j^p
\end{equation}
where ${\bf x}^p$ is naturally identified as the position of the $p$th
fundamental mono\-pole.

Because adjacent intervals have equal numbers of mono\-poles, we must
introduce jumping data.  At $s=s_p$, the boundary between the
$(p-1)$th and $p$th intervals, there is a two-component spinor $a^{(p)}$
that, according to the matching condition of
Eq.~(\ref{Nahmdiscontinuity}), must obey
\begin{equation}
     x_j^p - x_j^{p-1} =
     - \,{1 \over 2} a^{(p)\dagger} \, \sigma_j  \, a^{(p)} \, .
\label{oneoneoneMatching}
\end{equation}
Up to an irrelevant overall phase, the solution is
\begin{equation}
    a^{(p)} = \sqrt{2|{\bf x}^p -{\bf x}^{p-1}|}\, \left(
      \matrix{ \cos(\theta/2)\, e^{-i\phi/2} \cr \cr
             \sin(\theta/2)\, e^{i\phi/2} } \right)
\end{equation}
where $\theta$ and $\phi$ specify the direction of the vector
${\bf d}^p \equiv {\bf x}^{p-1} -{\bf x}^{p}$.

(It is a nontrivial result that jumping data satisfying the matching
condition can be found for arbitrary choices of the ${\bf x}^p$.  The
jumping data at the boundary between two intervals with the same value of
$k$ contain at most $4k$ real numbers, while the matching condition
imposes $3k^2$ constraints.  Hence, for $k>1$, the Nahm data within
the intervals must satisfy nontrivial restrictions for the matching to
be possible.)

Having obtained the Nahm data, the next step is to solve the
construction equation.  In order to obtain an SU($N$) solution, there
must be $N$ linearly independent solutions, with the $a$th such
solution consisting of $N-1$ functions $w_a^{(p)}(s)$, one for each
interval, and $N-2$ complex numbers $S_a^{(p)}$, one for each
inter-interval boundary.  Within each interval, the construction
equation is easily solved, giving
\begin{equation}
    w_a^{(p)}(s) = e^{(s-s_p)\,({\bf r}- {\bf x}_p)\cdot \sbsigma}
           \,   w_a^{(p)}(s_p) \, .
\end{equation}
The matching conditions at the boundaries then give
\begin{equation}
    w_a^{(p)}(s_{p}) =
        e^{(s_{p}-s_{p-1})\,{\bf d}_p \cdot
              \sbsigma}\, w_a^{(p-1)}(s_{p-1})
         - S_a^{(p)} a^{(p)}
\end{equation}
so that the entire solution is specified by giving its value at
$s=s_1$, together with the $S_a^{(p)}$ (p=2, 3, \dots, $N-1$).

For each solution, the two components of $w_a^{(1)}(s_1)$ contain four
real numbers and the $S_a^{(p)}$ give $2N-4$ more.  All together,
there should be $N$ such solutions obeying the orthonormality
condition of Eq.~(\ref{jumpingOrtho}).  This orthonormality condition
gives $N^2$ real constraints on the $2N^2$ numbers specifying the
solutions.  Of the remaining degrees of freedom, $N^2-1$ correspond to
the allowed changes of basis that are equivalent to SU($N$) gauge
transformations on the spacetime fields, while the last is an an
overall phase that has no effect on the spacetime fields.

It is a straightforward, albeit tedious, matter to obtain a complete
set of solutions and to then insert them into Eqs.~(\ref{jumpingPhi})
and (\ref{jumpingA}) to obtain $\Phi$ and $A_i$.  All of the required
integrals are readily evaluated, and the spacetime fields can be
expressed in closed (but not very compact) form in terms of elementary
functions.

\chapter{The moduli space of BPS monopoles}

Up to this point we have considered monopoles and dyons as classical
solitons of Yang-Mills-Higgs theory. While we started with general
theories, we saw how supersymmetry introduced many simplifications
into the study of solutions. The study of these BPS monopoles and
dyons has, in turn, contributed immensely toward our understanding of
SYM theories, especially in regard to the
nonperturbative symmetries of ${\cal N}\ge 2$
SYM theories known as dualities.

One important handle for studying the behavior and classification of
monopoles and dyons is
the low-energy moduli space approximation \cite{Manton:1981mp}.  In
this description, most of the field theoretical degrees of freedom are
ignored, leaving only a finite number of bosonic and fermionic
variables to be quantized.  The bosonic variables are the
collective coordinates that encode the positions and phases of the
individual monopoles, while the fermionic pieces are needed to complete certain
low-energy supersymmetries that are preserved by the monopole solutions.
Dyons arise in this description as excited states with
nonzero momenta conjugate to the phase coordinates.

The moduli space approximation ignores radiative interactions and is
relevant only when we ask questions suitable for the low-energy limit.
For instance, while one can study the scattering of monopoles within this
framework, the result is only reliable if none of monopoles are moving
rapidly or radiating a lot of electromagnetic energy. This can be
ensured by restricting to low velocity and by working in the regime
with small Yang-Mills coupling constant
\cite{Manton:bn,Stuart:tc}.\footnote{We are assuming here that the gauge
group has been broken to an Abelian subgroup.  Matters are more complicated
if there is an unbroken non-Abelian subgroup, as we will see in the
next chapter.}
This restriction is harmless when we
are investigating the possible types of low-energy monopole bound
states, which will be one of our main goals when we want to make
contact with the nonperturbative aspects of the underlying Yang-Mills
theories.

Although supersymmetry, specifically the supersymmetry that is left
unbroken by the monopoles, is important for understanding the
low-energy dynamics, we will start, in this chapter, with the purely
bosonic part of the theory.  When there is only one adjoint Higgs we
have the notion of fundamental monopoles, which was introduced in
Chap.~\ref{multimonChap}.  Each fundamental monopole carries four
collective coordinates, and thus a $4n$-dimensional moduli space
emerges as the natural setting for describing $n$ monopoles
interacting with each other. We will presently define, characterize,
and find explicit examples of such moduli spaces.

Of course, SYM theory with extended supersymmetry comes with either two or six
adjoint Higgs fields in the vector multiplet. Except in the
SU(2) theory, this feature turns out to qualitatively modify the
low-energy dynamics and is in fact quite crucial for recovering
most of the dyonic states in the theory.  However, by taking a suitable
limit in which one of the Higgs fields takes a dominant role in the
symmetry breaking, we can study monopole dynamics in such
multi-Higgs vacua with a simple and universal modification of the
moduli space dynamics. This modified moduli space dynamics will occupy
the second half of this review.  For now, we will concentrate on the
conventional moduli space dynamics, with only a single Higgs field.

We begin, in Sec.~\ref{moduliSpaceProperties}, by describing some
general properties of monopole moduli spaces.  We then go on to describe
how the moduli space metric can be determined in several special
cases.  In Sec.~\ref{Asymptotic}, we use the interactions between
well-separated monopoles to infer the metric for the corresponding
asymptotic regions of moduli space.  Next, in
Sec.~\ref{exactmodulispaces}, we show how these asymptotic results,
together with the general mathematical constraints on the moduli
space, determine the full moduli space for the case of two fundamental
monopoles.  If the two monopoles are of distinct types, it turns out
that the asymptotic form of the metric is actually the exact form for
the entire moduli space.  This result is extended to the case of an
arbitrary number of distinct monopoles in
Sec.~\ref{exactformanydistinct}.  Finally, in Sec.~\ref{trajectories},
we will illustrate the use of the moduli space approximation by using
the metrics we have obtained to discuss the scattering of two
monopoles.

\section{General properties of monopole moduli spaces}
\label{moduliSpaceProperties}

We recall from the discussion in Chap.~2 that in the moduli space
approximation the dynamics is described by a Lagrangian of the form
\begin{equation}
L = -\hbox{(total rest mass of monopoles)} +
\frac{1}{2} \,g_{rs}(z) \dot{z}^r\dot{z}^s
\label{modspaceLag}
\end{equation}
where the $z_r$ are the collective coordinates that parameterize the
monopole configurations, and the constant first term will usually be
omitted in our discussions.

\begin{figure}[t]
\begin{center}
\scalebox{1.2}[1.2]{\includegraphics{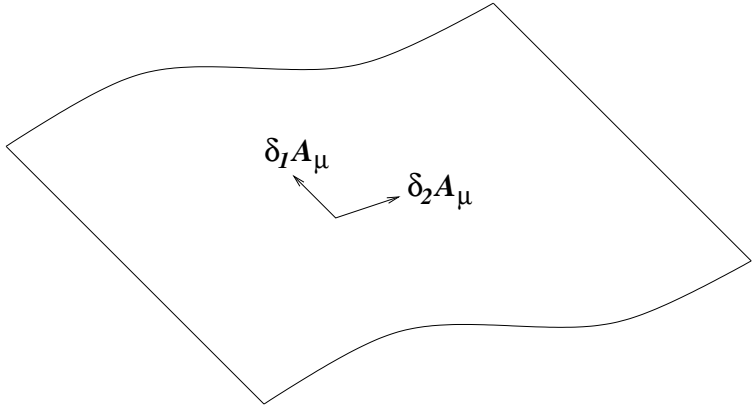}}
\par
\vskip-2.0cm{}
\end{center}
\begin{quote}
\caption{\small The monopole moduli space is a curved manifold
whose points correspond to monopole solutions. Thus, the tangent
vectors at any given point on the moduli space of a BPS
monopole encode the
infinitesimal deformations of the corresponding monopole
solution that preserve the BPS condition. \label{manifold}}
\end{quote}
\end{figure}

The moduli space is naturally viewed as a curved manifold with metric
$g_{mn}(z)$, as illustrated in Fig.~\ref{manifold}.  As was shown in
Sec.~\ref{chaptwoModspace}, the metric can be obtained from the
background gauge zero modes via
\begin{equation}
g_{rs}(z) = 2\int d^3 x \;\tr \left\{ \delta_r A_i \delta_s A_i
   + \delta_r \Phi \delta_s \Phi \right\}
    = 2\int d^3 x \;\tr \left\{ \delta_r A_a \delta_s A_a\right\} \, .
\label{metricdefinition}
\end{equation}
In the last integral we have used the convention, introduced
previously, of letting Roman indices from the beginning of the
alphabet run from 1 to 4, with $A_4 \equiv \Phi$.

This expression for the metric is anything but accessible.
The computation of the metric would seem to require that we know the
entire family of BPS monopole solutions, which remains a very difficult
task. Historically, moduli space metrics have been found by various
indirect methods that invoke the symmetries of the underlying gauge
theory and the moduli space properties that are derived from them.

One essential property of a monopole moduli space is its
hyper-K\"ahler structure.  In Sec.~\ref{indexDiracmodes} we found that
at each point on the moduli space there are three complex structures
$J^{(r)}$ that map the the tangent space onto itself and that obey the
quaternionic algebra
\begin{equation}
J^{(s)}J^{(t)}= -\delta^{st}+\epsilon^{stu}J^{(u)} \, .
\label{quaternionicAlgebra}
\end{equation}
Furthermore, as we will now show, it turns out that the manifold is
K\"ahler with respect to each of the $J^{(r)}$, which is equivalent to
saying that
\begin{equation}
\nabla J^{(s)}=0
\end{equation}
with respect to the Levi-Civita connection of the moduli space metric.
When a manifold possesses such a triplet of K\"ahler structures, it is
called a hyper-K\"ahler manifold.  This puts a tight algebraic
constraint on the curvature tensor and thus provides a differential
constraint on the moduli space metric.\footnote{A more detailed
discussion of complex structures, integrability, and K\"ahler
and hyper-K\"ahler geometry is given in Appendix~A.}

As explained in Sec.~\ref{KandHyperKappendix} of  Appendix~A, to
prove that a manifold is hyper-K\"ahler it is sufficient to show that
the three K\"ahler forms
\begin{eqnarray}
     \omega^{(s)}_{qr} &=& - g_{qn}J^{(s)n}_r  \cr
     &=& 2\int d^3 x \bar \eta^s_{ab} \Tr (\delta_q A_a \delta_r A_b)
\end{eqnarray}
are all closed.

We start by rewriting Eq.~(\ref{formofzeromode}) for the zero modes
as
\begin{equation}
     \delta_r A_a = \partial_r A_a - D_a \epsilon_r \, .
\end{equation}
It is is convenient to view $\epsilon_r$ as defining
a connection on the moduli space, and to define
the covariant derivative
\begin{equation}
     {\cal D}_r = \partial_r +ie[\epsilon_r, \,\,]
\end{equation}
and the field strength
\begin{equation}
  \phi_{rs} =  [{\cal D}_r, {\cal D}_s]
  = \partial_r \epsilon_s - \partial_s \epsilon_r
           + ie [\epsilon_r , \epsilon_s] \, .
\end{equation}

To show that the K\"ahler forms are closed, we must evaluate
\begin{eqnarray}
     \epsilon_{pqr} \partial_p \omega^{(s)}_{qr}
       &=&  2\epsilon_{pqr}\int d^3 x \,\bar \eta^s_{ab}
         \Tr [{\cal D}_p(\delta_q A_a \,\delta_r A_b)] \cr
      &=& 4 \epsilon_{pqr} \int d^3 x \,\bar \eta^s_{ab}
          \Tr[ ({\cal D}_p\delta_q A_a) \delta_r A_b] \cr
      &=& - 2 \epsilon_{pqr} \int d^3 x\, \bar \eta^s_{ab}
         \Tr[(D_a \phi_{pq}) \delta_r A_b] \, .
\end{eqnarray}
Using
Eq.~(\ref{quaternionicAction}), an integration by parts, and the background
gauge condition, Eq.~(\ref{chap2backgroundgauge}), we obtain
\begin{eqnarray}
     \epsilon_{pqr} \partial_p \omega^{(s)}_{qr}
     &=&  2 J^{(s)}_r{}^n \int d^3 x \Tr[(D_a \phi_{pq}) \delta_n A_a] \cr
     &=& - 2 J^{(s)}_r{}^n \int d^3 x \Tr[\phi_{pq} D_a \delta_n A_a] \cr
    &=& 0 \, ,
\end{eqnarray}
verifying that all three K\"ahler structures are closed, and thus that
any BPS monopole moduli space is hyper-K\"ahler
\cite{Atiyah:fd,Atiyah:dv,Atiyah-Hitchin}.

It is also important to take note of the isometries of the moduli space,
which reflect the underlying symmetries of the BPS monopole
solutions themselves. For instance, since we are
discussing monopoles in an $R^3$ space with rotational and translational
symmetries, the moduli space should possess corresponding
isometries. The translation isometry
shows up somewhat trivially in the center-of-mass part of the
collective coordinates and does not enter the interacting part of the
moduli space.

The SO(3) rotational isometry (which can, in general, be elevated to an SU(2)
isometry), turns out to be particularly useful, because it acts on
the relative position vectors of the monopoles.
Spatial rotation of a BPS solution always produces another BPS
solution. This takes one point on the moduli space to another, and
thus induces a mapping of the moduli space onto itself.  Because the
physics is invariant under such spatial rotations, this mapping
preserves the moduli space Lagrangian, and thus the metric.  

The infinitesimal generators of the isometries are realized as vector
fields on the moduli space.  We will denote the three generators of
the rotational isometry by $L^s$ with $s=1,2,3$. The statement that the
$L^s$ generate isometries is reflected in the fact that they are
Killing vector fields, whose components therefore satisfy
\begin{equation}
0=\left({\cal L}_{L^s}[g]\right)_{mn}\equiv
\nabla_m L^s_n+\nabla_n L^s_m
\end{equation}
where ${\cal L}_V$ denotes the Lie derivative with respect to the
vector field $V$.

The SU(2) structure of these isometries is in turn encoded in the commutators
of these vector fields,
\begin{equation}
[L^s,L^t]=\epsilon^{stu}L^u \, ,
\end{equation}
where the commutator of two vector fields, $X$ and $Y$, is defined as
\begin{equation}
[X,Y]^m\equiv X^n\partial_n Y^m-Y^n\partial_n Y^m \, .
\end{equation}
This SU(2) isometry does not leave the complex structures, $J^{(s)}$,
invariant. Instead, the complex structures transform as a triplet:
\begin{equation}
{\cal L}_{L^s}[J^{(t)}]=\epsilon^{stu}J^{(u)} \, .
\end{equation}
Equivalently, the three K\"ahler forms $w^s$ transform as
\begin{equation}
{\cal L}_{L^s}[w^t]=\epsilon^{stu}w^u \, .
\end{equation}
The fact that the $J^{(s)}$ transform as a rotational triplet can be
easily understood by recalling, from Eq.~(\ref{quaternionicAction}),
that their action originates from the action of the 't Hooft tensor
$\eta^s_{\mu\nu}$ on the zero modes. After carefully sorting through
how spatial rotation acts on the $\eta^s_{\mu\nu}$, one finds that the
$J^{(s)}$ form a triplet.

The unbroken gauge group, ${\rm U}(1)^r$, can also be used to transform a
BPS solution; this generates another set of isometries of the
moduli space.  (There are at most $r$ independent isometries of this
sort.) The zero modes associated with these gauge isometries take the
particularly simple form
\begin{equation}
\delta_A A_s= D_s\Lambda_A,\qquad \delta_A \Phi= ie[\Phi,\Lambda_A]
\end{equation}
with $A=1,2,\dots,r$ labelling the $r$ possible gauge rotations.
The zero mode equations then simplify to a single
second-order equation,
\begin{equation}
D^2\Lambda_A+e^2[\Phi,[\Phi,\Lambda_A]]=0 \, .
\end{equation}
Although the long-range part of the solution commutes with the
unbroken gauge group, the monopole cores, which contain charged
fields, are transformed.  Throughout this review, we will denote the
Killing vector fields associated with these U(1) isometries by $K^A$.
Returning to Eq.~(\ref{quaternionicAction}), we see that the effect of
a gauge transformation commutes with those of the $J^{(s)}$.  Hence,
these U(1) isometries, unlike the rotational isometry, preserve the
complex structures of the moduli space,
\begin{equation}
{\cal L}_{K_A}[J^{(s)}]=0 \, ,
\end{equation}
and are thus ``triholomorphic''.

In the following we will find it useful to have an explicit coordinate
system where the gauge isometries act as translations of the angular
coordinates. Generally, we may consider a coordinate system where
these Killing vectors are written as
\begin{equation}
K_A=\frac{\partial}{\partial \xi^A}
\end{equation}
for some angular coordinates $\xi^A$. The Lagrangian must then have no
explicit dependence on the $\xi^A$, other than via their velocities,
and so may be written most generally as
\begin{equation}
L = \frac{1}{2}\, h_{pq}(y)\dot{y}^p\dot{y}^q + \frac{1}{2} \,k_{AB}(y)\;
\left[\dot{\xi}^A + \dot{y}^pw^A_p(y)\right]
\left[\dot{\xi}^B + \dot{y}^qw^B_q(y)\right]
\label{generalexpectedform}
\end{equation}
where the $y^p$ are the other coordinates. In other words, the
$\xi^A$ are all cyclic coordinates whose conjugate momenta are
conserved quantities, just as in the case of SU(2) monopoles.  We can
identify these conjugate momenta as the electric charges that arise
when the monopole cores are excited in such a manner that the monopoles are
converted into dyons.

\section{The moduli space of well separated monopoles\label{Asymptotic}}

The metric on the moduli space determines the motion of slowly moving
dyons. Conversely, the form of the moduli space metric can be inferred
from a knowledge of the interactions between the dyons. In general,
this is not a simple task, since the complete interaction between the
dyons is no easier to understand than the complete form of the
classical Yang-Mills solitons.

On the other hand, a drastic simplification occurs when we restrict
our attention to cases where the monopole cores are separated by large
distances.  In this limit, the only interactions between the monopoles
come about by the exchange of massless fields, which are completely
Abelian \cite{Manton:hs,Gibbons:1995yw,Lee:1996kz}. In
other words, the interactions involved are simply the Maxwell forces
and their scalar analogue. By studying these interactions, then, we
will be able to recover those regions of the moduli space where the
intermonopole distances are all 
large.
In this section, we will show
how to do this.

\subsection{Asymptotic dyon fields and approximate gauge isometries}
\label{asymdyonfields}

Let us imagine that we have a set of $N$ fundamental monopoles, all
well separated from each other.  We label these by an index $j$.
Because only Abelian interactions are relevant at long distances, the
non-Abelian process of electric charge hopping from one monopole core
to another is extremely suppressed.  Consequently, in this regime we
have a larger number of ``gauge'' isometries than we have a right to
expect. Instead of having a conserved electric charge for each
unbroken U(1) gauge group, we effectively have a conserved electric
charge for each monopole core.  The $4N$ moduli of the
monopole solution are easily visualized as $3N$ position coordinates
${\bf x}_j$ and $N$ angular coordinates $\xi_j$, with $j$ labelling
the monopole cores. Translation along $\xi_j$ is then an
approximate symmetry of the moduli space metric, so we have an
approximate gauge isometry associated with each monopole.  The
effective Lagrangian of this approximate moduli space must be of the
form
\begin{equation}
L = \frac{1}{2}\, M_{ij}({\bf x}){\bf x}^i \cdot {\bf x}^j +
\frac{1}{2} \,K_{ij}({\bf x})\,
\left(\dot{\xi}^i +  {\bf W}^i_k({\bf x})\cdot \dot{\bf x}^k\right)
\left(\dot{\xi}^j +  {\bf W}^j_l({\bf x})\cdot \dot{\bf x}^l\right)
\label{expectedform}
\end{equation}
for some functions $M_{ij}$, $K_{ij}$, and ${\bf W}^i_j$ of the ${\bf
x}_k$.  This Lagrangian is similar in form to that displayed in
Eq.~(\ref{generalexpectedform}), but with the significant difference
that there is now a phase angle for every monopole, rather than just
one for each unbroken U(1) factor, no matter how many fundamental
monopoles of a given species are present.

Let us work in a gauge where the asymptotic Higgs field lies in the
Cartan subalgebra.  Then, as was described in
Sec.~\ref{generalgroupformalism}, the $j$th monopole, located at ${\bf
x}_j$, gives rise to an asymptotic magnetic field
\begin{equation}
  {\bf B}^{(j)} = g_j\,  ({\balpha}_j^*\cdot {\bf H})\,
  \frac{({\bf x}-{\bf x}_j)}{4\pi \,|{\bf x}-{\bf x}_j|^3}
\end{equation}
where $\balpha_j$ is one of the fundamental roots and $g_j = 4\pi/e$.
Exciting $Q_j$, the momentum conjugate to $\xi_j$,
gives rise to a long-range electric field
\begin{equation}
  {\bf E}^{(j)} = Q_j\,  ({\balpha}_j^*\cdot {\bf H})\,
   \frac{({\bf x}-{\bf x}_j)}{4\pi \,|{\bf x}-{\bf x}_j|^3}  \, .
\end{equation}
Because of the appearance of ${\balpha}_j^*$, instead of
${\balpha}_j$, the electric charge
$Q_j$ is quantized in integer units of $e\balpha_j^2$.

We will also need the long-range effects of these dyons on the Higgs
field.  Applying a Lorentz transformation to the solution of
Eq.~(\ref{fundmonosolution}), we see that the $j$th dyon induces a
deviation
\begin{equation}
 \Delta\Phi^{(j)} = - \frac{({\balpha}_j^*\cdot {\bf H})}
        {4\pi |{\bf x}-{\bf x}_j|}
  \sqrt{1- {\bf v}_j^2} \sqrt{g_j^2 +Q_j^2}+ O(r^{-2})
\label{higgs}
\end{equation}
from the vacuum value $\Phi_0$.

The interactions among these dyons are most easily described by a Legendre
transformation of the original monopole Lagrangian, in which we trade
off the $\xi_j$ in favor of their conjugate momenta $Q_j/e$. The
resulting effective Lagrangian is often called the Routhian, and has
the form
 \begin{equation}
\label{babeRouthian}
R=L-\frac{Q_j}{e} \,\dot \xi^j=\frac{1}{2}\, M_{ij}\;\dot{\bf x}^i\cdot \dot{\bf x}^j
-\frac{1}{2} \,(K^{-1})^{ij}\;\frac{Q_i}{e} \frac{Q_j}{e}
+ \frac{Q_i}{e} {\bf W}^i_j\cdot\dot{\bf x}^j \, .
\end{equation}
In the following section we will compute this Routhian directly from
the long-range interactions of dyons and then extract the asymptotic
geometry of the moduli space.

\subsection{Asymptotic pairwise interactions and the asymptotic metric}

We begin by considering a pair of well-separated dyons, and asking for
the effect of dyon 2 on the motion of dyon 1.  This has two parts ---
the long-range electromagnetic interaction and the scalar interaction.
The former is a straightforward generalization of the interaction
between a pair of moving point charges in Maxwell theory.
Given two U(1) dyons with electric and magnetic charges $Q_j$ and $g_j$, the
electromagnetic effects of the second on the first are described by the
Routhian
\begin{equation}
R^{(1)}_{\rm Maxwell} =
     Q_1 \left[ {\bf v}_1 \cdot {\bf A}^{(2)}({\bf x}_1)
    -  A_0^{(2)}({\bf x}_1)  \right]
    +  g_1 \left[ {\bf v}_1 \cdot \tilde{\bf A}^{(2)}({\bf x}_1)
    -  \tilde A_0^{(2)}({\bf x}_1)  \right] . \label{one}
\end{equation}
Here ${\bf A}^{(2)}$ and $A_0^{(2)}$ are the ordinary vector and scalar
electromagnetic potentials due to charge 2, while
$\tilde{\bf A}^{(2)}$ and $\tilde A_0^{(2)}$ are dual potentials defined
so that ${\bf E} = - {\bf \nabla \times \tilde A}$ and ${\bf B} = -{\bf
\nabla} \tilde A_0 + \partial \tilde{\bf A}/\partial t$.

Using standard methods to obtain these potentials, and
keeping only terms of up to second order in $Q_j$ or ${\bf v}_j$, we obtain
\begin{equation}
R^{(1)}_{\rm maxwell} =
   \frac{g_1g_2 }{ 4\pi r_{12} } \left[{\bf v}_1 \cdot{\bf v}_2
-  \frac{Q_1 Q_2}{g_1 g_2}  \right]
- \frac{1}{ 4\pi } (g_1Q_2 - g_2Q_1) ({\bf v}_2 -{\bf v}_1)\cdot
   {\bf  w}_{12}
\label{MaxwellRouth}
\end{equation}
where $r_{12}= |{\bf x}_1 - {\bf x}_2|$ and the Dirac monopole potential
\begin{equation}
{\bf w}_{12}={\bf w}({\bf x}_1-{\bf x}_2)
\end{equation}
obeys
\begin{equation}
    {\bf \nabla} \times {\bf w}({\bf r})  = -
     \frac{ {\bf r} }{ |{\bf r}|^3 } \, .
\end{equation}
In terms of the usual spherical coordinates for ${\bf r}$, we can
write
\begin{equation}\label{dirac}
{\bf w}({\bf r})\cdot d{\bf r}=\cos\theta \, d\phi
\end{equation}
locally.

These electromagnetic interactions can all be traced back to the
$F_{\mu\nu}^2$ term in the Maxwell Lagrangian.  In the Yang-Mills
case, the analogous term involves a trace over the group generators.
The result is that the right-hand side of Eq.~(\ref{MaxwellRouth})
must be multiplied by a factor of\footnote{The factor 2 arises
because our normalization convention, Eq.~(\ref{CartanStdNorm}),
replaces the usual 1/4 of the Maxwell Lagrangian by a 1/2, as in
Eq.~(\ref{lagrangian}).}
\begin{equation}
    2 \Tr [({\balpha}_1^* \cdot {\bf H})
        ({\balpha}_2^* \cdot {\bf H})]
       = {\balpha}_1^* \cdot {\balpha}_2^* \, .
\end{equation}

The scalar interaction is manifested as a position-dependent
modification of the dyon mass \cite{Gibbons:1995yw}.  The effective
mass of dyon 1 becomes
\begin{equation}
    m_1^{\rm eff} =  2 \sqrt{g_1^2 + Q_1^2} \, \,
    \Tr [({\balpha}_1^* \cdot {\bf H}) \, (\Phi + \Delta\Phi^{(2)}
 ({\bf x}_1))]
\end{equation}
and hence
\begin{eqnarray}
    R^{(1)}_{\rm scalar} &=&  m_1^{\rm eff} \sqrt{1 - {\bf v}_1^2 }  \cr\cr
      &=&  m_1 \left( 1 - {{\bf v}_1^2 \over 2} + {Q_1^2 \over g_1^2}
         \right)
      - {g_1g_2 \,{\balpha}_1^* \cdot {\balpha}_2^* \over 8\pi r_{12} }
       \left( {\bf v}_1^2 + {\bf v}_2 ^2 - {Q_1^2 \over g_1^2}
          - {Q_2^2 \over g_2^2} \right) \, .  \cr&&
\end{eqnarray}

Adding these contributions, subtracting the rest mass $m_1$, and
keeping terms up to second order in $Q_j$ or ${\bf v}_j$,
we obtain
\begin{eqnarray}
    R^{(1)} = &=& -m_1 \left( 1 - \frac{1}{2}{\bf v}_1^2 +
\frac{Q_1^2 }{2 g_1^2} \right) \nonumber\\
& -& \frac{g_1g_2 \, {\balpha}_1^* \cdot {\balpha}_2^*  }{ 8\pi r_{12} }
        \left[({\bf v}_1 -{\bf v}_2)^2
- \left(\frac{Q_1 }{g_1} - \frac{Q_2}{ g_2}\right)^2 \right] \nonumber \\
&-&\frac{{\balpha}_1^* \cdot {\balpha}_2^* }{ 4\pi }
       (g_1Q_2 - g_2Q_1) ({\bf v}_2 -{\bf v}_1)\cdot {\bf w}_{12} \, .
\end{eqnarray}
By interchanging particles 1 and 2,
a similar expression is obtained for $R^{(2)}$, the Routhian describing
the effects of particle 1 on particle 2.

The extension to an arbitrary collection of well-separated dyons
\cite{Lee:1996kz} is straightforward.  Since we are considering
fundamental dyons that all carry unit magnetic charges, we can set all
of the $g_j$ equal to $4\pi/e$.  The Routhian obtained by adding all
the pairwise interactions is of the form of Eq.~(\ref{babeRouthian}), with
\begin{equation}
M_{ij} =\cases{ m_i  -\high \sum_{k\ne i} \high
     {4 \pi {\balpha}_i^* \cdot {\balpha}_k^* \over e^2 r_{ik} }
     \, , \qquad  i=j           \cr \cr
    \high  {4\pi {\balpha}_i^* \cdot {\balpha}_j^* \over e^2 r_{ij} }
     \, , \qquad  i \ne j }
\label{Mform}
\end{equation}
\begin{equation}
   {\bf W}_i^j = \cases{ -\sum_{k\neq i}{\balpha}_i^*\cdot
         {\balpha}_k^*{\bf w}_{ik} \, , \qquad  i=j           \cr \cr
      {\balpha}_i^*\cdot  {\balpha}_j^*{\bf w}_{ij} \, , \qquad i \ne j }
\end{equation}
and
\begin{equation}
    K = {(4 \pi)^2 \over e^4} M^{-1} \, .
\label{Kform}
\end{equation}

The asymptotic moduli space metric is obtained by returning from the Routhian
back to the Lagrangian via a Legendre transform.  Substituting
Eqs.~(\ref{Mform})--(\ref{Kform})
into Eq.~(\ref{expectedform}), we obtain the desired asymptotic metric
\cite{Lee:1996kz},\footnote{Bielawski \cite{Bielawski,Bielawski:1998hj,Bielawski:1998hk} 
has shown rigorously that this asymptotic metric approaches the 
exact metric exponentially rapidly as the separations between monopoles are 
increased.}
\begin{equation}
    {\cal G}_{\rm asym}
    = M_{ij}d{\bf x}_i\cdot d{\bf x}_j+\frac{(4\pi)^2}{e^4}
   (M^{-1})_{ij}(d\xi_i+{\bf W}_{ik}\cdot d{\bf x}_k)(d\xi_j+{\bf W}_{jl}
  \cdot d{\bf x}_l) \, .
\label{metric}
\end{equation}

\subsection{Why does the asymptotic treatment break down?}

It is easy to see that this asymptotic approximation to the moduli
space metric cannot be exact for the case of two identical monopoles.
First of all, the $M_{jj}$ vanish and the asymptotic form becomes
singular if the intermonopole distance is too small.  Second, for the
case of two identical monopoles the approximate metric is independent
of the relative phase angle $\xi_1 -\xi_2$.  If this isometry were
exact, it would imply that the two-monopole solutions was axially
symmetric, which we know is not the case.  Furthermore, such an
isometry would correspond to an additional U(1) isometry, but for the
SU(2) case there is only one unbroken U(1) gauge group.

Neither of these problems would arise if we were considering a pair of
distinct monopoles \cite{Lee:1996if}.  Because $\balpha_1^* \cdot
\balpha_2^*$ is now negative, the $M_{jj}$ never vanish.  Also, for
two distinct monopoles there are always two different unbroken U(1)
isometries acting on the BPS solutions, so the appearance of an
additional U(1) is actually desired.  In fact, as will be shown in
detail in the next section, the asymptotic metric for a pair of
distinct monopoles can be extended without modification to all
distances and is identical to the exact moduli space metric found via
rigorous mathematical considerations.

The difference between these two cases can be understood by noting that
two fundamental monopoles of the same type can interact via
the exchange of a massive gauge boson.
This additional interaction is short-range, and so gives a correction to the
moduli space metric that falls exponentially with distance.  If the monopoles
are of different types, such gauge boson exchange is impossible, and
there is no modification to the metric.

\section{Exact moduli spaces for two monopoles}
\label{exactmodulispaces}

For a pair of monopoles, the moduli space is eight-dimensional. Of
these eight dimensions, three encode the center-of-mass motion of the
two-body system and must remain free, while at least one corresponds
to an exact gauge rotation. Thus the nontrivial part of the moduli
space is at most four-dimensional. With the various constraints on the
moduli space, in particular its hyper-K\"ahler property and the SO(3)
isometry from spatial rotations, not much choice is left. In fact, it
is via these abstract considerations that Atiyah and Hitchin
\cite{Atiyah:fd,Atiyah:dv,Atiyah-Hitchin}
were able to find the exact moduli space for two identical monopoles.
In this section, we will consider an arbitrary pair of monopoles,
identical or distinct, and find the exact moduli space thereof.

\subsection{Geometry of two-monopole moduli spaces}

Symmetry considerations tell us a great deal about the form of the
two-monopole moduli space $\cal M$. First of all, there must be three
directions, corresponding to overall spatial translations of the
two-monopole system, that are free of interaction.  In other words,
the metric components for these directions must be trivial.
Furthermore, the hyper-K\"ahler structure relates these three free
directions to a fourth one, at least locally, so that at least a
four-dimensional part of the moduli space comes with a flat
metric. This fourth direction must come from gauge rotations that are
a mixture of the two U(1) gauge angles associated with the
fundamental monopoles. This allows, in principle, a discrete mixing
between the free part of the gauge angles and the rest, and so we
conclude that the space must be of the form
\begin{equation}
{\cal M}=R^3\times \frac{R^1\times {\cal M}_0}{\cal D}
\label{manifoldWithDivisor}
\end{equation}
where ${\cal D}$ is a discrete normal subgroup of the isometry group
of $R^1\times{\cal M}_0$.

The isometry group of ${\cal M}_0$ is also easily determined.  Since
spatial rotation of a BPS solution about any fixed point yields
another BPS solution, ${\cal M}_0$ must possess a three-dimensional
rotational isometry.  As we noted in Sec.~\ref{moduliSpaceProperties},
this rotational isometry does not preserve the complex structures, but
rather mixes them among themselves.

If the two monopoles are of different types, there will be an
additional U(1) isometry. This is possible only if the gauge group is
rank 2 or higher, with at least two unbroken U(1) factors. One linear
combination of the two unbroken U(1) gauge degrees of freedom
generates the translational symmetry, alluded to above, along the
overall $R^1$.  The remaining generator must then induce a U(1)
isometry acting on ${\cal M}_0$.  Hence, ${\cal M}_0$ must be a
four-dimensional manifold that is equipped either with four Killing
vector fields that span an ${\rm su}(2)\times {\rm u}(1)$ algebra, or with three
Killing vectors that span ${\rm su}(2)$, depending on whether the monopoles
are distinct or identical.  Furthermore, the results of the previous
section show that the orbits of the rotational isometry on the
asymptotic metric are three-dimensional; clearly the exact metric must
also possess this property at large $r$.

For a four-dimensional manifold the fact that the moduli space is
hyper-K\"ahler implies that the manifold is a self-dual Einstein
manifold.  From this, together with the rotational symmetry properties
of the manifold, it follows that the metric can be written as
\begin{equation}
ds^2=f(r)^2\,dr^2+a(r)^2\sigma_1^2+b(r)^2\sigma_2^2+c(r)^2\sigma_3^2
\label{fabcmetric}
\end{equation}
where the metric functions obey
\begin{equation}
\frac{2bc}{f}\frac{da}{dr}=b^2+c^2-a^2-2\epsilon bc
     \label{abc}
\end{equation}
(and cyclic permutations thereof) with ${\epsilon}$ either 0 or 1,
while the three one-forms $\sigma_k$ satisfy
\begin{equation}
d\sigma_i=\frac{1}{2}\epsilon_{ijk}\sigma_j\wedge \sigma_k \, .
\end{equation}
An explicit representation for these one-forms is
\begin{eqnarray}
\sigma_1 &=& -\sin\psi d\theta +\cos\psi\sin\theta d\phi, \nonumber \\
\sigma_2 &=& \cos\psi d\theta +\sin\psi\sin\theta d\phi, \nonumber \\
\sigma_3 &=&  d\psi+\cos\theta d\phi
\end{eqnarray}
where the ranges of $\theta$ and $\phi$ are $[0,\pi]$ and $[0,2\pi]$,
respectively.

The function $f$ depends on the coordinate choice for $r$.  A
convenient choice for making contact with the results of the previous
section
is to take $\theta$ and $\phi$ to be the usual
spherical coordinates on $R^3$, $r$ to be a radial coordinate, and
$\psi$ to be a U(1) angle.  With this choice, it is easy to see that
\begin{equation}
\sigma_3=d\psi+{\bf w}({\bf r})\cdot d{\bf r}
\end{equation}
where ${\bf w}$ is the same Dirac potential as in Eq.~(\ref{dirac}).
In order that the metric tend to the asymptotic form
 ${\cal G}_{\rm rel}$, the range of $\psi$ must be $[0,4\pi]$.

\begin{figure}[t]
\begin{center}
\scalebox{1.1}[1.1]{\includegraphics{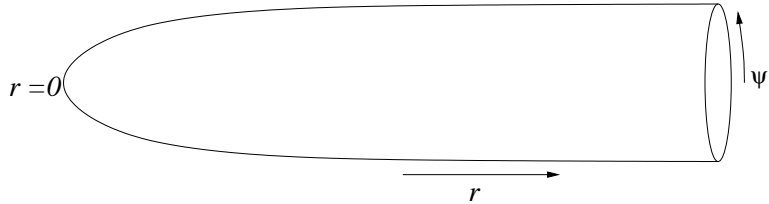}}
\par
\vskip-2.0cm{}
\end{center}
\begin{quote}
\caption{\small The Taub-NUT manifold with two of the three Euler
angles suppressed. The origin $r=0$ is a special point where one
circle collapses to a point. Everywhere else, we have a squashed $S^3$
at each fixed value of $r > 0$. \label{taub}}
\end{quote}
\end{figure}

We now quote the results of Atiyah and
Hitchin~\cite{Atiyah:fd,Atiyah:dv,Atiyah-Hitchin} and list all the smooth
geometries that are obtained from solutions to these conditions:

\begin{itemize}

\item
$\epsilon=0$  produces only one smooth solution with an
asymptotic region, the so-called Eguchi-Hanson gravitational
instanton \cite{Eguchi:1978gw}.
Its asymptotic geometry is $R^4/Z_2$ and does not have a compact
circle corresponding to a gauge U(1) angle.

\item
$ \epsilon=1$, $a=b=c$ gives a solution with
\begin{equation}
f=1 \, , \qquad a=b=c= - {r \over 2} \, .
\label{f2forflat}
\end{equation}
This corresponds to a flat $R^4$.
Dividing it by
$Z$ gives a cylinder, $R^3\times S^1$, which
would be ${\cal M}_0$ for a pair of noninteracting monopoles.  For an
interacting pair, however, this manifold is not acceptable, because it
has too much symmetry.

\item
$ \epsilon=1$, $a=b\neq c$ gives
\begin{equation}
f = \sqrt{1+\frac{2l}{r}} \, , \qquad a=b=-rf \, ,\qquad
       c=-\frac{2l}{f}
\label{f2forTaub}
\end{equation}
with $l>0$.  (A possible overall multiplicative constant has been
suppressed.)  This gives the Taub-NUT geometry with an SU(2) rotational
isometry \cite{Gibbons:xn}, which is illustrated in Fig.~\ref{taub}.
The range of $\psi$ is $[0,4\pi]$. Since $a=b$, the metric has no
dependence on $\psi$, and a shift of $\psi$ is a symmetry. This
generates an additional U(1) isometry, which is also triholomorphic
and thus could be associated with an unbroken U(1) gauge symmetry.

\item
$ \epsilon=1$, $a\neq b\neq c$ yields the Atiyah-Hitchin geometry with
an SO(3) rotational isometry and no gauge isometry
\cite{Atiyah:fd,Atiyah:dv,Atiyah-Hitchin}. There are two such smooth
manifolds, whose topology and global geometry are a bit involved.  We
will come back to them in Sec.~\ref{AtiyahHitchin}

\end{itemize}

\noindent

Thus, only two of the four cases, namely the Taub-NUT manifold
and the Atiyah-Hitchin manifold, can be part of the exact moduli
space for a pair of interacting monopoles.  These two geometries share
the same form for the asymptotic metric,
\begin{eqnarray}
ds^2
&=&\left(1+\frac{2l}{r}\right)\left(dr^2+r^2\sigma_1^2+r^2\sigma_2^2\right)+
\left(\frac{4l^2}{1+2l/r}\right)\sigma_3^2\nonumber \\
&=&\left(1+\frac{2l}{r}\right)\,d{\bf r}^2+\left(\frac{4l^2}{1+2l/r}\right)
(d\psi+ \cos\theta\,d\phi)^2
\label{twoAHsolutions}
\end{eqnarray}
up to an overall multiplicative constant.
The difference between the two is that the parameter $l$ is positive
for the Taub-NUT manifold and negative for the
Atiyah-Hitchin manifold.  With negative
$l$ this metric develops an obvious singularity at $r=2l$, signalling
that the Atiyah-Hitchin geometries must deviate from this
asymptotic form as $r$ become comparable to $2l$. On the other hand,
with positive $l$, this asymptotic form is exact for the Taub-NUT
geometry.

Finally, note that in the limit $l^2 \rightarrow \infty$, the metric
of Eq.~(\ref{twoAHsolutions}) becomes a flat metric with (after an
overall rescaling by $2l$) $a^2=b^2=c^2=f^{-2}=r$.  A coordinate
transformation with $r \rightarrow \tilde r =r^2/4$ brings this into
the form given in Eq.~(\ref{f2forflat}).

\subsection{Taub-NUT manifold for a pair of distinct monopoles}
\label{taubfortwo}

Let us now specialize the results of Sec.~\ref{Asymptotic} to the case
of two distinct fundamental monopoles.  If the corresponding simple
roots are orthogonal (i.e., if they are not connected in the Dynkin
diagram), then Eq.~(\ref{metric}) reduces to a flat metric, corresponding to
the fact that the monopoles do not interact with each other.  The more
interesting case is when $\balpha_1$ and $\balpha_2$ are connected in
the Dynkin diagram.  These may be roots of equal length; if not, we
can, without loss of generality, take $\balpha_2$ to be the shorter
root.  If we define
\begin{equation}
   \lambda = -2 {\balpha}_1^* \cdot {\balpha}_2^*
\end{equation}
then the general properties of Dynkin diagrams imply that $ \lambda
{\balpha}^2_2 =1$ and that
\begin{equation}
   p = \lambda {\balpha}^2_1 = { \balpha_1^2 \over \balpha_2^2}
\end{equation}
is an integer equal to 1, 2, or 3.

The first step is to convert from the original coordinates
to center-of-mass and relative variables.  For the spatial coordinates
we define the usual variables
\begin{equation}
 {\bf R} = \frac{m_1 {\bf x}_1 + m_2 {\bf x}_2}{m_1+m_2} \, ,\qquad
 {\bf r} = {\bf x}_1-{\bf x}_2 \, .
\end{equation}
To separate the phase variables, we first define a total charge
$q_\chi$ and a relative charge $q_\psi$ by
\begin{equation}
q_\chi = \frac{(m_1 Q_1 + m_2 Q_2)}{e\,(m_1+m_2)} \,,\qquad
q_\psi =  \frac{\lambda(Q_1-Q_2)}{2e} \,.\label{charge}
\label{CMandRelCharge}
\end{equation}
The coordinates conjugate to these charges are
\begin{equation}
\chi =  (\xi_1 + \xi_2) \,,\qquad
\psi = \frac{2(m_2\xi_1 - m_1\xi_2)}{\lambda \,(m_1+m_2)} \,.
\label{CMandRelAngles}
\end{equation}
When expressed in these variables, the metric of Eq.~(\ref{metric})
separates into a sum of two terms
\begin{equation}
   {\cal G}_{\rm asym} = {\cal G}_{\rm cm} + {\cal G}_{\rm rel}
\end{equation}
where
\begin{equation}
   {\cal G}_{\rm cm}=(m_1+m_2)\left[d{{\bf R}}^2
   + \frac{(4\pi)^2}{ e^4(m_1+m_2)^2 }d{\chi}^2\right]
\end{equation}
is a flat metric and \cite{Lee:1996if,Gauntlett:1996cw,Connell}
\begin{eqnarray}
   {\cal G}_{\rm rel} &=&
    \left( \mu + \frac{2\pi\lambda}{e^2 r} \right) \,d{\bf r}^2
   + \left(\frac{2\pi\lambda}{e^2}\right)^2
    \left( \mu+ \frac{2\pi \lambda}{e^2 r}\right)^{-1}
    \left[d{\psi}+ {\bf w}({\bf r})\cdot d{{\bf r}}\right]^2  \cr \cr
   &=& \left( \mu + \frac{2\pi\lambda}{e^2 r} \right) \,
   \left(dr^2+r^2\sigma_1^2+r^2\sigma_2^2\right)
  + \left(\frac{2\pi\lambda}{e^2}\right)^2
    \left( \mu+ \frac{2\pi \lambda}{e^2 r}\right)^{-1}
   \sigma_3^2  \, .\cr  &&
\label{taubnutrelativemetric}
\end{eqnarray}
Here $\mu$ is the reduced mass and ${\bf w}({\bf r})={\bf w}_{12}({\bf r})$.

Apart from an overall factor and a rescaling of $r$ by a factor of $\mu$,
this relative metric has the
same form as the Taub-NUT metric of Eq.~(\ref{f2forTaub}), with $l = \pi
\lambda/e^2$.
To verify that the manifold defined by the asymptotic metric is indeed
the Taub-NUT space, all that remains is to show that $\psi$ has
periodicity $4\pi$, which is required for the manifold to be
nonsingular at $r=0$.

We first recall, from the discussion in Sec.~\ref{asymdyonfields},
that $Q_j$ is quantized in units of $e \balpha_j^2$.  This implies
that $\xi_j$ has period $2\pi/\balpha_j^2$.  Hence, a shift of $\xi_1$
by $2\pi/{\mbox{\boldmath $\alpha$}}_1^2 $ implies the identification
\begin{equation}
(\chi, \psi) = \left(\chi +  \frac{2\pi}{{\mbox{\boldmath $\alpha$}}_1^2},
 \psi + \frac{4\pi m_2}{\lambda{\mbox{\boldmath $\alpha$}}_1^2
 (m_1+m_2)} \right),
\label{shift}
\end{equation}
while a $-2\pi/{\mbox{\boldmath $\alpha$}}_2^2$ shift of $\xi_2$ gives
\begin{equation}
(\chi, \psi) =  \left( \chi - \frac{2\pi}{ {\mbox{\boldmath $\alpha$}}_2^2},
\psi + \frac{4\pi m_1}{\lambda {\mbox{\boldmath $\alpha$}}_2^2
  (m_1+m_2)} \right)
\label{angles}.
\end{equation}
Combining $p$ steps of the first shift and one of the second then
gives
\begin{equation}
(\chi,\psi) = (\chi,\psi +4\pi) \, ,
\label{psiperiod}
\end{equation}
showing that $\psi$ has the required $4\pi$ periodicity.  The
identification in Eq.~(\ref{shift}) then defines the discrete
subgroup $\cal D$ that appears in Eq.~(\ref{manifoldWithDivisor})

As a consistency check, note that Eq.~(\ref{CMandRelCharge}) shows
that the quantization of $Q_j$ in units of $e \balpha_j^2$ implies
that $q_\psi$ has integer or half-integer eigenvalues, as is
appropriate for a momentum conjugate to an angle with periodicity
$4\pi$.  By contrast, $\chi$ is not periodic, and $q_\chi$ is not
quantized, unless the ratio of the monopole masses is a rational
number.

Thus, by simply continuing the asymptotic form of the moduli
space metric, we have found a smooth manifold that has all
the properties required of the exact moduli space. Not only have we
learned that the Taub-NUT manifold is the interacting part of the
exact moduli space, but we also learned that the naive asymptotic
approximation yields the exact metric for the case of a pair of
distinct monopoles \cite{Lee:1996if}.

\subsection{Atiyah-Hitchin geometry for two identical monopoles}
\label{AtiyahHitchin}

This brings us to the other possibility
for a pair of interacting monopoles. The decomposition of the full
moduli space into a free part and an interacting part should follow
from the asymptotic form of the metric. Since two identical
monopoles have exactly the same mass and the same magnetic charge, this
decomposition should be
\begin{equation}
{\cal M} = R^3 \times \frac{S^1 \times {\cal M}_0}{Z_{2}}.
\end{equation}
where, again, ${\cal M}_0$ is a four-dimensional hyper-K\"ahler space.
Proceeding as in the case of two distinct monopoles, but now with
$\lambda\balpha_1^2=\lambda\balpha_2^2 \equiv \lambda\balpha^2
=-2$, we find that the asymptotic form of the relative metric is
\begin{eqnarray}
   {\cal G}_{\rm rel}^{\rm asym} &=&
    \left( \mu - \frac{4\pi}{e^2 \balpha^2 r} \right) \,d{\bf r}^2
   + \left(\frac{4\pi}{e^2\balpha^2}\right)^2
    \left( \mu - \frac{4\pi }{e^2 \balpha^2 r}\right)^{-1}
    \left[d{\psi}+ {\bf w}({\bf r})\cdot d{{\bf r}}\right]^2 \cr \cr
   &=& \left( \mu - \frac{4\pi}{e^2 \balpha^2 r} \right)
  \, \left(dr^2+r^2\sigma_1^2+r^2\sigma_2^2\right)
  + \left(\frac{4\pi}{e^2\balpha^2}\right)^2
    \left( \mu - \frac{4\pi }{e^2 \balpha^2 r}\right)^{-1}\sigma_3^2
     \, .\cr  &&
\label{AHasymForm}
\end{eqnarray}
This has a singularity at $r= 4\pi/e^2 \balpha^2 \mu$, which tells us
that there must be some correction when the separation between the two
monopoles is small.

Up to a rescaling of $r$ and an overall factor, this asymptotic metric
has the form of Eq.~(\ref{twoAHsolutions}), with negative $l$.  As we
noted previously, this is the asymptotic metric for the Atiyah-Hitchin
geometry, the one remaining solution of Eq.~(\ref{abc}). In the
remainder of this section, we will characterize this geometry, with an
emphasis on its topology and its global geometry.

The general form of the metric given in Eq.~(\ref{fabcmetric}) leaves us the
freedom to redefine the radial coordinate $r$.  Following Gibbons and
Manton~\cite{Gibbons:df}, we fix this
freedom by setting
\begin{equation}
f=-\frac{b}{r} \, .
\end{equation}
We next parameterize the
radial coordinate by a variable $\beta$, defined by
\begin{equation}
r=2K(\sin(\beta/2))
\label{rAndBeta}
\end{equation}
where
\begin{equation}
K(x)\equiv \int_0^{\pi/2} dt \; \frac{1}{\sqrt{1-x^2\sin^2 t}}
\end{equation}
is the complete elliptic integral of the first kind.
As $\beta$ varies from 0 to $\pi$, the range of $r$ is
$[\pi,\infty)$. The Atiyah-Hitchin solution is then specified
by
\begin{eqnarray}
&&ab=-(\sin\beta) \frac{r\,dr}{d\beta} +\frac12(1-\cos\beta) r^2 \nonumber\\
&&bc=-(\sin\beta) \frac{r\,dr}{d\beta} -\frac12(1+ \cos\beta) r^2 \nonumber\\
&&ca=-(\sin\beta) \frac{r\,dr}{d\beta}
\end{eqnarray}
with $\beta$ determined as a function of $r$ by Eq.~(\ref{rAndBeta}).

This metric indeed asymptotes to Eq.~(\ref{twoAHsolutions}) (with
$l=-1$) as $r\rightarrow \infty$ ($\beta\rightarrow \pi$).  In order
to see how the singularity at small $r$ is replaced by a regular
geometry, we must also understand the metric near $r=\pi$.  Again
following Gibbons and Manton, we have
\begin{equation}
ds^2\simeq dr^2 +4(r-\pi)^2\sigma_1^2 +\sigma_2^2+\sigma_3^2 \, .
\label{metricnearcenter}
\end{equation}
In order that this metric give a smooth manifold near $r=\pi$, the
angle associated with $\sigma_1$ must have a period $\pi$
instead of the usual $2\pi$.
We can rephrase this by defining a new set of Euler angles by
\begin{eqnarray}
\sigma_1 &=& d\tilde \psi+\cos\tilde \theta d\tilde\phi  \nonumber \\
\sigma_2 &=& -\sin\tilde\psi d\tilde\theta
+\cos\tilde\psi\sin\tilde\theta d\tilde\phi \nonumber \\
 \sigma_3 &=&\cos\tilde\psi d\tilde\theta +
  \sin\tilde\psi\sin\tilde\theta d\tilde\phi
\end{eqnarray}
and imposing the identification
\begin{equation}
  I:\qquad \tilde\psi \rightarrow \tilde\psi+\pi\, .
\end{equation}
In terms of the original Euler angles, this is
\begin{equation}
I:\qquad \theta \rightarrow \pi-\theta\, ,\qquad
\phi \rightarrow \phi+\pi\, ,\qquad
\psi \rightarrow -\psi \, .
\end{equation}
From the viewpoint of the monopole solutions this
identification is quite natural, since it exchanges the positions of
the two identical monopoles, and thus maps any two-monopole solution
to itself.
The manifold that is obtained after making this identification is
known as the double-cover of the Atiyah-Hitchin
manifold \cite{Atiyah:fd}.  Near the ``origin" at $r=\pi$ its geometry is
that of $R^2\times S^2$.

A second smooth manifold can be obtained by making a further $Z_2$
division, defined by
\begin{equation}
I':\qquad \theta \rightarrow \theta\,,\qquad
\phi \rightarrow \phi\,,\qquad
\psi \rightarrow \psi+\pi\, .
\end{equation}
This is known as the Atiyah-Hitchin manifold, and is the manifold
denoted as $M_2^0$ in Ref.~\cite{Atiyah:fd}. Near $r= \pi$ it has the
geometry of $R^2\times RP^2$.

To decide which is the proper choice of ${\cal M}_0$, we need to
return to the definition of the center-of-mass and relative phase
angles $\chi$ and $\psi$.  We proceed as in Sec.~\ref{taubfortwo},
except that,
as noted above,
when $\balpha_1 = \balpha_2 = \balpha$ we have $\lambda \balpha^2
= -2$.  The analogues of the identifications in Eqs.~(\ref{shift}) --
(\ref{psiperiod}) tell us that $\psi$ has period $2\pi$, $\chi$ has
period $4\pi/\balpha^2$, and
\begin{equation}
(\chi, \psi) = (\chi +  \frac{2\pi}{\balpha^2},  \psi - \pi ) \, .
\label{atiyahtwisting}
\end{equation}
This identification corresponds to a $Z_2$ division on the product manifold, thus
yielding a manifold
\begin{equation}
{\cal M}=R^3\times \frac{S^1\times {\cal M}_0}{Z_2} \, .
\end{equation}

We now remember that the only role of the $R^3\times S^1$ in
monopole-monopole scattering is to supply a conserved total momentum and
total electric charge that are not affected by the scattering
process.  If we set these quantities equal to zero, then the
scattering is completely described by ${\cal M}_0/Z_2$; in order that
this be a smooth manifold, we must take ${\cal M}_0$ to be the
double-cover of the Atiyah-Hitchin manifold.

\section{Exact moduli spaces for arbitrary numbers of distinct monopoles}
\label{exactformanydistinct}

In the previous section, we saw that the asymptotic form of the moduli
space metric for a pair of distinct fundamental monopoles is in fact
the exact moduli space metric for all values of the monopole
separation.  The key to this surprising result lies in the gauge
isometry. As we noted at the very beginning of our discussion of the
asymptotic interactions between monopoles, the long-range interactions
involve only the interchange of photons and their scalar analogues,
because in the maximally broken phase all the other particles --- the
charged vector and scalar mesons --- are heavy and cannot propagate
over long distances.  The interactions mediated by these massive particles
fall exponentially with distance. Thus, the asymptotic form of the
metric for $k$ monopoles is always equipped with $k$ U(1) isometries.

For a pair of SU(2) monopoles, or for a pair of identical monopoles,
the two U(1) isometries cannot both be exact, since there is only one
U(1) gauge rotation acting on these monopoles. One might view the
short distance corrections in the Atiyah-Hitchin manifold as the
removal of the redundant gauge isometry. This is also reflected in the
fact that electric charge can hop from one monopole to the other.

For a pair of distinct monopoles, on the other hand, two U(1) gauge
isometries are, in fact, required. However small the impact parameter
is, the electric charges on the two monopole cores are separately
conserved.  If there were some short-distance correction to the
asymptotic metric, it would have to respect the additional constraint
of preserving two U(1) gauge isometries, in addition to all the usual
properties that are associated with monopole moduli spaces. In the
case of a two-monopole system, this constraint turns out to be
sufficiently stringent to fix the metric uniquely to be Taub-NUT.

What really happened here is that the only possible short-distance
correction comes from the exchange of heavy charged vector mesons, but
this is disallowed by the gauge symmetry combined with the BPS
equation.  Even with many distinct monopoles, this intuitive picture
of why the asymptotic form of the metric is actually the exact metric
should still work as long as no two monopoles are identical
\cite{Lee:1996kz}.  In this section, we will show that the asymptotic
metric for an arbitrary number of distinct monopoles is in fact the
exact moduli space metric. We will start by showing that it is smooth.

\subsection{The asymptotic metric is smooth everywhere}

We consider a system of $n$ fundamental monopoles with charges
$\balpha_i^*$, each corresponding to a different
simple root of the Lie algebra.  This set of simple roots defines a
subdiagram of the Dynkin diagram of the algebra.  If this subdiagram
has several disconnected components, the monopoles belonging to one
component will have no interactions with those belonging to others,
and the the total moduli space will be a product of moduli spaces
for each connected component.  It is therefore sufficient to to
consider the case where the $\balpha_j$ correspond to a connected
subset of simple roots, and thus to the full Dynkin diagram of a
(possibly smaller) simple gauge group.

There are several ways in which this moduli space could fail to be
smooth.  First, the $n \times n$ matrix $M$ would not be invertible if
$\det M$ vanished.  Second, the metric would be degenerate if
its determinant vanished; since
\begin{equation}
{\rm det}\;{\cal G_{\rm asym}} =  \left(\frac{4\pi}{e^2}\right)^{2n}
\left({\rm det}\;M\right)^{2} \, ,
\end{equation}
this possibility is equivalent to the first.  Finally, there could
be singularities when one or more of the $r_{ij}$ vanish.

We begin by showing that $\det M$ is nonzero whenever the $r_{ij}$ are
nonzero.  We start by recalling that its matrix elements are of the form
\begin{eqnarray}
M_{ii} &=& m_i  + \sum_{j\ne i} c_{ij} \nonumber \\
M_{ij} &=& -c_{ij}\, , \qquad
\hbox{\hskip 1cm  $i\neq j$},
\end{eqnarray}
where the $c_{ij}$ are all nonnegative functions of the $r_{ij}$ and
the $m_i$ are all positive definite.

It is trivial to see that ${\rm det}\, M >0$ for $n=2$.  We then
proceed inductively.  We note that the determinant vanishes if all of
the $m_i$ are zero, and that its partial derivative with respect to
any one of the masses is the determinant of the $(n-1)\times(n-1)$
matrix obtained by eliminating the row and column corresponding to
that mass.  The new matrix is of the same type as the first (but with
a shifting of the $m_j$), and so has a positive determinant by the
induction hypothesis.  If follows that ${\rm det}\, M >0$.

To study the behavior when some of the $r_{ij}$ vanish, it is more
convenient to switch to center-of-mass and relative coordinates.  To
do this, we observe that the Dynkin diagram contains $n$ links, which
we label by an index $A$.  Each of these is associated with a pair of
roots $\balpha_i$ and $\balpha_j$ for which $\lambda_A \equiv -2
\balpha_i^* \cdot \balpha_j^*$ is nonzero.  By analogy with our
treatment of the two-monopole case, we define center-of-mass and
relative coordinates
\begin{equation}
    {\bf R}=\frac{\sum m_i {\bf x}_i}{\sum m_i}\, ,\qquad
{\bf r}_A={\bf x}_i-{\bf x}_j \, , \label{split}
\end{equation} and
charges
\begin{equation}
   q_\chi=
  \frac{\sum m_i Q_i}{e\sum m_i}\, ,\qquad
   q_A=\frac{\lambda_A}{2e}(Q_i-Q_j) \, .
\end{equation} As before, the
$q_A$ have half-integer eigenvalues and their conjugate angles $\psi_A$
have period $4\pi$.

When rewritten in terms of these variables, the metric splits into the
sum of a flat metric for $\bf R$ and $\chi$ and a relative
moduli space metric,
\begin{eqnarray}\label{C}
{\cal G}_{\rm rel}
&=&C_{AB}\,d{\bf r}_A\cdot d{\bf r}_B\nonumber \\
&+&\frac{(2\pi)^2\lambda_A\lambda_B}{e^4}\,(C^{-1})_{AB}\,
[d\psi_A+{\bf w}({\bf r}_A)\cdot d{\bf r}_A]
[d\psi_B+{\bf w}({\bf r}_B)\cdot d{\bf r}_B] \, ,
\end{eqnarray}
where the $(n-1)\times (n-1)$ matrix $C_{AB}$ is
\begin{equation}
C_{AB}=\mu_{AB}+\delta_{AB}\,\frac{2\pi\lambda_A}{e^2 r_A}
\end{equation}
with $r_A=|{\bf r}_A|$ and $\mu_{AB}$ being a reduced mass matrix.

This relative metric is manifestly invariant under independent
constant shifts of the periodic coordinates $\psi_A$.  These
isometries, together with the isometry under uniform translation of
the global phase $\chi$, correspond to the action of the $n$
independent global U(1) gauge rotations, generated by the $\balpha_j
\cdot {\bf H}$, of the unbroken gauge group.

The ${\mbox{\boldmath $\alpha$}}_i$'s are connected and distinct.  It
is easy to see that the sum $\sum {\mbox{\boldmath $\alpha$}}_i^*$ is
then equal to $\mbox{\boldmath $\gamma$}^*$ for some positive root
$\bgamma$ of the group $G$.  Embedding of the SU(2) BPS monopole using
the subgroup generated by $\mbox{\boldmath $\gamma$}$ gives a solution
that is both spherically symmetric and invariant under the $n-1$ U(1)
gauge rotations orthogonal to $\mbox{\boldmath $\gamma$}\cdot {\bf
H}$.  It thus corresponds to a maximally symmetric point on the
relative moduli space that is a fixed point both under overall
rotation of the $n$ monopoles and under the $n-1$ U(1) translations.
This fixed point is clearly the origin, ${\bf r}_A=0$ for all $A$.
In the neighborhood of this point, the factors of $1/r_A$ are all
sufficiently large that the matrix $C_{AB}$ is effectively diagonal,
so that
\begin{equation}
{\cal G}_{\rm rel}\simeq \frac{2\pi}{e^2}\sum_A \lambda_A\left(\frac{1}{r_A}
d{\bf r}_A^2+r_A\,[d\psi_A+{\bf w}({\bf r}_A)\cdot d{\bf r}_A]^2\right),
\end{equation}
with the leading corrections being linear in the $r_A$.  Comparing
this with the results of Sec.~\ref{taubfortwo}, we see that the
manifold is nonsingular at the origin.

Finally, we consider the points where only some of the $r_A$'s vanish;
we use a subscript $V$ to distinguish those that vanish.  In inverting
$C_{AB}$ to leading order, it suffices to remove all components of
$\mu_{AB}$ in the rows or the columns labeled by the $V$'s.  The
matrix $C$ then becomes effectively block-diagonal, and consists of
the diagonal entries $2\pi\lambda_V/e^2 r_V$ and a number of smaller
square matrices. Looking for the part of metric along the ${\bf r}_V$
and $\psi_V$ directions, we find
\begin{equation} {\cal G}_{\rm rel}\simeq \frac{2\pi}{e^2}
  \sum_V \lambda_V\left(\frac{1}{r_V} d{\bf r}_V^2+r_V\,
  [d\psi_V+{\bf w}({\bf r}_V)\cdot d{\bf r}_V]^2\right)
+\cdots .
\end{equation}
The terms shown explicitly give a smooth manifold, as previously.  The
remaining terms, indicated by the ellipsis, consist of harmless finite
terms that are quadratic in the other $d{\bf r}_A$ and $d\psi_A$ as
well as mixed terms that involve a $d{\bf r}_V$ or a $d\psi_V$
multiplied by a $d{\bf r}_A$ or a $d\psi_A$.  The off-diagonal metric
coefficients corresponding to the latter vanish linearly near ${\bf
r}_V=0$, and hence cannot introduce any singular behavior at that
point. We thus conclude that the relative metric, and thus the total
metric, remains smooth as any number of monopoles come close together.

\subsection{The asymptotic metric is a hyper-K\"ahler quotient}

Actually, a cleaner way of showing that the asymptotic metric is
smooth (as well as that it is hyper-K\"ahler) is to show that it can
be obtained by a hyper-K\"ahler quotient procedure
\cite{Gibbons:1996nt}.  This alternate derivation is important not
only for showing the smoothness, but also for making contact with the
moduli space metric derived from the Nahm data, which should give the
exact form.  For simplicity, we take the case of an SU($n+2$) theory
broken to ${\rm U}(1)^{n+1}$, and consider a collection of $n+1$
distinct fundamental monopoles.

The hyper-K\"ahler quotient \cite{Hitchin:1986ea} procedure is more or
less the same as for a symplectic quotient, so let us briefly recall
the latter first. For more complete details, we refer readers to
Appendix~A.  Suppose that one is given a symplectic form $w$ (say on a
phase space) together with a symmetry coordinate $\xi$, or
equivalently a Killing vector $\partial/\partial\xi$ that not only
preserves the metric but also preserves the symplectic form $w$. A
symplectic quotient is a procedure for removing two dimensions
associated with such a cyclic coordinate. Formally, one does this by
first identifying a ``moment map" $\nu$ --- a function on the manifold
--- by
\begin{equation}
   d\nu =
  \left\langle \frac{\partial}{\partial \xi}\, ,w\right\rangle \, .
\end{equation}
The right hand side is
an inner product between the Killing vector field and the symplectic
2-form $w$, and the resulting 1-form is guaranteed to be closed
if
\begin{equation}
dw=0\,,\qquad {\cal L}_K w=0\, .
\end{equation}
Assuming trivial topology, the moment map $\nu$ is well-defined.

The submanifold on which $\nu$ takes a particular value, say $f$,
is a manifold $\nu^{-1}(f)$ with one fewer dimensions. One can
reduce by one more dimension by dividing  $\nu^{-1}(f)$ by
the group action $G$ of the Killing vector $\partial_\xi$. The
resulting manifold with two fewer dimensions,
$\nu^{-1}(f)/G$,
is the symplectic quotient of the original manifold, and is itself a
symplectic manifold. The symplectic quotient takes a more familiar
shape if we consider the manifold as the phase space for some
Hamiltonian dynamics. There, the quotient effectively corresponds to
restricting our attention to motions with a definite conserved
momentum, $\nu=f$, along a cyclic coordinate.

A hyper-K\"ahler manifold is essentially a symplectic manifold with
three symplectic forms, namely the three K\"ahler forms, defined
componentwise from the complex structure and the metric by
\begin{equation}
w_{mn}^{(s)}= g_{mk}(J^{(s)})^k{}_n \, .
\end{equation}
The
hyper-K\"ahler quotient reduces the dimension
by four, since we can now impose three moment maps for each
Killing vector field. We define the moment maps by
\begin{equation}
d\nu_s =\left\langle \frac{\partial}{\partial \xi}\,
 ,w^{(s)}\right\rangle
\end{equation}
where $\partial/\partial\xi$ preserves all three K\"ahler forms, and
consider the manifold
\begin{equation}
\left(\nu_1^{-1}(f_1)\cap
\nu_2^{-1}(f_2)\cap\nu_3^{-1}(f_3)\right) /G\, .
\end{equation}
This new manifold is also a hyper-K\"ahler manifold. If the initial
manifold was smooth the quotient is also smooth, provided that
the group action does not have a fixed submanifold, since the metric
on the quotient is inherited from the old manifold.

Consider a flat Euclidean space, $H^{n}\times H^{n}
=R^{4n}\times R^{4n}$, whose $8n$ Cartesian coordinates
are grouped into $2n$ quaternions $q^A$ and $t^A$ ($A= 1,2, \dots, n$).
We will assume a flat metric of the form
\begin{equation}
ds^2=\sum dq^A \otimes_s d\widehat q^A
+ \sum \mu_{AB}\, dt^A \otimes_s d\widehat t^B\, .
\end{equation}
(Here conjugation is denoted by a hat, and acts like Hermitian
conjugation,
\begin{equation}
\widehat{ab} =\widehat b \, \widehat a\, ,
\end{equation}
because quaternions do not commute.)

The three K\"ahler forms can be
compactly written as the expansion of
\begin{equation}
-\frac12 \left( \sum dq^A\wedge d\widehat q^A
+\sum \mu_{AB} dt^A\wedge d\widehat t^B\right)
=i w^{(1)}+ jw^{(2)}+kw^{(3)} \, ,
\end{equation}
which is necessarily purely imaginary since $\mu_{AB}$ is a symmetric
matrix.  The metric and the K\"ahler forms are nondegenerate as long
as the matrix $\mu$ is nondegenerate.

A useful reparameterization of the $q^A$  is obtained by introducing
$n$ three-vectors ${\bf r}_A$
such that
\begin{eqnarray}
\qquad q^A i\widehat q^A =ir^1_A+jr^2_A+k r^3_A\, ,
\label{phase}
\end{eqnarray}
and $n$ angular coordinates $\chi^A$ defined indirectly by
rewriting the first term in the metric as
\begin{equation}
\sum dq^A \otimes_s d\widehat q^A =
   \frac{1}{4}\sum\left[ \frac{1}{r_A}\,d{\bf r}_A^2+
r_A (d\chi^A+{\bf w}({\bf r_A})\cdot d{\bf r}_A)^2\right] \, .
\label{dqAtimesdqA}
\end{equation}
A shift of $\chi_A$ by $\Delta \chi_A$ is a multiplicative
map
\begin{equation}
q^A\rightarrow q^A e^{i\Delta \chi_A/2} \, .
\end{equation}
The reparameterization we want for $t^A$ is
\begin{equation}
t^A=\sum_B(\mu^{-1})_{AB}y^B_0+iy^A_1+jy^A_2+ky^A_3 \, ,
\end{equation}
from which it follows that the second term in the metric is
\begin{eqnarray}
\sum \mu_{AB}\, dt^A \otimes_s d\widehat t^B=
\sum \left[(\mu^{-1})_{AB}dy^A_0dy^B_0+\mu_{AB}\,
d{\bf y}^A\cdot d{\bf y}^B\right] \, .
\end{eqnarray}
In the new coordinates the K\"ahler forms are the three
imaginary parts of
\begin{equation}
\frac14 \sum_A d\chi^A\wedge (i\,dr^A_1+j\,dr^A_2+k\,dr^A_3)
+ \sum_A dy_0^A\wedge (i\,dy^A_1+j\,dy^A_2+k\,dy^A_3) +\cdots \, ,
\end{equation}
where the ellipsis denotes parts involving neither $\chi_A$
nor $y_0^A$.

We wish to start with this flat hyper-K\"ahler metric and use a
hyper-K\"ahler quotient to obtain a
$4n$-dimensional curved hyper-K\"ahler manifold.
To this end, consider the $n$ Killing vectors
\begin{equation}
    K_A =
2\frac{\partial}{\partial \chi^A} +\frac{\partial}{\partial y^A_0}
\label{isometryA}
\end{equation}
that generate
\begin{eqnarray}
   q^A &\rightarrow &q^Ae^{i\theta_A}\nn
   t^A &\rightarrow &t^A+\sum_B(\mu^{-1})^{AB}\,\theta_B \, .
\end{eqnarray}
The $3n$ moment maps are thus the $n$ purely
imaginary triplets in
\begin{equation}
  \frac12 \left( ir^1_A+jr^2_A+k r^3_A \right)
    + \left( iy^A_1+jy^A_2+ky^A_3 \right)
  = \frac12\left[q^A i\widehat q^A + (t^A-\widehat t^A)\right] \, .
\end{equation}
Setting these $3n$ moment maps
to zero, we may remove the ${\bf y}^A$ in favor of the ${\bf r}^A$,
\begin{equation}
{\bf y}^A=-\frac12 {\bf r}^A \, .
\end{equation}
This replacement gives us a $4n+n$ dimensional manifold
which can be further reduced by the symmetry action of
$R^n$.

The simplest method for doing this last step is to express the metric
in the dual basis in terms of some basis vector fields, instead of
one-forms, and set the generators of the isometry in
Eq.~(\ref{isometryA}) to zero. We will choose to work with the
coordinates defined by
\begin{equation}
\frac{\partial}{\partial \psi^A}=\frac{\partial}{\partial \chi^A} \, ,
\qquad
\frac{\partial}{\partial \theta^A}=2\frac{\partial}{\partial \chi^A}
+\frac{\partial}{\partial y^A_0}
\end{equation}
and set $\partial/\partial\theta^A$ to zero.
With this choice of coordinates, the metric of the quotient manifold
\begin{equation}
\left(\nu_1^{-1}(0)\cap
\nu_2^{-1}(0)\cap\nu_3^{-1}(0)\right) /R^n
\end{equation}
is
\begin{eqnarray}
ds^2
&=&\frac{1}{4}\,{\cal C}_{AB}\,d{\bf r}_A\cdot d{\bf r}_B\nonumber \\
&+&\frac{1}{4}\,({\cal C}^{-1})_{AB}\,
[d\psi_A+{\bf w}({\bf r}_A)\cdot d{\bf r}_A]
[d\psi_B+{\bf w}({\bf r}_B)\cdot d{\bf r}_B] \label{quotient}
\end{eqnarray}
where the matrix ${\cal C}_{AB}$ is
\begin{equation}
{\cal C}_{AB}=\mu_{AB}+\delta_{AB}\,\frac{1}{ r_A}  \, .
\end{equation}
Up to an overall factor of 1/4 and
a rescaling of distance by a factor of $2\pi/e^2{\balpha}^2$,
this is precisely the relative part of the asymptotic metric for a
chain of $n+1$ distinct monopoles in SU($n+2$) theory.
The reduced mass matrix $\mu_{AB}$ is a positive definite
matrix of rank $n$, as the construction here assumes.
Furthermore, its inverse $\mu^{-1}$ is also nondegenerate
as long as the monopoles are all of finite mass, and this
ensures that there is no fixed point under the $R^n$ action
used above. From this, we can conclude that this manifold is
free of singularities.

\subsection{The asymptotic metric is the exact metric}

While there is plenty of reason to believe that the asymptotic metric
for the case of all distinct monopoles is exact, there is as yet no
direct field theoretical proof of this assertion.\footnote{An
alternate approach to this proof can be found in
Ref.~\cite{Chalmers:1996jd}.}  However, very compelling support can be
found from the ADHMN construction. The Nahm data reproduces the
complete family of BPS monopoles and, furthermore, has its own
intrinsic definition of a moduli space metric. At first encounter,
this latter definition appears to have little to do with the field
theoretical definition of the moduli space metric, although for the
case of an SU(2) gauge group it has been shown mathematically
\cite{nakajima} that the two definitions give the same metric.

However, recent progress in string theory has given us a much better
understanding of the ADHMN construction in terms of D-branes. In
particular, it has become quite clear why the two definitions of the
moduli space metric should produce one and the same geometry; we refer
readers to Chap.~\ref{Brane} for more details.  Using this
knowledge, we show here that the asymptotic form of the metric is
precisely the same as the exact metric from the Nahm data
\cite{Murray:1996hi} and thereby prove the main assertion of this
section.

Before invoking the Nahm data, however, it is useful to generalize
slightly the hyper-K\"ahler quotient construction above. Instead of
using $H^n\times H^{n}$ as the starting point, we want to start with
$H^{n}\times H^{n+1}$, where the $H^n$ is to be taken the same as the
first factor in the previous construction. We have $2n+1$ quaternionic
variables, $q^A$ ($ A=1,2,\dots, n$) and
\begin{equation}
  T^i=\frac{1}{m_i}\;x^i_0+ ix^i_1+jx^i_2+kx^i_3
       \, ,\qquad i=0,1,2,\dots, n  \, .
\end{equation}
We introduce the flat metric
\begin{equation}
   ds^2 =
\sum_A dq^A \otimes_s d\widehat q^A
   +\sum_i m_i\, dT^i\otimes_s d\widehat T^i \, .
\end{equation}
As the notation suggests, the $m_i$ will later be identified with the
masses of individual monopoles.

Let us take a hyper-K\"ahler quotient with the action
\begin{equation}
T^i\rightarrow T^i+\eta
\end{equation}
for any real number $\eta$. The three moment maps are the
imaginary parts of
\begin{equation}
  \nu = \frac12\;\sum_i m_i (T^i-\widehat T^i) \, .
\end{equation}
The subsequent hyper-K\"ahler quotient
reduces the $H^{n+1}$ factor to $H^n$ with the metric
\begin{equation}
   d\hat s^2 =
\sum_{A,B} \left[(\mu^{-1})_{AB}dy^A_0dy^B_0+\mu_{AB}\,
d{\bf y}^A\cdot d{\bf y}^B\right]
\end{equation}
where the reduced mass matrix $\mu$ is associated with the $m_i$
and the $y$ coordinates are constructed from the $x$
coordinates by writing
\begin{equation}
{\bf y}^A={\bf x}^{A-1}-{\bf x}^{A},\qquad
\frac{\partial}{\partial y_0^A} =
\frac{\partial}{\partial x_0^{A-1}}-\frac{\partial}{\partial x_0^{A}}
\end{equation}
while setting
\begin{equation}
  0 =\sum m_i{\bf x}^i, \qquad
   0=\sum m_i\frac{\partial}{\partial x_0^{i}} \, .
\end{equation}

From this, then, we can proceed as before to produce the relative
part of the smooth asymptotic metric by a hyper-K\"ahler quotient.
Since the two quotient operations commute, we conclude that our moduli
space metric can be thought of as the hyper-K\"ahler quotient of
$H^n\times H^{n+1}$ with respect to the $n+1$ isometries
generated by the Killing vectors
\begin{equation}
  K_0= \sum_{i=0}^n m_i\,\frac{\partial}{\partial x_0^{i}}
\end{equation}
and
\begin{equation}
   K_A =
\frac{\partial}{\partial x_0^{A-1}}-
\frac{\partial}{\partial x_0^{A}}
    +2\frac{\partial}{\partial\chi^A} \, ,
\label{KAisometries}
\end{equation}
where the $\chi^A$ are certain phases of the $q^A$, as in
Eq.~(\ref{dqAtimesdqA}).  In fact, the role of the first isometry is not
difficult to guess. Its associated moment maps are $\sum_i m_i{\bf
x}^i$, so the quotient due to this simply removes the center-of-mass
part of the moduli space. We leave it to interested readers to verify
that the quotient of $H^n\times H^{n+1}$ by $R^{n}$, with only the $n$
isometries of Eq.~(\ref{KAisometries}),
reproduces our asymptotic form for the total moduli space metric, up
to a periodic identification of one free angular coordinate.  The
condition that the $3n$ moment maps vanish can be written more
suggestively in terms of the coordinates of $H^{n}\times H^{n+1}$,
\begin{equation}
\frac12\, q^A i\widehat q^A = {\rm Im}(T^A-T^{A-1}) \, ,
\end{equation}
where ${\rm Im}(T)\equiv (T-\widehat T)/2$.

There is a very obvious correspondence with the Nahm data for this
system, which were discussed in Sec.~\ref{oneoneonecase}.  Because we
are considering a chain of $n+1$ distinct ${\rm SU}(n+2)$ monopoles,
we need $n+1$ contiguous intervals, of lengths proportional to the
$m_p$.  Since there is only one monopole of each type, the Nahm data
on the $p$th interval includes a triplet of functions $T^{(p)}_i(s)$
that, by the Nahm equation, are equal to a constant, $x_i^p$, on the
interval, together with $T^{(p)}_0(s)$, which we are not assuming to
have been gauged away.  We can identify the former with the imaginary
part of a quaternion $\tilde T^p$, with the real part being
\begin{equation}
  {1 \over m_p} x_0^p \equiv \int ds T^{(p)}_0(s)  \, .
\end{equation}
A natural metric for this part of the Nahm data is then
\begin{equation}
  \sum_p \frac{1}{m_p}(dx_0^i)^2+m_p\left(d{\bf x}^p
    \cdot d{\bf x}^p\right)
   =\sum_p m_i\, d\tilde T^p\otimes_s d\widehat{\tilde T}^p \, .
\end{equation}

In this trivial example of the ADHMN construction, the only subtle
part was obtaining the jumping data at the boundaries.  It is not hard
to see that the matching condition of Eq.~(\ref{oneoneoneMatching}) is
equivalent to requiring that there be quaternions $\tilde q^A$ such
that
\begin{equation}
\frac12\, \tilde q^A i\widehat{\tilde q}^A = {\rm Im}(\tilde T^A-\tilde
T^{A-1})  \, .
\end{equation}
The natural metric for these is $\tilde q^A$ is the canonical one,
\begin{equation}
  d\tilde s^2 = \sum_A d\tilde q^A\otimes_s d\widehat{\tilde q}^A \, .
\end{equation}

When we studied this example in Sec.~\ref{oneoneonecase}, we worked in
a gauge where the $T_0^{(p)}$ were identically zero.  Had we not done
so, we would have found that the gauge action of
Eqs.~(\ref{gaugeaction}) and (\ref{gaugeactiononW}) also acts on the
jumping data, with the effect being that the phase $\tilde\chi^A$
associated with $\tilde q^A$ is shifted by an amount that is
determined by the transformations of the $T_0^{(p)}$ in the adjacent
intervals.  The invariance under this local gauge action is then
equivalent to the isometry generated by
\begin{equation}
  \tilde K_A =
\frac{\partial}{\partial \tilde x_0^{A-1}}-
\frac{\partial}{\partial \tilde x_0^{A}}
    +2\frac{\partial}{\partial\tilde\chi^A} \, .
\end{equation}

The correspondence with the moduli space metric is clear.  We simply
drop the tildes and associate the Nahm data and the jumping data with
the $H^{n+1}$ and $H^n$ factors, respectively.  The
vanishing of the moment maps is the matching
condition on the Nahm data, while the division by $R^n$ is the
identification due to the gauge action on the
Nahm data. With this mapping of variables, the metric derived from the
Nahm data is exactly equal to the asymptotic form of the metric that
we found by considering only the long-range interactions. This
concludes the proof.

\section{Monopole scattering as
trajectories in moduli spaces\label{trajectories}}

In the moduli space approximation, one assumes that the low-energy
dynamics of the full field theory can be reduced to that of the zero
modes, and can therefore be described by the Lagrangian of
Eq.~(\ref{modspaceLag}).  Time-dependent solutions are then given by
geodesics on the moduli space, with open geodesics corresponding to
monopole scattering and closed geodesics to bound states.  The
essential justification for this approximation is energetic.  The
excitation of a mode of oscillation is greatly suppressed if the
available energy is small compared to the scale set by the
eigenfrequency of the mode.  Hence, for a system with no massless
fields, the dynamics at sufficiently low energy should involve only
the zero modes.

However, our situation is not quite so simple, because the theories we
are considering all have massless U(1) gauge fields.  Excitation of
these fields, in the form of radiation, is always energetically
allowed.  To establish the validity of the moduli space approximation
\cite{Manton:bn,Stuart:tc}, one must show that such radiation is
suppressed when the monopole velocities are small.  For the case of
two monopoles of masses $\sim M$ with relative velocity $v$, this can
be done by treating the monopoles as point sources moving along a
geodesic trajectory.  Standard electromagnetic techniques then show
that the total dipole radiation is proportional to $Mv^3$, with higher
multipoles suppressed by additional powers of $v$.  (For two identical
monopoles, the dipole radiation vanishes and the quadrupole
contribution, proportional to $Mv^5$, dominates.)  This argument
breaks down when the cores overlap.  However, the modes significantly
affected by the core overlap are those with wavelengths comparable to
the core radius $\sim e^2 M^{-1}$. These modes have quanta with
energies $\sim e^{-2} M$, and so their excitation is energetically
suppressed for slowly moving monopoles.

As an illustration, let us consider the geodesics for a two-monopole
system, whose relative moduli space metric has the form shown in
Eq.~(\ref{fabcmetric}).  This system can be viewed as a top, with
``body-frame'' components of the angular velocity defined by $\sigma_j
= \omega_j\, dt$ and $a^2$, $b^2$, and $c^2$ being position-dependent
principal moments of inertia \cite{Atiyah:fd,Atiyah:dv,Gibbons:df}.
The quantities
\begin{eqnarray}
    J_1 = a^2 \omega_1  \cr
    J_2 = b^2 \omega_2  \cr
    J_3 = c^2 \omega_3
\end{eqnarray}
are then the body-frame components of angular momentum.  Unlike the
``space-frame'' components, these are not separately conserved,
although the sum of their squares,
\begin{equation}
    J^2 = J_1^2 + J_2^2 + J_3^2 \, ,
\end{equation}
is.  After converting from the
angular velocities to the angular momenta by means of
a Legendre transformation, we can describe the dynamics by
means of the Routhian
\begin{equation}
    R = {1 \over 2} f^2 \dot r^2 - {J^2_1 \over 2 a^2}
      - {J^2_2 \over 2 b^2} - {J^2_2 \over 2 c^2}  \, .
\end{equation}

If the two monopoles are distinct, we have the Taub-NUT metric
with $a^2 = b^2$, and the system is a symmetric top.
This additional symmetry implies that $J_3$ is conserved, from which
it follows that $J_1^2 + J_2^2$ is also constant.  The latter quantity
is (up to a multiplicative constant) the ordinary orbital angular
momentum, while $J_3$ is proportional to the relative U(1) charge.
If, instead, the monopoles are identical, the moduli space is the
Atiyah-Hitchin space with $a^2 \ne b^2$, and so $J_3$ is not
separately conserved; trajectories with a net change in $J_3$
correspond to scattering processes in which U(1) charge is exchanged
between the two monopoles \cite{Atiyah:fd,Atiyah:dv}.

Both cases allow open trajectories corresponding to nontrivial
scattering.  For distinct monopoles, Eq.~(\ref{taubnutrelativemetric})
shows that the principal moments of inertia are all increasing
functions of $r$.  It follows that all geodesics begin and end at $r =
\infty$, so there are no closed geodesics and no bound states.
For identical monopoles, on the other hand, $c^2$ --- but not $a^2$ or
$b^2$ --- is a decreasing function of $r$, making it possible to have
bound orbits.

We will not discuss the scattering trajectories in detail, reserving
our comments for one particularly interesting case.\footnote{Further
discussions of scattering and bound trajectories can be found in
\cite{Houghton:1995uj,Feher:1986ib,Gibbons:1986hz, Gibbons:1987sp,
Bates:1988ta, Temple-Raston:1988xs, Temple-Raston:1988bc,Cordani:1989ip,
Schroers:1991tf,Bielawski:1996}.}
If the monopoles approach each
other head-on, with vanishing impact parameter, all three $J_i$
vanish.  The trajectories are then purely radial.  With our
conventions for the principal moments of inertia, the line of approach
is along the 3-axis. [To see this, note that at large $r$, with units
restored as in Eqs.~(\ref{taubnutrelativemetric}) and
(\ref{AHasymForm}), $a^2 \approx b^2 \approx \mu r^2$, while $c^2$
tends to a constant.]  It is a straightforward matter to integrate the
geodesic equations to obtain $r$ as a function of time.  The only
subtlety occurs at the point of minimal $r$.

For distinct monopoles, this minimal value is $r=0$.  In the
neighborhood of this point, the Taub-NUT metric approximates that of
flat four-dimensional Euclidean space.  It is then clear that the
geodesic we want passes straight through the origin without bending.
Thus, the two monopoles pass through each other without any
deflection.  Indeed, the only other possibility allowed by the axial
symmetry of the problem would have been a complete reversal of
direction, with the monopoles receding along their initial paths of
approach.

The situation is different when the two monopoles are identical.
Equation~(\ref{metricnearcenter}) shows that the minimum value, $r=
\pi$, corresponds to a two-sphere rather than a point.  In this region
the manifold is approximately the product of a flat two-dimensional
plane, with polar coordinates $\tilde r = r - \pi$ and $\tilde \psi$,
and a two-sphere spanned by $\tilde \theta$ and $\tilde \phi$.  As an
incoming radial trajectory passes through $\tilde r =0$, $\tilde \psi$
increases by $\pi/2$ (i.e., half of its total range), while $\tilde
\theta$ and $\tilde \phi$ are unchanged.  This shift in $\tilde \psi$
corresponds to an interchange of $\sigma_2$ and $\sigma_3$.  Hence,
the monopoles approach head-on, merge and cease to be distinct objects
as $r$ approaches $\pi$, and then re-emerge and recede back-to-back
along a line perpendicular to their line of approach
\cite{Atiyah:fd,Atiyah:dv}.  This $90^\circ$ scattering gives a rather
dramatic demonstration of the lack of axial symmetry in the
two-monopole system.

\chapter{Nonmaximal symmetry breaking}
\label{unbrokenNonAb}

We have focussed up to now on monopoles in theories where the adjoint
Higgs field breaks the gauge group maximally, to a product of U(1)'s.
However, monopoles can also occur when there is a larger, non-Abelian,
unbroken symmetry, as long as it contains at least one U(1) factor.  This
brings in a number of interesting features, which we will describe in
this chapter.

The case of non-Abelian unbroken symmetry can be viewed as a limiting
case of maximal symmetry breaking, corresponding to a special value of
the Higgs vacuum expectation value [degenerate eigenvalues, in the
case of SU($N$)].  In this limit, additional gauge bosons (and their
superpartners) become massless.  The formulas obtained in
Chap.~\ref{multimonChap} imply that, correspondingly, some of the
fundamental monopoles should also become massless.  On the one hand,
such massless monopoles are to be expected from a duality symmetry, to
be the duals of the gauge bosons of the unbroken non-Abelian group.
From another viewpoint, however, they seem problematic, since it is
clear that the theory cannot have a nontrivial massless classical
solution.

The resolution is found by looking at solutions containing both
massive and massless monopoles, with the constituents chosen so that
the total magnetic charge is purely Abelian.  At the level of
classical solutions, the massless monopoles are then realized as one
or more clouds of non-Abelian field that enclose the massive monopoles
and shield their non-Abelian magnetic charge.  Turning to their
dynamics, one finds that the collective coordinates that described the
massive fundamental monopoles survive even when some of these monopoles
become massless, and that the moduli space Lagrangian has a smooth limit as
the unbroken symmetry is enlarged.

We begin, in Sec.~\ref{masslessIntroSection}, by adapting the
formalism and results of Chap.~\ref{multimonChap} to the case where
the symmetry breaking is no longer maximal.  Next, in
Sec.~\ref{masslessExamples}, we describe several classical solutions
containing both massive and massless monopoles.  We discuss the moduli
space and its metric in Sec.~\ref{masslessmodulimetric}, focussing in
particular on the examples described in the previous section.
Finally, in Sec.~\ref{cloudtrajectories}, we discuss the use of this
metric to treat the scattering of the massive monopoles and massless
clouds.  In the course of this discussion, we will see that the range of
validity of the moduli space approximation is more limited than when
the symmetry breaking is maximal, and we
will discuss the conditions under which it can be considered reliable.

\section{Simple roots, index calculations, and massless monopoles}
\label{masslessIntroSection}

As we saw in Eq.~(\ref{defOFh}), the vacuum expectation value of the
Higgs field defines a vector $\bf h$ whose properties determine the
nature of the symmetry breaking.  Maximal symmetry breaking occurs
when ${\bf h}\cdot \balpha$ is nonzero for all roots $\balpha$.  If,
instead, there are some roots orthogonal to $\bf h$, then these are
the roots of some non-Abelian semisimple group $K$, of rank $r'$, and
the unbroken subgroup is $K \times {\rm U}(1)^{r-r'}$.  One
consequence for monopole solutions is that the homotopy group is now
smaller, since $\Pi_2(G/H) = \Pi_1[K \times {\rm U(1)}^r]= Z^{r-r'}$.
As a result, there are only $r-r'$ integer topological charges.

As in the maximal symmetry breaking
(MSB) case, it is useful to define a set of simple roots
${\bbeta}_a$.  However, we can no longer require that these satisfy
Eq.~(\ref{hdefinessimple}), but rather can only impose the weaker
condition
\begin{equation}
    {\bf h} \cdot {\bbeta}_a \ge 0  \, .
\label{NUShdefinessimple}
\end{equation}
We will sometimes need to distinguish the simple roots
that are orthogonal to $\bf h$.  We will denote these by
$\bgamma_j$ ($j=1, 2, \dots, r'$), and the remaining simple
roots (possibly renumbered) by
$\bbeta_p$ ($p=1, 2, \dots, r-r'$).  Note that the $\bgamma_j$
form a set of simple roots for the subgroup $K$.

\begin{figure}[t]
\begin{center}
\scalebox{1.0}[1.0]{\includegraphics{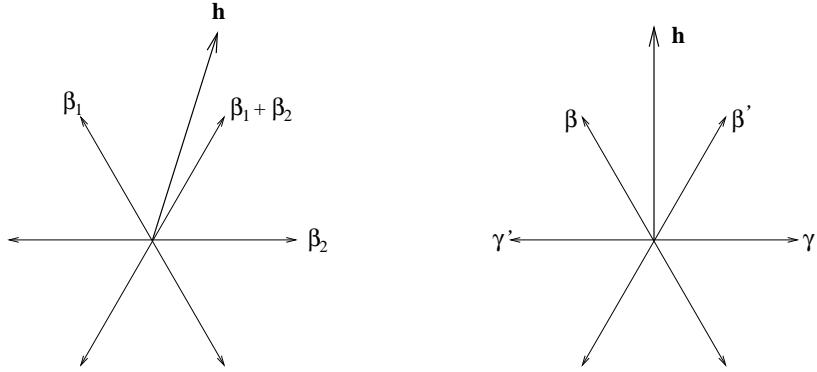}}
\par
\vskip 0.0cm{}
\hskip-5.0cm{}
\end{center}
\begin{quote}
\caption{\small Two symmetry breaking patterns of SU(3).  With generic
symmetry breaking, as on the left, the unbroken gauge group is ${\rm
U}(1)\times{\rm U}(1)$ and $\bbeta_1$ and $\bbeta_2$ are the simple
roots that define the fundamental monopoles. When ${\bf h}$ is
orthogonal to $\bgamma$, as on the right, the unbroken group is ${\rm
SU}(2) \times {\rm U}(1)$ and the simple roots can chosen to be
either ${\bbeta}$ and ${\bgamma}$ or ${\bbeta}'$ and
${\bgamma}'$.
\label{su3figure}}
\end{quote}
\end{figure}

Equation~(\ref{NUShdefinessimple}) does not uniquely determine the
simple roots.  There will be several possible choices, related to each
other by gauge transformations in the unbroken group $K$.  This can be
illustrated by considering the case of SU(3), whose root diagram is
shown in Fig.~\ref{su3figure}.  With $\bf h$ oriented as in the left-hand
diagram, the unbroken subgroup is U(1)$\times$U(1), and
Eq.~(\ref{hdefinessimple}) fixes the simple roots to be the ones
denoted ${\bbeta}_1$ and ${\bbeta}_2$.  When the symmetry breaking is
to SU(2)$\times$U(1), as shown on the right-hand side, the simple
roots can be chosen to be either ${\bbeta}$ and ${\bgamma}$ or
${\bbeta}'$ and ${\bgamma}'$, with the two pairs related by a
rotation by $\pi$ in the unbroken SU(2).

For the MSB case, the magnetic charge quantization, the BPS mass
formula, and the counting of zero modes all suggested that a general
BPS solution should be viewed as being composed of a number of
fundamental monopoles, each associated with a particular simple root.
The situation is a bit more complex now.
The arguments that led to Eq.~(\ref{bfGdecomp}) go through essentially
unchanged, and imply that the magnetic charge vector $\bf g$ defined
in Eq.~(\ref{defOFg}) must be of the form
\begin{equation}
   {\bf g} = {4\pi \over e} \sum_{a=1}^r n_a \bbeta_a^*
     = {4\pi \over e} \left(\sum_{p=1}^{r-r'}  n_p\bbeta_p^*
       + \sum_{j=1}^{r'} k_j \bgamma_j^* \right)
\label{NUScharge}
\end{equation}
where the $n_p$ and the $k_j$ are all integers.  (This will, in general,
entail a renumbering of the $\bbeta_a$.)  In general, the $k_j$
depend on which set of simple roots was chosen and are not even gauge
invariant.  The remaining coefficients, $n_p$, on the other hand, are
gauge invariant and do not depend on the particular choice of simple
roots.  They are the topological charges.

For the MSB case, there are solutions corresponding to any set of
positive $n_a$.  This might lead one to expect that with nonmaximal
symmetry breaking there would be a solution for any choice of positive
$n_p$ and $k_j$.  However, with a different set of simple roots some
of these would correspond to negative $k_j$, and thus would not be expected
to give rise to classical solutions.  Thus, for SU(3) broken to
SU(2)$\times$U(1) one would only expect to find solutions with $k \le
n$.  Outside the BPS limit the restrictions on the $k_j$ are even more
severe, because solutions for which the non-Abelian component of the
magnetic charge is nonminimal [e.g., $|n-k| > 1/2$ for this SU(3)
example] are unstable \cite{coleman-fifty,Brandt:1979kk}.  These
instabilities are absent in the BPS limit, because of the effects of
the long-range massless Higgs fields.

The BPS mass formula of Eq.~(\ref{massAsSum}) becomes
\begin{equation}
   M= \sum_{a=1}^r n_a m_a = \sum_{p=1}^{r-r'} n_p  m_p
\label{NUSmass}
\end{equation}
where the second equality uses the fact that the orthogonality of
$\bgamma_j$ to $\bf h$ implies the vanishing of the corresponding mass.

When we turn to the index theory calculations, matters become somewhat
more complicated.  Two separate issues arise.  The first concerns the
calculation of $\cal I$.  The derivation used in the MSB case goes
through unchanged up to Eq.~(\ref{intermediateindexcalc}), but the next
step in the derivation used the fact that the ${\bf h}\cdot \balpha$
were all nonzero, which is no longer the case.  The terms arising from
the roots orthogonal to $\bf h$ (i.e., the roots of the unbroken
subgroup) make no contribution to the sum, with the result that
Eq.~(\ref{finalIndex}) is replaced by
\begin{equation}
   {\cal I} = 4\sum_{a=1}^r n_a
   -{e \over \pi}  {\sum_{{\sbalpha}\in K}}' {\bf g} \cdot {\balpha}
\label{NUSfirstindexformula}
\end{equation}
with the sum in the second term being over the positive roots of $K$.

The second issue relates to the possible continuum contribution,
${\cal I}_{\rm cont}$.  A nonzero contribution of this type can only
arise from the large-$r$ behavior of the terms in $\cal D$ and ${\cal
D}^\dagger$ that affect the massless fields.  With maximal symmetry
breaking there can be no such contribution, since the long-distance
behavior of $\cal D$ and ${\cal D}^\dagger$ is determined by the
massless fields and these fields, being Abelian, do not interact with
themselves or each other.  This simple argument for the vanishing of
${\cal I}_{\rm cont}$ clearly fails when there are non-Abelian
massless fields.  In fact, one can show explicitly that a nonzero
${\cal I}_{\rm cont}$ can actually occur.  Returning to the SU(3)
example above, we can use the root $\bbeta$ to obtain an embedding of
the SU(2) unit monopole via Eq.~(\ref{fundmonosolution}).  Because of
the spherical symmetry of this solution, the zero mode equations can
be explicitly solved \cite{Weinberg:1979zt}.  There turn out to be
precisely four normalizable zero modes, whereas evaluation of the
right-hand side of Eq.~(\ref{NUSfirstindexformula}) gives ${\cal I} =
6$.  The difference is due to a nonzero ${\cal I}_{\rm cont}$.

These difficulties in counting the zero modes disappear if the
magnetic charge is purely Abelian; i.e., if \cite{Weinberg:1982ev}
\begin{equation}
    {\bf g} \cdot \bgamma_j =0
\label{purelyabelian}
\end{equation}
for all $j$, so that the long-range magnetic field is invariant under
the subgroup $K$.  (Note that this does not imply that the $k_j$
vanish.)  First, the $k_j$ are now gauge-invariant and independent of
the choice of simple roots.  [Thus, for our example of SU(3) broken to
SU(2)$\times$U(1) the magnetic charge is
purely Abelian when $\bf g$ is of the form
\begin{equation}
   {\bf g} = {4\pi \over e} \left( 2n \bbeta^* + n \bgamma^* \right)
      ={4\pi \over e}  \left( 2n {\bbeta'}^*
        + n {\bgamma'}^* \right) \, .
\end{equation}
As we see, the coefficients are the same for either choice of simple
roots.]  Next, since $\bf g$ is orthogonal to all the roots of $K$,
the second term in Eq.~(\ref{NUSfirstindexformula}) vanishes,
so\footnote{In Ref.~\cite{Weinberg:1982ev} an equivalent expression
was given in which only the $n_p$ appeared, but with coefficients that
depended on the particular $\bbeta_p$.  This turns out
\cite{Bowman:1985kh} to not be as useful in elucidating the structure
of these configurations.}  \cite{Lee:1996vz}
\begin{equation}
   {\cal I} = 4\sum_{a=1}^r n_a = 4\sum_{p=1}^{r-r'} n_p
    + 4\sum_{j=1}^{r'} k_j \, .
\label{NUSindex}
\end{equation}
Finally, the vanishing of the non-Abelian components of the magnetic
charge implies a faster falloff for the non-Abelian fields.  A detailed
analysis shows that this falloff is rapid enough to guarantee the
vanishing of ${\cal I}_{\rm cont}$, so that the number of normalizable
zero modes is correctly given by Eq.~(\ref{NUSindex}).

With these results in mind, we will restrict our considerations to
solutions that obey\footnote{This condition can always be satisfied by
adding an appropriate of collection of monopoles at a large distance
from the configuration of interest.  The fact that adding distant
monopoles makes a difference reflects the fact that the difficulties
associated with solutions that violate Eq.~(\ref{purelyabelian}) are
all due to their slow long-range falloff.}$^,$\footnote{For more on
solutions with non-Abelian magnetic charge that violate
Eq.~(\ref{purelyabelian}), including a discussion of the dimensions of
the spaces of solutions, see
Refs.~\cite{Murray:1989zk,Bais:1997qy,Schroers:1998pg}.}
Eq.~(\ref{purelyabelian}).  For such solutions,
Eqs.~(\ref{NUScharge}), (\ref{NUSmass}), and (\ref{NUSindex}) suggest
an interpretation in terms of fundamental monopoles, each
corresponding to a simple root, and each with four degrees of freedom.
However, Eq.~(\ref{NUSmass}) would imply that the fundamental
monopoles corresponding to the $\bgamma_j$ --- which would have purely
non-Abelian magnetic charges --- would be massless.  As we have
already noted, this seems somewhat problematic, since it is easy to
show that the theory cannot have any massless classical solitons.
Indeed, using Eq.~(\ref{fundmonosolution}) to construct the
fundamental monopole solution corresponding to the one of the
$\bgamma_j$ simply yields the pure vacuum.  Nevertheless, we will see
that it can be meaningful to speak of such ``massless monopoles'',
which can be viewed as the counterparts of the massless elementary
``gluons'' carrying electric-type non-Abelian charge.  Note however
that, in contrast with the massive fundamental monopoles, the massless
monopoles do not carry topological charges.

\section{Classical solutions with massless monopoles}
\label{masslessExamples}

One way to gain insight into the massless monopoles is to examine some
classical solutions with nonzero values for the $k_j$.  In this
section, we will examine three of these in some detail.  One, arising
in an SO(5) model, is comprised of just two monopoles, one massive and
one massless, and is the simplest possible solution containing a
massless monopole \cite{Weinberg:1982jh}.  In fact, it is sufficiently
simple that it can be obtained by direct solution of the Bogomolny
equations.  We will then use the ADHMN construction to study two
solutions that each contain one massless and two massive monopoles ---
an SU($N$) solution with two distinct massive monopoles
\cite{Weinberg:1998hn}, and an SU(3) solution in which the massive
monopoles are identical \cite{dancer-nonlin}.

\subsection{One massive monopole and one massless monopole in SO(5)
broken to SU(2)$\times$U(1)}
\label{so5solution}

\begin{figure}[t]
\begin{center}
\scalebox{1.0}[1.0]{\includegraphics{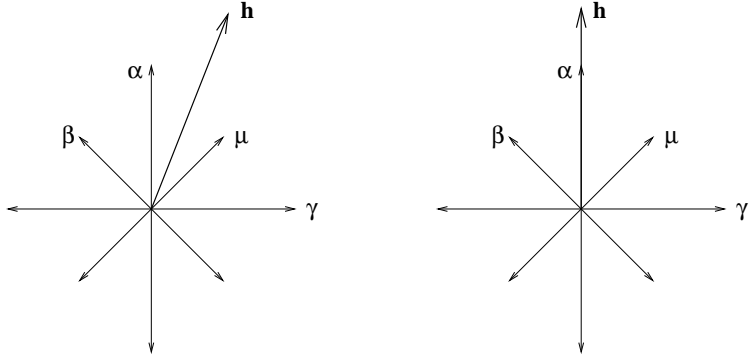}}
\par
\vskip-2.0cm{}
\end{center}
\begin{quote}
\caption{\small Two symmetry breaking pattern of SO(5).  With generic
symmetry breaking, as on the left, the unbroken gauge group is ${\rm
U}(1)\times{\rm U}(1)$ and $\bbeta^*$ and $\bgamma^*$ are the two
fundamental monopole charges. When ${\bf h}$ is orthogonal to
$\bgamma$, as on the right, the unbroken group is ${\rm SU}(2) \times
{\rm U}(1)$, and the $\bgamma^*$ monopole becomes massless.
\label{so5figure}}
\end{quote}
\end{figure}

The simplest example \cite{Weinberg:1982jh} containing a massless
monopole, but with a purely Abel\-ian total magnetic charge, occurs in a
theory with SO(5) broken to SU(2)$\times$U(1) as illustrated by the
root diagram in Fig.~\ref{so5figure}.  A solution with
\begin{equation}
   {\bf g} = {4\pi \over e}\left(\bbeta^* + \bgamma^* \right)
\end{equation}
would correspond to one massive $\bbeta$-monopole and one massless
$\bgamma$-monopole and, according to Eq.~(\ref{NUSindex}), should
have eight normalizable zero modes.  Three of these must
correspond to spatial translations of the solution, and four others
must be global gauge modes corresponding to the generators of the
unbroken SU(2)$\times$U(1).  While the origin of the last zero mode
may not be immediately apparent, it certainly cannot be a rotational mode,
because any solution that is not rotationally invariant must have at
least two rotational zero modes.  Hence, this zero mode must correspond to
the variation of a parameter that has no direct interpretation in
terms of a symmetry.

Since there are no rotational zero modes, the solution must be
spherically symmetric.  The resulting simplifications make it possible
to directly solve the Bogomolny equations.\footnote{For the
construction of this solution by the ADHMN method, see
Ref.~\cite{Bowman:1984vm}} We begin by noting that any element of the
Lie algebra of SO(5) can be written as
\begin{equation}
     P = {\bf P}_{(1)} \cdot {\bf t}(\balpha)
      + {\bf P}_{(2)} \cdot {\bf t}(\bgamma)
      + \tr P_{(3)} M
\end{equation}
where ${\bf t}(\balpha)$ and ${\bf t}(\bgamma)$ are defined as in
Eq.~(\ref{embeddingGenerators}), and
\begin{equation}
     M = {i \over \sqrt{{\bbeta}^2}} \left(
      \matrix{E_{\sbbeta} & -E_{-\sbmu}  \cr \cr
              E_{\sbmu} & E_{-\sbbeta} } \right) \, .
\end{equation}
We then consider the spherically symmetric ansatz
\begin{eqnarray}
     A_{i(1)}^a  &=& \epsilon_{aim} \hat r^m A(r) \, , \qquad
     \phi_{(1)}^a  =  \hat r^a H(r) \, ,  \cr\cr
     A_{i(2)}^a &=& \epsilon_{aim} \hat r^m G(r) \, ,\qquad
     \phi_{(2)}^a =  \hat r^a K(r)    \, ,  \cr\cr
     A_{i(3)}^a &=&  \tau_i F(r) \, ,\qquad
     \phi_{(3)}^a =  i J(r)  \, .
\end{eqnarray}
Substituting this into the Bogomolny equation gives
\begin{eqnarray}
     0 &=& A' + {A \over r} +e\left(A +{1 \over er}\right)H
              + 2eF(F+J)    \cr \cr
     0 &=& H' + e\left(A +{2 \over er}\right)A  + 2eF(F+J)  \cr \cr
     0 &=& G' + {G \over r} + e\left(G +{1 \over er}\right)K
                     + 2eF(F-J) \cr \cr
     0 &=& K' + e\left(G +{2 \over er}\right) G + 2eF(F-J)  \cr \cr
     0 &=&  F' +{e \over 2}(H-A-G+K)F +{e \over 2}(A-G)J   \cr \cr
     0 &=& J' + {e \over 2}(2A -H +K -2G)F \, .
\label{so5eqs}
\end{eqnarray}
In order that the solutions be nonsingular, $A$, $G$, $H$, and $K$ must all
vanish at the origin; $F(0)$ and $J(0)$ are unconstrained.  As $r$
tends to infinity, all of the functions except for $H$ must vanish; to
get the desired symmetry breaking, we must require that
$H(\infty) \equiv v$ be nonzero.

If we try setting $F=-J$, the first two lines in Eq.~(\ref{so5eqs})
give a pair of equations involving only $A$ and $H$.  These are the
same as would be obtained for the unit SU(2) monopole.  Referring to
the results in Eq.~(\ref{monoPSsoln}), and converting from the
conventions of Eq.~(\ref{hedghogansatz}) to those used here, we obtain
\begin{eqnarray}
     A(r) &=&  {v \over \sinh evr} - {1 \over er}   \cr \cr
     H(r) &=&  v \coth evr - {1 \over er} \, .
\end{eqnarray}
The remaining four lines of Eq.~(\ref{so5eqs}) then imply that
$G = K$, and that
\begin{eqnarray}
     0 &=& G' + e\left(G +{2 \over er}\right) G + 4eF^2    \cr\cr
     0 &=& F' + {e \over 2}(H - 2A +G) F  \, .
\end{eqnarray}
These are solved by
\begin{eqnarray}
    F &=& {v \over \sqrt{8} \cosh (evr/2)} L(r, a)^{1/2}    \cr\cr
    G &=& A(r) L(r, a)
\end{eqnarray}
where
\begin{equation}
   L( r, a) = { a \over a + r \coth(evr/2)}  \, .
\end{equation}
We see that there is a core region, of radius $\sim 1/ev$, outside of
which the massive fields fall exponentially.

The quantity $a$, which enters here as a constant of integration, can
take on any positive real value.  It has no effect on the energy, and
so the eighth zero mode evidently corresponds to variation of $a$.
Some physical understanding of $a$ can be obtained by examining the
fields outside the core region.  Let us assume, for the sake of
simplicity, that $a$ is much greater than the core radius.  We see
that $L\approx 1$ in the region $1/ev \ll r \ll a$, so that both $A$
and $G$ fall as $1/r$.  The $1/r^2$ part of the magnetic field is then
just that which would be produced by an isolated $\bbeta$-monopole,
corresponding to a magnetic charge with both Abelian and non-Abelian
components.  On the other hand, when $r \gg a$ we find that $L \sim
a/r$.  It follows that $G \sim 1/r^2$ and that the Coulomb part of the
magnetic field comes only from $A$ and is purely Abelian.  Thus, we
can view the solution as being composed of a massive
$\bbeta$-monopole, with a fixed core radius $\sim 1/ev$, that is
surrounded by a cloud of non-Abelian field of radius $\sim a$ that
shields the non-Abelian part of the magnetic charge
\cite{Abouelsaood:1983ft}.  This cloud, whose radius is apparently
arbitrary, can be seen as the manifestation of the massless monopole.

It is instructive to look at this solution from another point of view.
The case of SU(2)$\times$U(1) breaking can be viewed as a limit of the
maximally broken theory in which the Higgs vacuum expectation value
has been varied so that one of the fundamental monopole masses goes to
zero.  Thus, we can imagine starting with a solution of maximally
broken SO(5) containing two monopoles, one of each type, separated by
a distance $R$.  As we begin to restore SU(2) gauge symmetry, one of
the two monopoles begin to decrease in mass and grow in size. However,
the growth of this
would-be massless monopole is affected by its nonlinear coupling to
the other monopole.  When the radius of the lighter monopole becomes
of order $R$, this interaction prevents any further increase in size,
and the monopole evolves into the non-Abelian cloud \cite{Lu:1998br}.

A curious feature is that the limiting solution depends only on the
initial monopole separation, and not on the relative spatial
orientation of the two massive monopoles.  We will encounter this from
another viewpoint when we study the moduli space metric in
Sec.~\ref{so5metric}, where we will find that the angular spatial
coordinates of the maximally broken case are replaced by internal
symmetry variables when the symmetry breaking is nonmaximal.

As a final remark, note that we could also imagine starting with a (1,
1) solution of maximally broken SU(3) and taking a similar limit.  In
this case, which does not satisfy Eq.~(\ref{purelyabelian}), the
growth of the lighter monopole is not cut off by the presence of the
massive monopole, but instead continues until, in the massless limit,
the monopole has infinite radius but is essentially indistinguishable
from the vacuum \cite{Lu:1998br}.  Indeed, the limiting (1, [1])
solution that one obtains in this fashion is gauge-equivalent to the
(1, [0]) massive monopole.

\subsection{(1, [1], \dots, [1], 1) monopole solutions in SU($N$) broken to
U(1)$\times$SU($N-2$)$\times$U(1)}
\label{masslessoneoneone}

A somewhat more complicated example \cite{Weinberg:1998hn} is obtained
by considering SU($N$) broken to U(1)$\times$SU($N-2$)$\times$U(1),
with our notation indicating that the unbroken U(1)'s correspond to
the simple roots at the ends of the Dynkin diagram.  We will use the
notation introduced below Eq.~(\ref{suNasymBfield}) to indicate the
magnetic charges of a solution, with the only modification being that
massless monopoles will be indicated with a square bracket.  Thus, a
solution composed of one monopole of each type --- two massive and
$N-3$ massless, in all --- would be a (1, [1], \dots, [1], 1)
solution.

This solution is a limiting case of the (1, 1,
\dots, 1) solution whose Nahm data was obtained in
Sec.~\ref{oneoneonecase}.  For that solution, the range $s_1 < s <
s_N$ was divided into $N-1$ subintervals.  The $T_j$ were constant on
each of these intervals, with their values giving the locations ${\bf
x}^p$ of the corresponding monopoles.  There were also jumping data at
each of the interval boundaries, with the data at the boundary between
the $(p-1)$th and $p$th boundaries obeying
\begin{eqnarray}
      a^{(p)\dagger} \,  \bsigma \, a^{(p)} &=&
         2( {\bf x}^{p-1} - {\bf x}^p) \cr
  a^{(p)\dagger} \, a^{(p)}  &=& 2|{\bf x}^p - {\bf x}^{p-1}| \, .
\label{jumpAndX}
\end{eqnarray}
The (1, [1], \dots, [1], 1) solution corresponds to the limit in which
all but the first and last subintervals have zero width, so that
$s_2 = s_3 = \dots = s_{N-1}$.  The previously obtained Nahm data
are unaffected by this limit.

Going from the Nahm data to the spacetime fields involves solving for
the $w_a^{(p)}(s)$ within each interval, and then finding $S_a^{(p)}$
that satisfy the condition
\begin{equation}
      w_a^{(p)}(s_p) -  w_a^{(p-1)}(s_p) = -S_a^{(p)} a^{(p)} \, .
\label{zerowidthW}
\end{equation}
With the intermediate intervals reduced to zero width, the
corresponding $w_a^{(p)}(s)$ become simply numbers, rather than
functions.  Furthermore, they do not contribute to the spacetime
fields, since they only enter through integrals over a zero range.
Thus, the scalar field is given by
\begin{equation}
   \Phi^{ab}
       = \int_{s_1}^{s_2} ds \, s w^{(1)\dagger}_a (s) w^{(1)}_b(s)
 + \int_{s_2}^{s_N} ds \, s w^{(N-1)\dagger}_a (s) w^{(N-1)}_b(s)
          + s_2 \sum_{p=1}^{N-2} S_a^{(p)}{}^* S_b^{(p)} \, ,
\end{equation}
with similar simplifications occurring in the normalization integral,
Eq.~(\ref{jumpingOrtho}), and in the expression for the gauge field,
Eq.~(\ref{jumpingA}).  Hence, the $N-2$ constraints implied by
Eq.~(\ref{zerowidthW}) effectively reduce to the single constraint
\begin{equation}
      w_a^{(N-1)}(s_2) -  w_a^{(1)}(s_2)
           = - \sum_{p=1}^{N-2} S_a^{(p)} a_p \, .
\end{equation}

\begin{figure}[t]
\begin{center}
\scalebox{1.0}[1.0]{\includegraphics{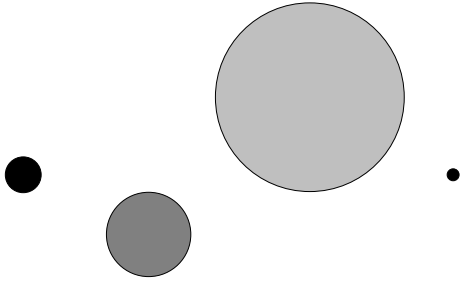}}
\par
\vskip-2.0cm{}
\end{center}
\begin{quote}
\caption{\small As the unbroken symmetry becomes enlarged, some of
the fundamental monopoles become very light and grow in its size.  In
this figure, we show four distinct monopoles for SU(5) maximally
broken to ${\rm U}(1)^4$. As the limit of unbroken ${\rm U}(1)\times
{\rm SU}(3)\times {\rm U}(1)$ is approached, the two middle monopoles
become light, large and dilute. Eventually the massless monopoles lose
their individual identities and merge into a single monopole cloud, as
illustrated in Fig.~\ref{cloud-ellipse}.
\label{notyetcloud}}
\end{quote}
\end{figure}

\begin{figure}[t]
\begin{center}
\scalebox{1.0}[1.0]{\includegraphics{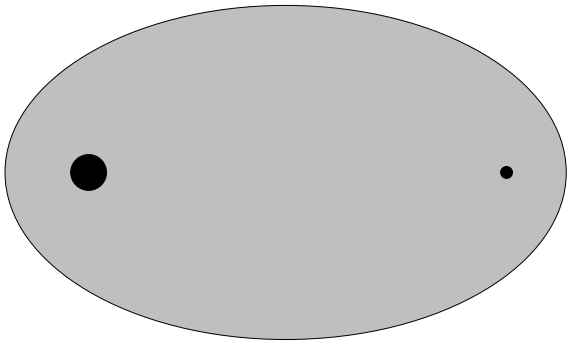}} \par
\vskip-2.0cm{}
\end{center}
\begin{quote}
\caption{\small This figure shows the monopoles of
Fig.~\ref{notyetcloud} in the limit where the unbroken symmetry is
${\rm U}(1)\times {\rm SU}(3)\times {\rm U}(1)$. The two remaining
massive monopoles are denoted by black circles. The two massless
monopoles have turned into a cloud of non-Abelian field that surrounds
the two massive monopoles and screens their non-Abelian magnetic
charge.  The cloud parameter $b$ measures the size of this ellipsoidal
cloud. When $b$ has the minimum allowed value, equal to the the
separation between the two massive monopoles, the cloud merges
completely into the massive monopoles.
\label{cloud-ellipse}}
\end{quote}
\end{figure}

Examining these last two equations, we see that the
substitution
\begin{eqnarray}
        S_a^{(p)}  &\rightarrow& \tilde S_a^{(p)}
               = U_{pq} S_a^{(q)}   \cr
        a_p  &\rightarrow& \tilde a_p
               = U_{pq} a_q
\end{eqnarray}
with $2 \le p,q \le N-1$ and $U$ an $(N-2)\times(N-2)$ unitary matrix,
has no effect on the spacetime fields. [This is a reflection
of the additional unbroken SU($N-2$) gauge symmetry.]  However, the
changes in the $a^{(p)}$ would, through Eq.~(\ref{jumpAndX}), imply
changes in the ${\bf x}_p$ for $2 \le p \le N-2$, leaving invariant
only the quantities
\begin{equation}
    \sum_{p=2}^{N-2} {\bf x}^{p-1} - {\bf x}^p
                = {\bf x}^1 - {\bf x}^{N-1} \equiv {\bf R}
\end{equation}
and
\begin{equation}
    \sum_{p=2}^{N-2} |{\bf x}^{p-1} - {\bf x}^p|
        \equiv b       \, .
\label{bDef}
\end{equation}
Thus, the individual massless monopole positions lose their
significance and together yield a single gauge invariant quantity,
$b$.  Note that $b \ge R$, where $R=|{\bf R}|$ is the distance between
the two massive monopoles.  (See Figs.~\ref{notyetcloud} and
\ref{cloud-ellipse}).

The physical significance of $b$ becomes clear once the spacetime
fields are obtained from the Nahm data.  Let $y_L$ and $y_R$ be the
distances from a given point to the two massive monopoles.  In the
region outside the massive monopole cores, but with $y_L + y_R \ll b$,
the long-range parts of the magnetic and scalar fields have both
Abelian and non-Abelian components and are just what would they would
have been if only the two massive monopoles were present.  On the
other hand, in the region where $y_L + y_R \gg b$ only the Abelian
parts of the long-range fields survive.  Thus, the effect of the
massless monopole(s) is to create an ellipsoidal cloud that, like the
cloud in the
SO(5) example, shields the non-Abelian magnetic charges.  The size of
this cloud is measured by the ``cloud parameter'' $b$.

In the SU(4) case, there are twelve zero modes.  Six correspond to the
massive monopole positions and five to the global gauge modes of the
unbroken subgroup.  The one remaining zero mode corresponds to
variations of $b$.  More generally, for $N > 4$ there are $4(N-1)$
zero modes, with six again corresponding to the massive monopole
positions and one to the cloud parameter.  The leaves $4N-11$, which
is smaller than the dimension of the unbroken gauge group.  This is
explained by realizing that the solutions for $N>4$ are actually
embeddings of SU(4) solutions.  Hence, for any given solution
there is a U($N-4$) subgroup that leaves the solution invariant and
does not give rise to any zero modes.  The number of global gauge
modes is therefore ${\rm dim}\,[{\rm U(1)}\times{\rm
SU}(N-2)\times{\rm U(1)}] - {\rm dim}\,[{\rm U}(N-4)] = 4N-11$, which
accounts for all the remaining zero modes.

Thus, although going to a larger group brings in additional massless
monopoles, it does not give any additional gauge invariant parameters.
Indeed, the spacetime fields themselves are essentially unchanged as
the group is enlarged.  We see one cloud, even though there are $N-3$
massless monopoles.

\subsection{(2, [1]) solutions in SU(3) broken to
SU(2)$\times$U(1)}
\label{DancerSoln}

The last example \cite{dancer-nonlin} we will consider in detail is
that of one massless and two massive monopoles\footnote{For earlier
work on this SU(3) case, see Ref.~\cite{Ward:1981qq}.  The closely related
Nahm construction of Sp(4) solutions with one massless and two massive monopoles
is described in Ref.~\cite{Lee:1997ny}.}
 for SU(3) broken to
SU(2)$\times$U(1).  The Nahm data for these can be obtained directly
from the results in Sec.~\ref{twoOneInSU3}, where we treated the (2,
1) solutions of maximally broken SU(3).  For the latter case, the Nahm
data consists of a triplet of $2\times 2$ matrices $T_i^L(s)$ on the
interval $s_1<s<s_2$ and a triplet of constants $t_i^R$ corresponding
to the interval $s_2 < s <s_3$.  The form of the $T_i^L(s)$ was given
in Eq.~(\ref{twoPlusOneT}), and the matching conditions at $s=s_2$
required that $t_i^R$ be equal to the 22 component of $T_i^L(s_2)$.

These data continue to satisfy the Nahm equation when the interval
$(s_3 - s_2) \rightarrow 0$.  However, a new symmetry appears in the
limit.  When the interval has zero width, the construction equation
solutions on that interval, $w_a^{(2)}(s)$, make no contribution to
the spacetime fields.  Now suppose that we were to apply an SU(2)
gauge action to the $T_i^L(s)$.  The resulting redefinition of basis
would change their 22 components, and thus the $t_i^R$, and so would
not be an invariance of the maximally broken theory.  However, since
the $t_i^R$ only enter the construction in the determination of the
$w_a^{(2)}(s)$, this gauge action has no effect on the spacetime
fields when the breaking is to SU(2)$\times$U(1).  Hence, there is no
loss of generality in using this gauge action to rotate the $\tau_j'$
of Eq.~(\ref{twoPlusOneT}) into the standard Pauli matrices $\tau_j$,
and writing
\begin{equation}
    T_i^L(s) = {1 \over 2}
           \sum_{j} A_{ij} f_j(s -s_1; \kappa, D) \tau_j
           + R_i {\rm I}_2 \, .
\end{equation}
The vector $\bf R$ and the orthogonal matrix $A_{ij}$ correspond to
spatial translations and spatial rotations, respectively, of the
solution.  They contain six independent parameters which, when taken
together with the four global gauge parameters that do not enter the
Nahm data, leave only two non-symmetry parameters.  These are the
quantities $\kappa$ and $D$, which must satisfy $0 \le \kappa \le 1$
and
\begin{equation}
      0 \le D(s_2 - s_1) <  2 K(\kappa) \, .
\label{boundonDandKappa}
\end{equation}

Our experience with the (2, 1) solution suggests that, for values
large compared to the massive monopole core radius, $D$ should
correspond to the separation between the massive monopoles.  Further,
we might guess that
\begin{equation}
    r = {D \over 2[2 K(\kappa) - D (s_2- s_1)]} \, ,
\end{equation}
which gave the distance of the $\bbeta_2$-monopole from the
center-of-mass of the $\bbeta_1$-monopoles in the (2, 1) case, would
specify the size of a non-Abelian cloud similar to that found in the
previous two examples.  In the limit where $D$ and $r$ are both large,
these interpretations are borne out by analysis of asymptotic cases
and examination of numerical solutions \cite{Dancer:zx,Irwin:1997ew}.
(In particular, note that for limiting case $r \rightarrow \infty$ the
Nahm data has a pole at $s_2$, and the solution is an embedding of the
SU(2) two-monopole solution into SU(3), as should be expected when the
non-Abelian cloud becomes infinite in size.)

Although the generic (2, [1]) solution has no rotational symmetry,
there are two special cases that are axially symmetric.  In both, the
spacetime fields can be obtained explicitly \cite{dancer-nonlin}. If
$\kappa=1$, the elliptic functions become hyperbolic functions, and
\begin{eqnarray}
   f_1(s -s_1; \kappa, D) &=& f_2(s -s_1; \kappa, D)
        = - {D \over \sinh(Ds) }  \cr
   f_3(s -s_1; \kappa, D) &=& - D \coth(Ds) \, .
\label{hyperbolicSolution}
\end{eqnarray}
For large $D$ these ``hyperbolic solutions'' correspond to a pair of
massive monopoles, separated by a distance $D$, that are surrounded by
a massless monopole cloud of minimum size.

The ``trigonometric solutions'' are obtained by setting
$\kappa=0$, so that
\begin{eqnarray}
   f_1(s -s_1; \kappa, D) &=&  - D \cot(Ds)   \cr
   f_2(s -s_1; \kappa, D) &=& f_3(s -s_1; \kappa, D)
      = - {D \over \sin(Ds) } \, .
\label{trigSolution}
\end{eqnarray}
Because $K(0) = \pi/2$, Eq.~(\ref{boundonDandKappa}) implies that $D <
\pi/(s_2-s_1)$, so the cores of the two massive monopoles must overlap in
this case.  In fact, examination of the solutions suggests that they
can be interpreted as two coincident massive monopoles
surrounded by a
massless cloud that varies from minimal size to infinite radius as
$D$ ranges over its allowed values.

Finally, if $D=0$ the elliptic functions become
independent of $\kappa$.  The hyperbolic and trigonometric solutions
then coincide and yield a spherically symmetric solution with
$f_1=f_2=f_3 = 1/s$.

\subsection{Multicloud solutions}

The three examples above all had a single non-Abelian cloud.  This
remained true even if there were several massless monopoles, as in the
SU($N$) solutions of Sec.~\ref{masslessoneoneone} with $N>4$.
However, this is not necessarily the case.  Solutions with multiple
clouds can be obtained \cite{Houghton:2002bz} by considering the same
breaking of SU($N$) as in Sec.~\ref{masslessoneoneone}, but choosing
the magnetic charges to be (2, [2], \dots, [2], 2); for
simplicity,\footnote{The solutions for $N > 6$ are essentially
embeddings of those for SU(6), while those for SU(4) and SU(5) can be
viewed as constrained SU(6) solutions.} we will assume that $N \ge 6$.
These solutions include, as a special case, ones that are essentially
combinations of two disjoint (1, [1], \dots, [1], 1) solutions, each
with its own massless cloud enclosing a pair of massive monopoles.
However, in the generic solution the massive monopoles are not paired
up in this fashion.  Instead, there is a somewhat more complex
structure.  For each of the massive species of monopole, there is a
massless cloud surrounding two identical monopoles [essentially, a
copy of the (2, [1]) solution of Sec.~\ref{DancerSoln}].  These two
(2, [1]) structures can either overlap or be disjoint, but in either
case are enclosed by two other clouds, one nested within the other.
There are thus a total of four clouds (although there are at least six
massless monopoles).  There is an independent size parameter for each
of these clouds, and in addition there are parameters that specify the
relative group orientations of the various clouds.  For a more
detailed description of these solutions, see
Ref.~\cite{Houghton:2002bz}.

\section{Moduli space metrics with massless monopoles}
\label{masslessmodulimetric}

Just as in the case of maximal symmetry breaking, one can define a
metric on the moduli space. Provided that the net magnetic charge is
purely Abelian and satisfies Eq.~(\ref{purelyabelian}), this metric is
a smooth limit of the moduli space metric for the corresponding
solutions with maximal symmetry breaking.  As examples of such
metrics, we will consider in this section the three
single-cloud solutions with clouds that were described in detail in
the previous section.  In the next section we will discuss the
application of these metrics to the study of monopole dynamics.

\subsection{SO(5) solutions with one massive monopole and one massless
monopole}
\label{so5metric}

We start by returning to the SO(5) example considered in
Sec.~\ref{so5solution}.  Regardless of whether the unbroken group is
${\rm U}(1)\times{\rm U}(1)$, with two massive monopoles, or ${\rm
SU}(2)\times{\rm U}(1)$, with one massless monopole and one
massive monopole, the moduli space is eight-dimensional.  In the
maximally broken case, the moduli space splits into a flat
four-dimensional space, spanned by the three center-of-mass
coordinates and the overall U(1) gauge angle, and a four-dimensional
relative moduli space whose metric ${\cal G}_{\rm rel}$ takes the
Taub-NUT form given in Eq.~(\ref{taubnutrelativemetric}).  Three of
the coordinates on this relative moduli space are naturally
interpreted as specifying the relative positions of the two monopoles,
while the fourth can be taken to be the U(1) angle $\psi$ defined by
Eq.~(\ref{CMandRelAngles}).  We can now go to the nonmaximally broken
case by taking one of the monopole masses to zero.  In this limit the
reduced mass $\mu$ vanishes and the metric becomes
\begin{equation}
{\cal G}_{\rm rel} \longrightarrow {\cal G}_{\rm rel}(\mu=0)=
\frac{2\pi\lambda}{e^2} \left(\frac{1}{r}\,d{{\bf r}}^2
+ r\,[d{\psi}+ {\bf w}({\bf r})\cdot d{{\bf r}}]^2\right) \, .
\label{relativeflatmetric}
\end{equation}

Alternatively, we can exploit the fact that this is the one nontrivial case
where we have a complete family of explicit classical solutions.  From
these we can obtain the background gauge zero modes and then use the
defining Eq.~(\ref{metricdefinition}) to get the metric \cite{Lee:1996vz}.  Thus,
varying the cloud parameter $a$ in the expressions given in
Sec.~\ref{so5solution} gives a zero mode, which happens to already
satisfy the background gauge conditions.  The three other zero modes
can be obtained from infinitesimal SU(2) transformations or,
more easily, by utilizing the local quaternionic symmetry on the
moduli space and applying the transformations of
Eq.~(\ref{transformingzeromodes}).  The metric obtained by this
procedure has precisely the same form as that in
Eq.~(\ref{relativeflatmetric}), but with $r$ replaced by $a$ and the
three gauge SU(2) Euler angles replacing the spatial angles $\theta$
and $\phi$ associated with $\bf r$ and the U(1) phase angle $\psi$.

Note that this limit of the Taub-NUT manifold is actually
a flat $R^4$. Mapping to the usual Cartesian
coordinates via
\begin{equation}
w+iz= \sqrt{r}\;\cos(\theta/2) \,e^{-i(\phi+\psi)/2}, \qquad
x+iy= \sqrt{r}\;\sin(\theta/2)\, e^{-i(\phi-\psi)/2} \, ,
\end{equation}
transforms the metric of Eq.~(\ref{relativeflatmetric}) to the
manifestly flat form
\begin{equation}
  {\cal G}_{\rm rel} =
  \frac{8\pi\lambda}{e^2}\left(dw^2+ dx^2+dy^2+dz^2\right) \, .
\end{equation}
The isometry of the Taub-NUT metric is enhanced to ${\rm SO}(4)= {\rm
SU}(2)\times {\rm SU}(2)$ in this limit.  The first SU(2) is the
rotational isometry that was always there, whose action on the
classical solution becomes trivial in the massless limit, while the
second SU(2) is the gauge isometry, enhanced from the U(1) triholomorphic
isometry of Taub-NUT.  This is consistent with the the well-known fact
that when we pick a particular hyper-K\"ahler structure on $R^4$, only
one of the two SU(2)'s becomes triholomorphic, while the other rotates
the three K\"ahler structures.\footnote{This SO(5) (1, [1]) solution
can be extended, by embedding, to a (1, [1], \dots [1])
solution with one massive and $N$
massless monopoles in a theory with Sp($2N+2$) broken to
U(1)$\times$Sp($2N$).  The above derivation of the moduli space metric
is readily generalized, and one finds that the relative moduli space is
$R^{4N}$ \cite{Lee:1996vz}.}

It was important here that we were dealing with a system whose
magnetic charge was purely Abelian.  If we had started out with the
(1, 1) solutions of maximally broken SU(3), we would have obtained the
same flat $R^4$ relative moduli space in the massless limit.  However,
as we have already noted, the classical (1, 1) solutions do not behave
smoothly in this limit: the massless monopole expands without bound
\cite{Lu:1998br}, and the (1, [1]) solution that is obtained in
the limit is gauge equivalent to the (1, [0]) solution which, having
only a single monopole, has no relative moduli space.

Furthermore, the triholomorphic SU(2) isometry of the moduli space
metric cannot possibly correspond to the enhanced unbroken gauge
symmetry.  The first indication of this is the fact that the
long-range tail of the solution, which is not invariant under the
SU(2), would naively seem to give an infinite moment of inertia for
SU(2) gauge rotations.  (On closer inspection
\cite{Abouelsaood:1982dz}, one finds that the moment of inertia
actually vanishes, an equally troubling result.)  A deeper problem
emerges on closer inspection.  The long-range tails of the non-Abelian
components of the fields produce a topological obstruction that makes
it impossible to find a basis for the unbroken SU(2) that is smooth
over all of space (or even over the sphere at spatial infinity).  As a
result, one cannot even define an action of this gauge SU(2) on the
moduli space
\cite{Nelson:bu,Balachandran:1982gt,Abouelsaood:1983gw,Nelson:1983fn}.
This obstruction, which is sometimes referred to as the global color
problem, only arises when the magnetic charge has a non-Abelian component, and
is absent when Eq.~(\ref{purelyabelian}) is satisfied.

\subsection{SU($N$) (1, [1], \dots [1], 1) solutions}

We next turn to the case of the (1, [1], \dots [1], 1) SU($N$)
solutions that were described in Sec.~\ref{masslessoneoneone}.  As was
noted there, these solutions can be obtained as a limiting case of the
(1, 1, \dots, 1) solutions of the maximally broken theory.  Thus, we
should be able to obtain the moduli space metric by taking the
appropriate limit of the metric of Eq.~(\ref{C}).

The $(4N-4)$-dimensional moduli space
splits into a four-dimensional center-of-mass part and a
$(4N-8)$-dimensional
relative part.
As in the SO(5) example, only the latter part is
affected by the enhanced symmetry.  In fact, the only effect on the
metric comes through the reduced mass matrix $\mu_{AB}$.  Computing
this first with nonvanishing masses $m_i$, and
then taking the middle $N-3$ masses to zero, we find that all of its
components are equal; i.e.,
\begin{equation}
  \mu_{AB}=\bar \mu\equiv \frac{m_1m_{N-1}}{m_1+m_{N-1}}
\end{equation}
for all $A$ and $B$ from 1 to $N-2$. If we set the root lengths to
unity, so that $\lambda_A=1$ for all pairwise interactions in the
SU($N$), and then eliminate the coupling constant factors by rescaling
the intermonopole separations ${\bf r}_A$ and the metric itself, the
metric of Eq.~(\ref{C}) becomes \cite{Lee:1996vz}
\begin{eqnarray}
   {\cal G}_{\rm rel}
   &=& C_{AB}\,d{\bf r}_A\cdot d{\bf r}_B\nonumber \\
  && + (C^{-1})_{AB}\,
   \left[d\psi_A+{\bf w}({\bf r}_A)\cdot d{\bf r}_A \right]
  \left[d\psi_B+{\bf w}({\bf r}_B)\cdot d{\bf r}_B \right]
\label{taubiancalabi}
\end{eqnarray}
with
\begin{equation}
{\cal C}_{AB}=\bar \mu+\delta_{AB}\,\frac{1}{ r_A}  \, .
\end{equation}

We now want to show that this metric remains well-behaved in the limit
of enhanced symmetry breaking, even though the reduced mass matrix is
now degenerate.  Again, the simplest way to show this is by realizing
the metric via a hyper-K\"ahler quotient. We start with the $H^{N-2}\times
H$ spanned by the quaternions
\begin{equation}
    q^A=a^Ae^{i\chi^A/2}\, , \quad A=1,2,\dots , N-2 \, ;
    \qquad t=\frac{1}{\bar \mu}\,y_0+iy_1+jy_2+ky_3
\end{equation}
and with the flat metric
\begin{equation}
   ds^2 =  \bar\mu\, dt\otimes_s d\widehat t
         +\sum_A dq^A\otimes_s d \widehat q^A  \, .
\end{equation}
We take the quotient using the symmetry
\begin{equation}
    \chi^A\rightarrow \chi^A+2\theta \, ,
      \qquad y_0\rightarrow y_0+\theta \, ,
\end{equation}
whose moment map is
\begin{equation}
  \nu = \frac12\left[\sum_A q^Ai\widehat q^A
           + (t-\widehat t )\right] \, .
\end{equation}
Proceeding as in Sec.~\ref{exactformanydistinct}, we identify
\begin{equation}
       q^Ai\widehat q^A=ir^A_1+jr^A_2+kr^A_3
\end{equation}
and
\begin{equation}
\frac{\partial}{\partial\psi^A}=\frac{\partial}{\partial\chi^A}
\end{equation}
and set both the moment map and
\begin{equation}
    K = 2\sum_A\frac{\partial}{\partial\chi^A}
     +\frac{\partial}{\partial y_0}
\end{equation}
to zero.  This produces the smooth metric of
Eq.~(\ref{taubiancalabi}), known as the Taubian-Calabi
metric, as the hyper-K\"ahler quotient.

We expect this geometry to have both an SU(2) rotational isometry and
a ${\rm U}(N-2) = {\rm SU}(N-2)\times {\rm U}(1)$ gauge isometry. The
rotational isometry was already present in the maximally broken case,
and so should remain in the massless limit as well.  To see how the
triholomorphic gauge isometry emerges in
the massless limit, we note that
the hyper-K\"ahler structure of $H^{N-2}\times H$
is invariant under right multiplication,
\begin{equation}
    q^A\rightarrow q^Bp_B{}^A  \, ,
\end{equation}
by any quaternionic matrix $p$ such that
\begin{equation}
   \sum_C p_A{}^C\hat p_B{}^C =\delta_{AB} \, .
\end{equation}
Of this invariance, only the part involving matrices that
commute with the action of the
hyper-K\"ahler quotient procedure survives as a
triholomorphic isometry of the Taubian-Calabi metric.
This eliminates the matrices that involve either $j$ or $k$,
leaving only complex unitary matrices $p$, and thus
a ${\rm U}(N-2)$ triholomorphic isometry, just as expected.

This ${\rm U}(N-2)$  leaves the separation vector
\begin{equation}
    {\bf R} = \sum_A {\bf r}_A = \sum_A q^Ai\hat q^A = -t+\hat t
\end{equation}
invariant.  Further, by using the fact that $Q\hat Q= \hat Q Q$
is real for any quaternion $Q$, it is easy to show that $|Qi \hat Q|^2
= |Q \hat Q|^2 = (Q \hat Q)^2$.  It then follows that the cloud
parameter $b$ defined in Eq.~(\ref{bDef}) can be written as
\begin{equation}
   b= \sum_A |{\bf r^A}| = \sum_A |q^Ai\hat q^A|
            = \sum_A q^A \hat q^A  \, .
\end{equation}
It is clear from the last expression that $b$ is also invariant under
the ${\rm U}(N-2)$ isometry.

In this language the rotational SU(2) is realized in terms
of unit quaternions $u$ ($u\hat u=\hat u u = 1$) via
\begin{equation}
  q^A\rightarrow uq^A \, ,\qquad t\rightarrow ut\hat u  \, .
\end{equation}
Under this SU(2), $ {\bf R}$ rotates as
a triplet, while $b$ is invariant, just as expected.

The relative metric of Eq.~(\ref{taubiancalabi}) can be rewritten in
an alternative form, expressed in terms of $b$, $R=|{\bf R}|$, and
$4N-10$ angular and group orientation variables, that proves to be
quite useful for studying the actual dynamics of the monopoles
\cite{Chen:2001ge}.  The solutions with fixed $b$ and $R$ lie on
$(4N-10)$-dimensional orbits in the relative moduli space that are
defined by the action of the rotational and gauge symmetries of the
theory.  Locally, these orbits are
\begin{equation}
   M_{4N-10} = \frac{{\rm SU}(2)\times {\rm U}(1)
    \times {\rm SU}(N-2)}{{\rm SO}(2)\times {\rm U}(N-4)}  \, .
\end{equation}
Here the SU(2) is the rotational symmetry, while the ${\rm U}(1)
\times {\rm SU}(N-2)$ is the unbroken gauge symmetry with the overall
center-of-mass U(1) symmetry factored out.  As we noted previously,
for any given solution
there is a U($N-4$) subgroup of the gauge symmetry that leaves the
solution invariant.  In addition, the action of the rotational SU(2)
mixes with the gauge symmetry (something that is not unusual for
monopoles) in such a way that there there is one combination of
a rotation about $\bf R$ and a gauge rotation that leaves the solution
invariant; this leads to the SO(2) factor in the denominator.

This suggests defining a natural basis as follows.
We can always
compute a one-form $\lambda_v$ dual to any isometry generator
$v$ by contracting with the metric
\begin{equation}
   \lambda_v =ds^2(\cdot,v)  \, .
\end{equation}
This $\lambda_v$ can be thought of as the associated conserved
momentum, in the sense that the time-derivative $\lambda_v/dt$ is the
conserved quantity.  In this way we can construct the ``conserved"
one-forms from the ${\rm SU}(2)\times {\rm U}(1)\times {\rm
SU}(N-2)$ symmetry.
As in the simpler two-monopole case we considered in
Sec.~\ref{trajectories}, what we actually need are not the ``space
components'' of these quantities, defined relative to axes that are fixed in
(real or gauge) space, but rather the ``body-frame'' components that
are defined with respect
to axes that move with the monopole configuration.  These can
be organized nicely as follows:
\begin{itemize}
\item{} Rotational symmetry gives three angular momentum  components $J_s$.
Although these body-frame components are not individually conserved
in general,
angular momentum conservation does imply that
\begin{equation}
    J^2 = J^2_1 + J^2_2 + J^2_3
\end{equation}
is constant.
Furthermore, if the body axes are chosen so that  ${\bf R}=(0,0,R)$,
the moments of inertia for $J_1$ and $J_2$ are equal so, as in
a symmetric top, $J_3$ is conserved.

\item{} The U(1) gauge isometry leads to a conserved
quantity $Q$ that is the relative electric charge of the two massive monopoles.

\item{} The unbroken gauge group gives both a triplet $T_s$,
corresponding to the SU(2) subgroup defined by the decomposition ${\rm
SU}(N-2)\rightarrow {\rm SU}(2)\times {\rm U}(N-4)$, and a set of
$2N-8$ complex (or $4N-16$ real)
components $\tau^1_\alpha$ and
$\tau^2_\alpha$ ($\alpha=1,2,\dots, N-4$) that correspond to the off-diagonal
components in the same decomposition.  The conserved quadratic Casimir
is
\begin{equation}
    T^2 = T^2_1 + T^2_2 + T^2_3
    + \sum_\alpha [\tau^1_\alpha (\tau^1_\alpha)^* +
               \tau^2_\alpha (\tau^2_\alpha)^* ]  \, ,
\end{equation}
where the U($N-4$) terms that vanish identically have been omitted.

\end{itemize}

Finally, because of the mixing between the gauge rotation and spatial
rotation, there is one identity among the above,
\begin{equation}
     J_3 = T_3  \, ,
\end{equation}
leaving us with a total of $4N-10$ basis one-forms, as required. These,
together with $dR$ and $db$, constitute a complete basis.

In terms of these quantities, and with the rescaling of lengths undone,
the metric takes the form
\begin{equation}
ds^2= \bar\mu\, dR^2 + \frac\kappa2\,\left[\frac{(db+dR)^2}{(b+R)}
+\frac{(db-dR)^2}{(b-R)}\right]+ ds^2_{\rm angular}
\label{bRanglemetric}
\end{equation}
where $\kappa = 2\pi/e^2$ and
\begin{eqnarray}
ds^2_{\rm angular}&=&\frac{1}{R^2(\kappa+\bar \mu b)}\sum_{s=1,2}
   \left[bJ_s^2+ \left(b+{\bar \mu R^2\over \kappa} \right) T_s^2
          -2\sqrt{b^2-R^2}J_sT_s\right]
     \nonumber \\
    &+&{\bar\mu\over \kappa^2}  Q^2+\frac{1}{\kappa(b^2-R^2)}
        \left[b(J_3^2+Q^2)+2RJ_3Q\right]
    \nonumber \\
     &+& {4\over \kappa}\sum_{\alpha=1}^{N-4}
  \left[\frac{\tau^1_\alpha (\tau^1_\alpha)^*}{b+R}+
    \frac{\tau^2_\alpha (\tau^2_\alpha)^*}{b-R}\right]   \, .
\label{fullsuNcloudmetric}
\end{eqnarray}
This representation of the metric is useful because, by fixing
the conserved quantities, one can obtain an
effective Lagrangian involving only $R$ and $b$ only.

In particular, trajectories with $J^2 = T^2 = Q$ lie on the
two-dimensional quotient space $\cal Y$ that is obtained by dividing
the relative moduli space by the group of rotations and unbroken gauge
symmetries.  The metric $ds_{\cal Y}^2$ on this space, which is given by the
first two terms in Eq.~(\ref{fullsuNcloudmetric}), has an apparent
singularity at $b=R$.  This singularity is not physical and
can be eliminated by defining
\begin{equation}
   \matrix{ x = \sqrt{\kappa} \left[ \sqrt{b+R} + \sqrt{b-R}\right]
\cr\cr
    y = \sqrt{\kappa} \left[ \sqrt{b+R} - \sqrt{b-R}\right] }
                \, \qquad\qquad  0 \le y \le x \,  ,
\label{firstoctant}
\end{equation}
in terms of which the metric is
\begin{equation}
   ds^2_{\cal Y} = dx^2 + dy^2
     + {\mu \over 4 \kappa^2} (x\,dy + y\, dx)^2  \, .
\label{xymetric}
\end{equation}
This definition maps the entire physical range $0 \le R \le b <
\infty$ to the octant $0 \le y \le x < \infty$.  This octant is
bounded by the $x$-axis, corresponding to $R=0$ (i.e., solutions in
which the massive monopoles coincide) and by the line $x=y$,
corresponding to $b=R$ (i.e., solutions with minimal cloud size).
It is not
geodesically complete, because geodesics can reach the
boundaries in finite time.\footnote{This is a reflection of the fact
that this quotient space is not a manifold, because some solutions ---
those lying on the boundaries of the octant --- have an enlarged
invariance group.}  A geodesically complete space can be obtained
by extending the definitions
of $x$ and $y$ outside their original range by appropriate changes in
signs.  For example, in the octant $ 0 \le -y \le x < \infty$ we define
\begin{equation}
   \matrix{ x = \sqrt{\kappa} \left[ \sqrt{b-R} + \sqrt{b+R}\right]
    \cr\cr
    y = \sqrt{\kappa} \left[ \sqrt{b-R} - \sqrt{b+R}\right] }
     \, \qquad\qquad  0 \le -y \le x \, \,.
\end{equation}
A trajectory crossing the $x$-axis then corresponds to one in which
the two massive monopoles approach head-on, meet, and then pass
through each
other.  Proceeding in a similar fashion in the remaining octants
gives an eightfold mapping of the $b$-$R$ moduli space onto the
$x$-$y$ plane.

\subsection{SU(3) (2, [1]) solutions}

In the previous two examples the moduli space metric was obtained
either directly from the explicit solutions or by taking the massless
limit of a known metric that had previously been obtained by more
indirect means.  Neither of these options is available to us when we
turn to the SU(3) solutions with one massless and two massive
monopoles that we described in Sec.~\ref{DancerSoln}.  Instead, we
will quote the results of Dancer \cite{Dancer:kn}, who obtained the
metric as the metric on the space of Nahm data.\footnote{The metric
for the maximally broken (2, 1) solutions was subsequently determined
\cite{Houghton:1999qu}, but again by approaching the problem through
the Nahm data.}

The relative moduli space is eight-dimensional, with solutions of
fixed $\kappa$ and $D$ lying on six-dimensional orbits generated by the
rotational SO(3) and unbroken gauge SU(2) groups.  Rather than display
the full expression for the metric, we will focus on the
two-dimensional space, which we will again denote ${\cal Y}$, that is
obtained by taking the quotient by these symmetry groups.  It is
convenient to replace $\kappa$ and $D$ by the variables
\begin{eqnarray}
      x &=& (2 - \kappa^2)D^2
      = f_3^2(u; \kappa, D) + f_2^2(u; \kappa, D)
                - 2 f_1^2(u; \kappa, D) \cr \cr
    y &=& - \sqrt{3} \kappa^2 D^2
      =  - \sqrt{3}\, \left[f_3^2(u; \kappa, D) - f_2^2(u; \kappa, D)
     \right]
\end{eqnarray}
where the $f_j(u; \kappa, D)$ are the top functions defined in
Eq.~(\ref{topfunctions}).  (Note that the combinations of these functions
that appear on the right-hand side are independent of $u$.)

\begin{figure}[t]
\begin{center}
\hskip -1cm\scalebox{0.4}[0.4]{\includegraphics{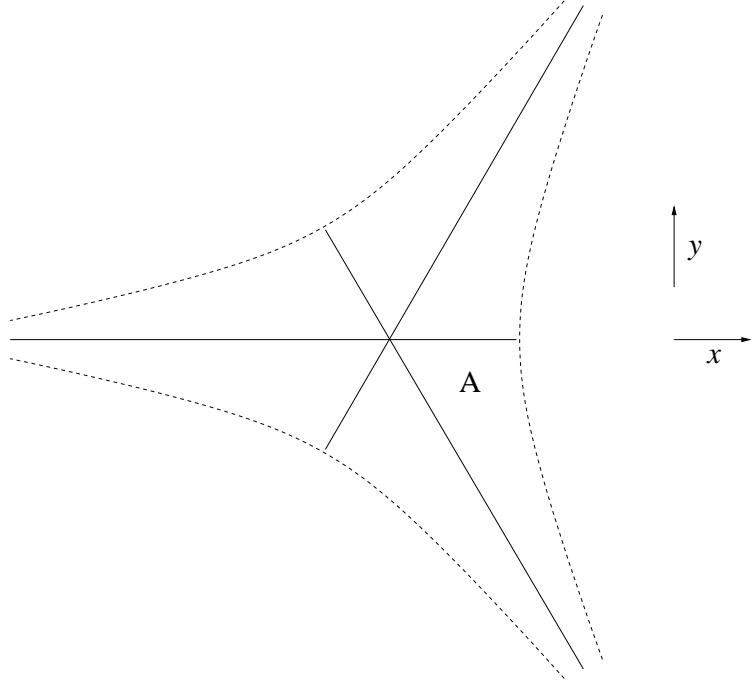}} \par
\vskip-2.0cm{}
\end{center}
\begin{quote}
\caption{\small The geodesically complete space containing six copies
of the quotient space $\cal Y$ for the (2, [1]) solutions for SU(3)
broken to SU(2)$\times$U(1).
The long straight lines correspond to the axially symmetric hyperbolic solutions,
and the short line segments to the trigonometric solutions.
The curved boundaries, which are not included in the space, correspond to the
solutions with two massive, but no massless, monopoles.}
\label{DancerFig}
\end{quote}
\end{figure}

The allowed values of $\kappa$ and $D$ are then mapped onto
region A in
Fig.~\ref{DancerFig}, and the metric is (up to an overall constant)
\begin{equation}
    ds^2_{{\cal Y}} = H
        \left[\sqrt{3} (g_1 + g_2) dx + (g_1 - g_2) dy \right]^2
      + g_1 (\sqrt{3}\, dx + dy)^2 +  g_2 (\sqrt{3}\, dx - dy)^2
\end{equation}
where
\begin{eqnarray}
   H(x,y) &=& f_1(s_2-s_1; \kappa, D) \, f_2(s_2-s_1; \kappa, D)
     \,f_3(s_2-s_1; \kappa, D)  \cr \cr
  g_1(x,y) &=&  \int_0^{s_2-s_1} du \, {1 \over f_2^2(u; \kappa, D) }
          \cr \cr
  g_2(x,y) &=&  \int_0^{s_2-s_1} du \, {1 \over f_3^2(u; \kappa, D) }  \, .
\end{eqnarray}

The long and short straight lines bounding
region A correspond to the axially symmetric hyperbolic and
trigonometric solutions of Eqs.~(\ref{hyperbolicSolution}) and
(\ref{trigSolution}), respectively.
The curved boundaries corresponds to the (2, 0)
solutions, which are actually embeddings of the SU(2) two-monopole
solutions of Atiyah and Hitchin.  This boundary is geodesically
infinitely far from any point in the interior and so is not actually
part of ${\cal Y}$.

A geodesically complete manifold can be
obtained, in a procedure similar to that used in
the previous example,
by mapping six copies of ${\cal Y}$ (corresponding to the six
possible orderings of the $f_j$) onto the $x$-$y$ plane, as shown in
Fig.~\ref{DancerFig}.  Points far out on the legs correspond to
configurations with two well-separated monopoles, with the three legs
corresponding to three perpendicular axes of separation.  The boundary
curves are the geodesics for the SU(2) two-monopole solutions, and
thus illustrate the $90^{\circ}$ scattering angle for head-on
collisions that was discussed in Sec.~\ref{trajectories}.

\section{Geodesic motion on the moduli space}
\label{cloudtrajectories}

Having found the metric on the moduli space, we can now investigate
the interactions of the massive and massless monopoles by studying
the geodesic motions.

We start with a particularly simple case, the SO(5) example that was
discussed in Sec.~\ref{so5solution}.  We saw in Sec.~\ref{so5metric}
that the moduli space was flat four-dimensional Euclidean space,
$R^4$.  If this is described by spherical coordinates, the radial
distance is proportional to the square root of the cloud parameter,
$\sqrt{a}$, while the three angular coordinates correspond to the
Euler angles that specify the orientation in the unbroken SU(2).

The geodesics are straight line motions with constant velocity.
Purely radial geodesics correspond to fixed SU(2) orientation
with $\sqrt{a}$ varying linearly with time; i.e., to solutions
whose cloud parameters obey
\begin{equation}
     a = k (t-t_0)^2
\end{equation}
where $k$ and $t_0$ are constants.

Nonradial geodesics correspond to motions that include excitation of
the SU(2) gauge zero modes, and thus to time-dependent solutions with
nonzero SU(2) electric charge.  It is evident that at large times the
cloud size in these solutions also grows quadratically with time.

The solutions with two massive monopoles and a single cloud provide
less trivial examples.  In the SU($N$) solutions of
Sec.~\ref{masslessoneoneone} the massive monopoles correspond to two
nodes in the Dynkin diagram that are not joined by a common link.
Hence, these monopoles do not interact directly with each other and so
can only affect each other through their mutual interactions with the
massless monopoles that form the cloud.  This suggests that nontrivial
scattering effects should only happen when the massive monopoles are
near the cloud.

These expectations can be tested by analyzing \cite{Chen:2001ge} the
geodesics of the metric of Eq.~(\ref{bRanglemetric}).  When $b \gg R$,
these behave as
\begin{eqnarray}
    R &=& v |t| + \dots  \cr
    b &=& k t^2 + \dots
\label{asymSU4geod}
\end{eqnarray}
where $v$ and $k$ are constants and the ellipsis represents subdominant
terms.  Thus, asymptotically the massive monopoles move at constant
velocity on straight lines, while the cloud (which is almost spherical
when $b \gg R$) behaves like the cloud of the SO(5) example.  In this
regime the energy associated with the geodesic motion,
\begin{equation}
    E \approx  {\mu \over 2} \dot R^2
      + {\kappa\over 2}{\dot b^2 \over  b} \, ,
\label{energydivision}
\end{equation}
can be separated into two parts, associated with the massive monopoles and
the cloud, respectively, that are each approximately constant.

One can go beyond this asymptotic analysis by numerically integrating
the geo\-desic equations of motion.  First, suppose that the angular
momentum and the gauge charges all vanish, so that the motion is
described by a trajectory on the two-dimensional space spanned by $x$
and $y$, with the metric given by Eq.~(\ref{xymetric}).  At large
negative time all solutions have $b \gg R$, with the cloud contracting
and the massive particles approaching each other at constant velocity,
in accord with Eq.~(\ref{asymSU4geod}).  This behavior continues until
$b \approx R$; i.e., until the cloud collides with the massive
monopoles.  At this point the cloud and massive monopoles interact, as
evidenced by a change in the velocity of the massive monopoles; in
some cases, this interaction can be strong enough to reverse the
direction of motion of the massive monopoles.  All trajectories have
at least one such interaction. (Some have two points with $b= R$, but
none have more than two.)  A striking fact about these interactions is
that they are of short duration, restricted to the time when $b$ is
very close to $R$.  At least in its interactions, the cloud behaves as
if it were a thin shell.

The overall effect of the monopole-cloud interaction can be measured
by comparing the division of energy between the cloud and the massive
monopoles at $t = -\infty$ and at $t= \infty$.  This effect turns out to be
greatest if the collision(s) between the cloud and the massive monopoles
[i.e., the point(s) where $b=R$] occurred at small values of $b$.

On every trajectory there is a point where $R=0$; i.e., where
the massive particles pass through each other.  (If the interaction
with the cloud is strong enough to reverse the massive monopole
directions, then they pass through each other twice.)  There is no
change in the motion, either of the massive monopoles or of the cloud,
when this happens.  This is in agreement with expectations, since
there is no direct interaction between the massive monopoles.

The main qualitative effect of having nonzero angular momentum and
gauge charges can be seen by examining the angular part of the metric,
given in Eq.~(\ref{fullsuNcloudmetric}).  Because of the factors of
$R$ and $b-R$ that appear in the denominators of the various terms,
there are effective potential barriers that prevent the system from
reaching either $R=0$ or $b=R$.  As a result, the motion is restricted
to a single octant of the $x$-$y$ plane.

The other possibility with two massive monopoles and a single cloud is
the SU(3) solution described in Sec.~\ref{DancerSoln}.  In this case
the massive monopoles are of the same type and so can interact
directly as well as through their mutual interactions with the cloud.
The geodesics for the case of vanishing angular momentum and SU(2)
charge lie on the two-dimensional surface shown in
Fig.~\ref{DancerFig}.  Numerical studies of these \cite{Dancer:1992hf}
show that their behavior is consistent with that found when the
massive monopoles are distinct.  As with distinct massive monopoles,
the cloud starts large, contracts to a minimum size, and then expands
indefinitely.  Again, the interaction of the massive monopoles with
the massless one is significant only when the massive monopoles are
close to the cloud and is strongest if this happens when the cloud is
small.  The main difference from before is that the massive monopoles
interact with each other even when the cloud is far in the distance.
Instead of passing undeflected through each other, as the distinct
massive monopoles do, they scatter by 90$^\circ$, just as a pair of
SU(2) monopoles would.

There is, however, a problem with these results.  In all three of
these cases the geodesics on the moduli space predict a cloud whose
size grows quadratically with time at large $t$, which means that its
expansion velocity eventually exceeds the speed of light.  This
strongly suggests that there is a breakdown of the moduli space
approximation.

When we discussed the validity of this approximation previously, in
Sec.~\ref{trajectories}, we noted that there is a potential problem
when massless fields are present, because excitation of these is
always energetically allowed.  However, we saw that when the symmetry
breaking is maximal and the massless fields are Abelian the radiation is
sufficiently suppressed at low monopole velocities to preserve the
validity of the approximation.  An essential ingredient in this
argument was the fact that the charged massive fields, which are the
potential sources for the radiation, are
confined to the fixed-radius monopole cores.

The situation is rather different now. The possible sources for the
radiation now include the non-Abelian gauge fields, which extend
throughout the core, and even beyond.  Some insight can be gained from
an analysis that compared numerical solutions of the spacetime field
equations with the predictions of the moduli space approximation for
the SO(5) example \cite{Chen:2001qt}.  The two agree well until roughly
the time, $t_{\rm cr}$, when the moduli space approximation predicts
that the cloud velocity should reach the speed of light.  Beyond this
time, the field profiles in the cloud region are no longer well
approximated by simply allowing the collective coordinates of the BPS
solution to be time-dependent.  Instead, the expanding cloud
essentially becomes a wavefront moving outward at constant velocity
$c$.  In the regions well inside the cloud, however, the fields
continue to be well approximated by the moduli space approximation,
suggesting that the predictions for the asymptotic motion of the
massive monopoles in the (1, [1], \dots, 1) and (2, [1]) solutions
remain reliable.

It is interesting to note that the duration of the period when the
moduli space approximation is valid is inversely proportional to the
energy.  Using Eqs.~(\ref{asymSU4geod}) and (\ref{energydivision}),
for example, one finds that
\begin{equation}
    t_{\rm cr} \sim {\kappa \over E} = {2\pi \over e^2 E}  \, .
\label{tHeis}
\end{equation}
As we noted in Sec.~\ref{sec2-quant}, weak coupling ensures that the
radius of a massive monopole is much greater than its Compton
wavelength, so that the classical field profile remains meaningful
even in the quantum theory.  Equation~(\ref{tHeis}) gives a
complementary result for massless monopoles.  For weak coupling, the
period in which the moduli space description of the cloud gives a good
approximation is much longer than the uncertainty in time set by the
uncertainly principle.  Hence, the classical description should be
reliable on the times scales relevant for this analysis.

There is one last topic we should address in this section.  So far, we
have only considered trajectories in which the gauge orientation
collective coordinates remain constant, so that the gauge charges
vanish.  Relaxing this condition might be expected to give dyonic solutions
carrying both magnetic and electric charges.  Of particular interest
would be solutions in which the electric charge was in the non-Abelian
unbroken subgroup; because of the analogy with QCD, such objects have
been termed ``chromodyons''.

These were first investigated in the context of an SU(5) grand unified
model.  It was soon found that the fundamental massive monopoles in
this theory \cite{Dokos:1979vu} cannot give rise to chromodyons
\cite{Abouelsaood:1982dz}, because of the topological obstruction,
noted at the end of Sec.~\ref{so5metric}, to globally defining a basis
for the unbroken gauge group
\cite{Nelson:bu,Balachandran:1982gt,Abouelsaood:1983gw,Nelson:1983fn}.
Because the examples we have considered in this chapter have purely
Abelian magnetic charges, they have no such obstruction and so one
might ask if they could be used to construct chromodyons.

The obvious starting point would be the SO(5) example with a single
massive monopole.  A stable chromodyon would correspond to a geodesic
trajectory with fixed cloud parameter $a$ and one (or more) of the
gauge orientation angles varying periodically with time.  Since
we already know that the geodesics are all straight lines in $R^4$,
such trajectories are clearly excluded.  They would be allowed, at least
within the moduli space approximation, if the cloud size could somehow
be held fixed.  This can be done by going beyond the BPS regime and
adding an appropriate potential.  However, a new difficulty, again
associated with massless radiation, arises.  Not only are the
non-Abelian gauge bosons massless, but they also carry non-Abelian
electric charge.  This opens up the possibility that the would-be
chromodyon could radiate away its
electric charge.  Numerical studies of the SO(5) example \cite{guothesis} suggest that
it suffers from precisely this affliction, and there seems little
reason to believe that the difficulty would be absent in other cases.
Hence, it appears that even when topology allows chromodyons, dynamics
may not.

\chapter{Multi-Higgs vacua in SYM theory and multicenter dyons}
\label{multihiggsChap}

Up to now, we have concentrated on the physics of
monopoles and dyons when only one adjoint Higgs field
acquired a vacuum expectation value.  For the simplest
gauge group with monopoles, SU(2), this restriction
hardly matters, because one can always use the global $R$ symmetry
of SYM theory to remove all but one
of the vevs. This appears to be one reason why the rich
new physics of multi-Higgs vacua had been neglected
for a long time.

For larger gauge groups and generic Higgs vevs,
this possibility is no longer available. The reason is simple: the
expectation value of an adjoint Higgs field entails $r$ mass scales,
corresponding to the generators of the Cartan subalgebra. If there are
two adjoint Higgs expectation values, there are $2r$ independent mass
scales. On the other hand, the global $R$ symmetry is independent of
the rank of the gauge group, and so in general cannot rotate $2r$
masses into $r$ masses. For the classification of generic monopoles
and dyons in a generic vacuum, we can no longer stick to the
single-Higgs model.

We have already noted, in Sec.~\ref{susyconnectionSec}, that when both
magnetic and electric charges are present the conditions for
maintaining some unbroken supersymmetry are a bit involved, and we saw
the possibility of 1/4-BPS dyons \cite{Bergman:1997yw,Bergman:1998gs}
in ${\cal N}=4$
theories. Yet, all the monopoles and dyons we have discussed so far
have been 1/2-BPS from the ${\cal N}=4$ viewpoint. In this chapter, we
will see how 1/4-BPS dyons arise in the context of generic vacua of
${\cal N}=4$ SYM theory, and will explore the nature of these solitons. In
the process, we will learn that the modified BPS equations involve at
most two independent adjoint Higgs fields, and are thus directly
applicable to the ${\cal N}=2$ case as well; the only difference is
the amount of supersymmetry that is preserved. Any given 1/4-BPS
soliton solution of ${\cal N}=4$ SYM theory can be thought of as a solution
to ${\cal N}=2$ SYM theory with the same gauge group. The supersymmetry
properties of the latter are a more subtle issue, to which we will
return in later chapters.

In the first half of this chapter, Sec.~\ref{newbps}, we will show how
the BPS energy bound and equations are modified in the presence of
additional Higgs fields.  Then, in Sec.~\ref{addlHiggsSec}, we will
specialize to the case where all but one of the Higgs fields can be
treated as small, in a sense that we will make more precise, and show
that their effects can be described by adding a potential energy to
the moduli space Lagrangian.

\section{Generalized BPS equations \label{newbps}}

It turns out that when more than one adjoint Higgs field has a nonzero
expectation value the BPS equations are modified in an essential way,
leading to a new class of dyonic BPS solutions.  One unexpected and
important characteristic of these new solutions is that they should be
really regarded as composites of two or more solitonic cores balanced
against each other by long-range static forces. These static forces
can be derived rigorously from the Yang-Mills-Higgs Lagrangian and are
a combination of long-range Coulomb forces and forces mediated by
scalar particle exchange.

\subsection{Energy bound}
\label{chap7energyboundsec}

We start by recalling the purely bosonic part of the Lagrangian
for SYM theory with extended supersymmetries that was
given in Eq.~(\ref{susyLag}).  The corresponding energy density
is \cite{Lee:1998nv}
\begin{equation}\label{ef}
   {\cal H} = \Tr \left\{ E_i^2 + B_i^2 + \sum_P(D_0\Phi_P)^2+
 \sum_P (D_i\Phi_P)^2 - {e^2\over 2}\sum_{P,Q}[\Phi_P,\Phi_Q]^2 \right\} \,.
\end{equation}
For ${\cal N}=4$, there are six adjoint scalar fields.  We choose two
arbitrary six-dimensional unit vectors $\hat m_P$ and $\hat n_P$ that
are orthogonal to each other and decompose the scalar fields as
\begin{equation}
   \Phi_P = b\hat m_P + a\hat n_P + \bphi_P  \, .
\end{equation}
Here $\bphi_P$ is orthogonal to both $\hat m_P$ and $\hat n_P$,
in the sense that
\begin{equation}
\hat m_P\bphi_P=\hat n_P\bphi_P =0 \, ,
\end{equation}
and represents four independent adjoint scalar fields.
We may regard (\ref{ef}) as the energy density of ${\cal N}=2$
SYM theory by restricting the $\Phi_P$ to just
$\Phi_1$ and $\Phi_2$ or, equivalently, to $a$ and $b$.
In the latter case, the four adjoint scalar fields
associated with $\bphi_P$ would be absent.

Using this decomposition, we rewrite the energy density as
\begin{eqnarray}
   {\cal H} &=& \Tr \biggl\{
        \left[ B_i^2 + (D_i b)^2 \right]
      + \left[ E_i^2 + (D_i a)^2 \right]
      + \left[ (D_0 b)^2  +e[a,b]^2 \right]
       \cr\cr && \qquad
      + \sum_P \left[ (D_0 \bphi_P)^2 + e[a, \bphi_P]^2 \right]
      + (D_0 a)^2
      + e^2 \sum_P[b ,\bphi_P]^2
       \cr\cr && \qquad
      + e^2 \sum_{P,Q}[\bphi_P, \bphi_Q]^2
      + \sum_P(D_i \bphi_P)^2   \biggr\}\cr\cr
     & =& \Tr \biggl\{
       ( B_i - D_i b)^2
      + (E_i -D_i a)^2
      + (D_0 b - ie [a,b])^2
       \cr\cr && \qquad
      + \sum_P \left(D_0 \bphi_P - ie [a, \bphi_P] \right)^2
      + (D_0 a)^2
      + e^2 \sum_P[b ,\bphi_P]^2
       \cr\cr && \qquad
     + e^2 \sum_{P,Q}[\bphi_P, \bphi_Q]^2
      + \sum_P(D_i \bphi_P)^2
      \cr\cr && \qquad
      + 2 B_i \, D_i b
      + 2 E_i \, D_i a
      + 2ie [a,b] \, D_0 b
      + 2ie \sum_P [a, \bphi_P]\, D_0 \bphi_P  \biggr\} \, .
\label{calHasSquares}
\end{eqnarray}
The last four cross terms can be rewritten with the
aid of the Bianchi identity, $D_i B_i=0$, and Gauss's law,
\begin{equation}
D_i E_i - ie[\Phi_P, D_0\Phi_P] = 0 \, ,
\label{gaussIn7}
\end{equation}
to give
\begin{eqnarray}
   {\cal H}&=& \Tr \biggl\{
       ( B_i - D_i b)^2
      + (E_i -D_i a)^2
      + (D_0 b - ie [a,b])^2
      + \sum_P \left(D_0 \bphi_P - ie [a, \bphi_P] \right)^2
      \cr\cr && \qquad
      + (D_0 a)^2
      + e^2 \sum_P[b ,\bphi_P]^2
      + e^2 \sum_{P,Q}[\bphi_P, \bphi_Q]^2
      + \sum_P(D_i \bphi_P)^2
      \cr\cr && \qquad
      + 2 \partial_i (B_i b) + 2 \partial_i (E_i a)
  \biggr\}  \, .
\end{eqnarray}

Every term is nonnegative, except for the last two, which are total
derivatives.  The surface terms arising from the latter then give the
bound
\begin{equation}
{\cal E}=\int d^3x \;{\cal H}  \; \ge \; \hat n_P {\cal Q}^E_P
  + \hat m_P {\cal Q}^M_P
\label{bps1}
\end{equation}
where
\begin{eqnarray}
{\cal Q}_P^M &=& 2\int d^3x \;\partial_i\,( \Tr \Phi_P B_i )  \\
{\cal Q}_P^E &=&  2\int d^3x\; \partial_i\, ( \Tr \Phi_P E_i )
\end{eqnarray}
are defined in a manner analogous to the ${\cal Q}_M$ and ${\cal Q}_E$
of Eq.~(\ref{higgscharge}).  However, while the latter two were
proportional to the actual electric and magnetic charges, differing
from them only by a common factor of the SU(2) Higgs vev $v$, the
situation now is a bit more complicated.  At large distance, the
asymptotic magnetic, electric, and scalar fields must all commute.
Therefore, in any fixed direction the asymptotic forms of these fields
can be simultaneously rotated into the Cartan subalgebra.  By analogy with
Eqs.~(\ref{defOFh}) and (\ref{defOFg}), they can then
be represented by vectors $\bf g$, $\bf q$, and the eigenvalue vectors
${\bf h}_P$ of the expectation values $\langle\Phi_P\rangle$.  We then
have
\begin{equation}
     {\cal Q}_P^M = {\bf h}_P \cdot {\bf g}   \, ,\qquad
     {\cal Q}_P^E = {\bf h}_P \cdot {\bf q}  \, ,
\end{equation}
while the bound of Eq.~(\ref{bps1}) can be rewritten as
\begin{equation}
   {\cal E} \ge {\bf a} \cdot {\bf q} + {\bf b} \cdot {\bf g}
\label{chap7Ebound1}
\end{equation}
where ${\bf a} = \hat n_P{\bf h}_P$ and ${\bf b} = \hat m_P{\bf h}_P$.

The most stringent bound is obtained by varying $\hat n_P$ and $\hat
m_P$ so as to maximize the right-hand side of
Eq.~(\ref{bps1}). This requires, first of all, that $\hat m_P$ and
$\hat n_P$ both lie on the plane spanned by ${\cal Q}^M_P$ and ${\cal
Q}^E_P$.  Next, the directions of $\hat m_P$ and $\hat n_P$ should be
chosen so that $\hat n_P {\cal Q}^E_P$ and $\hat m_P {\cal Q}^M_P$ are
both positive.  Assuming both of these conditions to hold, let
$\theta$ be the angle between $\hat m_P$ and ${\cal Q}^M_P$, and
$\alpha < \pi$ the one between between ${\cal Q}^M_P$ and ${\cal
Q}^E_P$.  One then finds that the right-hand side of Eq.~(\ref{bps1})
is maximized when
\begin{equation}
\tan \theta = \frac{ {\cal Q}^E|\cos\alpha|}{{\cal Q}^M
        + {\cal Q}^E\sin\alpha}
\end{equation}
where ${\cal Q}^M$ and ${\cal Q}^E$ are the magnitudes of the vectors
${\cal Q}^M_P$ and ${\cal Q}^E_P$.  (Note that this implies that
${\bf b} \cdot {\bf q} = {\bf a} \cdot {\bf g}$.)
This gives the bound
\begin{equation}
   {\cal E} \ge
     \sqrt{({\cal Q}^M)^2 +({\cal Q}^E)^2
   +2 {\cal Q}^M {\cal Q}^E \sin \alpha} \, .
\label{chap7Ebound2}
\end{equation}

\subsection{Primary and secondary BPS equations}

The lower bound on $\cal E$ is saturated when the bulk terms in the
energy density
all vanish.  {}From this we obtain a total of eight sets of equations. The
first is the most familiar,
\begin{equation}
   B_i= D_i b \, .
\label{bbps}
\end{equation}
This is the usual Bogomolny equation, which admits magnetic monopole
solutions.  Note that this magnetic equation can be solved
independently of the remaining equations. The other BPS equations
influence only the choice of the unit vector $\hat m_P$. This fact is
of crucial importance when we construct the BPS solution later. For
this reason, we call Eq.~(\ref{bbps}) the primary BPS equation.

The other BPS equations are to be solved in the
background of this purely magnetic BPS solution.  They are
\begin{eqnarray}
&& E_i= D_i a  \label{ebps} \\
&& D_0 b=-ie[b,a] \label{bphi} \\
&& D_0 \bphi_P=-ie[\bphi_P,a] \label{zetaphi} \\
&& D_0 a=0 \label{aphi}
\end{eqnarray}
and
\begin{eqnarray}
&& [b,\bphi_P]=0  \label{commbzeta}\\
&& [\bphi_P,\bphi_Q]=0 \label{commzetazeta}\\
&& D_i \bphi_P =0 \, .\label{Dizeta}
\end{eqnarray}
In addition, we must impose Gauss's law,
\begin{equation}
D_i E_i = ie\,\left( [b, D_0b ]+[a,D_0 a]
+[\bphi_P,D_0\bphi_P]\right)  \, .
\label{gauss1}
\end{equation}
Inserting Eqs.~(\ref{ebps}) -- (\ref{aphi}) in Gauss's law
gives a linear equation for $a$,
\begin{equation}
D_i D_i\, a  = e^2\,[b, [ b,
a]]+ e^2[\bphi_P,[\bphi_P,a]]  \, .
\label{gauss15}
\end{equation}

Matters can be simplified further by writing
the solution to the primary
equation in a form where the nontrivial fields occupy irreducible
blocks, and working in the unitary, or string, gauge where $b$
is diagonal and time-independent. With this gauge choice, $\partial_0 A_i$
is also zero and Eq.~(\ref{ebps}) is solved by
\begin{equation}
A_0 = -a
\end{equation}
while $D_0\bphi_P-ie\,[a,\bphi_P]=\partial_0\bphi_P=0$ requires that
$\bphi_P$ also be time-independent.  In the background of a generic
monopole solution, the last three equations, (\ref{commbzeta}),
(\ref{commzetazeta}), and (\ref{Dizeta}), imply that $\bphi_P$ is a
constant times the identity in each of the irreducible blocks occupied
by the monopole solution.\footnote{In the language of string web, to
be discussed in chapter \ref{Brane}, this translates to the
requirement that the string web be planar.}

Now Eq.~(\ref{gauss15}) is a zero-eigenvalue problem for a nonnegative
operator acting linearly on $a$.  In order to have the bosonic
potential vanish at
infinity, $a(\infty)$ must commute
with  $b(\infty)$ and $\bphi_P(\infty)$. Furthermore,
the actual solution can have nontrivial behavior only inside
each of the irreducible blocks, defined by $b$, where the $\bphi_P$
are just numbers times the identity matrix.
Thus the $\bphi_P$ must commute with $a$ everywhere and the last
term in Eq.~(\ref{gauss15}) drops out, yielding
\cite{Lee:1998nv,Hashimoto:1998zs}
\begin{equation}
\qquad D_i D_i\, a  = e^2\,[b, [ b, a]]  \, ,
\label{gauss2}
\end{equation}
which we call the secondary BPS equation.

Finally, recall that in Sec.~\ref{susyconnectionSec} we showed that a
1/4-BPS solution of the ${\cal N}=4$ theory was obtained by requiring
that all but two scalar fields vanish and that the remaining two
satisfy Eq.~(\ref{quartertwo}).  These requirements are identical
to\footnote{We could have obtained, instead, the equivalent of
Eq.~(\ref{quarterone}) if we had made a different choice of sign when
completing the squares in Eq.~(\ref{calHasSquares}).  In this case, we
would have found that the most stringent energy bound was obtained by
requiring $\hat n_P Q_P^E$ to be negative, and so would have been led
to the same solutions, but with $a$ redefined in such a way that its sign
was reversed.}  Eqs.~(\ref{bbps}) -- (\ref{Dizeta}), thus verifying
that solutions obeying the primary and secondary BPS equations are
indeed 1/4-BPS.  (For the special case $\alpha=0$, these solutions
can be rotated into a form with only a single nontrivial scalar,
and are actually 1/2-BPS.)

\subsection{Multicenter dyons are generic}

Now that we have generalized the BPS equations, let us characterize
the solutions.  We saw above that the BPS equations split into two
groups, one involving the original Bogomolny equation for the magnetic
sector, and the other leading to the second-order Eq.~(\ref{gauss2})
to be solved in the background of a purely magnetic solution to the
first. Because of this, the solutions are parameterized by the same
monopole moduli space.  The new story is that for any given BPS
monopole solution the electric sector is uniquely determined, because
the solution to the second-order equation is completely fixed by the
Higgs expectation values and the moduli coordinates that characterize
the BPS monopole.  Note in particular that the electric sector must,
in general, be nontrivial; with gauge groups larger than SU(2), it is
only in special cases that a purely magnetic solution is possible.

A somewhat unexpected consequence of this result is
that, if we fix the asymptotic Higgs field and the electric
charge, the relative positions of the monopole
cores are constrained and generically
lead to a collection of well-separated
dyonic cores \cite{Lee:1998nv,Bak:1999hp}. Unlike the case with only
one nontrivial Higgs field, these cores cannot be moved
freely relative to one another, unless we also change the
electric charge or the Higgs vevs. In the next section
we will study this odd behavior in more detail,
and find that there is really nothing mysterious
about it; it is simply a result of classical forces
generated by the Yang-Mills-Higgs system on these solitonic
objects.

To illustrate the general structure of these solutions, it
is instructive to consider the secondary BPS equation
(\ref{gauss2}) when we have a single fundamental monopole.
Since the latter
is an embedded SU(2) monopole solution, we have
\begin{equation}
D^2_{\rm SU(2)}\, a  =
e^2\,[\Phi_{\rm SU(2)}, [ \Phi_{\rm SU(2)}, a]] \, .
\label{gauss'}
\end{equation}
For this somewhat degenerate
case, there is really only one solution for $a$, which
can be written as
\begin{equation}
  a=c \,\Phi_{\rm SU(2)} + \hbox{constant}\label{14SU(2)}
\end{equation}
where $c$ is an integration constant and the last term must commute
with the magnetic part of the solution everywhere.  Thus, we also have
\begin{equation}
  E_i=c D_i\Phi_{\rm SU(2)}= c B_i^{\rm SU(2)}\label{chargeSU(2)} \, .
\end{equation}
Note that the electric field is proportional to the
magnetic field.

For a collection of well-separated fundamental monopoles, this form of
the solution is a good approximation near each of the monopole
cores. Thus, turning on the vacuum expectation value $\langle
a\rangle$ endows each core with an electric charge in the
corresponding SU(2) subgroup. The amount of electric charge is
determined by $\langle a\rangle$ and by the particulars of the
magnetic solution.  Since the general magnetic solution to the primary
BPS equation consists of separated fundamental monopoles, the generic
dyonic solution in a SYM theory looks like a collection of many
embedded SU(2) dyons whose relative positions are determined by the
balance between various long-range forces.  An explicit solution
involving two such dyonic cores in SU(3) gauge theory can be found in
Ref.~\cite{Lee:1998nv}.\footnote{A somewhat special solution of 1/4
BPS dyons with a single spherically symmetric core can be found in
Refs.~\cite{Hashimoto:1998zs,Hashimoto:1998nj,Kawano:1998bp}. For this
solution, however, the electric charge is fixed and incompatible with
the charge quantization condition. See also Ref.~\cite{Houghton:1999cm},
which considers 1/4-BPS dyons in the case where the $b$ field only
breaks the SU(3) symmetry to SU(2)$\times$U(1), with a further
breaking to U(1)$\times$U(1) arising from the second Higgs field.}

\section{Additional Higgs expectation values as perturbations}
\label{addlHiggsSec}

Understanding this new breed of solution becomes a little easier,
however, when we approach these solutions from a different
perspective.  In this section we will try to construct these dyons as
classical bound states of monopoles or, equivalently, as static orbits
in the moduli space of monopoles.  To do this, we will assume that the
additional Higgs expectation values are much smaller than the first,
and show that the perturbation due to the additional Higgs fields
generates an attractive bosonic potential energy between the monopole
cores.

We thus start with a Yang-Mills theory with a single adjoint
Higgs field $\Phi$ and solve its Bogomolny equation,
\begin{equation}
  B_i=D_i \Phi  \, . \label{first}
\end{equation}
We then begin to turn on expectation values of additional adjoint
Higgs fields $a_I$ where  $I=1,2,3,4,5$ for ${\cal N}=4$ SYM theory and
$I=1$ for ${\cal N}=2$ SYM theory. In terms of the decomposition of the six (or two)
adjoint scalar fields in the previous section, we are assigning
\begin{eqnarray}
b=\hat m_P\Phi_P\quad\rightarrow \quad \Phi \, ,
\end{eqnarray}
and treating the other five scalar field on an equal footing,
\begin{eqnarray}
\bphi_P\:\hbox{and}\:a=\hat n_P\Phi_P\quad\rightarrow \quad  a_I, \:I=1,2,3,4,5.
\end{eqnarray}
Because of the quartic commutator term in the Lagrangian,
the vacuum condition on the $\langle a_I\rangle$ requires that
they commute with the expectation value of $\Phi$.
With an SU(2) gauge group, this uniquely fixes the direction of
the vevs, which then allows one to use a global $R$-symmetry
to remove all but one vacuum expectation value. This is no
longer true for gauge groups of rank $\ge 2$.

Note that we did not need to make any assumption about the relative
sizes of these Higgs
expectation values when finding the Bogomolny bound
in the previous section. In contrast, here we need to assume that
that the mass scales in $\langle a_I\rangle$ are much smaller than
those in $\langle \Phi\rangle=\langle b\rangle$.
One immediate effect of turning on
such extra expectation values of the $a_I$'s is that the BPS monopole
solutions $(\bar A_a,\bar \Phi)$ of the magnetic BPS equation(\ref{first})
are not, in general, solutions to the full field equations
when the expectation values $\langle a_I\rangle$ are turned on
\cite{Fraser:1997xi}.
As a result, the monopoles exert static forces on each other.
In this language the electric charge behaves as an angular
momentum and generates a repulsive angular momentum barrier. The
resulting BPS dyons are then obtained via the balance between the
potential energy and the angular momentum barrier.

\subsection{Static forces on monopoles}

For sufficiently small $\langle a_I\rangle$, we should be
able to treat these forces as arising from an extra potential energy
due to the nontrivial $a_I$ fields in the background of the monopole
solution. In other words, when $\langle a_I\rangle\neq 0$, the monopole
background induces a nontrivial behavior in the $a_I$ that
``dresses'' the monopoles and contributes to the energy of the system
in a manner that depends on which monopole solution was used for
the background.

Let us parameterize the size of the additional Higgs expectation values
by assuming that
\begin{equation}
|\langle a_I\rangle|/|\langle\Phi\rangle| = O(\eta)
\end{equation}
where $\eta$ is a small dimensionless number.
To find the effect to leading order in $\eta$, we imagine a static
configuration of monopoles that satisfies the Bogomolny equation. Let us
try to dress this configuration with a time-independent $a_I$ field, at the
smallest possible cost in energy. The strategy is a two-step
process. First, we find the minimum energy due to this
additional Higgs vev for a given monopole configuration,
and then incorporate it into the low-energy monopole dynamics.
Second, we solve this modified dynamics to find out how the monopoles
react to the additional Higgs vev.

With some hindsight we will call this new interaction energy
${\cal V}$, for it will prove to be a potential energy
term. ${\cal V}$ is obtained by
using the $a_I$ field equations to minimize \cite{Bak:1999sv}
\begin{equation}
 \Delta E=  \int d^3 x \; \Tr\left\{    ( {\bar D}_j a_I)^2
 - e^2\left([a_I,\bar\Phi\,]\right)^2  \right\}
\label{energy}
\end{equation}
with the $\langle a_I \rangle$ held fixed.
(We will ignore terms,
such as $[\,a_I,a_J]^2$, that are higher order in $\eta$.)
Thus, we solve
\begin{equation}
\bar D^2_j a_I - e^2[\bar \Phi,[\bar \Phi,a_I]] = 0
\label{laplace}
\end{equation}
and insert the result back into $\Delta E$ to
obtain the minimum energy needed to
maintain the monopole configuration in the presence of the
$\langle a_I\rangle$.

A crucial point to note here is that the equation for the $a_I$ is
identical to that obeyed by the gauge zero modes
\cite{Tong:1999mg}. The gauge zero modes are always of the
form\footnote{We are using here the four-dimensional Euclidean
notation in which $a$ runs from 1 to 4, with $\delta A_4 = \delta
\Phi$.}  $\delta A_a= \bar D_a\epsilon$ and must obey the background
gauge condition, Eq.~(\ref{backgroundgauge}).  The latter implies that
the gauge function $\epsilon$ must satisfy
\begin{equation}
  \bar D^2_j\epsilon-e^2[\bar\Phi,[\bar\Phi,\epsilon]]=0 \, .
\end{equation}
We notice that the ${\bar D}_a a_I$ have exactly the same form as the
global gauge zero modes, $\delta A_a=\bar D_a\epsilon$, with the gauge
function $\epsilon=a_I$. Thus, it must be true that we can express the
$\bar D_a a_I$ as linear combinations of gauge zero modes.
Consequently, each ${a}_I$ picks out a linear combination
\begin{equation}
K^r_A \frac{\partial}{\partial z_r} = \frac{\partial}{\partial
\xi_A}
\end{equation}
of U(1) Killing vector fields on the moduli space.  More precisely,
each $K_A$ corresponds to a linear combination $K_A^r \delta_r A_a$ of
gauge zero modes and each $D_a a_I$ is a linear combination of these,
\begin{equation}
    \bar D_a a_I = a_I^A K_A^r \delta_r A_a
            \equiv G_I^r \delta_r A_a \, ,
\end{equation}
where we have expanded the Cartan-valued vev as
\begin{equation}
{\bf a}_I = \sum_A a^A_I \label{defa}
\lambdabf_A
\end{equation}
with the $\lambdabf_A$ being the fundamental weights, which obey
$\lambdabf_A \cdot \betabf_B = \delta_{AB}$.

We then express the potential energy ${\cal V}$, obtained by minimizing
the functional $\Delta E$ in Eq.~(\ref{energy}) in the monopole
background, in terms of the monopole moduli parameters \cite{Bak:1999da,Bak:1999vd} as
\begin{equation}
{\cal V} = \int d^3 x \:{\rm Tr}\:\left\{(a_I^A K_A^r\delta_r A_a)
(a_I^B K_B^s\delta_s A_a)\right\}
=\frac{1}{2}\, g_{rs}\,a_I^A K_A^r  a_I^B  K_B^s
= \frac{1}{2}\, g_{rs} G_I^r G_I^s \, .
\end{equation}
The value of this potential energy depends on the monopole
configuration we started with.
The low-energy effective Lagrangian, which was
purely kinetic when the $a_I$ were absent,
picks up a potential energy term that lifts some of the moduli,
and takes the form
\begin{equation}
{\cal L}= \frac{1}{2}\,g_{rs}\dot z^r\dot z^s
        - \frac{1}{2}\, g_{rs} G_I^r G_I^s \, .
\label{modspaceLwithV}
\end{equation}
In the
current approximation, where the additional Higgs fields are treated as
perturbations, the mass scale introduced by the potential energy is much
smaller than that of the charged vector mesons, and we can still
consistently truncate to this moduli space mechanics. The procedure we
employed here should be a very familiar one. When we talk about, say,
the Coulombic interactions among a set of charged particles, we also
fix the charge distribution by hand, and then estimate the potential
energy that it costs.  Using this potential energy, we then find how the
charged particles interact at slow speed.

Of course, there is the possibility of interaction terms involving the
moduli velocities as well as the $a_I$ fields, but in the low-energy
approximation used here the only relevant such terms would be of order
$v\eta$. However, it is clear that neither the back-reaction of the
$a_I$ on the magnetic background nor the time-dependence of the $a_I$
can produce such a term. Thus, to leading order, the Lagrangian of
Eq.~(\ref{modspaceLwithV}) captures all of the bosonic interactions
among the monopoles in the presence of the nonzero ${\bf a}_I$.

A special solution to the $a_I$ equation deserves further attention. Since
\begin{equation}
\nabla^2\bar\Phi-e^2[\bar \Phi,[\bar\Phi,\bar\Phi]]=\nabla^2\bar\Phi
=\nabla\cdot \bar B=0
\end{equation}
one can always separate from $a_I$ the part proportional
to $\bar\Phi$ by writing
\begin{equation}
a_I=c_I\bar \Phi +\Delta a_I
\end{equation}
and requiring ${\rm Tr}\left(\langle \Delta a_I\rangle
\langle\bar\Phi\rangle\right)=0$. The U(1) Killing vector associated
with the gauge function $\epsilon=\bar \Phi$ is
the free U(1) angle that is one of the center-of-mass degrees of
freedom.   The square of this Killing vector is
independent of the moduli, and the potential energy term in question
simply adds a positive constant to the energy of the system.

This ``extra" energy can be easily understood by going back to
the field theory and reanalyzing the BPS equation. As was mentioned at
the beginning of the section, an $a_I$ vacuum expectation value
proportional to that of $\Phi$ can be rotated away by a redefinition
of $\Phi$. Once this is done, we can make the replacement
\begin{eqnarray}
\bar\Phi &\rightarrow&
\left(1+\sum_I c_I^2\right)^{1/2}\,\bar\Phi \nn
a_I &\rightarrow & \Delta a_I \, .
\end{eqnarray}
Expanding the mass formula in terms of the small $c_I$, we get back
the constant energy terms $\sim c_I^2/2$.
Thus, we could have started with these rotated Higgs fields and
regarded the $\Delta a_I$, instead of the $a_I$, as the
perturbation. The potential energy would then be generated entirely by
the $\Delta a_I$, and there would be no constant energy shift from the
center-of-mass part of the moduli space.  For this reason, the part of
the $a_I$ proportional to $\bar \Phi$ will be ignored for most of this
review.

\subsection{Dyonic bound states as classical orbits}
\label{dyonorbitssubsec}

In the classical moduli space approximation bound dyons should appear
as closed, stationary orbits along U(1) phase angles. Let us consider
now the effect on the existence of such closed orbits of adding the
potential energy $\cal V$ generated by one\footnote{As
will become clear in Chap.~\ref{modspacefermionchap}, dyons such as
these classical monopole bound states can become BPS only if just one
such bosonic potential energy is turned on; i.e., only one of the
$a_I$ can be excited (up to an orthogonal transformation among the
$a_I$). This corresponds to having only two adjoint Higgs fields
participating in the low-energy dynamics and, in the language of the
classical BPS equations of Sec.~\ref{newbps}, corresponds to the
decoupling of the $\bphi_P$. This motivates removing all but one of
the $a_I$. } additional Higgs field $a$ \cite{Lee:2000rp}. It is immediately
clear that one will generically find many more closed orbits in the
presence of $\cal V$ than otherwise. For example, if one considers the
case of $n$ distinct monopoles, it can be shown rigorously that no
closed orbits are possible in the absence of such a potential energy.
The existence of a potential energy will, understandably, change this
completely.

As a special case, let us take a pair of distinct monopoles in a
theory with SU(3) broken to ${\rm U}(1)\times {\rm U}(1)$. Before turning
on the additional Higgs fields, the purely kinetic interaction
Lagrangian of the pair can be distilled down to
\begin{equation}
  L_0=\frac12\left(1+\frac{1}{r}\right)\dot{\bf r}^2
  +\frac12\left(1+\frac{1}{r}\right)^{-1}
  \left[\dot \psi+{\bf w}({\bf r})\cdot \dot {\bf r}\right]^2
\end{equation}
where, for the sake of simplicity, we have started with the Taub-NUT
relative metric of Eq.~(\ref{taubnutrelativemetric}) and transformed
to dimensionless quantities defined by the rescalings
\begin{equation}
{\bf r}\rightarrow \frac{2\pi{\bf r}}{e^2\mu} \, ,\qquad
t\rightarrow \frac{(2\pi)^2t}{e^4\mu } \, ,\qquad
L \rightarrow {e^4 \mu L\over (2\pi)^2}  \, .
\label{randtandLrescaling}
\end{equation}
We have also taken the root lengths to be equal to unity, so that
$\lambda=1$.

A dyonic state with a relative
electric charge $q$ would be governed by the Routhian
\begin{equation}
R_0=\frac12\left(1+\frac{1}{r}\right)\dot{\bf r}^2
-\frac{q^2}{2}\left(1+\frac{1}{r}\right)
+q{\bf w}({\bf r})\cdot \dot {\bf r} \, .
\end{equation}
This has three interaction terms: one that
modifies the inertia as a function of the separation $r$; a repulsive
potential energy; and a velocity-dependent coupling that generates a
Lorentz force, due to a unit monopole sitting at the origin ${\bf
r}=0$, on a particle of charge $q$. Despite the various interaction
terms, the (rescaled) conserved energy takes the simple form
\begin{equation}
E=\frac12\left(1+\frac{1}{r}\right)\left({\bf v}^2+q^2\right) \, .
\end{equation}
{}From the form of the effective potential energy, which is
monotonically decreasing toward $r=\infty$, it is fairly clear
that, as we saw in Sec.~\ref{trajectories},
no bound orbits are possible with this classical dynamics.

A more complete characterization of the classical trajectories
is possible if we utilize an additional conserved quantity.
The conserved angular momenta has the familiar form
\begin{equation}
{\bf J}= \left(1+\frac{1}{r}\right){\bf r}\times{\bf v}+ q\hat {\bf r}
\end{equation}
with the last term being characteristic of charged particles in a
monopole background. This severely restricts the possible trajectories
because
\begin{equation}
{\bf J}\cdot \hat {\bf r}=q
\end{equation}
is also a conserved quantity. This says that the trajectories lie along a
cone going through the origin ${\bf r}=0$, with an opening angle
$\cos^{-1}(q/J)$ around ${\bf J}$. Also note the inequality
\begin{equation}
J^2-q^2\ge 0 \, ,
\end{equation}
which is saturated only when the cone collapse to a line.

One more conserved vector is known to exist.  It is of
the Runge-Lenz type \cite{Lee:2000rp},
\begin{equation}
{\bf K}= \left(1+\frac{1}{r}\right){\bf v}\times {\bf J}
-(E-q^2)\hat {\bf r} \, .
\end{equation}
The linear combination
\begin{equation}
{\bf N}\equiv
q{\bf K}+ (E-q^2){\bf J}
\end{equation}
of these two conserved vectors gives us another conserved inner product,
\begin{equation}
[q{\bf K}+ (E-q^2){\bf J}]\cdot {\bf r}=q(J^2-q^2) \, .
\end{equation}
Thus, the trajectories
also must lie on a plane which is orthogonal to ${\bf N}$ and
displaced from the origin by
\begin{equation}
\Delta{\bf r}=\frac{q(J^2-q^2)}{N^2}\,{\bf N}  \, .
\end{equation}
Combined with the previous result, this shows that the trajectories
are always conic sections.

Now let us consider what happens when we turn on a second Higgs field
as a perturbation. The only U(1) Killing vector on the Taub-NUT
manifold is $\partial_\psi$, and the effect of turning on a small,
second Higgs expectation value $a$ should show up as a potential
energy term.  The unbroken gauge group is ${\rm U}(1) \times {\rm
U}(1)$, with one factor acting on the center-of-mass part.  Because of
this, there is only one independent component in the expectation value
$\langle \Delta a_1\rangle$ that generates a nontrivial potential
energy term; we denote this value by $a$.  Keeping in mind that $\psi$
has period $4\pi$, we see that this generates a potential
energy $g_{\psi \psi}(ea)^2/2$.  After introducing the
dimensionless combination
\begin{equation}
\tilde a\equiv \frac{(2\pi)^2a}{e^3\mu} 
\end{equation}
and rescaling as in Eq.~(\ref{randtandLrescaling}), we have
\begin{equation}
{\cal V}= \frac{1}{2}\frac{\tilde a^2}{1+1/r} \, .
\label{potentialfortwodistinct}
\end{equation}
A remarkable fact is that this potential gives a Lagrangian,
\begin{equation}
L=\frac12\left(1+\frac{1}{r}\right)\dot{\bf r}^2
+\frac12\left(1+\frac{1}{r}\right)^{-1}
\left[\dot \psi
 + {\bf w}({\bf r})\cdot \dot {\bf r}\right]^2-{\cal V} \, ,
\end{equation}
whose dynamics admits exactly the same forms for the conserved vectors
${\bf J}$, ${\bf K}$ and ${\bf N}$, provided that when writing $\bf K$
and $\bf N$ we keep in mind that the conserved energy
\begin{equation}
E=\frac12\left(1+\frac{1}{r}\right)\left({\bf v}^2+q^2\right)+
\frac{1}{2}\left(1+\frac{1}{r}\right)^{-1}\tilde a^2
\end{equation}
now includes an additional contribution from the potential energy.
Thus, after we take into account the additional Higgs field,
all trajectories are still conic sections.

Of the five kinds of conic sections, only circles and ellipses
correspond to bound trajectories. The condition for a closed trajectory
is  then expressible in terms of the angle between ${\bf N}$ and
${\bf J}$ in the following manner.\footnote{We thank Choonkyu Lee for
useful conversations on this classical dynamics.}  Given the angular momenta
${\bf J}$, the cone encloses ${\bf J}$ with an opening angle
$0\le \alpha=\cos^{-1}q/J \le \pi $. Let $\beta$ be
the angle between ${\bf J}$ and ${\bf N}$. {}From the explicit
form of the conserved vectors, it is a matter of
straightforward computation to show that
\begin{equation}
\cos\beta=\frac{\sqrt{J^2-q^2}}{J}\times
\frac{E-q^2}{\sqrt{E^2-\tilde a^2q^2}}
\end{equation}
while
\begin{equation}
\cos(\pi/2-\alpha)=\sin\alpha=\frac{\sqrt{J^2-q^2}}{J} \, .
\end{equation}
In addition to the inequality $J^2\ge q^2$, the fact that ${\bf
N}^2\ge 0$ gives another constraint,
\begin{equation}
E\ge |\tilde aq| \, .
\end{equation}

For the sake of simplicity, we will assume that $q\ge 0$ so that
$\alpha <\pi/2$.  Then, the trajectory will be an ellipse (or a
circle) if $\alpha+\beta $ is smaller than $\pi/2$, a parabola if
$\alpha+\beta=\pi/2$, and a hyperbola if $\alpha+\beta $ is larger
than $\pi/2$.  Hence, the trajectory is bound and closed if and only
if the ratio
\begin{equation}
\frac{\cos\beta}{\cos(\pi/2-\alpha)}=\frac{E-q^2}{\sqrt{E^2-\tilde a^2q^2}}
\end{equation}
is strictly larger than 1.  This is equivalent to requiring that
\begin{equation}
    \frac{\tilde a^2+q^2}{2} > E > q^2
\end{equation}
which, in turn, implies that $|q| < |\tilde a|$.  The same result is
obtained for negative $q$.

One simple corollary is that if the potential energy term $\sim \tilde
a^2$ is absent, no bound orbit at all is possible in this two-body
problem. This last statement also holds in the many-body problem with all
distinct monopoles, as was shown by Gibbons \cite{Gibbons:1996wc}.
Without the potential energy, all classical orbits are hyperbolic.

\begin{figure}[t]
\begin{center}
\scalebox{0.9}[0.9]{\includegraphics{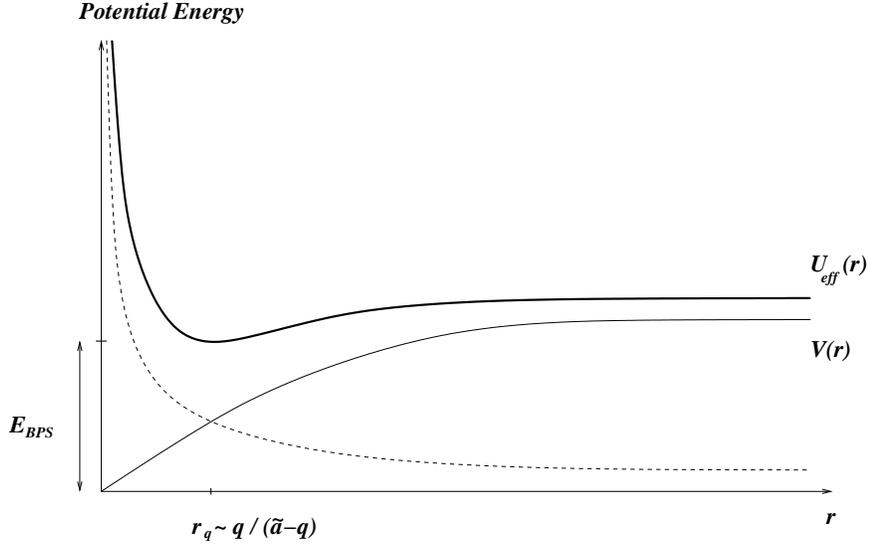}}
\par
\vskip-2.0cm{}
\end{center}
\begin{quote}
\caption{\small Potential energies between a pair of distinct
monopoles as a function of separation. The solid line is the potential
energy ${\cal V}$ between a pair of bare monopoles, while the dotted
line is an angular momentum barrier generated by assigning a relative
electric charge $q$. The thick line is the effective potential energy
$U_{\rm eff}$ between such a pair of dyons, which has a minimum at a
separation $r = q/(\tilde a -q)$, with the excitation energy
saturating the BPS bound.
\label{potential}}
\end{quote}
\end{figure}

\subsection{Static multicenter dyons and balance of forces}
\label{staticmulticentersec}

An interesting limiting case of a bound orbit is found when the cone
collapses to a line, so that $\alpha=0$ with positive $q$, in which
case the entire angular momentum comes from the $q\hat r$ piece. In
this case the energy must saturate its lower bound, $E=\tilde a q$,
and the ``orbit'' is simply a stationary point at a fixed
distance. With the two monopoles as above, this static configuration
is easy to understand.  The effective potential energy in the charge
$q$ sector is
\begin{equation}
\frac{q^2}{2}\left(1+\frac{1}{r}\right)+{\cal V}
=\frac{q^2}{2}\left(1+\frac{1}{r}\right)+
\frac{\tilde a^2}{2}\left(1+\frac{1}{r}\right)^{-1}
\end{equation}
which, for $\tilde a>q$, has a global minimum at
\begin{equation}
r= \frac{q}{\tilde a-q}
\label{qOveraminusq}
\end{equation}
with the minimum energy being $E=\tilde aq$. The contribution from the
charge $q$ to the effective potential energy behaves exactly like an
angular momentum barrier that balances against the
attractive potential energy ${\cal V}$.

Restoring the physical units is easy; we only need to reverse the
rescaling performed above, so that Eq.~(\ref{qOveraminusq}) becomes
\begin{equation}
\frac{e^2}{2\pi}r=\frac{q}{4\pi^2a/e^3- \mu q} \, .
\end{equation}
Because the time must be also rescaled back, the physical energy
receives an additional multiplicative factor and becomes
\begin{equation}
E=\frac{e^4\mu}{(2\pi)^2}\,\tilde aq=eaq \, .
\end{equation}
For a larger collection of distinct monopoles,
the general form of the effective potential energy in physical units
is
\begin{equation}
U_{\rm eff} = \frac12\left(\frac{4\pi^2}{e^2} \sum_{A,B}
(C^{-1})_{AB}\,\lambda_Aa^A\, \lambda_B a^B +\frac{e^4}{4\pi^2}\sum_{A,B}
C_{AB} \,\frac{q_A}{\lambda_A}\,\frac{q_B}{\lambda_B} \right)
\end{equation}
with $a^A$ defined as in Eq.~(\ref{defa}). $C_{AB}$ is the matrix in
Eq.~(\ref{C}) that characterized the relative moduli space metric,
while the dimensionless number that encodes the strength of the
interaction between a pair of monopoles is
$\lambda_A=-2\bbeta^*_{j}\cdot\bbeta^*_{k}$, with $\bbeta_j$ and
$\bbeta_k$ being the simple roots joined by link $A$ of the Dynkin
diagram.  The static minimum energy configuration is found when
\begin{equation}
  \frac{e^3}{4\pi^2}\sum_B C_{AB} \frac{q_B}{\lambda_B}
          = \lambda_A a^A \, .
\end{equation}
(Note that there is no sum over $A$ on the right-hand side.)
The solution is
\begin{equation}
\frac{e^2}{2\pi}r_A=\frac{q_A}{4\pi^2\lambda_A
  a^A/{e^3}-\sum_B\mu_{AB}\,q_B/\lambda_B} \, .
\end{equation}

We thus find a static dyon solution involving interacting cores
separated by finite distances. In particular, the distance
$r_A$ becomes infinite as $a_A$ approaches the critical value
\begin{equation}
   a_A^c = \frac{e^3\mu}{4\pi^2}
      \sum_B\frac{\mu_{AB}}{\lambda_A\lambda_B}\,q_B \, .
\end{equation}
Note that, although the distances are thus fixed, there are
some moduli that remain massless. For instance, the distance between
the first and second dyon cores is fixed, as is the distance between
the second and third, but the distance between the first and third is
not. When a finite size bound state of this type exists, its energy is
\begin{equation}
E=e\,\sum_Aa^A\,q_A
\end{equation}
regardless of the $\lambda_A$.

Such a dyonic configuration, with several soliton
cores balanced against each other at fixed separations, is
quite typical of dyons that preserve at most four supersymmetries.\footnote{
The same phenomenon has been observed in various different regimes.
In the strongly coupled description of ${\cal N}=2$ SYM
theories, this was observed by solving for approximate solutions based
on the Seiberg-Witten geometry of the vacuum moduli space
\cite{Ritz:2000xa,Argyres:2000xs,Ritz:2001jk,Argyres:2001pv}.
These solutions resemble a stringy picture that had been
investigated earlier \cite{Bergman:1998br,Mikhailov:1998bx,DeWolfe:1998zf,DeWolfe:1998bi},
although the latter authors apparently did not realize the
multicentered nature of these states. Interestingly, the multicentered
nature of the dyons persists when strong gravity is introduced and
the dyons are hidden behind extremal horizons. See Refs.~\cite{Denef:2000nb,Denef:2001xn}.}
In Chap.~\ref{Bound} we will study the quantum counterparts of these dyons
by  realizing dyons as quantum bound states of monopoles.
To do this, we must first derive the most general monopole moduli space
dynamics, which will be the subject of the next chapter.
As we saw above, the existence of more than one adjoint Higgs field
changes the traditional moduli space dynamics by adding
a potential energy term. The main objective of the next chapter is to
determine how this modifies the complete moduli space dynamics, with
fermionic contributions included.

\chapter{Moduli space dynamics from SYM theories}
\label{modspacefermionchap}

In the earliest examples of monopole moduli space dynamics, for
theories with just a single nontrivial adjoint Higgs vev, one finds a
purely kinetic supersymmetric quantum mechanics on a smooth
hyper-K\"ahler manifold.\footnote{Although the general features had
been known for some time, a precise derivation of the supersymmetric
low-energy dynamics from the SYM theory was only given in the early
1990's.  General issues concerning the treatment of the fermionic
collective coordinates were addressed in Ref.~\cite{Manton:aa}, while
a full-fledged derivation of the low-energy effective actions for pure
${\cal N}=2$ SYM \cite{Gauntlett:1993sh} and for ${\cal N}=4$ SYM
\cite{Blum:1994wd,Cederwall:1995bc} followed shortly after. In all of
these, only a single adjoint Higgs field was included in the
analysis.}  This is sufficient for studying certain BPS states, such
as the dyons in SU(2) SYM theories and the purely magnetic states in
an arbitrary ${\cal N}=4$ SYM.  However, for gauge groups of rank two
or higher this approach misses a vast class of BPS states.  These
dyonic states, which preserve only one-fourth of the supersymmetry,
require the presence of two nontrivial Higgs fields, and for this case
the purely kinetic low-energy dynamics is no longer valid.  Although
the low-energy dynamics can still be described in terms of the moduli
space, it is governed by a Lagrangian that is considerably more
complex.  In this chapter we will derive this low-energy Lagrangian
for both ${\cal N}=2$ and ${\cal N}=4$ SYM theories.

In the general Coulombic vacuum of ${\cal N}=2$ or ${\cal N}=4$ SYM theory, where we
have two or more adjoint Higgs vevs turned on, many of the
would-be moduli of a BPS monopole solution are lifted by a
potential energy \cite{Lee:1998nv,Bak:1999da}.
As we saw in the previous chapter, this phenomenon is
crucial in the construction of 1/4-BPS dyons in ${\cal N}=4$ SYM
theory, since the static multiparticle nature of these
dyons is tied to the long-range behavior of this potential energy
\cite{Lee:1998nv}.
Similarly, many of the fermionic zero modes are lifted by the same
mechanism. The mass scale of these lifted modes is
proportional to the square of the additional Higgs expectation
values. As long as the latter are sufficiently small, we may be able
to describe monopole interactions in terms of these light, would-be
moduli parameters.  The usual moduli space dynamics is a
nonrelativistic mechanics where the bare masses of the monopoles are
far larger than the typical kinetic energy scale, and the slow motion
justifies ignoring radiative interactions with the massless
fields \cite{Manton:1981mp}. Adding a similarly small potential
energy to the system is not likely to upset this approximation scheme
\cite{Bak:1999sv}.

In the previous chapter, we discovered the general structure of this
potential energy for a purely bosonic theory with only bosonic moduli.
However, supersymmetry requires that fermionic fields also be present.
In this chapter we will see that these fermionic fields lead to the
introduction of fermionic moduli.  In fact, the low-energy moduli
space dynamics possesses a supersymmetry that is inherited from the
supersymmetry of the underlying field theory
\cite{Gauntlett:1993sh,Blum:1994wd,Cederwall:1995bc,Bak:1999da,
Gauntlett:1999vc,Bak:1999vd,Gauntlett:2000ks}.
Understanding this low-energy supersymmetry will pave
the way for searching for quantum BPS states in general SYM
theories, which will be the topic of the next chapter.

We begin in Sec.~\ref{modspacegeom} with a brief discussion of the
bosonic and fermionic zero modes and the geometry of the moduli space;
much of this is a collection of results obtained earlier in this
review.  Next, in Sec.~\ref{lowELagInSYM}, we derive the low-energy
effective Lagrangians for both pure ${\cal N}=2$ and ${\cal N}=4$
SYM theories. We borrow in part some notation
and tools introduced in Refs.~\cite{Gauntlett:1993sh,Blum:1994wd,Cederwall:1995bc}
but go beyond these papers in that we consider the effects of multiple
adjoint Higgs fields, and therefore include potential energy terms in the
low-energy effective Lagrangian. Similar but independent derivations,
in the context of the most general ${\cal N}=2$ SYM
theories, are also given in Refs.~\cite{Gauntlett:2000ks,Kim:2006tq}.

Then, in Sec.~\ref{lowESUSYandquant}, we discuss the supersymmetry
properties and quantization of these low-energy Lagrangians.  We begin
this discussion in Sec.~\ref{general}, where the main features of the
low energy superalgebra for a sigma model with a potential energy are
illustrated with the examples of $N=1$ real and complex supersymmetry
with a flat target manifold. The key difference from the usual
nonlinear sigma model is the emergence of a central charge associated
with an isometry.  Such central charges will eventually contribute to
the central charges of ${\cal N}=2$ and ${\cal N}=4$ SYM dyons as an
extra energy contribution due to the electric fields.  In
Sec.~\ref{n=2q}, we summarize the supersymmetry transformation rules
on the moduli space for the case of pure ${\cal N}=2$ SYM theory, and give
the quantum supersymmetry algebra.  Section~\ref{n=4q} repeats this
exercise for the moduli space dynamics arising from ${\cal N}=4$ SYM
theory.

These low-energy Lagrangians have solutions that saturate
Bogomolny-type bounds and preserve some of the supersymmetry of the
moduli space dynamics.  As one might expect, these moduli space BPS
solutions are closely related to the BPS solutions of the full quantum
fields.  In Sec.~\ref{bpsAndBPS} we discuss and clarify this
relationship.  Finally, in Sec.~\ref{seiberg-section}, we discuss the
connections with Seiberg-Witten theory.

This moduli space dynamics with a potential energy was developed first
in Refs. \cite{Bak:1999da} and \cite{Bak:1999ip}, where the authors
found a supersymmetric mechanics that reproduced the known 1/4-BPS
dyon spectra of ${\cal N}=4$ theories.  This effort was later extended
to the ${\cal N}=2$ pure SYM case in
Ref.~\cite{Gauntlett:1999vc}. These papers were, however, based on
constraints from the anticipated spectrum and low-energy
supersymmetry.  A field theoretical derivation of the moduli dynamics
was carried out in related papers.  The derivation of the bosonic part
of the potential was first developed in
Refs.~\cite{Bak:1999sv} and \cite{Bak:1999vd} and this was later
generalized in Ref.~\cite{Gauntlett:2000ks} to include the full set of
bosonic and fermionic degrees of freedom for a general ${\cal N}=2$
SYM theory with hypermultiplets. We will concentrate in this
chapter on the cases
of pure ${\cal N}=2$ and ${\cal N}=4$ SYM theories, and postpone the case of
${\cal N}=2$ SYM theory with hypermultiplets to
Appendix~\ref{nequal2appendix}.

\section{Moduli space geometry and adjoint fermion zero modes}
\label{modspacegeom}

We recall that, given a family of BPS solutions $A_a({\bf x}, z)$, the
bosonic zero modes can be written in the form\footnote{Here we have
again adopted the four-dimensional Euclidean notation, to be used
throughout this chapter, in which Roman letters at the beginning of
the alphabet run from 1 to 4, with $A_4 \equiv \Phi$.  We will also
use the convention that partial derivatives with indices $q,r, \dots$
from the middle of the alphabet
are derivatives with respect to the moduli space coordinates.}
\begin{equation}
    \delta_r A_a  = {\partial A_a \over \partial z^r} -D_a \epsilon_r
     \equiv \partial_r A_a - D_a \epsilon_r
\label{chap8deltaA}
\end{equation}
where $\epsilon_r$ is chosen so that the background gauge condition
\begin{equation}
   0 = D_a \delta_r A_a
\label{chap8background}
\end{equation}
is satisfied.
As we have noted previously, one
can view $\epsilon_r$ as defining a connection on
the moduli space, with a corresponding gauge covariant derivative
\begin{equation}
   {\cal D}_r  = \partial_r + ie\,[\epsilon_r, \,\,]
\end{equation}
and a field strength
\begin{equation}
    \phi_{rs} = \partial_r \epsilon_s - \partial_s \epsilon_r
           + ie [\epsilon_r , \epsilon_s]  \, .
\end{equation}
We note, for later reference, that
\begin{equation}
    D_a^2 \, \phi_{rs} = 2ie [\delta_r A_a, \delta_s A_a] \, .
\label{D2Phi}
\end{equation}

The moduli space has a naturally defined metric
\begin{equation}
   g_{rs} =  2 \int d^3x \, \Tr \,\delta_r A_a \, \delta_s A_a
     \equiv \langle \delta_r A \, ,  \delta_s A \rangle
      \, .
\label{chap8metric}
\end{equation}
As we showed in Sec.~\ref{manifold}, this metric is hyper-K\"ahler, with a
triplet of complex structures $J^{(i)}_r{}^s$ that obey the
quaternionic algebra, Eq.~(\ref{quaternionicAlgebra}), and act on the
zero modes by
\begin{equation}
    J^{(i)}_r{}^s \, \delta_s A_a  =
          - \bar\eta^i_{ab} \, \delta_r A_b \, .
\label{JdeltaA}
\end{equation}
A straightforward (although somewhat tedious) calculation using
Eqs.~(\ref{chap8deltaA}) and (\ref{chap8background}) shows that the
Christoffel connection associated with this metric is given by
\begin{equation}
  \Gamma_{prs}=g_{pq}\Gamma^q_{rs}
   = \langle \delta_p A \, , {\cal D}_r\delta_s A \rangle
   = \langle \delta_p A \, , {\cal D}_s\delta_r A \rangle \, .
\end{equation}

The calculation of the Riemann tensor is somewhat more complex.
Straightforward calculation yields
\begin{eqnarray}
   R_{pqrs} &=& g_{st} \left[ \partial_q \Gamma^t_{pr}
      -\partial_p \Gamma^t_{qr}  + \Gamma^u_{pr} \Gamma^t_{qu}
      - \Gamma^u_{qr} \Gamma^t_{pu}  \right]  \cr\cr
     &=& ie \langle \phi_{qp} \, [\delta_r A, \delta_s A] \rangle
    + \langle {\cal D}_q \delta_s A\, ,
            {\cal D}_p \delta_r A \rangle
     - \langle {\cal D}_p \delta_s A\, ,
         {\cal D}_q \delta_r A \rangle
     \cr\cr &&
    - \langle {\cal D}_q \delta_s A\, , \delta_t A \rangle
     g^{tu} \langle \delta_u A\, , {\cal D}_p\delta_r A \rangle
    + \langle {\cal D}_p \delta_s A\, , \delta_t A \rangle
     g^{tu} \langle \delta_u A\, , {\cal D}_q\delta_r A \rangle
     \, .  \cr &&
\end{eqnarray}
To proceed further, we use the facts that the $\delta_p A$ are
annihilated by the zero mode operator $\hat{\cal D}$, and that these
together with the nonzero eigenmodes of
$\hat{\cal D} \hat{\cal D}^\dagger$ form a complete set, to
write
\begin{equation}
     \langle {\cal D}_q \delta_s A\, ,
            {\cal D}_p \delta_r A \rangle
  -\langle {\cal D}_q \delta_s A\, , \delta_t A \rangle
     g^{tu} \langle \delta_u A\, , {\cal D}_p\delta_r A \rangle
   = - \langle {\cal D}_q \delta_s A\, ,
         \Pi {\cal D}_p \delta_r A \rangle
\end{equation}
where
\begin{equation}
   \Pi = \hat {\cal D}^\dagger
      \left(\hat{\cal D} \hat{\cal D}^\dagger \right)^{-1}
       \hat{\cal D}
      = - \hat {\cal D}^\dagger \left(D_aD_a\right)^{-1}
       \hat{\cal D}
\end{equation}
projects onto the space orthogonal to the zero modes.

The evaluation of the expression on the right-hand side is simplified
by working in the equivalent quaternionic formulation defined by
Eqs.~(\ref{bosetofermi}) and (\ref{Diraczeromodeeq}); in this
language,
\begin{equation}
    \hat{\cal D}_{ab} \, \delta_p A_b
       = {1\over 2} \tr \, e_a e_c D_c e_b^\dagger \, \delta_p A_b
\end{equation}
where the $e_a$, defined by $e_j= - i \sigma_j$, $e_4 = I$, obey
\begin{equation}
    e_a e_b^\dagger
      = \delta_{ab} + i \sigma_k \bar \eta^k_{ab} \, .
\end{equation}
Making use of Eqs.~(\ref{D2Phi}) and (\ref{JdeltaA}), and
defining $J^{(4)q}_p \equiv \delta_p{}^q$, one eventually obtains
\begin{eqnarray}
   R_{pqrs} &=& ie \left\{
     \langle \phi_{qp}, [\delta_r A, \delta_s A] \rangle
      - {1\over 2} \langle J^{(c)t}_q J^{(c)u}_p \phi_{ur},
         [\delta_t A, \delta_s A] \rangle
      \right. \cr \cr  && \qquad \qquad  \left.
      +{1\over 2} \langle J^{(c)t}_p J^{(c)u}_q \phi_{ur},
         [\delta_t A, \delta_s A] \rangle \right\}  \, .
\label{riemanntensor}
\end{eqnarray}

In this chapter we will be concerned solely with fermion fields that
transform under the adjoint representation of the gauge group.  The zero
modes of such fermions are closely related to the bosonic zero
modes.\footnote{The more complex issues that arise with fermions in
other representations are discussed in
Appendix~\ref{nequal2appendix}.}  In Sec.~\ref{indexsection}, we
showed that the $\delta_r A_a$ could be obtained by first seeking
solutions of the Dirac equation
\begin{equation}
     \Gamma^a D_a \chi = 0
\label{chap8eucdirac}
\end{equation}
for an adjoint representation $\chi$, with $\Gamma_i = \gamma_0
\gamma_i$ and $\Gamma_4 = \gamma_0$ being a set of Hermitian Euclidean
gamma matrices.  These fermionic solutions are all antichiral with
respect to $\Gamma_5 = \Gamma_1 \Gamma_2 \Gamma_3 \Gamma_4$.  This
construction can be inverted to express the fermionic zero modes in
terms of the bosonic ones via
\begin{equation}
   \chi_r = \delta_r A_a \, \Gamma^a \, \zeta
\label{fermionzeromodesubr}
\end{equation}
where $\zeta$ is a c-number spinor.  Without loss of generality,
we can require that $\zeta$ satisfy
\begin{equation}
    \zeta^\dagger \Gamma^a \Gamma^b \zeta
         = \delta_{ab} + i \bar\eta^3_{ab} \, .
\label{zetaconvention}
\end{equation}

The actions of the complex structures on the bosonic zero modes have
counterparts on the fermionic modes.  If $\zeta$ is chosen to obey
Eq.~(\ref{zetaconvention}), then
\begin{eqnarray}
    J^{(1)}_r{}^s \chi_s &=& i \Gamma^0 \,\chi^c_r
  \label{J1onPsi}\\
    J^{(2)}_r{}^s \chi_s &=&  -\Gamma^0 \,\chi^c_r
   \label{J2onPsi}\\
    J^{(3)}_r{}^s \chi_s &=& i    \,\chi_r
   \label{J3onPsi}
\end{eqnarray}
where the charge conjugate spinor is defined by
\begin{equation}
    \chi^c = C (\bar \chi)^T
\end{equation}
with\footnote{Equations~(\ref{J1onPsi}) and (\ref{J2onPsi}) assume a
particular choice of the arbitrary phase in $C$.  This choice, of
course, has no effect on our final results.}  $C^{-1} \gamma_\mu C = -
(\gamma_\mu)^T$.  These equations reflect the fact that the mapping
from bosonic to fermionic modes is two-to-one, as was first noted in
Sec.~\ref{indexsection}.  Thus, while the hyper-K\"ahler structure
relates four bosonic zero modes to each other, on the fermionic side
it only couples pairs of charge conjugate zero modes.

\section{Low-energy effective Lagrangians from
SYM theories}
\label{lowELagInSYM}

Here, and for the remainder of this review, we will take a low-energy
approach to dyonic BPS states.  Instead of solving for solitonic
solutions, we will realize dyons as excited states of monopoles in the
moduli space dynamics, much as we did, in the purely bosonic context,
in Sec.~\ref{addlHiggsSec}.  To this end, we assume that one linear
combination of the scalar fields, $b$, has a vev that is much greater
than that of all the others, and that to lowest order $b$ satisfies
the primary BPS Eq.~(\ref{bbps}).  Furthermore, we recall from
Sec.~\ref{chap7energyboundsec} that the magnetic charge, the electric
charge, and the scalar field vacuum expectation values can all be
simultaneously rotated into the Cartan subalgebra and then represented
by vectors in the root space.  We assume that when this is done $\bf
b$ is parallel to $\bf g$, and that the remaining scalar fields all have
vevs that are orthogonal to $\bf g$.

\subsection{ ${\cal N}=2$ SYM}
\label{N2lowenergy}

We first
consider pure ${\cal N}=2$ SYM theory, whose
Lagrangian\footnote{This can be obtained from the ${\cal N}=4$
Lagrangian of Eq.~(\ref{Nequal4Lag}) by defining $\chi = \chi_1 + i\chi_2$,
$G_{12}=b$, $H_{12}=a$, and setting the remaining fermionic and
scalar fields to zero.}
we write as
\begin{eqnarray}
   {\cal L} &=& \Tr \left\{ -{1 \over 2} F_{\mu\nu}^2
     + (D_\mu b)^2 +  (D_\mu a)^2 +e^2 [b,a]^2
    \right. \cr && \qquad\qquad \left.
    +  i \bar \chi \gamma^\mu D_\mu \chi -e \bar \chi [b, \chi]
    + ie \bar \chi \gamma^5 [a, \chi] \right\} \, .
\end{eqnarray}
As discussed above, we
fix the U(1) R-symmetry by requiring that to lowest order
$b \equiv A_4$ obey the purely magnetic primary BPS equation, and that
$a$ (which is assumed to be small) be orthogonal to the magnetic
charge (i.e., that $\bf a$ and $\bf g$ be orthogonal).

We construct our low-energy approximation by supplementing our
previous requirement that the motion of $A_a$ be restricted to the
moduli space with the assumption that only the zero modes of $\chi$
are excited.  Thus, we have
\begin{eqnarray}
    A_a &=& A_a({\bf x}, z(t))       \\
    \chi &=&  \chi_r({\bf x}, z(t)) \, \lambda^r(t) =
        \delta_r A_a({\bf x}, z(t)) \, \Gamma^a \zeta
             \, \lambda^r(t) \, .
\label{N2SYMansatz}
\end{eqnarray}
The $\lambda^r$ are Grassmann-odd collective coordinates.
Ordinarily,
one might have allowed these coefficients to be complex.  However,
Eq.~(\ref{J3onPsi}) would then imply that they were not all
independent.  Instead, we obtain a complete set of independent
variables by taking the $\lambda^r$ to be all real.

The $z^r$ and $\lambda^r$ are the only independent dynamical
variables.  The other bosonic fields, $A_0$ and $a$, are to be viewed
as dependent variables to be expressed in terms of these collective
coordinates.  As previously, we require these fields to be small, so
that our procedure will lead to a valid expansion.  More precisely, if
we denote the order of our expansion by $n$, then the velocities $\dot
z^r$ and the field $a$ are both of order $n=1$, and each fermionic
variable is of order $n=1/2$.  We include terms up to order $n=2$ in
the Lagrangian, and so only need the lowest order ($n=1$) approximation
to $A_0$ and $a$.  To this order, $A_0$ and $a$ are determined by
solving their static field equations in a fixed background.  The
solutions of these equations must then be substituted back into the
Lagrangian to yield an effective action for the collective
coordinates.  This generalizes the procedure by which we obtained the
moduli space potential energy from the low-energy dynamics of the
bosonic fields in Chap.~\ref{multihiggsChap}.

Substituting our ansatz for $\chi$, and using
Eq.~(\ref{zetaconvention}) and the Grassmann properties of the
$\lambda^r$, we find that Gauss's law can be written as
\begin{equation}
   D_a \left(D_a A_0 - \dot A_a \right)
           = 2ie Y^{rs} [\delta_r A_a, \delta_s A_a]
\label{chap8gauss}
\end{equation}
where
\begin{equation}
   Y^{rs} = -{i \over 4}\, \lambda^r  \lambda^s \, .
\end{equation}
The static field equation for $a$ is
\begin{equation}
   D_a^2 a = 2ie Y^{rs}   [\delta_r A_a, \delta_s A_a] \, .
\label{chap8Da2}
\end{equation}
(In obtaining this equation, we have used the fact that the
chirality properties of the fermion zero modes imply that
$\gamma^0\gamma^5 \chi = - i \chi$.)
Recalling Eq.~(\ref{D2Phi}), we see that these equations are solved by
\begin{equation}
    A_0 = \dot z^r \epsilon_r + Y^{rs}\phi_{rs}\label{A0Solution}
\end{equation}
\begin{equation}
    a = \bar a +  Y^{rs}\phi_{rs}
\end{equation}
where
$\bar a$ is a solution of the homogeneous equation $D^2_a \bar a =0$.
In fact, from the discussion in Sec.~\ref{addlHiggsSec}, we know that
\begin{equation}
     D_a \bar a = G^r \, \delta_r A_a \label{chap8Dabar}
\end{equation}
where
\begin{equation}
      G^r = a^A K_A^r
 \end{equation}
is a linear combination of the triholomorphic Killing vector fields
corresponding to U(1) gauge transformations.  The factor of $a^A$, the
coefficient that appears when the expectation value of $a$ is expanded
in terms of fundamental weights, arises from the requirement that $a$
attain its vacuum expectation value at spatial infinity.

We must now substitute our results back into the Lagrangian.  The
lowest order term,
\begin{equation}
   L_0 = \int d^3x \,\Tr \left\{ -{1 \over 2} F_{ab}^2
          +i \bar \chi \gamma^a D_a \chi \right\}
     =  - {\bf b}\cdot {\bf g}
\end{equation}
is just minus the energy of the static purely magnetic solution,
with the fermion term giving a vanishing contribution.
The leading nontrivial part of the dynamics arises from the next
contribution,
\begin{eqnarray}
   L_1  &=& \int d^3x\, \Tr \left\{  F_{a0}^2 - (D_a a)^2
     + i \bar \chi \gamma^0 D_0 \chi
        + ie \bar \chi \gamma^5 [a, \chi] \right\} \cr \cr
   &=&  \int d^3x \, \Tr \left\{
          (-\dot z^r \delta_r A_a +  Y^{rs} D_a  \phi_{rs}  )^2
        - (D_a \bar a + Y^{rs} D_a  \phi_{rs})^2
         \right. \cr \cr && \qquad \left.
       + i  \chi^\dagger \dot \chi
       + e \chi^\dagger [(-\dot z^r \epsilon_r + \bar a), \chi]
      \right\}  \, ,
\end{eqnarray}
whose evaluation we must now undertake.

We start with the first two, purely bosonic, terms.  We note that:
\begin{itemize}
\item{} The square of the first part of the $F_{a0}$ term gives the
  usual bosonic collective coordinate kinetic energy, $(1/2)
  g_{rs} \dot z^r \dot z^s$.

\item{} The square of the first part of the $D_a a$ term gives
  the bosonic potential energy of Chap.~\ref{multihiggsChap},
  $-(1/2) g_{rs}G^r G^s$.

\item{} There is no contribution from the cross-terms linear in $Y$.
  This follows by integrating by parts and using the background gauge
  condition (in the first term) and the fact that $\bar a$ solves the
  homogeneous equation (in the second term).

\item{} The terms quadratic in $Y$ cancel.  When we turn to the ${\cal
  N}=4$ case, we will find that the analogous terms survive and lead
  to four-fermion contributions to the Lagrangian.
\end{itemize}

This leaves us with the fermionic terms.  After integration over the
spatial coordinates, the terms independent of
$\bar a$ give
\begin{eqnarray}
  &&  i \int d^3x \Tr\left( \chi^\dagger \dot \chi
         + ie \chi^\dagger [ \dot z^p \epsilon_p , \chi] \right)
    \cr && \qquad\qquad
    =   i \int d^3x \, \zeta^\dagger  \Gamma^a \Gamma^b \zeta \,
           \lambda^r
   \left[  \Tr \left(\delta_r A_a \delta_s A_b \right) \dot \lambda^s
       + \dot z^p \lambda^s \, \Tr \left(\delta_r A_a
         {\cal D}_p \delta_s A_b \right) \right]
  \cr && \qquad\qquad = {i \over 2}\, g_{rs} \lambda^r \dot \lambda^s
     +{i \over 2}\, g_{rs} \Gamma^s_{pq} \dot z^p \lambda^r\lambda^q
  \cr   && \qquad\qquad
          ={i \over 2}\, g_{rs} \lambda^r D_t \lambda^s \, .
\end{eqnarray}
Here $D_t$ is the covariant time derivative along the trajectory
$z(t)$, and we have dropped a total time derivative term.
The Yukawa term leads to
\begin{eqnarray}
   4ie Y^{rs}\int d^3x\, \Tr \delta_r A_a \, [\bar a, \delta_s A_a]
  &=& 4   Y^{rs}  \int d^3x\,
      \Tr \left[ \delta_r A_a \left( D_a {\cal D}_s \bar a
           - {\cal D}_s D_a \bar a \right) \right]  \cr
  &=& -4 Y^{rs}  \int d^3x\,
    \Tr \left[ \delta_r A_a {\cal D}_s \left( G^u \delta_u A_a
           \right) \right]  \cr
  &=&  -2 Y^{rs}\, g_{rq} \left(\partial_s G^q
        + \Gamma^q _{us} G^u\right)  \cr
  &=&  2 Y^{rs} \, \nabla_r G_s \, .
\end{eqnarray}

Adding all these pieces together, we obtain the low-energy effective
Lagrangian
\begin{equation}
   L  = {1 \over 2}\left[g_{rs} \,\dot z^r \dot z^s
    + ig_{rs} \lambda^r D_t \lambda^s - g_{rs} G^r G^s
   -i\lambda^r\lambda^s \nabla_r G_s   \right]  - {\bf b}\cdot {\bf g} \, .
\label{N2lowenergyLag}
\end{equation}

\subsection{ ${\cal N}=4$ SYM}

In Eq.~(\ref{Nequal4Lag}) we wrote the ${\cal N}=4$ SYM Lagrangian
in a form that made the SU(4) = SO(6) R-symmetry
manifest.  A form that is more convenient for our present purposes
is obtained by defining
\begin{eqnarray}
    \Phi_1 = \eta_{rs}^2 G_{rs}  \, ,  \qquad
    \Phi_2 &=& \eta_{rs}^1 G_{rs}  \, ,  \qquad
    \Phi_3 = \bar\eta_{rs}^2 H_{rs} \, ,   \cr\cr
    \Phi_4 = \bar\eta_{rs}^3 H_{rs}   \, ,  \qquad
    \Phi_5 &=& -\bar\eta_{rs}^1 H_{rs}   \, ,  \qquad
    \Phi_6 = \eta_{rs}^3 G_{rs} \, ,    \cr\cr
    \chi =  \chi_1 + i \chi_2  \, ,  &&\qquad
    \xi  =  \chi_3 + i \chi_4 \, .
\end{eqnarray}
This gives
\begin{eqnarray}
   {\cal L} &=& \Tr \left\{ -{1\over 2}F_{\mu\nu}^2
       + \sum_P(D_\mu \Phi_P)^2
       + {e^2 \over 2}\sum_{P,Q}[\Phi_P, \Phi_Q]^2
 + i \bar\chi \gamma^\mu D_\mu \chi
      + i \bar\xi \gamma^\mu D_\mu \xi
      \right. \cr\cr &&   \left.
      - e \bar \chi[\Phi_6,\chi] - e \bar \xi [\Phi_6,\xi]
     +ie \bar \chi \gamma^5 [\Phi_4, \chi]
     -ie \bar \xi \gamma^5 [ \Phi_4, \xi]
      \right. \cr\cr  &&  \left.
      +i e\bar \chi [ (\Phi_1 - i \Phi_2),\xi^c]
      -i e \bar\xi^c [ (\Phi_1 + i \Phi_2),\chi]
      \right. \cr\cr &&  \left.
      + e \bar \chi \gamma^5 [(\Phi_3 + i\Phi_5),\xi]
      - e \bar \xi \gamma^5 [ (\Phi_3 - i\Phi_5), \chi]
      \vphantom{{e^2 \over 4}} \right\} \, .
\end{eqnarray}

We now choose $\Phi_6 $ to be the primary BPS field $b$ which
plays the role of $A_4$ in the Euclidean four-dimensional notation, in
whose background $\chi$ and $\eta$ both have zero modes given by
Eq.~(\ref{fermionzeromodesubr}).  We generalize the ${\cal N} =2$
low-energy ansatz of Eq.~(\ref{N2SYMansatz}) to
\begin{eqnarray}
    A_a &=& A_a({\bf x}, z(t))         \cr\cr
    \chi &=&   \delta_r A_a({\bf x}, z(t)) \, \Gamma^a \zeta
            \, \, \lambda^r_\chi(t)    \cr\cr
    \xi &=&   \delta_r A_a({\bf x}, z(t)) \,  \Gamma^a \zeta
             \,\, \lambda_\xi^r(t)
\label{N4SYMansatz}
\end{eqnarray}
with $\lambda_\chi$ and $\lambda_\xi$ both real.  Again, we
must determine the remaining fields in terms of the collective
coordinates.

Proceeding as in the ${\cal N}=2$ case, we write Gauss's
law as
\begin{equation}
   D_a (D_a A_0 - \dot A_a) = 2ie Y_0^{rs} [\delta_r A_b , \delta_s A_b]
\label{N4gauss}
\end{equation}
and the equations for the remaining scalar fields $a_I=\Phi_I$ ($I=1,2,\dots,5$)
as
\begin{equation}
   D_a^2 a_I = 2ie Y_I^{rs} \,[\delta_r A_b , \delta_s A_b]
\label{N4scalarEq}
\end{equation}
where
\begin{eqnarray}
   Y_0^{rs} &=&  -{i\over 4}\left( \lambda^r_\chi \lambda^s_\chi
         + \lambda^r_\xi \lambda^s_\xi \right)\cr \cr
   Y_1^{rs} &=&   -{i\over 4}\left(
           \lambda^q_\chi J^{(1)}_q{}^r  \lambda^s_\xi
        -   \lambda^q_\xi J^{(1)}_q{}^r  \lambda^s_\chi \right)
                 \cr\cr
   Y_2^{rs} &=&   -{i\over 4}\left(
           \lambda^q_\chi J^{(2)}_q{}^r  \lambda^s_\xi
        -   \lambda^q_\xi J^{(2)}_q{}^r  \lambda^s_\chi \right)
          \cr\cr
   Y_3^{rs} &=&   -{i\over 4}\left(
           \lambda^q_\chi J^{(3)}_q{}^r  \lambda^s_\xi
        -   \lambda^q_\xi J^{(3)}_q{}^r  \lambda^s_\chi \right)
          \cr\cr
   Y_4^{rs} &=& -{i\over 4} \left( \lambda^r_\chi \lambda^s_\chi
         - \lambda^r_\xi \lambda^s_\xi \right) \cr\cr
   Y_5^{rs} &=&  - {i\over 4}\left( \lambda^r_\chi \lambda^s_\xi
         + \lambda^r_\xi \lambda^s_\chi   \right) \, .
\end{eqnarray}
If we combine $\lambda^r_\chi$ and $\lambda^r_\xi$ into a single two-component
spinor
\begin{equation}
   \eta^r = \left(\matrix{\lambda^r_\chi \cr\cr \lambda^r_\xi }
           \right)
\end{equation}
and write $\bar\eta = \eta^T\sigma_2$,
then $Y_0^{rs}$ and the $Y_I^{rs}$ can be written more compactly as
\begin{equation}
    Y_0^{rs} = -{i \over 4} \bar\eta^r \sigma_2  \eta^s
\end{equation}
and
\begin{equation}
    Y_I^{rs} = -{i \over 4} \bar\eta^r (\Omega_I)^s{}_q \eta^q
\end{equation}
where
\begin{eqnarray}
    (\Omega_j)^s{}_q &=&  i J^{(j)}_q{}^s \, , \qquad j=1,2,3 \cr \cr
    (\Omega_4)^s{}_q &=& i\sigma_1 \delta^{qs} \cr \cr
    (\Omega_5)^s{}_q &=& -i\sigma_3 \delta^{qs} \, .
\end{eqnarray}

Equations~(\ref{N4gauss}) and (\ref{N4scalarEq}) are solved by
\begin{eqnarray}
    A_0 &=& \dot z^r \epsilon_r + Y_0^{rs} \phi_{rs}  \cr
   \Phi_I = a_I &=& \bar a_I + Y_I^{rs} \phi_{rs}
\end{eqnarray}
with the $\bar a_I$ being solutions of $D_a^2 \bar a_I=0$ that give
the expectation values of the $\Phi_I$ at spatial infinity.  These
solutions must then be substituted back into the Lagrangian.  Most of
the manipulations are completely analogous to those used in the ${\cal
N}=2$ case.  The only new feature is that the four-fermion terms terms
no longer cancel.  Instead, using
the identity
\begin{equation}
    (\sigma_j)_{\alpha\beta} (\sigma_j)_{\gamma\delta}
   = 2\delta_{\alpha\delta} \delta_{\beta\gamma}
       - \delta_{\alpha\beta} \delta_{\gamma\delta}
\end{equation}
and recalling Eq.~(\ref{riemanntensor}), we find that these are
equal to
\begin{eqnarray}
   -2ie \left[Y_I^{pq} Y_I^{rs} - Y_0^{pq} Y_0^{rs} \right] \Tr \int
     d^3x \, \phi_{rs}\, [\delta_p A_a, \delta_q A_a]
     = -{1\over 12}
     R_{rstu}\, \bar\eta^r \eta^t \bar\eta^s \eta^u
     \, .
\end{eqnarray}
The final form of the low-energy effective Lagrangian is then
\begin{eqnarray}
  L &=& {1 \over 2}\left[g_{rs} \,\dot z^r \dot z^s
            + ig_{rs} (\eta^r)^T D_t \eta^s
     -  g_{rs} G^r_I G^s_I
    -i \bar \eta^r (\Omega_I \eta)^s\,\nabla_r G_{Is} \right]
        \cr &&
        -{1\over 8}
     R_{rstu}\, (\eta^r)^T \eta^s (\eta^t)^T \eta^u
        - {\bf b}\cdot {\bf g} \, ,
\label{N4lowenergyLag}
\end{eqnarray}
We have used here the identity
\begin{equation}
     R_{rstu}\, \bar\eta^r \eta^t \bar\eta^s \eta^u
     = {3\over 2}
     R_{rstu}\, (\eta^r)^T \eta^s (\eta^t)^T \eta^u\, ,
\end{equation}
which can be derived by using the cyclic symmetry of the Riemann
tensor together with the symmetry properties of the products of four
Grassmann variables.

Finally, we note that
the structure of the $\Omega_I$, which may at first seem a bit
strange, can be understood in terms of an R-symmetry.  The original
field theory possessed an SO(6) R-symmetry.  This was broken to SO(5)
when we singled out one of the scalar fields as part of the solution to
the primary BPS equation. This SO(5) symmetry is inherited by the
moduli space Lagrangian, and acts on the fermion variables by
\begin{eqnarray}
&&\eta \rightarrow
e^{{1\over 2}\theta_{KL}{\cal J}_{KL}}\,\,\eta
\label{phase1}
\end{eqnarray}
where $\theta_{KL}= -\theta_{LK}$ is a real parameter and
the SO(5) generators are given by
\begin{equation}
{\cal J}_{ij}=1\otimes \epsilon_{ijk}\,J^{(k)}\,, \ \ \
{\cal J}_{45}=i\sigma_2\otimes 1\,,
\ \ \ {\cal J}_{4i}=\sigma_1 \otimes J^{(i)}\,,\ \ \
{\cal J}_{5i}=-\sigma_3 \otimes J^{(i)}
\label{generator1}
\end{equation}
with $s,t,u=1,2,3$.  The definition of the $\Omega_I$ is such that the
$Y^{rs}_I$ transform as five-vectors under this transformation.
The ten generators of SO(5) in (\ref{generator1}) exhaust all the possible
covariantly constant, antisymmetric structures present in the {\cal N}=4
supersymmetric sigma model without potential, so this
realization of the R-symmetry is rather unique.

\section{Low-energy supersymmetry and quantization }
\label{lowESUSYandquant}

The low-energy Lagrangians that we have obtained in the previous
section possess supersymmetries that are inherited from those of the
underlying quantum field theories.  In this section we will describe
these moduli space supersymmetries in detail and then discuss
the quantization of these theories.

\subsection{Superalgebra with a central charge\label{general}}

An important feature of these low-energy effective theories is
the presence of a bosonic potential energy term of the
form $|G|^2$, where $G$ is a triholomorphic Killing vector
field built out of generators of gauge isometries.
It is instructive to examine how such a
potential can modify, but still preserve, a sigma-model supersymmetry.
Of
particular importance is to realize how a central charge emerges from
$G$, and how a quantum state with this moduli-space dynamics can preserve
all or part of the supersymmetries of the quantum mechanics.

A classic paper that dealt with potentials in supersymmetric sigma
models is Ref.~\cite{Alvarez-Gaume:1983ab}, where many of the
mathematical structures we shall see in this chapter were
discussed in the context of two-dimensional supersymmetric sigma models.
A further generalization of this formal approach was
given in Ref.~\cite{Hull:1993ct}, and the general structure of the
latter accommodates well the low-energy dynamics of monopoles from
SYM. Since we are dealing with
quantum mechanics instead of a two-dimensional field theory, our
discussion is related to discussions in these two references by
a dimensional reduction, but the basic supersymmetry structures
remain the same.\footnote{More recent papers which considers
massive quantum mechanical non-linear sigma-model with extended
supersymmetries include Refs.~\cite{Peeters:2001np,Bellucci:2001ax}.}

\subsubsection{Complex superalgebra}

The conventional examples of moduli
space dynamics are
all sigma models whose quantum mechanical degrees
of freedom live freely on some smooth manifold.  With a
flat target space, say $R^n$,
the simplest sigma model Lagrangian with a complex
supersymmetry is
\begin{equation}
{ L}={1\over 2}   \dot{z}_q \dot{ z}^q +
i \varphi_q^* \partial_t \varphi^q \, .
\end{equation}
Upon canonical quantization,
\begin{equation}
   [z^q, p_r]=i\delta^q_r\, ,\qquad
   \{\varphi^q,\varphi_r^*\}=\delta^q_r \, ,
\end{equation}
one finds a Hamiltonian
\begin{equation}
{ H}=\frac12\,p_q p^q
\end{equation}
and a complex supercharge
\begin{equation}
S=\varphi^q p_q
\end{equation}
that satisfy
\begin{equation}
\{ S,S^\dagger\}=2{H}  \, ,\qquad \{S,S\}
             = \{S^\dagger,S^\dagger\} = 0 \, .
\end{equation}
All of this can be generalized to the more general case where the
target manifold is curved.  However, if there is more than one
possible supersymmetry, there will often be restrictions on the target
manifold.  Thus, the fact that the monopole moduli space is hyper-K\"ahler
may be understood as being due to the existence of four
supersymmetries.  When the monopoles are from an ${\cal N}=4$ SYM theory,
these supersymmetries are complex, as in the
trivial example above, while ${\cal N}=2$ theories lead to real
supersymmetries, which we will examine shortly.

Now let us imagine inventing a new supercharge of the form
\begin{equation}
{  Q}\equiv S-\varphi^*_qG^q=\varphi^q p_q-\varphi^*_qG^q
\end{equation}
where $G^r$ is a vector field on the target manifold, and
see if a sensible Lagrangian exists and is invariant under
such a supercharge. Actually, there is more than
one way of adding such an additional term associated with
a vector field.  The simplest variation would be to rotate the
second term by a phase; e.g.,
\begin{equation}
{  Q}\equiv S+i\varphi^*_qG^q=\varphi^q p_q+i\varphi^*_qG^q  \, .
\end{equation}
In terms of the moduli dynamics we found for monopoles in ${\cal N}=4$
SYM, the first choice corresponds to turning on only $G_4$, while the
second choice corresponds to turning on only $G_5$.  For the sake of
simplicity, we will consider here only the first choice.

The anticommutator of the modified supercharges defines a Hamiltonian
\begin{equation}
   { H}= {1 \over 2}\{  Q,  Q^\dagger\}
   = {1 \over 2} p_q p^q +{1 \over 2} G_q G^q
+i\partial_qG_r \varphi^{q*}\varphi^{r*}
+i \partial_qG_r\varphi^q\varphi^r
\end{equation}
that can be easily derived
from the Lagrangian
\begin{equation}
  {L}={1\over 2} \,  \dot{z}_q \dot{ z}^q -\frac12\,  G_qG^q +
i \varphi_q^* \partial_t \varphi^q -i\, \partial_qG_r \varphi^{q*}
\varphi^{r*}-i\, \partial_qG_r\varphi^q\varphi^r  \, .
\end{equation}
Note the appearance of both the potential term, $G^2/2$, and its
superpartner.

If the supercharge is to generate
a symmetry, then $[{  Q}, {H}]$ must vanish, which
in turn implies that $[  Q^\dagger,  Q^2]=0$. Computing this last
identity, one finds that it is satisfied if and only if
\begin{equation}
\partial_qG_r+\partial_r G_q=0 \, .
\end{equation}
In other words, $G$ must be a Killing vector field. It then follows
that
\begin{equation}
{Z}\equiv
\left(G^q p_q-i\partial_rG_q\varphi^r\varphi^{q*}\right)
=-\frac12\,\{  Q^\dagger,   Q^\dagger\}=
-\frac12\,\{  Q,  Q\}
\end{equation}
is a conserved quantity and becomes the central charge
of the superalgebra.  Thus, the
modified superalgebra closes as long
as $G$ generates an isometry on the target manifold. In terms of
\begin{equation}
Q_\pm=Q\pm Q^\dagger
\end{equation}
the superalgebra become
\begin{equation}
\{Q_\pm,Q_\pm\}= \pm 2{H}-2{ Z} \, ,\qquad
\{Q_\pm,Q_\mp\}= 0 \, .
\end{equation}
A BPS state of such a quantum mechanics typically preserves half of
the supersymmetry, either $Q_+$ or $Q_-$, depending on whether the
eigenvalue of ${ Z}$ is positive or negative. This is how the 1/4-BPS
dyons we encountered in Chaps.~\ref{BPSchap} and \ref{multihiggsChap} are
realized in the moduli space dynamics of monopoles in ${\cal N}=4$
SYM.

Having a potential energy term in a supersymmetric theory is hardly
new; the important point here is that this particular form extends
naturally to cases with four supersymmetries.  In the case of sigma
models with four (real or complex) supercharges, the appropriate
constraints on $G$ are that it should be a Killing vector field, as
above, and that it should also be triholomorphic. That is, the
diffeomorphism flow induced by $G$ should preserve not only the metric
but also all three complex structures. Thus, the modified moduli space
dynamics we have found fits quite naturally with this deformation of
supersymmetry.

Having an electric charge means that a state is
not invariant under the gauge isometries, but rather has
nonzero momenta conjugate to the associated cyclic coordinates.
This generically translates to having a nontrivial
eigenvalue of the central charge ${ Z}$, resulting
in a state that preserves at most half of the
supersymmetries. The moduli space dynamics inherits four
complex supercharges from the field theory; from these,
four real supercharges are preserved by the special states
whose energies equal the absolute value of the central
charge. These states preserve $4=1/4\times 16$ supercharges, just
as 1/4-BPS states in ${\cal N}=4$ SYM theory should.

\subsubsection{Real superalgebra}

A curious variation on this happens if we
consider real supersymmetry. Returning to the
example of a flat target manifold as above, but
with real fermions obeying
\begin{equation}
   \{\lambda^q, \lambda^r\} = \delta^{qr}  \, ,
\end{equation}
we have a Lagrangian
\begin{equation}
 { L} ={1\over 2}   \dot{z}_q \dot{ z}^q +
\frac{i}{2} \lambda_q\partial_t \lambda^q
\end{equation}
and a sigma-model supercharge
\begin{equation}
S = \lambda^q p_q \, .
\end{equation}
Twisting gives
\begin{equation}
Q= S-\lambda_q G^q= \lambda^q (p_q-G_q)  \, .
\end{equation}
This would normally be regarded as the introduction
of a gauge field $G$ on the target manifold. However,
we must try a different interpretation here.

Motivated by the
appearance of a central charge in the case of complex
supersymmetry, let us split $Q^2$
into two pieces as
\begin{equation}
  Q^2={ H}-{Z}
\end{equation}
where
\begin{eqnarray}
{H}&=& \frac12\,(p_q p^q+G_qG^q)
-\frac{i}{4} (\partial_q G_r -\partial_r G_q)
\lambda^r\lambda^q \nonumber\\
{ Z}&=& G^q p_q+
   \frac{i}{4}(\partial_q G_r -\partial_r G_q)\lambda^r\lambda^q \, .
\end{eqnarray}
Again, the superalgebra closes, with two separately
conserved quantities ${ H}$ and ${Z}$, as long
as $G$ is a
Killing vector field.

In actual examples, the identification of ${H}$ (as opposed to ${H}-{
Z}$) as the energy has to be made by examining how the energy defined
by the field theory propagates to the moduli space dynamics. Once this
is done, we are left with the possibility of positive energy BPS
states that preserve all the supercharges of the moduli space
dynamics.  Again, four supersymmetries are preserved by such BPS
states.  The dyons of ${\cal N}=2$ SYM theory generically arise from the
moduli space dynamics, and the BPS dyons obtained in this manner would
be 1/2-BPS with respect to ${\cal N}=2$.

\subsection{Low-energy superalgebra from pure ${\cal N}=2$ SYM}
\label{n=2q}

The low-energy dynamics for ${\cal N}=2$ SYM theory that we derived in
Sec.~\ref{N2lowenergy} provides an example with real supersymmetry on
a curved target space.  The effective Lagrangian,
Eq.~(\ref{N2lowenergyLag}), gives an action that is invariant under
the supersymmetry transformations \cite{Gauntlett:1999vc,Gauntlett:2000ks}
\begin{eqnarray}
\delta z^q   &=& -i\ep\lambda^q -i\sum_{j=1}^3 \ep_{(j)}  \lambda^{r}{J_{\;r}^{(j)q}}
\nn
\delta \lambda^q &=&\ep(\dot z^q -G^q) +i\ep \lambda^r\Gamma^q_{rs}\lambda^s
\nn
&+&\sum_{j=1}^3\ep_{(j)}\left[-(\dot
z^r- G^r){J_{\;r}^{(j)q}} + i \lambda^r {J_{\;r}^{(j)t}}\Gamma^q_{ts}\lambda^s
 \right]\label{N2transformation}
\end{eqnarray}
with Grassmann-odd parameters $\ep$ and $\ep_{(j)}$.
The term containing  $\Gamma^q_{rs}\lambda^r\lambda^s$
in the second line vanishes on its own due to the symmetry property of
$\Gamma^q_{rs}$, but we have kept it to give the formula a more balanced appearance.
The action is also invariant under the
symmetry transformation
\begin{eqnarray}
 \delta z^q &=&k G^q  \nn
\delta\lambda^q &=&k{G^q}_{,r}\lambda^r
\end{eqnarray}
that is related to the requirement that $G$ be a triholomorphic
Killing vector field.

To quantize this effective action, we first introduce an orthonormal
frame $e^E_q$ and define new fermionic variables $\lambda^E=\lambda^q
e_q^E$ that commute with all bosonic variables.  The remaining
canonical commutation relations are then given by
\begin{eqnarray}
    [z^q,p_r]&=&i\delta^q_r  \cr \cr
    \{\lambda^E,\lambda^F\}&=&\delta^{EF} \, .
\end{eqnarray}
Note that, to have consistent canonical commutators, we have to use
the $\lambda^E$ as the canonical variables instead of the $\lambda^q$.
If we insisted on starting with the $\lambda^q$, the canonical
conjugate pair of fermions would be $\lambda^q$ and
$\lambda_q=g_{qr}\lambda^r$, but these two cannot simultaneously
commute with $p_q$ unless the metric is constant. By using
$\lambda^E$, which is conjugate to itself, we neatly avoid this
problem and, as a bonus, find a natural map $\lambda^E=\gamma^E/{\sqrt
2}$ of the fermions to the Dirac matrices on the moduli space.

The supercovariant momentum operator, defined by
\begin{equation} \label{spinconnection}
\pi_q=p_q-{i\over 4}\omega_{q EF}[\lambda^E,\lambda^F]  \, ,
\end{equation}
where $\omega_{q\, F}^{\, \, E}$ is the spin connection,
appears in the supercharge.  If we make the above
identification between $\lambda^E$ and $\gamma^E/{\sqrt 2}$,
the wave function is interpreted as a spinor on the moduli space.
In this picture, $i\pi_q$ behaves exactly as the
covariant derivative on a spinor, so that we may identify
\begin{equation}
\pi_q =-i\nabla_q  \, .
\end{equation}
Other useful identities are
\begin{eqnarray}
{[}\pi_q,\lambda^r{]}&=&i\Gamma^r_{qs}\lambda^s ,\nn
{[}\pi_q,\pi_r{]}&=&
-{1\over 2}R_{qrst}\lambda^s\lambda^t
\end{eqnarray}
where we have chosen to present the action of $\pi_q$ on $\lambda^q$.
These identities can be derived from Eq.~(\ref{spinconnection}) after taking
into account the fact that the $\lambda^E$ commute with $p_q$.

The four supersymmetry charges take the form
\begin{eqnarray}
Q&=&\lambda^q(\pi_q-G_q) \nn
Q_{j}&=&\lambda^q {J^{(j)r}_{\,\, q}}(\pi_r-G_r)  \qquad j=1,2,3
\label{N2supercharges}
\end{eqnarray}
with $J^{(j)}_{\;rs}=-J^{(j)}_{\;sr}$,
and their algebra  is given by
\begin{eqnarray}
\{Q,Q\}&=&2({H}-{ Z})   \nn
\{Q_j,Q_k\}&=&2\,\delta_{jk}({H}-{ Z})  \nn
\{Q,Q_j\}&=&0
\end{eqnarray}
where the Hamiltonian ${ H}$ and the central charge ${ Z}$ are
\begin{eqnarray}
{H}&=&
{1\over 2\sqrt{g}}\pi_q \sqrt{g }g^{qr}\pi_r
+ {1\over 2}G_q G^q  + {i\over 2}\lambda^q\lambda^r \nabla_q G_r \nn
{ Z}&=& G^q \pi_q -{i\over 2}  \lambda^q\lambda^r(\nabla_q G_r) \, .
\label{hamiltonian1}
\end{eqnarray}
Note that the operator $i{Z}$ is  the Lie derivative ${\cal L}_G$
acting on spinors.

We see that, as in the real supersymmetry example of the previous
section, the states either preserve all four supersymmetries (if
${H}={ Z}$), or else none at all.  A simple, yet unexpected, corollary
is that the dyon spectrum of the parent ${\cal N}=2$ SYM theory
is asymmetric under a change in sign of the electric charges.  If we
flip the sign of all electric charges (while maintaining those in the
magnetic sector), the central charge ${ Z}$ flips its sign as well,
making $Q^2=0$ impossible.  With complex supersymmetry, we can
preserve either $Q+Q^\dagger$ or $Q-Q^\dagger$, depending on the sign
of the central charge.  With real supersymmetry, we do not have this
luxury.  We will see in the next chapter how asymmetric spectra emerge
in specific examples, but it is important to remember that this
asymmetry is largely a consequence of the general form of the
superalgebra.

\subsection{Low-energy superalgebra from ${\cal N}=4$ SYM theory\label{n=4q}}

The low-energy effective Lagrangian for ${\cal N}=4$ SYM theory, given in
Eq.~(\ref{N4lowenergyLag}), is invariant under the complex
superalgebra defined by the transformations \cite{Bak:1999da,Bak:1999vd,Gauntlett:2000ks}
\begin{eqnarray}
\delta z^q &=& \bar\epsilon \eta^q +\sum_{j=1}^3
{\bar\epsilon}_{(j)} (J^{(j)}\eta)^q \cr \cr
\delta \eta^q  &=& -i\dot{z}^q\sigma_2 \epsilon- \bar\epsilon \eta^r\Gamma^q_{rs}  \eta^s
-i (G^{\bI}\Omega^\bI)^q \epsilon \nn
&+&\sum_{j=1}^3 \left[i(J^{(j)}\dot{z})^q \sigma_2\epsilon_{(j)}
-\bar\epsilon_{(j)}(J^{(j)}\eta)^s\Gamma^q_{st}
\eta^t
-i(G^{\bI}J^{(j)}\Omega^\bI)^q\epsilon_{(j)}\right]
\label{transformation}
\end{eqnarray}
where $\epsilon$ and $\epsilon_{(j)}$ are two-component spinor
parameters.\footnote{As in the previous section, we define
$\bar\epsilon=\epsilon^T\sigma_2$
and  $\bar\epsilon_{(j)} =\epsilon^T_{(j)}\sigma_2$.} Thus, there are
eight real (or four complex) supersymmetries.

When the theory is quantized, the spinors $\eta^E = e_q^E \eta^q$
commute with all the bosonic dynamical variables.  The
remaining fundamental commutation relations are
\begin{eqnarray}
&&[z^q, p_r ] = i\delta^q_r \cr\cr
&&\{\eta^E_\alpha, \eta^F_\beta\} = \delta^{EF}\delta_{\alpha\beta}\,.
\label{commutators}
\end{eqnarray}
If we define  supercovariant momenta by
\begin{eqnarray}
&& \pi_q \equiv p_q -{i\over 2}\omega_{EF\,q}
(\eta^E)^T \eta^F
\label{cov}
\end{eqnarray}
where $\omega_{EF\,q}$ is again the spin connection,
the supersymmetry generators
can be written as
\begin{eqnarray}
&&Q_\alpha = \eta^q_\alpha \pi_q
 - (\sigma_2\,\Omega^\bI \eta)^q_\alpha G^{\bI}_q \label{generator0} \\
  \cr
&&Q^{(j)}_\alpha = (J^{(j)}\eta)^{q}_\alpha \pi_q
-(\sigma_2 J^{(j)}\Omega^\bI \eta)^q_\alpha G^{\bI}_q \,.
\label{generator}
\end{eqnarray}
These charges satisfy the {\cal N}=4 superalgebra
\begin{eqnarray}
 \{Q_\alpha,Q_\beta  \} &=& \{Q^{(1)}_\alpha,Q^{(1)}_\beta\}
 =\{Q^{(2)}_\alpha,Q^{(2)}_\beta\}
   =\{Q^{(3)}_\alpha,Q^{(3)}_\beta\}  \cr
&=& \delta_{\alpha\beta} \; {H}
-2(\sigma_3)_{\alpha\beta} \; {Z}_{4}
-2(\sigma_1)_{\alpha\beta} \; { Z}_{5} \cr
  \{Q_\alpha,\ Q^{(i)\,}_\beta  \} &=&
2(\sigma_2)_{\alpha\beta} \; { Z}_{i}  \cr
\{Q^{(1)}_\alpha,Q^{(2)}_\beta  \} &=&
2(\sigma_2)_{\alpha\beta} \; { Z}_{3} \cr
 \{Q^{(2)}_\alpha,Q^{(3)}_\beta  \} &=&
2(\sigma_2)_{\alpha\beta} \; { Z}_{1} \cr
\ \ \  \{Q^{(3)}_\alpha,Q^{(1)}_\beta  \} &=&
2(\sigma_2)_{\alpha\beta} \; { Z}_{2}
\label{algebra4}
\end{eqnarray}
where the Hamiltonian is
\begin{equation}
{ H}=
{1\over 2} \left[ {1\over \sqrt{g}}\pi_q \sqrt{g }g^{qr}\pi_r
+{1\over 4}R_{qrst}(\eta^q)^T \,
\eta^r (\eta^s)^T  \eta^t
 + g^{qr}G^{\bI}_q G^{\bI}_r
+i D_q G^{\bI}_r \, \bar\eta^q
\Omega^\bI \eta^r \right]
\label{hamiltonian4}
\end{equation}
and the
\begin{eqnarray}
{Z}_{I}= G_{\bI}^{\bmu} \pi_\bmu -
{i\over 2}  \nabla_\bmu G^{\bI}_\bnu \,(\eta^\bmu)^T
\eta^\bnu
\end{eqnarray}
are central charges.

It is often useful to write these supercharges in a complex form.
This can be done by defining
\begin{equation}
    \varphi^q\equiv{1\over \sqrt{2}} (\eta_1^q-i\eta_2^q)
    ={1\over \sqrt{2}}(\lambda_\chi^q - i\lambda_\xi^q)
\end{equation}
and then combining the supercharges of Eq.~(\ref{generator0}) to
give
\begin{eqnarray}
    Q&\equiv& {1\over \sqrt{2}}(Q_1-iQ_2)
   = \varphi^q \pi_q
 -\varphi^{*q} (G^4_q-iG^5_q)-i\sum_{j=1}^3
G^j_q (J^{(j)}\varphi)^q
\cr
   Q^\dagger &\equiv& {1\over \sqrt{2}}(Q_1+iQ_2)
= \varphi^{*q} \pi_q
 -\varphi^{q} (G^4_q +iG^5_q)+
i\sum_{j=1}^3 G^j_q (J^{(j)}\varphi^*)^q
\,.  \cr &&
\label{generators}
\end{eqnarray}
with analogous definitions for
$Q^{(i)}$ and ${Q^{(i)}}^\dagger$.
The positive definite nature of the Hamiltonian can be seen easily
in the anticommutators
\begin{equation}
\{Q,Q^\dagger\}=\{Q^{(1)},Q^{(1)\dagger}\}
 =\{Q^{(2)},Q^{(2)\dagger}\}
   =\{Q^{(3)},Q^{(3)\dagger}\}  =2{H}
\end{equation}
while the central charges appear in other parts of superalgebra;
e.g.,
\begin{eqnarray}
\{Q,Q\}=& \{Q^{(j)},Q^{(j)}\} &=-2{Z}_4+2i{ Z}_5 \cr \cr
\{Q^\dagger,Q^\dagger\} =& \{Q^{(j)\dagger},Q^{(j)\dagger}\} &=
-2{ Z}_4-2i{Z}_5
\end{eqnarray}
for $j=1$, 2, or 3.
Once we adopt this complex notation, it is natural to introduce an
equivalent geometrical notation for realizing the fermionic part of
the algebra of Eq.~(\ref{commutators}).  Defining the vacuum state
$|0\rangle$ to be annihilated by the $\varphi^q$ gives the
one-to-one correspondence
\begin{equation}
(\varphi^{*q_1}\varphi^{*q_2}\cdots\varphi^{*q_k})|0\rangle
\quad\leftrightarrow\quad dz^{q_1}\wedge
dz^{q_2}\wedge \cdots \wedge dz^{q_k}  \, ,
\end{equation}
in terms of which we can reinterpret $\varphi^{*q}$ as the exterior
product with $dz^q$ and $\varphi^q$ as the contraction with
$\partial/\partial z^q$.

\section{BPS trajectories and BPS dyons}
\label{bpsAndBPS}

The purely bosonic part of the low-energy Lagrangian obtained from
${\cal N}=2$ SYM theory,
Eq.~(\ref{N2lowenergyLag}),
corresponds to a classical energy
\begin{eqnarray}
    {\cal E}_2 &=& {1\over 2}g_{qr} (\dot z^q \dot z^r +G^q G^r)
         +{\bf b} \cdot {\bf g}   \cr
   &=& {1\over 2} (\dot z^q \mp G^q) (\dot z_q \mp G_q)
           \pm \dot z^q G_q +{\bf b} \cdot {\bf g}   \cr
         &\ge& | \dot z^q G_q | +{\bf b} \cdot {\bf g} \, .
\end{eqnarray}
Using the expression for the metric, Eq.~(\ref{chap8metric}), as well as
Eqs.~(\ref{A0Solution}) and (\ref{chap8Dabar}), and setting all fermionic
quantities to zero, we obtain
\begin{eqnarray}
      g_{qr} \dot z^q G^r &=& 2 \Tr \int d^3x \,
               (\dot z^q \delta_q A_a) (G^r \delta_r A_a) \cr
     &=& 2 \Tr \int d^3x \, F_{0a} D_a a \cr
     &=& 2 \Tr \int d^3x \, \partial_i(a F_{0i}) \cr
     &=& {\bf a} \cdot {\bf q}  \, .
\end{eqnarray}
This gives the Bogomolny-type bound
\begin{equation}
   {\cal E}_2 \ge |{\bf a} \cdot {\bf q}| + {\bf b} \cdot {\bf g}
\end{equation}
that is saturated if and only if
\begin{equation}
     \dot z^q = \pm G^q  \, ,
\label{N2bpsTrajectory}
\end{equation}
with the upper or lower sign being chosen according to whether
${\bf a} \cdot {\bf q}$ is positive or negative.

Similarly, for the ${\cal N}=4$ low-energy Lagrangian,
Eq.~(\ref{N4lowenergyLag}), we have
\begin{equation}
    {\cal E}_4 = {1\over 2}g_{qr} (\dot z^q \dot z^r +G^q_I G^r_I)
         +{\bf b} \cdot {\bf g}  \,.
\end{equation}
If $\hat n_I$ is any unit vector and $G^{\perp q}_I$ is the part of
$G^q_I$ orthogonal to $\hat n_I$, then
\begin{eqnarray}
   {\cal E}_4 &=& {1\over 2}
      (\dot z^q \mp \hat n_I G_I^q) (\dot z_q \mp \hat n_I G_{Iq})
           \pm \dot z_q \hat n_I G_I^q
        + {1 \over 2} G_I^{\perp q} G_{Iq}^\perp
           +{\bf b} \cdot {\bf g}   \cr\cr
     &\ge&  |(\hat n_I {\bf a}_I) \cdot {\bf q} |
           +{\bf b} \cdot {\bf g}  \, .
\end{eqnarray}
For this bound to be saturated, the $G_I^{\perp q}$ must all vanish.
Using the SO(5) invariance of the low-energy theory, we can then
rotate the $G_I$ so that only a single one, say $G_5$, is nonvanishing.
The solutions satisfying the Bogomolny bound will then have
\begin{equation}
   {\cal E}_4 = |{\bf a}_5 \cdot {\bf q}| + {\bf b} \cdot {\bf g}
\end{equation}
and satisfy
\begin{equation}
      G^q_I = \pm \delta_{I5} \, \dot z^q
\label{N4bpsTrajectory}
\end{equation}
with the upper or lower sign being chosen according to the sign of
${\bf a}_5 \cdot {\bf q}$.

Our experience with the monopole solutions that satisfy the Bogomolny
bound in the context of the classical field theory suggests that the
moduli-space trajectories satisfying Eqs.~(\ref{N2bpsTrajectory}) or
(\ref{N4bpsTrajectory}) should preserve some of the supersymmetry of
the low-energy dynamics.  Surprisingly, this is only partially true.
The ${\cal N}=2$ theory has four real supersymmetries, whose actions
were given in Eq.~(\ref{N2transformation}).  Trajectories with $G^q =
\dot z^q$ preserve all of these.  However, those with $G^q = -\dot
z^q$ preserve none.  Thus, despite saturating the energy bound at the
classical level, the latter trajectories are not BPS and do not lead
to BPS dyons in the full quantum theory.  As we will see in the next
chapter, this leads to an essential asymmetry in the dyon spectrum of
the ${\cal N}=2$ theory.

By contrast, for ${\cal N}=4$ the trajectories saturating the Bogomolny
bound all preserve half of the eight real supersymmetries of the
low-energy theory.  With conventions chosen so that these solutions
obey Eq.~(\ref{N4bpsTrajectory}), the unbroken supersymmetries are
given by Eq.~(\ref{transformation}), with $\epsilon$ and the
$\epsilon_{(j)}$ required to be
eigenvectors of $\sigma_3$ with eigenvalue
1 or $-1$ according to the sign of ${\bf a}_5 \cdot {\bf q}$.

The BPS moduli-space trajectories have a natural correspondence with
the BPS dyons of the full field theories.  Indeed, the lower bounds on ${\cal
E}_2$ and ${\cal E}_4$ coincide with the dyon mass bounds,
Eqs.~(\ref{chap7Ebound1}) and (\ref{chap7Ebound2}), that we obtained
in the previous chapter.  Moreover, the fact that only one choice of
sign in Eq.~(\ref{N2bpsTrajectory}) gives a true BPS solution has a
counterpart in the ${\cal N}=2$ SYM theory.  The latter theory has only a
single central charge and so, as we saw in
Sec.~\ref{susyconnectionSec}, the multi-Higgs dyon solutions
saturating the classical energy bound are either 1/2-BPS (for one
choice of sign in $E_i = \pm D_i a$) , or not BPS at all (for the
other choice of sign).  In ${\cal N}=4$ SYM theory, both choices of sign give
solutions that are 1/4-BPS.

There is one subtlety in this correspondence that must be pointed out.
Achieving the energy bound of Eq.~(\ref{chap7Ebound1}) requires that
${\bf a} \cdot {\bf g} = {\bf b} \cdot {\bf q}$.  Our treatment in
this chapter has been based on the assumption that ${\bf a} \cdot {\bf
g} =0$, thus implying ${\bf b} \cdot {\bf q} = 0$.  For generic
electric charge, the latter need not vanish and, in fact, has a simple
interpretation. The electric excitation energy $ \pm {\bf a}\cdot{\bf
q}$ captures only the energy due to relative electric charges. The
center-of-mass part of the electric charge, which is necessarily
parallel to $\bf g$ and thus orthogonal to $\bf a$, gives an electric
energy that arises as a term $({\bf b}\cdot{\bf q})^2/2{\bf
b}\cdot{\bf g}$ in the kinetic energy of the center-of-mass sector
moduli.

\section{Making contact with Seiberg-Witten theory}
\label{seiberg-section}

Before closing, we would like to comment on how the central
charge ${Z}$ of the supersymmetric quantum mechanics relates to the
Seiberg-Witten \cite{Seiberg:1994rs,Seiberg:1994aj} central charge
${\cal Z}_{SW}$ for ${\cal N}=2$ SYM theory.
The Seiberg-Witten description of ${\cal N}=2$ SYM mainly concerns the
vacuum structure of the theory in the Coulomb phase. The Coulomb phase,
where the gauge symmetry is broken to the Cartan subgroup, comes
with a set of  massless U(1) vector multiplets whose kinetic terms 
specify completely the low-energy dynamics.
Since
an ${\cal N}=2$ vector multiplet contains a complex scalar field, it
is then a matter of specifying the geometry of the vacuum moduli
space spanned by these scalar fields. One way to represent the
massless scalar fields associated with the U(1)'s is to 
assemble $r$ complex fields as the components
of a column vector ${\bf A}$ so that
the massive vector meson with integer-quantized electric charge ${\bf n}_e$
has a mass $|{\bf n}_e \cdot {\bf A}|$.
In the weak coupling limit, ${\bf A}=e({\bf h}_1+i{\bf h}_2)$.

Because of the extended supersymmetry, the
vacuum moduli space has a very restrictive kind of geometry,
known as special K\"ahler geometry, and the low-energy dynamics of the ${\cal N}=2$
SYM in the Coulomb phase is completely determined by the knowledge of
a single locally holomorphic function, ${\cal F}$, termed the prepotential.
In the weak coupling limit, this prepotential takes a universal
form,
\begin{equation}
{\cal F}\simeq \frac{\tau({\bf A})}{2}{\bf A}\cdot {\bf A} \, ,
\end{equation}
with
\begin{equation}
\tau\equiv \frac{\theta}{2\pi}+\frac{4\pi i}{e^2} \, .
\end{equation}
The prepotential is not a single-valued function, but instead transforms
nontrivially as we move around the vacuum moduli space.  This
turns out to be a blessing, because its transformation properties
are entirely determined by which BPS particles become massless, and at 
which points;
the knowledge of these transformation properties is
often enough to fix the entire prepotential exactly.

In fact,
${\bf A}$ also transforms nontrivially as we move around the moduli space.
The transformation properties of ${\bf A}$ and ${\cal F}$ are tied in the
following sense.  One can define
a magnetic version of ${\bf A}$ by
\begin{equation}
{\bf A}_D=\frac{\partial}{\partial {\bf A}}{\cal F}
\end{equation}
and, with this, form a $2r$-dimensional column vector
\begin{equation}
{\cal P}\equiv \left(
\begin{array}{c}
{\bf A}\\
{\bf A}_D
\end{array}\right) \, .
\end{equation}
The vacuum moduli space of ${\cal N}=2$ SYM is riddled with singular
points and cuts, but these are all associated with some BPS particles
becoming massless in some vacuum. The transformations of ${\cal F}$
and ${\cal P}$ occur as one moves around such a singularity
or passes over a cut, and may be expressed generally as
\begin{equation}
{\cal P}\quad \rightarrow \quad U{\cal P} \, ,
\end{equation}
where $U$ is an element of the infinite discrete group Sp$(2r,Z)$.
The set of $U$'s are called the monodromy group.

In the Coulomb phase, there are charged particles that are electrically
and magnetically charged with respect to the unbroken U(1)'s above.
In terms of ${\cal P}$, the central charge of these particles is
written as
\begin{equation}
{\cal Z}_{SW}={\bf n}_m \cdot {\bf A}_D + {\bf n}_e \cdot {\bf A} +
\sum_f N_f(m_R^f+im_I^f) \,,
\end{equation}
where ${\bf n}_e$ and ${\bf n}_m$ are the electromagnetic charge
vectors. The last term is a contribution from massive
hypermultiplets with complex masses $m^f$, with $N_f$ being the
fermion excitation number for the $f$th flavor. While the quantities
${\bf A}$ and ${\bf A}_D$ transform under the $U$'s, the central charge
itself gives the masses of physical particles, and so must be
invariant.  This means that
one must also transform the charges as
\begin{equation}
({\bf n}_e, {\bf n}_m) \quad \rightarrow  \quad ({\bf n}_e, {\bf n}_m)U^{-1}
\end{equation}
so that
\begin{equation}
{\cal Z}_{SW} \quad \rightarrow \quad
{\cal Z}_{SW}
\end{equation}
as we move around singularities and cross cuts in the vacuum moduli space.

In the weak coupling regime, assuming $\theta=0$ for further simplicity,
${\bf A}$ and ${\bf A}_D$ are given by
${\bf A}=e({\bf h}_1+i{\bf h}_2)$ and
${\bf A_D}=4\pi i{\bf A}/e^2$.  Rewriting the above expression for
${\cal Z}_{SW}$ in terms of $\bf a$ and $\bf b$ gives
\begin{equation}
{\cal Z}_{SW}=i({\bf b}\cdot{\bf g} + {\bf a}\cdot{\bf q} + N_fm_I)
+({\bf b}\cdot{\bf q}  + N_fm_R) \; ,
\end{equation}
where we have used ${\bf a\cdot g}=0$. Because ${\bf b\cdot g}$ is
large, the BPS mass $|{Z}_{SW}|$ is approximately
\begin{equation}
M=|{\cal Z}_{SW}|\simeq {\bf b}\cdot{\bf g} + {\bf a}\cdot{\bf q} +
\sum_f N_fm_I^f +\frac{({\bf b}\cdot{\bf q}  +\sum_f
N_fm_R^f)^2}{2{\bf b}\cdot{\bf g}} \;.
\end{equation}
Already we begin to recognize individual contributions to the BPS
mass from the low-energy dynamics. The first term is the rest mass
of the monopoles. The last term would be the energy from the
center-of-mass electric charge if we could ignore $m_R^f$. On the
other hand, the approximation we have adopted demands that the bare
fermion mass is at most of the same order of magnitude as $e{\bf
a}$, which implies that $m^f_R$ should be much smaller than $e{\bf
b}$. The second and third terms are the central charges of the
low-energy dynamics,\footnote{See Appendix B.1.3 for discussion of
massive fermion contributions to the low-energy dynamics.}
\begin{equation}
Z={\bf a}\cdot{\bf q} + \sum_fN_fm_I^f \; .
\end{equation}
Therefore, the mass formula is approximated by
\begin{equation}
M=|{\cal Z}_{SW}|\simeq {\bf b}\cdot{\bf g} + {Z} +\frac{({\bf
b}\cdot{\bf q})^2}{2{\bf b}\cdot{\bf g}} \;,
\end{equation}
which relates the central charge, $Z$, of the low-energy dynamics to
that of the ${\cal N}=2$ SYM theory.

\chapter{BPS dyons as quantum bound states\label{Bound}}

We will now use the low-energy dynamics developed in the previous
chapter to explore the spectrum of states in SYM theory. We will focus
in particular on the BPS states; the fact that these preserve some of
the supersymmetry of the theory makes them particularly amenable to
analysis.

An important motivation for studying these states is to understand
the duality symmetries of these theories.  As we first noted in
Sec.~\ref{OM}, Montonen and Olive \cite{Montonen:1977sn} conjectured
that there might be a symmetry that exchanged weak and strong
coupling and electric and magnetic charges.  In the SU(2) theory the
masses of the elementary charged gauge boson and the monopole are
consistent with this duality, but the full multiplet structure of
spin states with unit electric or magnetic charge is only invariant
when there is ${\cal N}=4$ supersymmetry.

The existence of dyonic states leads to further requirements.  We
have seen that excitation of the U(1) gauge zero mode about the unit
monopole leads to states whose electric and magnetic charges (in
units of $e$ and $4\pi/e$, respectively) are $(n, 1)$, with $n$ any
integer. The Montonen-Olive duals of these would have to have unit
electric charge and multiple magnetic charge, and so should be bound
states containing $n$ monopoles.  As we will see, not only do these
states exist in the ${\cal N}=4$ SYM
theory, but they also imply the existence of still further states.
In fact, the Montonen and Olive duality extends to an SL(2,$Z$)
symmetry that requires states with charges $(n,m)$
for all coprime integers $n$ and $m$ \cite{Sen:1994yi}

When the gauge group is larger than SU(2), duality requires bound
states even for the purely magnetic parts of the spectrum.  Recall
that the basic building blocks for the BPS states are the fundamental
monopoles. The number of types of these is equal to the rank of the
gauge group which, for all groups larger than SU(2), is smaller than
the number of massive gauge bosons carrying (positive) electric-type
charges.  Thus, for maximally broken SU($N$) there are $N-1$
fundamental monopoles, but $N(N-1)/2$ massive vector mesons.  The
extra monopoles required for duality can only arise as bound states.
We will show that such bound states exist and are BPS, but only in
the ${\cal N}=4$ theory.  In the ${\cal N}=2$ theory there are no
corresponding BPS bound states and, at least for some ranges of
parameters, no such bound states at all.  This is despite the fact
that there are classical solutions that might seem to give the duals
of the gauge mesons.  

Although ${\cal N}=4$ supersymmetry tends to make the spectrum of
BPS states relatively simple and easy to determine, the explicit
construction of these states is rarely trivial and becomes
increasingly cumbersome, or even effectively impossible, for larger
charges.  However, there are some additional tools that we can use.
Some ${\cal N}=4$ theories can be easily obtained as theories of
D3-branes in type IIB string theory, as we will discuss in the next
chapter. In this approach, the ${\rm SL}(2,Z)$ duality of ${\cal
N}=4$ SYM theory follows from the ${\rm SL}(2,Z)$ duality of the type IIB
theory; once the latter is accepted as a fact, an ${\rm
SL}(2,Z)$-invariant spectrum is automatic.  A more conservative
point of view might be to say that the ${\rm SL}(2,Z)$ invariance
found in the ${\cal N}=4$ field theory is strong evidence for the
corresponding invariance of the string theory.  Either way, the
stringy construction allows an easy generalization to a large class
of gauge groups and provides easy pictorial hints to novel BPS
states.

A case in point is the 1/4-BPS dyons \cite{Lee:1998nv}, whose
existence was first realized in type IIB theory
\cite{Bergman:1997yw}, where these states are constructed by having
a web of fundamental strings and D-strings with ends on three or
more D3-branes \cite{Dasgupta:1997pu,Sen:1997xi}.  These
objects had not been recognized from the field theory approach
because the conventional treatment of low-energy monopole dynamics
had been based on models with a single adjoint Higgs, which
necessarily excluded all such-1/4 BPS states,

At the same time, this is not to say that type IIB theory is more
powerful in counting and isolating precise BPS spectra. It is
important to remember that the correspondence between the two
theories is not at the classical level, but rather at the quantum
level.  Just as we must quantize the moduli space dynamics on the
field theory side, the string web must also be quantized. In
particular, the moduli space of the string web has little to do with
that of the field theory dyons and is, in fact, more difficult to
quantize.  Thus, the field theory side may give us better control
for addressing some of the more precise and specific questions
concerning the spectrum.  In the later part of this chapter we will
demonstrate the existence and determine the degeneracy of some of
the simpler 1/4 BPS dyons.

For the case of ${\cal N}=2$ SYM theories, both approaches
tend to be more difficult to handle. {}From the string theory
side, there are diverse constructions of the gauge theories, but in
all of them it is quite nontrivial to find the corresponding BPS
spectrum. In the elegant formulation of Seiberg-Witten theory
\cite{Seiberg:1994rs,Seiberg:1994aj} as a theory of wrapped
M5-branes in M-theory \cite{Witten:1997sc,Klemm:1996bj}, we know how
to realize BPS dyons as open membranes. Nevertheless, establishing
the existence of a given dyon is all but impossible, except at
particular points of the moduli space
\cite{Henningson:1997hy,Mikhailov:1997jv}.  From the field theory
side also, the constraints \cite{Ferrari:1996sv,Bilal:1996sk} coming
from the Seiberg-Witten description and S-duality are difficult to
analyze beyond the simple rank 1 case of SU(2) theories
\cite{Fraser:1996pw,Taylor:2001hg}. The main culprit is the
extremely interesting phenomena that the BPS spectrum can change as
we change the vacuum of the theory along the Coulomb phase
\cite{Seiberg:1994aj,Henningson:1995hj}.  Understanding the spectrum
in this approach requires understanding the latter phenomena
everywhere on the Seiberg-Witten vacuum moduli space.

However, even if one managed to understand the structure of 
the vacuum moduli space
completely and explicitly, this would be only
the beginning of the problem. The reason is that this approach is
basically a bootstrap where one tries to find a solution to a set of
consistency conditions that becomes intractable as the rank of the
gauge group increases beyond unity. For practical purposes, one
typically needs additional input, such as the BPS spectrum in some corner
of the vacuum moduli space. An obvious place to look for BPS
spectra is, of course, the weak coupling regime, which is the main
focus of this chapter. In this regard,
some good news is that the semiclassical approach involving the
moduli space description remains more or less manageable
\cite{Lee:1996if,Sen:1994yi,Sethi:1995zm,Gauntlett:1995fu}, and does
not get significantly worse than in the ${\cal N}=4$ case.

We start, in Sec.~\ref{modspaceboundsec}, with some generalities
concerning moduli space bound states, especially those BPS states
that preserve part of the supersymmetry.  In
Sec.~\ref{twobodyboundsection} we explicitly construct two-body
bound states.  We first consider the case of two identical
monopoles, showing how Montonen-Olive duality is naturally extended
to an SL(2,$Z$) duality, and then turn to the case of two distinct
monopoles.  Then, in Sec.~\ref{manybodyboundsection}, we consider
the problem of many-body bound states.  Although the explicit
construction of these states is much more difficult than for the
two-body case, their number can be determined by using index theory
methods.  We conclude the chapter, in Sec.~\ref{boundstabilitysec},
with a brief discussion of the difficulties in finding bound states
with four supercharges.

\section{Moduli space bound states}
\label{modspaceboundsec}

The fundamental degrees of freedom for the low-energy Hamiltonian
are the bosonic collective coordinates $z^r$ that span the moduli
space, together with their fermionic counterparts that arise from
the fermion zero modes.  Because each complex fermionic variable
corresponds to a two-state system (the zero mode being either
occupied or unoccupied), the states of the system are naturally
described by a multicomponent wave function in which each component
is a function of the $z^r$.  A $k$-monopole system in ${\cal N}=2$
SYM theory is described by $4k$ bosonic and $2k$ complex fermionic
variables.  Its wave function has $2^{2k}$ components.  As we saw in
Sec.~\ref{n=2q}, these are most naturally viewed as a column vector,
with the action of the fermionic variables being represented by
Dirac matrices of appropriate dimension.  For ${\cal N}=4$ SYM theory,
there are $4k$ complex fermionic variables and a $2^{4k}$-component
wave function that can be conveniently written as a linear
combination of differential forms, as described in Sec.~\ref{n=4q}.

The moduli space is locally a product of a flat center-of-mass
manifold, spanned by the center-of-mass position and an overall U(1)
phase angle, and a nontrivial relative moduli space.  This implies
that the Hamiltonian can be written as the sum of center-of-mass and
relative pieces, and that its eigenstates can be described by wave
functions that are products of center-of-mass and relative wave
functions.  Because of identifications, such as those in
Eqs.~(\ref{shift}) and (\ref{psiperiod}), that follow from the
periodicities of certain phases angles, this factorization of the
moduli space is only local, and not global.  Although this has no
effect on the form of the Hamiltonian, it does, as we will see,
produce an entanglement between the quantization of the overall U(1)
charge and the state of the relative variables.

The eigenstates of the center-of-mass Hamiltonian are labelled by
the total momentum and a quantized U(1) charge.  The corresponding
fermionic variables do not enter the Hamiltonian at all, and thus
affect neither the energies nor the form of the wave functions.
Instead, their only effect is to generate a supermultiplet of
degenerate states.  For ${\cal N}=2$ SYM theory, these supermultiplets
contain $2^2=4$ four states, while for ${\cal N}=4$ there are $2^4 =
16$ states, exactly matching the charged vector meson
supermultiplet; as we saw in Sec.~\ref{OM}, this ${\cal N}=4$
structure is precisely what is required to have a Montonen-Olive
duality between the states with unit magnetic and unit electric
charges in SU(2).

Thus, the nontrivial part of the spectrum of states is associated
with the relative moduli space.  Our interest here is in bound
states, which correspond to normalizable wave functions on the
relative moduli space.  When combined with the states of the
center-of-mass Hamiltonian, each such bound state will yield a tower
of states of increasing overall U(1) charge, with each state in the
tower having a degeneracy of 4 (for ${\cal N}=2$) or 16 (for ${\cal
N}=4)$ arising from the center-of-mass fermionic zero modes.

In our analysis of the relative moduli space Hamiltonian, we will
focus in particular on the BPS states that preserve some of the
supersymmetry.  There are two cases to consider:

\subsubsection{${\cal N}=4$ SYM}

States preserving one-half of the supersymmetries of the low-energy
dynamics, and thus one-fourth of the field theory supersymmetries,
are only possible if there is just a single nonzero $G_I$, which we
may take to be $G_5$, and only one nonvanishing central charge
($Z_5$ in this case).  Such states are annihilated by one of the
operators
\begin{equation}
    {\cal D}_\pm \equiv   \sqrt{i}Q \pm \sqrt{-i} Q^\dagger \, ,
\end{equation}
which obey
\begin{equation}
    {\cal D}_\pm^2 = \pm 2(H \mp Z_5) \, .
\end{equation}
If the state is represented by a differential form $\Omega$, as
described in Sec.~\ref{n=4q}, then this BPS condition becomes
\begin{eqnarray}
    0 &=& {\cal D}_\pm \Omega  \cr
      &\equiv &(\sqrt{i}\varphi^m \pm \sqrt{-i} \varphi^{*m})
          (\pi_m  \mp G^5_m) \Omega  \cr
     &=& \sqrt{-i} (d - \iota_{G^5})
     \pm \sqrt{i}(d^\dagger - \iota_{G^{5\dagger}}) \, ,
\label{N4Dplusminus}
 \end{eqnarray}
where $\iota_K$ denotes the contraction with a vector field $K$ and
$\iota_K^\dagger$ is the exterior product by the one-form obtained
from $K$ by lowering its index.

A complex conjugation that keeps operators, such as $\varphi$ and
$d$, untouched will transform ${\cal D}_+$ into $i{\cal D}_-$. This
conjugation thus pairs every state with nonvanishing $Z_5$ with a
state carrying the opposite sign for this central charge.

Once a state solves one of these equation, it also solves three
other similar equations where $Q$ and $Q^\dagger$ are replaced by
$Q_{(k)}$ and $Q_{(k)}^\dagger$, for $k=1$, 2, or 3, since the
superalgebra is such that we have the identity
\begin{equation}
\left(\sqrt{i}Q_{(k)} \pm \sqrt{-i} Q^\dagger_{(k)}\right)^2
  =\pm 2H-2Z_5={\cal D}_\pm^2\;.
\end{equation}
Thus, solutions to Eq.~(\ref{N4Dplusminus}) with a particular sign
choice actually preserve four real supersymmetries, or half of the
low-energy supersymmetry. These are 1/4-BPS in ${\cal N}=4$ SYM theory.

In the special case where the central charges all vanish, the states
annihilated by ${\cal D}_+$ are also annihilated by ${\cal D}_-$,
and so preserve all of the low-energy supersymmetry. These states
are 1/2-BPS in  ${\cal N}=4$ SYM theory.

\subsubsection{${\cal N}=2$ SYM}

In ${\cal N}=2$ SYM theory, the BPS states preserve all of the low-energy
supersymmetry, and are 1/2-BPS with respect to the full field
theory.  These states have energy equal to the central charge $Z$.
{}From Eq.~(\ref{N2supercharges}) we see that these can be
represented by spinors $\Psi$ obeying
\begin{equation}
    0 = {\cal D} \Psi \equiv \gamma^m(-i\nabla_m-G_m)\Psi \; .
\label{N2Doperator}
\end{equation}
where we have used the map $\sqrt{2}\lambda^m=\gamma^m$ between the
real fermions and the Dirac matrices on the target manifold. A
crucial difference from the ${\cal N}=4$ case is that there is no
pairing of BPS states; these states exist only for one sign of the
moduli space central charge.

As before, once a state solves this equation, it also solves three
other similar equations in which ${\cal D}=\sqrt{2}Q$ is replaced by
$\sqrt{2}Q_{(k)}$ with $k=1$, 2, or 3, since
\begin{equation}
2Q_{(k)}^2=2H-2Z={\cal D}^2\; .
\end{equation}
Thus, solutions to Eq.~(\ref{N2Doperator}) preserve all four
supersymmetries. These are  BPS states in  ${\cal N}=2$ SYM
theory.

\section{Two-body bound states}
\label{twobodyboundsection}

When there are only two monopoles involved, the bound state problem
is simple enough to allow an explicit construction of states.  We
will start with the case of two (necessarily identical) SU(2)
monopoles and then consider that of two distinct monopoles in a
larger group.

\subsection{Two identical monopoles}
\label{twoidentbound}

Since for SU(2) there is (up to a rotation) only a single nonzero
Higgs vev, the low-energy Hamiltonian has no potential energy term
and is purely kinetic.  The low-energy dynamics then has no central
charges, and so all ${\cal N}=4$ BPS states are 1/2-BPS within the
full field theory.  Any such bound states correspond to square
normalizable harmonic forms on the relative moduli space.

In particular, Montonen-Olive duality would require a two-monopole
bound state, carrying one unit of electric charge, to provide the
supermultiplet dual to the dyonic supermultiplet with one unit of
magnetic and two units of electric charge.\footnote{SL(2,$Z$)
duality requires many additional bound states; we will return to
this point shortly.} Because the electric-magnetic mapping is to be
one-to-one, this bound state should be unique.

We recall, from Sec.~\ref{AtiyahHitchin}, that the relative moduli
space ${\cal M}_0$ for two SU(2) monopoles is the double cover of
the Atiyah-Hitchin manifold.  Its metric may be written as
\begin{equation}
ds^2=\sum_m \omega^m\otimes \omega^m
\end{equation}
where we have defined a basis of one-forms
\begin{eqnarray}
\omega^0&=&f(r)\,dr \nonumber \\
\omega^1&=&a(r)\,\sigma_1\nonumber \\
\omega^2&=&b(r)\,\sigma_2\nonumber \\
\omega^3&=&c(r)\,\sigma_3
\end{eqnarray}
and the functions $a$, $b$, and $c$ are those described in
Sec.~\ref{AtiyahHitchin}.  Our conventions will be such that these
functions are all positive, and that $a$ is the function that
vanishes at the ``origin'', $r=\pi$.

For the bound state to be unique, it must be represented by a form
that is either self-dual or anti-self-dual. If it were not, another
normalizable harmonic form could be generated by a Hodge dual
transform.  Also, since the Hamiltonian (in the absence of a
potential energy) is really a Laplace operator and so does not mix
forms of different degree, the wave function should correspond to a
form of definite degree. This, combined with the uniqueness,
restricts us to middle-dimensional forms (i.e., two-forms).
Furthermore, the uniqueness also requires that the state be a
singlet under the SO(3) isometry, so we discover that the wave
function must be one of the six possibilities 
\begin{equation}
          \Omega^{(s)}_\pm =N^{(s)}_\pm(r)
      \left(\omega^0\wedge \omega^s \pm \frac12 \,
       \epsilon_{stu}\,\omega^t\wedge\omega^u\right) \, ,
\end{equation}
where $s=1$, 2, or 3 and the summation is only over $t$ and $u$.
Harmonicity follows if the two-form is closed,
$d\Omega^{(s)}_\pm=0$. The latter condition gives a first-order
equation for the $N^{(s)}_\pm$, which is solved by
\begin{eqnarray}
N^{(1)}_\pm &=&\frac{1}{bc}\exp\left[\mp\int dr\,\frac{fa}{bc}
\right] \nn N^{(2)}_\pm &=&\frac{1}{ca}\exp\left[\mp\int
dr\,\frac{fb}{ca} \right] \nn N^{(3)}_\pm
&=&\frac{1}{ab}\exp\left[\mp\int dr\,\frac{fc}{ab} \right]  \, .
\end{eqnarray}
Substituting the form of Atiyah-Hitchin metric, detailed in
Sec.~\ref{AtiyahHitchin}, we find that only one of these six
possibilities leads to a wave function that is normalizable and yet
nonsingular at $r= \pi$, namely $N^{(1)}_+$.  The only possible
ground state is therefore
\begin{equation}
\Omega^{(1)}_+ = N^{(1)}_+(r) \left(\omega^0\wedge \omega^1 +
   \omega^2\wedge\omega^3\right) \, .
\end{equation}

The physical wave function on the entire moduli space is the product
of $\Omega^{(1)}_+$ with a form on the center-of-mass moduli space.
Now recall\footnote{Here, and below, we have set $\balpha^2=1$,
which is the standard convention for SU($N$)} from
Eq.~(\ref{atiyahtwisting}) that the angle $\psi$ (with range $2\pi$)
of the approximate U(1) on ${\cal M}_0$ must be twisted with the
angle $\chi$ (with range $4\pi$) of the exact U(1) on the
center-of-mass moduli space in such a way as to give the
identification
\begin{equation}
(\chi,\psi)\sim (\chi+2\pi,\psi-\pi) \, .
\end{equation}
Since increasing $\psi$ by $\pi$ flips the signs of
\begin{eqnarray}
&&\sigma_1 = -\sin\psi d\theta +\cos\psi\sin\theta d\phi \, , \nn
&&\sigma_2\wedge \sigma_3=d\sigma_1 \, ,
\end{eqnarray}
and thus of $\Omega^{(1)}_+$, we can only obtain a single-valued
total wave function if the center-of-mass part also changes sign
when $\chi \rightarrow \chi + 2\pi$.  Thus, the total wave function
must correspond to a form that can be written as
\begin{equation}
   \Omega = \Omega^{(1)}_+ \otimes e^{i(k+1/2)\chi} \otimes
        \Omega_{\rm CM}
\end{equation}
with $k$ any integer.

It takes  more care to show that the value $k+1/2$ for the momentum
conjugate to $\chi$ translates to an electric charge $n=2k+1$. The
key point is to recall that in all two-body cases, whether the
monopoles are identical or distinct, the momentum conjugate to
$\chi$ is always related to the (approximately) conserved electric
charges $Q_1$ and $Q_2$ on the individual monopoles by
\begin{equation}
   q_\chi=\frac{m_1Q_1+m_2Q_2}{m_1+m_2}  \, .
\end{equation}
Hence, the total U(1) electric charge of a pair of identical
monopoles is the simple sum
\begin{equation}
Q_1+Q_2 =2q_\chi \, .
\end{equation}

Montonen-Olive duality required a unique supersymmetric bound state
dyon with one unit of electric and two units of magnetic charge.  We
have found not only that (1,2) state, but a full tower of states
with with electric and magnetic charges $(2k+1,2)$ for arbitrary
integer $k$.  Applying Montonen-Olive duality to these would require
additional states with charges $(2,2k+1)$, and one might well expect
that the construction of these states would lead to still further
states, thus continuing the process.

The explanation for this is that the Montonen-Olive duality naturally
extends to an SL(2,$Z$) symmetry \cite{Sen:1994yi,Sen:1992ch, Sen:1993zi,
Schwarz:1993mg, Sen:1994fa}.  To understand this symmetry,
recall that the Yang-Mills Lagrangian can be extended to include a
topological ``$\theta$-term'' that has no effect on the classical
field equations but that leads to important quantum effects.  If we
rescale the gauge potential $A_\mu$ by a factor of $e$, so that the
coupling does not appear explicitly in the field strength, the pure
gauge part of the extended Lagrangian then takes the form
\begin{equation}
    {\cal L}_{\rm gauge} = -{1\over 2e^2} \Tr F_{\mu\nu} F^{\mu\nu}
         + {\theta\over 16\pi^2} \Tr F_{\mu\nu} \tilde F^{\mu\nu}
\end{equation}
where $\tilde F^{\mu\nu} = (1/2) \epsilon^{\mu\nu\alpha\beta}
F_{\alpha\beta}$.  The variable $\theta$ is periodic, with a shift
$\theta \rightarrow \theta + 2\pi$ having no effect on the physics.

When working with these rescaled gauge fields, it will also be
convenient to define rescaled charges $\hat Q_M = e Q_M$ and $\hat
Q_E = e Q_E$.  The $\theta$-term has no effect on the quantization
of the magnetic charge, so
\begin{equation}
   \hat Q_M = 4\pi m
\end{equation}
with $m$ any integer.  However, the quantization of electric charge
is modified \cite{Witten:1979ey}, so that now
\begin{equation}
   \hat Q_E = e^2 \left(n - {\theta\over 2\pi}m \right)
\end{equation}
with $n$ an integer.

If we combine $\theta$ with $e$ to define a complex coupling
constant
\begin{equation}
    \tau = {\theta \over 2\pi} + {4\pi i \over e^2}
\end{equation}
and define a complex charge
\begin{equation}
     q =  \hat Q_M -i \hat Q_E  \, ,
\end{equation}
the charge quantization conditions can be written more compactly as
\begin{equation}
    Re\,q = 4\pi m \, , \qquad  Re\,q\tau = 4\pi n  \, .
\end{equation}

For $\theta=0$, Montonen-Olive duality replaces $e$ by $e'=4\pi/e$
and interchanges electric and magnetic charges, corresponding to the
transformations
\begin{equation}
    \tau \rightarrow \tau'= -{1 \over \tau}\, , \qquad
     q\rightarrow q'= \tau q  \, .
\label{dualitytransf}
\end{equation}
It is natural to extend the duality conjecture by assuming that the
transformations of Eq.~(\ref{dualitytransf}) leave the theory
invariant even when $\theta \ne 0$.  We then have
\begin{eqnarray}
     4\pi m' &=& Re\,q' = Re\, \tau q = 4\pi n  \cr \cr
     4\pi n' &=& Re\,\tau' q' = - Re\, q  = -4\pi m \, ;
\end{eqnarray}
i.e., the vector $(n,m)^t$ is transformed by the matrix
\begin{equation}
     S = \left( \matrix{0 & 1 \cr -1& 0} \right) \, .
\end{equation}

A second invariance, under $ \tau \rightarrow \tau - 1$, follows
from the periodicity of $\theta$.  This corresponds to the
transformation $(n,m) \rightarrow (n + m,n)$, which can be
represented by
\begin{equation}
     T = \left( \matrix{1 & 1 \cr 0& 1} \right) \, .
\end{equation}
The matrices $S$ and $T$, when multiplied in an arbitrary sequence,
generate SL(2,$Z$), the group of $2\times 2$ matrices with integer
elements and unit determinant.  We thus are led to expect that the
spectrum of states should be invariant under the action of
SL(2,$Z$).\footnote{Although such an SL(2,$Z$) electromagnetic duality
is a hallmark of ${\cal N}=4$ SYM theories, there is a class of ${\cal
N}=2$ SYM theories that also possess BPS spectra that respect an
SL(2,$Z$) duality \cite{Lee:1997fy}. These theories have gauge group
Sp($2k$) with four hypermultiplets in the fundamental representation
and one hypermultiplet in antisymmetric tensor representation, and
include the SU(2) theory of Ref.~\cite{Sethi:1995zm,Gauntlett:1995fu}
as a special case.  Duality-invariant spectra
for small magnetic charges were demonstrated in
these three references.  However, in
general ${\cal N}=2$ SYM theories have
complicated vacuum moduli spaces, plagued by marginal stability domain walls,
and are not expected to admit duality-invariant spectra; see
Ref.~\cite{Cederwall:1996hy} for an explicit example of this.}

It is not hard to show that any pair of coprime integers $(n,m)$
can be obtained by acting on (1,0) with an SL(2,$Z$) matrix.
Conversely, if $n$ and $m$ have a common factor, then $(n,m)$ cannot
be obtained from (1,0).  Hence, the generalized Montonen-Olive
duality, together with the existence of the unit charged vector
meson state, requires that there be states corresponding to all
coprime integers $m$ and $n$.  With only two monopoles present,
these are just the states with charges $(2k+1,2)$ that we have
found.\footnote{For the counting of all the $(q,p)$ towers of BPS
states, see Refs.~\cite{Porrati:1995ge,Segal:1996eb,Ferrari:1997gu}.}

In fact, one can extend the two-body computation above a little
further and establish a vanishing theorem stating that no other
ground state exists on ${\cal M}_0$.  This reveals something that we
might not have known a priori just from the ${\rm SL}(2,Z)$
invariance. Not only does it show that the requisite $(2k+1, 2)$ BPS
states exist with the right supermultiplet structure, but it would
also be a strong indication that the spurious states with charges
$(2q,2p)$ are all absent.

In contrast with the rich spectrum of bound states in the ${\cal N}
=4$ theory, there are no dyonic bound states of two or more
monopoles in ${\cal N} =2$ pure SU(2) SYM theory
\cite{Ferrari:1996sv,Gauntlett:1995fu}. If there
were any such bound states, they would imply the existence of
additional dyonic states in the ${\cal N} = 4$ theory, in conflict
with the uniqueness results cited above.

This SL(2,$Z$) action naturally extends to ${\cal N}=4$ SYM for
other simply laced gauge groups, since the SL(2,$Z$) acts on each
root of the gauge algebra equally and simultaneously. As in
Montonen-Olive duality, the subtleties arise in the case of
non-simply laced gauge groups, namely Sp($2N$), SO(2$N$+1), $F_4$,
and $G_2$, where the electric-magnetic duality interchanges long roots
and short roots.
Recall that this, in particular, exchanges
SO($2N$+1) with Sp($2N$), where the magnetic charge associated with
a short root of SO(2$N$+1) is actually the long root of Sp($2N$)
and vice versa. 

The SL(2,$Z$) is generated by two generators, $S$ and $T$, and this
exchange of odd-dimensional orthogonal gauge group and symplectic
gauge group happens under $S$. $T$, on the other hand, shifts electric
charge by a quantized amount proportional to the magnetic charge, and
does not by itself change the gauge group.  The full SL(2,$Z$) action
can be reconstructed from these two generators, and mixes these two
gauge groups. For this pair the SL(2,$Z$) action can be also easily
understood by realizing the ${\cal N}=4$ SYM theory in terms of
D3-branes and orientifold 3-planes, whereby the SL(2,$Z$) duality of
type IIB string theory is inherited by the worldvolume SYM theory
\cite{Witten:1998xy,Hull:1994ys,Schwarz:1995dk}. For the other
non-simply laced cases, $F_4$ and $G_2$, the SL(2,$Z$) does not change
the gauge group but involves a shift of vacuum in addition to a change
of coupling constants, since the long roots and short roots can be
interchanged. We refer the reader to Ref.~\cite{Argyres:2006qr} for
these two exceptional cases.

\subsection{Two distinct monopoles}
\label{twodistinctbound}

\subsubsection{${\cal N}=4$ SYM}

We now consider a larger gauge group and turn to the case of two
distinct fundamental monopoles, with masses $m_1$ and $m_2$, that
are associated with the simple roots $\bbeta_1$ and $\bbeta_2$
(which are assumed to be connected in the Dynkin diagram).  We saw
in Sec.~\ref{taubfortwo} that the relative moduli space is the
Taub-NUT manifold with rotational SU(2) isometry and a
triholomorphic U(1) isometry.  If we rescale coordinates as in
Eq.~(\ref{randtandLrescaling}), its metric can be written as
\begin{eqnarray}
    ds^2 &=& \left(1 +\frac{1}{r}\right)\, d{\bf r}^2
   + \left(1+\frac{1}{r}\right)^{-1}
     (d\psi+\cos\theta d\phi)^2  \cr \cr
    & = & \sum_m \omega^m\otimes \omega^m
\end{eqnarray}
where the basis forms $\omega^m$ are now given by
\begin{eqnarray}
   \omega^0&=&\sqrt{1+1/r}\,dr \nonumber \\
   \omega^1&=&\sqrt{r^2+r}\,\sigma_1\nonumber \\
   \omega^2&=&\sqrt{r^2+r}\,\sigma_2\nonumber \\
   \omega^3&=&\sqrt{\frac{r}{1+r}}\,\sigma_3 \, .
\end{eqnarray}
Note that $\omega^1 + i\omega^2$ transforms as a unit charge state
under the U(1) gauge isometry, while $\omega^0$ and $\omega^3$ are
neutral.

Again, the full moduli space is the product of the relative moduli
space and a center-of-mass moduli space, with identifications on
$\chi$ and $\psi$ that are now given by Eqs.~(\ref{shift}) and
(\ref{psiperiod}).  As we saw in Sec.~\ref{taubfortwo}, these imply
that $q$, the momentum conjugate to $\psi$, must be an integer or
half-integer.  The condition on the momentum conjugate to $\chi$ is
such that the total electric charge corresponds to a root space
vector
\begin{equation}
{\bf q}=e(n/2+q)\bbeta_1+e(n/2-q)\bbeta_2
\end{equation}
where the integer $n$ is odd (even) whenever $2q$ is odd (even).

In contrast with the SU(2) case, there will in general be additional
Higgs vevs and, therefore, a potential energy $\cal V$ on the moduli
space that is obtained from
\begin{equation}
   \tilde G_5 =\tilde a\frac{\partial}{\partial \psi} \label{tildeG}
\end{equation}
where
\begin{equation}
   \tilde a \equiv \frac{4\pi^2 a}{e^3\mu}\, .
\end{equation}
In the sector with fixed relative charge $q$, there is a repulsive
``angular momentum'' barrier that combines with $\cal V$ to produce
an effective potential energy whose form was given in
Eq.~(\ref{potentialfortwodistinct}).  As was noted then, this has a
minimum at a finite value of $r$ if and only if $q^2 < \tilde a^2$.
Otherwise, the minimum moves out to infinity, implying that a dyonic
bound state cannot form.

We first look for BPS bound states in the ${\cal N}=4$ theory, which
must satisfy Eq.~(\ref{N4Dplusminus}).  Without any loss of
generality, we can assume that $\tilde a \ge 0$, and look for bound
states obeying\footnote{As noted below Eq.~(\ref{N4Dplusminus}), the
states annihilated by ${\cal D}_-$ can be obtained by complex
conjugation of those annihilated by ${\cal D}_+$.  To obtain the
bound states for $\tilde a< 0$, let us define a kind of Hodge star
operator, $\star\equiv\prod_E(\varphi^E-\varphi^{*E})$.  The
identity $\star\, {\cal D}_-(\tilde a) = i {\cal D}_+(-\tilde
a)\star$ then implies that $\Omega'_{q}(-\tilde a) =
\star\;\Omega_q(\tilde a)$.}
\begin{equation}
   {\cal D}_+\Omega_q=0  \, .
\end{equation}
These bound state wave functions can be chosen to carries three
conserved quantum numbers: the relative electric charge, $q$, the
total angular momentum, $j$, and the third component of the angular
momentum, $m$.  All of these are quantized to be an integer or
half-integer.  We will denote the BPS wave functions with these
quantum numbers as $\Omega^j_{m;q}$.

For the moment, we will put aside the special case case $q=0$ (i.e.,
no relative electric charge), and assume that $q\ne 0$.  In this
case there is always a nonvanishing low-energy central charge, and
so the BPS bound states only preserve 1/4 of the field theory
supersymmetry.  When $q\ge 1$, these states come in four distinct
angular momentum multiplets, of total angular momenta $j=q$,
$q-1/2$, $q-1/2$, and $q-1$, giving a total of $8q$ wave functions.
When $q=1/2$, only the first three multiplets are present, but these
by themselves give a degeneracy of $8q$.  In either case, each of
these $8q$ states acquires an additional factor of 16 degeneracy
from the center-of-mass fermion zero modes.  Taken all together,
these $16\times 8q$ degenerate states form a single 1/4-BPS
supermultiplet with highest angular momentum $q+1$.

The wave functions for these states are most easily written in terms
of the spherical harmonics on $S^3$, which are usually denoted by
$D^j_{mk}$.  A unit $S^3$ has ${\rm SO}(4)={\rm SU}(2)_L\times {\rm
SU}(2)_R$ isometry.  The spherical harmonics, $D^j_{mk}$, have the
same quadratic Casimir, $j(j+1)$, for the two SU(2)'s but
independent values $m$ and $k$ for the third component eigenvalues
for ${\rm SU}(2)_L$ and ${\rm SU}(2)_R$.  However, because we only
have an ${\rm SU}(2)_L\times {\rm U}(1)_R$ isometry, our multiplets
have a definite eigenvalue $k$, which is to be identified with the
electric charge contribution to $q$.  In other words, in a given
multiplet $m$ ranges over $-j,-j+1,\dots,j$, while $k$ takes a fixed
value in that range.

After some trial and error, one finds that for $q \ge 1$ the
simplest angular momentum multiplet, with $j=q-1$, takes the
form\footnote{It is important to recognize that the SU(2) rotational
isometry we rely on here is not quite the physical angular momentum.
Because of the triplet of complex structures, it turns out that a
spin contribution must be added to $j$ to give the actual angular
momentum. We refer the reader to Ref.~\cite{Bak:1999ip} for complete
details.}
\begin{equation}
\Omega^{q-1}_{m;q}= {r^{q-1}e^{-(\tilde a-q)r}}\, (\omega^0 +
i\omega^3)\wedge(\omega^1 +i\omega^2)D^{q-1}_{m(q-1)} \, ,
\label{firstmultiplet}
\end{equation}
with $m$ taking values $-q+1,-q+2,\dots,q-1$, and that the largest
multiplet, with $j=q$, is given by
\begin{eqnarray}
\Omega^q_{m;q} &=& {r^qe^{ - (\tilde a-q)r}\over 1+r}\times \nn
&&\Biggl\{ \left[ \tilde a \left( 1 +
  \omega^0\wedge\omega^1\wedge\omega^2\wedge\omega^3 \right)
   +  \left(\tilde a+ {1\over 1+r}\right)( \omega^0\wedge\omega^3 +
\omega^1\wedge\omega^2)
\right] D^q_{mq}\nonumber\\
 &&-\sqrt{q/2}\,
(\omega^0 + i\omega^3)\wedge (\omega^1
+i\omega^2)D^q_{m(q\!-\!1)}\Biggr\} \, , \label{secondmultiplet}
\end{eqnarray}
with $m$ taking values $-q,-q+1,\dots,q$. [Note how the U(1) charge
is a combination of the charge on the spherical harmonics and that
on the forms.]

The remaining wave functions, with angular momentum $q-1/2$, can be
found most easily by acting with ${\cal D}_-$ on those found above.
This gives $2q-1$ states
\begin{eqnarray}
\label{oddsolb}
  {\cal D}_- \Omega^{q-1}_{m;q} =
     {r^qe^{ - (\tilde a-q)r}\over\sqrt{r+r^2}}\,(\omega^1 +
i\omega^2)\wedge(1+\omega^0\wedge\omega^3)D^{q-1}_{m(q-1)}
\end{eqnarray}
and $2q+1$ states
\begin{eqnarray}
\label{oddsola}
     {\cal D}_- \Omega^q_{m;q}
&&{r^qe^{ - (\tilde a-q)r}\over \sqrt{r+r^2}}\,\Bigl[ (\omega^0
\!+\! i\omega^3)\wedge(1\!+\!\omega^1\wedge\omega^2)
\sqrt{2q}D^q_{mq}\nn &&\hskip 1.8cm+ i(\omega^1 \!+\!
i\omega^2)\wedge(1\!+\!\omega^0\wedge\omega^3)D^q_{m(q\!-\!1)}
\Bigr].
\end{eqnarray}
The states in the two $j=q-1/2$ multiplets are obtained from linear
combinations of these $4q$ wave functions.

There is a slight modification if $q=1/2$.  In this case the
expressions in Eqs.~(\ref{firstmultiplet}) and (\ref{oddsolb}) are
undefined, and the entire set of $8q=4$ states is given by
Eqs.~(\ref{secondmultiplet}) and (\ref{oddsola}). Note that in both cases
the $4q$ states with charge $j=q -1/2$ are given by forms
of odd degree, while the remaining $4q$ wave functions are composed
of forms of even degree. Finally, all of these $q>0$ wave functions
are normalizable only if $q<\tilde a$, providing a natural cut-off
for the existence of a bound state. Recall that this criterion was
also present for the classical bound states discussed in chapter 7.

The case $q=0$ (i.e., no relative electric charge) is special.
There is a unique state, with $j=0$.  The supercharges of the
low-energy supersymmetry are all preserved, and this state is
1/2-BPS from the viewpoint of the full field theory.  Its wave
function is
\begin{equation}
\Omega^0_{0;0}={e^{ - \tilde a r}\over 1+r}\,\left[ \tilde a +
(\tilde a+ {1\over 1+r})( \omega^0\wedge\omega^3 +
\omega^1\wedge\omega^2) + \tilde
a\,\omega^0\wedge\omega^1\wedge\omega^2\wedge\omega^3) \right].
\end{equation}
In the limit of aligned vacua ($a=0$), this state becomes a
threshold bound state of two monopoles with the drastically simpler
form \cite{Gauntlett:1996cw,Lee:1996if},
\begin{equation}
\Omega^0_{0;0}=d\,\left(\frac{\sigma_3}{1+1/r}\right) =
d\,\left(\frac{\omega^3}{\sqrt{1+1/r}}\right)  \, .
\end{equation}
Note that the solutions in this last case can be (and, in fact,
first were) obtained by the same line of attack as the SU(2)
solutions of Sec.~\ref{twoidentbound}.

Note that all of these wave functions are chiral with respect to the
natural chirality operator of ${\cal D}_+$, namely the product of
all the $(\sqrt{i}\varphi^m + \sqrt{-i} \varphi^{*m})$. In Sec.~9.3,
this chirality operator is denoted as $\tau_s^+$. With the
wave function represented  as a differential form, this chirality
translates to self-duality for even forms and imaginary
anti-self-duality for odd forms. Later we will count the dyonic
bound states of many distinct monopoles by computing the index of
$\tau_s^+$, i.e., the difference between the numbers of chiral and
of antichiral solutions to the ${\cal D}_+$ equations. This
explicit construction of bound states, where all of them come out to
be chiral, suggests that such an index counting will actually count
the number of bound states, and not just a difference.

\subsubsection{${\cal N}=2$ SYM}

The main difference for monopoles in ${\cal N}=2$ SYM theory
is that the wave function $\Omega$ is now represented by a Dirac
spinor on the moduli space, with a BPS state obeying the Dirac
Eq.~(\ref{N2Doperator}).  With a spinorial $\Omega$, writing down
the explicit form of the wave function is more cumbersome, and so we
will just summarize the results \cite{Gauntlett:1999vc}.

In the relative moduli space, the bound state wave functions exist
only if $1/2\le q<\tilde a$ or $\tilde a<q\le -1/2$. These wave
functions are organized into a single angular momentum multiplet
with angular momentum $j=|q|-1/2$, and are all of the same
chirality.  When combined with the half-hypermultiplet structure
from the center-of-mass fermions, they form a single BPS multiplet
with highest spin $|q|$ and total degeneracy $4\times 2|q|$.  Note
that the dyons with large $|q|$ are in multiplets with large highest
spin.\footnote{ Such high-spin dyons remain massive everywhere on
the vacuum moduli space, and do not enter the Seiberg-Witten
description of ${\cal N}=2$ theories in any crucial way. In
particular, the states with $|q|>1$, and possibly those with
$|q|=1$, would be completely missed if we were to use a bootstrap
argument to generate dyons by acting with monodromies on simple
elementary particles or fundamental monopoles. In order to
understand the complete BPS spectrum, one must at least start with
the above weak coupling spectrum as an input to the bootstrap.}

Perhaps the most important, yet very counterintuitive aspect of the
${\cal N}=2$ dynamics is that is that, in stark contrast with the
${\cal N}=4$ case, BPS bound states with $q=0$ are nowhere to be
found.  The absence of these states is a dramatic illustration of
the fact that the relation between classical solutions and quantum
states is more subtle than is often appreciated.

Note first that if two monopoles are associated with simple roots
$\bbeta_1$ and $\bbeta_2$ (and thus have magnetic charges
proportional to $\bbeta_1^*$ and $\bbeta_2^*$), they will interact
only if $\bbeta_1$ and $\bbeta_2$ are linked in the Dynkin diagram.
Given any two such linked simple roots, there is always a composite
root $\balpha$ whose dual is $\bbeta_1^* + \bbeta_2^*$.  We now show
that it is always possible to construct a classical solution whose
magnetic charge is the sum of these two monopole charges.

When there is only a single nonvanishing Higgs field, this solution
is obtained by using $\balpha$ to embed the SU(2) solution via
Eq.~(\ref{fundmonosolution}).  This result is easily extended to the
case with two nontrivial Higgs fields by using the R-symmetry to
rotate the Higgs fields, $\phi_1$ and $\phi_2$, into a new pair,
$\phi'_1$ and $\phi'_2$, such that $\phi'_2$ is in the Cartan
subalgebra and of the form ${\bf h}'_2 \cdot {\bf H}$, with ${\bf
h}'_2$ orthogonal to $\balpha$.  The desired solution is then
obtained by using the above embedding to generate the gauge fields
and $\phi'_1$, and then adding a spatially constant $\phi'_2$.  In
both of these cases, it is easy to show that the energy of the
classical solutions is related to the mass of the corresponding
gauge bosons by the replacement $e \rightarrow 4\pi/e$.

We showed in Sec.~\ref{indexsection} that for a single Higgs field
the classical solutions built from composite roots are actually just
special multimonopole solutions in which the noninteracting
monopoles happen to be coincident.  Hence, one might reasonably
expect the corresponding quantum state to be a two-particle state.
Once this is realized, the absence of a BPS bound state in the
${\cal N}=2$ theory should not be surprising.  Rather, it is the
fact that the ${\cal N}=4$ theory contains such a bound state,
{\it in addition} to the two-particle states, that should be seen as
remarkable and as a nontrivial test of the duality conjecture.

On the other hand, when there are two nonvanishing Higgs fields the
classical BPS solution obtained by embedding via a composite root has a
mass that is less than the sum of the masses of its components.  Since
the component monopoles cannot be separated by a small perturbation,
one is justified in interpreting this solution as a classical bound
state.  This does not, however, guarantee the existence of a BPS
quantum bound state (or even, if the potential is too shallow, any
quantum bound state at all).  This can be understood by noting
that the classical solution
corresponds to a fixed value of the intermonopole separation.  In the
quantum state, the wave function has a finite spread about this
value, to values of the separation for which there is no classical BPS
solution.  For the state to preserve some supersymmetry requires a
delicate interplay between the fermionic degrees of freedom and the
quantum fluctuations of the bosonic degrees of freedom.  This interplay
turns out to be possible only in the ${\cal N}=4$ theory.

\section{Many-body bound states and index theory methods}
\label{manybodyboundsection}

We need more a systematic approach to the problem to generalize the
bound state counting to the many-body case, since the explicit
construction of bound states becomes much more difficult beyond the
two-body case.  Instead of the direct construction of bound states,
we will proceed by using index theory methods.  The index
calculations can be quite involved, given that the quantum mechanics
involves many degrees of freedom with complicated interaction terms.
However, we will later see that it is precisely these interaction
terms that simplify the index calculations enormously.  Let us start
with some generalities, following Ref.~\cite{Stern:2000ie}.

We will define three different indices, each of which will be useful
for counting one type of BPS state.  In each case, there is a $Z_2$
grading $\tau$ that anticommutes with the supercharges that
annihilate the states in question.  The index counts the difference
between the numbers $n_+$ and $n_-$ of ground states with $\tau$
eigenvalues of 1 and $-1$.  We are actually interested in the sum,
$n_+ + n_-$, for which one needs a more refined understanding of the
dynamics, such as a vanishing theorem.  We will assume that such a
vanishing theorem does exist, so that either $n_+=0$ or $n_-=0$, and
assume that the absolute value of the index equals the number of
ground states of interest.

Finally, we note that in all of these cases we can calculate the
indices separately for each subspace of fixed central charges; in
our problems these central charges are completely determined by the
electric charges.

\begin{itemize}


\item 1/2-BPS states in ${\cal N}=4$ SYM

These states are annihilated by all of the supercharges of the
low-energy theory, which is only possible if the central charges all
vanish.  There is a canonical $Z_2$ grading, which in the geometric
language is defined on $k$-forms by
\begin{equation}
   \tau_4\equiv(-1)^k
\end{equation}
or, equivalently, by
\begin{equation}
   \tau_4\equiv \prod 2\eta^E_1\eta^E_2 = \prod 2\lambda_\chi^E
   \lambda_\xi^E =
   \prod (\varphi^{*E}\varphi^{E}-\varphi^{E}\varphi^{*E}) \, ,
\end{equation}
which anticommutes with all of the supercharges.  Thus, the sign of
$\tau_4$ is determined by whether the state is bosonic and
fermionic, and so the associated index, ${\cal I}_4$, is just the
usual Witten index.  For the two-monopole example of
Sec.~\ref{twodistinctbound}, our explicit construction of the bound
states shows that ${\cal I}_4 = 1$ if $q=0$, and vanishes otherwise.


\item 1/4-BPS states in ${\cal N}=4$ SYM

As we have seen, these can only occur if the Higgs vevs are such
that there is only a single nonvanishing $G_I$, which we can choose
to be $G_5$.  The 1/4-BPS states are annihilated by one (or both, if
the state is actually 1/2-BPS) of the operators ${\cal D}_\pm$
defined in Eq.~(\ref{N4Dplusminus}).  With only a single $G_I$,
there is a second type of $Z_2$ grading, defined by the operators
\begin{equation}
   \tau_s^\pm \equiv\prod (\sqrt{i}\,\varphi^E\pm \sqrt{-i}\,\varphi^{*E})
\end{equation}
that anticommute with ${\cal D}_\pm$.  We will denote the
corresponding indices by ${\cal I}_s^\pm$.  This generalizes the
signature index that counts the difference between the numbers of
self-dual and anti-self-dual wave functions.  From the results of
Sec.~\ref{twodistinctbound}, we see that for $0 < |q| < |\tilde a|$
and $\pm \tilde a q > 0$ we have ${\cal I}_s^\pm = 8|q|$ and that
${\cal I}_s^+ = {\cal I}_s^- = 1$ if $q=0$.  For all other cases,
${\cal I}_s^\pm = 0$.


\item 1/2-BPS states in ${\cal N}=2$ SYM

These are annihilated by the Dirac operator $\cal D$ of
Eq.~(\ref{N2Doperator}).  This anticommutes with the $Z_2$ grading
defined by the operator
\begin{equation}
    \tau_2=\prod \sqrt{2}\,\lambda^E =\prod \gamma^E  \, .
\end{equation}
The sign of $\tau_2$ is determined by whether the state is bosonic
and fermionic, and so the associated index ${\cal I}_2$, like ${\cal
I}_4$ above, is the usual Witten index.  For the two-monopole
examples of the previous section, ${\cal I}_2 = 2|q|$ if $aq > 0$
and $1/2 \le |q| < |a|$.

\end{itemize}

\subsection{Bound states of many distinct ${\cal N}=4$ monopoles}

We will consider specifically the bound states of many distinct
SU($N$) monopoles, corresponding to fundamental roots $\bbeta_1$,
$\bbeta_2$, \dots, $\bbeta_{k+1}$.  The moduli space potential
energy is derived from a single combination of the triholomorphic
Killing vector fields, $G=e\,\sum_A a^A K_A$.  From the analysis of
Sec.~\ref{staticmulticentersec}, we find that the effective
potential has a nontrivial minimum, thus allowing a classical bound
state, if
\begin{equation}
|q_A| < |\tilde a_A|
    \equiv \frac{4\pi^2}{e^3}  \sum_B (\mu^{-1})_{AB}\,a_B \, .
\end{equation}
This condition also guarantees the existence of a mass  gap in the
system, and allows us to compute the index using the index theorem
\cite{Stern:2000ie}.  Otherwise, there is a net repulsive force
between some of the monopoles, and there cannot be any bound state,
classical or quantum. The marginal case of $|q_A|=|\tilde a_A|$ is
more subtle; we will ignore this case except for some special
limits.

A standard theorem asserts that a Dirac operator ${\cal D}$ can be
deformed continuously without changing its index, as long as the
deformation does not destroy an existing mass gap.  Thus, as long as
we start with a case that has a mass gap as above, we can safely
multiply $G$ by a large number $T$ to find another Dirac operator
with an even larger mass gap, but with the same index. On the other
hand, a larger coefficient of $G$ means that the potential energy
gets stiffer and the low-energy motion gets confined closer to the
zeros of $G$ or, equivalently, nearer to the fixed points of $G$. In
this special set of examples, the one and only fixed point of $G$ is
the origin, $\vec r_A=0$, so it suffices to solve a local index
problem near the origin. Furthermore, the finite curvature at the
origin is overwhelmed by the ever-increasing scale associated with
the rescaled Killing vector $T\,G$.

For sufficiently large $T$, the problem reduces to one where
the geometry is a flat $R^{4k}$, and $G$ is a linear combination of
certain rotational vectors from each $R^4$ factor. The problem then
decomposes into many $R^4$ problems.  On the other hand, we may use
the same kind of deformation of the two-monopole problem to reduce
it to a flat $R^4$ problem as well.  The two-monopole problem has
been solved explicitly, so we already know the value of the index
for the $R^4$ problem. Then, since the multimonopole index problem
factorizes into many $R^4$ problems, all we need to do to recover
the value of the index for the multimonopole case is to take the
product of the known two-monopole indices for each interacting pair
of monopoles within the group.

Thus, when we consider a bound state with relative charges
$q_1,q_2,\dots,q_k$, we can count the number of states by
considering successive pairs $(\bbeta_A, \bbeta_{A+1})$ with
relative charges $q_A$.  Counting the degeneracy $d_A$ of each pair
as if no other monopoles were present, the degeneracy of the bound
state wave function involving all $k+1$ monopoles would be simply
the product of all the $d_A$.\footnote{Of course, to get the true
degeneracy, at the end of the day one must multiply by the factor of
either 16 or 4 from the center-of-mass part of the moduli space.} In
the remainder of this subsection, we will write out the resulting
index formulae explicitly, and make some contact with physics.

\subsubsection{1/2-BPS bound states}

Of the three indices, only ${\cal I}_4$ is robust against turning on
more than one of the $G^I$. Turning on an additional $G^I$ always
increases the mass gap, and is a Fredholm deformation that preserves
${\cal I}_4$. The index computation~\cite{Stern:2000ie} yields
\begin{equation}
{\cal I}_4 = \left( \prod_A\left\{\begin{array}{ll} 1 & \qquad q_A=0
\\ 0 & \qquad q_A \neq 0 \end{array} \right\}\right)  \, .
\end{equation}
Since the central charge of the state that contributes to the index
is zero, the state must be annihilated by all of the supercharges of
the quantum mechanics and be 1/2-BPS in the ${\cal N}=4$ SYM
theory.  This is consistent with the existence of a unique purely
magnetic 1/2-BPS bound state of monopoles in a generic Coulomb
vacuum, as is expected from the ${\rm SL}(2,Z)$ electromagnetic
duality. One of the generators of the ${\rm SL}(2,Z)$ maps the
massive charged vector supermultiplets to purely magnetic bound
states in a one-to-one fashion.  After taking into account the
automatic degeneracy of 16 from the free center-of-mass fermions,
the total degeneracy of these bound states is always 16, which fits
the ${\cal N}=4$ vector multiplet nicely. This purely magnetic bound
state was previously constructed by Gibbons in special vacua where
all the $G^I$ vanish.\footnote{One might think that the existence of
this bound state is obvious, since the potential energies are all
attractive and there exists a classical BPS monopole with the same
magnetic charge. However, none of these guarantees the existence of
a BPS bound state at the quantum level. In fact, the same set of
facts are true for a pair of distinct monopoles in ${\cal N}=2$
SU(3) SYM theory, but we know that a purely magnetic bound
state does not exist as a BPS state in that theory.}

\subsubsection{1/4-BPS bound states}

The existence of 1/4-BPS states requires that the relevant parts of
the Higgs expectation values be such that only one linearly
independent $G^I$ is present, which is just the condition that is
needed to make ${\cal I}_s^\pm$ available. In addition, the
effective potential energy in the charge eigensector must be
attractive along all asymptotic directions
for a bound state to exist. This condition takes the simple form
\begin{equation}
   |q_A| < |\tilde a_A| \, .
\end{equation}

Given the mass gap, the index ${\cal I}_s^\pm$ was computed and the
result \cite{Stern:2000ie} is
\begin{equation}
{\cal I}_s^\pm = \left(\prod_A\left\{\begin{array}{ll} 8|q_A| &
\qquad \pm \tilde a_Aq_A>0
\\ 1 & \qquad \tilde a_Aq_A=0 \\ 0 & \qquad \pm \tilde a_Aq_A <0
\end{array} \right\}\right).
\end{equation}
Note that the index is nonvanishing only if each of the $\pm \tilde
a_Aq_A$ is nonnegative. This is in addition to the usual requirement
that
\begin{equation}
\pm \, \sum_A \tilde a_Aq_A >0 \, ,
\end{equation}
which is necessary for the states to be annihilated by ${H}\mp {Z}$
with ${Z}= {Z}_5= \sum_A \tilde a_Aq_A $ being the central charge.
The index indicates that the degeneracy of such a 1/4-BPS state is
\begin{equation}
16\times \prod_A {\rm Max}\,\{8|q_A|,1\}.
\end{equation}
with the factor of 16 arising from the free center-of-mass fermions.

In the two-monopole bound states, the number $8|q|$ is accounted for
by four angular momentum multiplets with $j=|q|$, $|q|-1/2$,
$|q|-1/2,$ and $|q|-1$ (except for $|q|=1/2$, where the first three
suffice). The top angular momentum $|q|$ in the relative part of the
wave function has a well-known classical origin: When an
electrically charged particle moves around a magnetic object, the
conserved angular momentum is shifted by a factor of $eg/4\pi$.
While the fermions can and do contribute, the number of fermions
scales with the number of monopoles, and not with the charge $q_A$.
In fact, for large charges the top angular momentum of such a dyonic
bound state wave function is
\begin{equation}
j_{\rm top}=\sum_A |q_A|,
\end{equation}
so that the highest spin of a dyon would be
\begin{equation}
1+j_{\rm top}=1+\sum_A |q_A|
\end{equation}
after taking into account the universal vector supermultiplet
structure from the free center-of-mass part. The actual multiplet
structure is not difficult to derive, and we find
\begin{equation}
V_4\otimes \biggl(\otimes_A \biggl\{[|q_A|]\oplus[|q_A|-1/2]
\oplus[|q_A|-1/2]\oplus[|q_A|-1]\biggr\}\biggr)  \, .
\end{equation}
Here $V_4$ denotes the vector supermultiplet of ${\cal N}=4$
superalgebra, and $[j]$ denotes a spin $j$ angular momentum
multiplet.

The largest supermultiplet contained in this has highest spin
$j_{\rm top}+1$; such a supermultiplet has a degeneracy (including
the factor from the center-of-mass fermionic zero modes) of
\begin{equation}
16\times 8\sum_A |q_A| \, .
\end{equation}
Unless all but one of the $q_A$ vanishes, this is much less than the
number of states we found above. Thus, this implies that there are
many 1/4-BPS, and thus degenerate, supermultiplets of dyons for a
given set of electromagnetic charges. For large electric charges
$q_A$, thus, the number of dyon supermultiplets scales
as,\footnote{It has been conjectured that for large electric and
magnetic charges the degeneracy will eventually scale exponentially,
with the exponent being linear in the charges \cite{Kol:1998zb}.
There is, to date, no field theoretical confirmation of this,
although in the supergravity regime such large degeneracies are
implied by black hole entropy functions.}
\begin{equation}
\left(\prod_A {\rm Max}\,\{8|q_A|,1\}\right)/\left(\sum_A
8|q_A|\right).
\end{equation}
While one would expect to find degenerate states within a
supermultiplet, there is no natural symmetry that accounts for the
existence of many supermultiplets with the same electromagnetic
charges and the same energy.

\subsection{Bound states of many distinct ${\cal N}=2$ monopoles}

In ${\cal N}=2$ SYM theories, a state can be either BPS or
non-BPS. There is no such thing as a 1/4-BPS state. Dyons that would
have been 1/4-BPS when embedded in an ${\cal N}=4$ theory are
realized as either 1/2-BPS or non-BPS, depending on the sign of the
electric charges.

Whenever there is a mass gap, the index ${\cal I}_2$
is~\cite{Stern:2000ie}
\begin{equation}
{\cal I}_2 = \left(\prod_A\left\{\begin{array}{ll} 2|q_A| & \qquad
\tilde a_Aq_A > 0
 \\ 0 & \qquad \tilde a_Aq_A \le 0 \end{array} \right\}\right) \, ,
\end{equation}
which gives us a possible criterion for BPS dyons to
exist.\footnote{This field theory counting was precisely reproduced
later by a string theory construction using D-branes wrapping
special Lagrange submanifolds in a Calabi-Yau manifold
\cite{Denef:2002ru}.} This condition is similar to the condition for
BPS dyons or monopoles to exist in ${\cal N}=4$ SYM theories,
but differs in two aspects. The first is that, given a set of $a_A$,
all of which are positive (negative), the electric charges $q_A$
must be all positive (negative). The overall sign of the electric
charge matters.

The second difference from the ${\cal N}=4$ case is that, as we have
already noted for the two-monopole case, there is no purely magnetic
BPS bound state of monopoles, even though there exists a classical
BPS solution with such a charge.  In fact, the index indicates that
all relative $q_A$ must be nonvanishing for a BPS state to exist.
Assuming the vanishing theorem, the number of BPS dyonic bound state
under the above condition is \cite{Stern:2000ie}
\begin{equation}
4\times \prod_A 2|q_A| \, ,
\end{equation}
with the overall factor of $4$ coming from the quantization of the
free center-of-mass fermions. The actual multiplet structure is
\begin{equation}
C_2\otimes \biggl(\otimes_A \;[|q_A|-1/2]\biggr)
\end{equation}
where $C_2$ denotes the half hypermultiplet of the ${\cal N}=2$
superalgebra.

For large electric charges we again observe the proliferation of
supermultiplets. The top angular momentum, and thus the size of the
largest supermultiplet, can only grow linearly with $\sum |q_A|$.
This means that the number of supermultiplets with the same electric
charges scales at least as
\begin{equation}
\left(\prod_A 2|q_A|\right) /\left(\sum_A 2|q_A|\right)
\end{equation}
for large $q_A$.

\section{Difficulties in finding BPS states with four supercharges}
\label{boundstabilitysec}

Much of this chapter has been devoted to counting dyonic states that
are either 1/4-BPS in ${\cal N}=4$ theories or 1/2-BPS in ${\cal
N}=2$ theories. In either case, the BPS states in question preserve
four supercharges. We succeeded in counting the dyons made out of a
chain of distinct monopoles, and also found that their existence
depends sensitively on the choice of the vacuum and the coupling
constant.

Just as we found in our earlier classical analysis, these dyons are
typically loosely bound states of more than one charged particle.
The size of the wave function grows indefinitely as we increase
certain electric charges beyond critical values or as we move the
vacuum toward some limiting values. This behavior of a bound state
breaking up into two infinitely separated dyons is at the heart of
the marginal stability that became familiar from the study of ${\cal
N}=2$ SYM theory.\footnote{As was mentioned in Chap.~7,
essentially the same phenomenon has been found as well in the
opposite limit of the strongly coupled regime
\cite{Ritz:2000xa,Argyres:2000xs,Ritz:2001jk,Argyres:2001pv}.}

Although the same sort of marginal stability mechanism exists for
the 1/4-BPS dyons in ${\cal N}=4$ and the usual BPS dyons in ${\cal
N}=2$, there are further subtleties in ${\cal N}=2$ SYM theory.  In
particular, unlike in ${\cal N}=4$ SYM theory, the existence of a classical
BPS solution does not guarantee the existence of its quantum
counterpart, even when the low-energy effective theory has a mass
gap. The clearest example of this can be found in the ${\cal N}=2$ pure
SU(3) theory. As we saw in the previous sections, a purely magnetic
bound state of the two monopoles does not exist as a quantum bound
state, even though classically it would be on equal footing with the
other two, lighter monopoles. This absence of the third, heaviest
monopole was shown first for the moduli dynamics without a potential
energy \cite{Fraser:1997nd} and then more recently for the case
where a potential energy is present \cite{Gauntlett:1999vc}.

Overall, cataloging states with four unbroken supercharges turns out
to be a rather difficult task, not only in SYM theories, but
also in superstring theories. In fact, the two problems are often
closely related. For instance, if one realizes ${\cal N}=2$ SYM theory
as the dynamics of a wrapped M5 brane \cite{Witten:1997sc}, the BPS
states correspond to supersymmetric open membranes with boundaries
circling specific combinations of topological cycles on the wrapped
M5 brane \cite{Henningson:1997hy,Mikhailov:1997jv}. If one
realizes these theories by Calabi-Yau compactification of type II
string theories \cite{Klemm:1996bj}, the BPS states are D-branes
completely wrapped on supersymmetric cycles of the Calabi-Yau
manifold. The problem of finding such states manifests itself in
many diverse mathematical forms in string theory, of which we have
just mentioned two.

Other approaches to this general class of problem have been attempted.
One method involves a worldvolume approach, in which
one tries to determine the existence and the degeneracy by studying
boundary conformal field theories \cite{Lerche:2000uy,Lerche:2000iv}
or a topological version thereof
\cite{Hori:2004ja,Hori:2004zd,Walcher:2004tx,Hori:2006ic}. A more
geometrical approach\footnote{See, for example,
Refs.~\cite{Douglas:1999vm,Douglas:2000ah,Douglas:2000qw,Diaconescu:2000ec,Denef:2001ix}.
} led later to an attempt to encase the problem in a new
mathematical framework called a ``derived category''; for the latter,
see Refs.~\cite{Douglas:2000gi,Lazaroiu:2003md}. However, these problems remain
largely unsolved.

In this respect, the counting of BPS dyons in this chapter
represents one of the most concrete and successful programs we know
of. It is true that our approach here is applicable for only a small
corner of the entire landscape of this class of problem, but it is
also one where computation can be performed explicitly and with a
well-defined approximation procedure. The hope is that one can
eventually find ways to connect our findings to other regimes and
find useful information about the behavior of BPS spectra in other
regimes.

\chapter{The D-brane picture and the ADHMN construction\label{Brane}}

D-branes are nonperturbative objects, found in some string theories,
that accommodate the endpoints of open strings
\cite{Polchinski:1995mt}.  A D$p$-brane is a D-brane that has $p$
spatial dimensions.  A flat, infinitely extended D-brane in
$R^{9+1}$ preserves half of the 32 supercharges of the spacetime, so
the dynamics of the D-brane itself must respect 16 supercharges.
This fact restricts the possible form of the low-energy dynamics
quite severely and naturally gives it a gauge theory structure.
Furthermore, a stack of many identical D$p$-branes is associated
with a Yang-Mills dynamics with 16 supersymmetries.  A soliton of
the SYM theory is then transformed into a local deformation
of this stack of D$p$-branes, and we can ``view'' such solitons by
seeing how the D$p$-branes are deformed locally.

What makes this representation of the SYM theory especially
useful for the study of solitons is that there is an alternate
picture of these Yang-Mills solitons in terms of lower dimensional
D-branes. For the BPS monopoles with which we are concerned, the
relevant picture is a segment of D1-brane stretched orthogonally
between a pair of D3-branes.  The dynamics of the pair of D3-branes
is exactly that of an ${\cal N}=4$ ${\rm U}(2)$ SYM theory
that is spontaneously broken to ${\rm U}(1)\times {\rm U}(1)$ by the
separation between the two D3-branes.

Thus, the motion of the monopole/D1-segment can be described from
two completely different viewpoints --- either as a trajectory on
the moduli space or as a motion in the space of the classical vacua
of a ($1+1$)-dimensional SYM theory compactified on an
interval.  From the latter viewpoint, the Nahm equation emerges as
the supersymmetric vacuum condition on the ($1+1$)-dimensional
theory \cite{Diaconescu:1996rk,Kapustin:1998pb,Tsimpis:1998zh}.  This is the
underlying physics behind the Nahm data, and gives us a rationale
for identifying the geometry of the Nahm data moduli space with that
of the monopole moduli space.\footnote{For a
review of other topological solitons from the D-brane viewpoint, see
Ref.~\cite{Tong:2005un}.}

We will start our discussion, in Sec.~\ref{DbraneYangMills}, with a
overview of D-branes and their relation to SYM theories.
Next, in Sec.~\ref{Dbranesolitons}, we will describe in more detail
how solitons --- and monopoles in particular --- fit into this
picture.  T-duality and the relationship between monopoles and
instantons are discussed in Sec.~\ref{instantonpartons}.  Finally,
the connection between the Nahm data and D-branes is explained in
Sec.~\ref{DbraneNahm}.

\section{D-branes and Yang-Mills dynamics}
\label{DbraneYangMills}

D-branes are extended objects that are charged with respect to the
so-called Ramond-Ramond tensor fields.  Historically, these objects
were first found as black $p$-brane solutions; i.e., as charged
black-hole-like objects of an extended nature. A classic paper by
Polchinski \cite{Polchinski:1995mt} showed how to realize these
objects in terms of conformal field theory as boundaries on which a
string can end. This latter characterization provides a very
powerful tool for studying D-branes. In this review, however, we do
not have space for a systematic introduction to open string
theories. Rather, we will approach D-branes heuristically and borrow
key results from string theory whenever convenient.

\subsection{D-brane as a string background}

The D-branes that we will be interested in are those found in type
IIA and type IIB string theories. The Ramond-Ramond tensor fields
$C^{(p+1)}$ are antisymmetric tensor fields, or equivalently
$(p+1)$-forms, living in the ten-dimensional spacetime. There is a
gauge transformation involving a $p$-form $\Lambda^{(p)}$,
\begin{equation}
C^{(p+1)}\rightarrow C^{(p+1)}+d\Lambda^{(p)} \, ,
\end{equation}
in complete parallel with the case of the usual vector gauge fields.
The invariant field strength is thus
\begin{equation}
H^{(p+2)}=dC^{(p+1)}
\end{equation}
and a typical equation of motion takes the form
\begin{equation}
\nabla\cdot H^{(p+2)}= \cdots
\end{equation}
with electric sources and interaction terms on the right hand side.
The case of $p=0$  corresponds to the usual Abelian gauge field.

One way to think about a D-brane is as a supersymmetric background
for type II superstrings. The action for the low-energy effective
theory, type II supergravity, is
\begin{eqnarray}
   S&=&
   \int_{\rm spacetime}\sqrt{g}e^{-2\phi}\left(R+4(\nabla\phi)^2
   -\frac{1}{2\cdot 3!}|dB|^2 \right)\nn
   &+&
   \int_{\rm spacetime} \sqrt{g} \sum_p \frac{-1}{2\cdot (p+2)!}
    |dC^{(p+1)}|^2 +\cdots \, .
\end{eqnarray}
This contains terms with a dilation $\phi$ and a Kalb-Ramond 2-form
field $B$. The ellipsis represents various interaction terms as well
as those required for the supersymmetric completion of the theory.
Just as an electrically charged particle couples minimally to a
vector gauge field through
\begin{equation}
   S_{\rm int}  =    \int_{{\rm worldline}} C^{(1)}
\end{equation}
and enters the equation of motion for the latter via
\begin{equation}
\nabla\cdot H^{(2)}= *\delta_{\rm worldline}+\cdots  \, ,
\end{equation}
we may imagine extended objects with $p$ spatial dimensions that
couple minimally to these higher rank tensor fields via
\begin{equation}
  S_{\rm int}  = \int_{{\rm worldvolume}} C^{(p+1)}
\end{equation}
and provide electric source terms of the form
\begin{equation}
\nabla\cdot H^{(p+2)}= *\delta_{\rm worldvolume}+\cdots  \, .
\end{equation}
Here $\delta_{\rm worldvolume}$ is the $(9-p)$-form delta function
supported on the worldvolume and $*$ is the Hodge dual operation.

The coupling to gravity allows us to find solutions with finite
energy per unit volume that carry such electric charges. These are
typically gravitational solitons of an extended nature, which are
generically black $p$-brane solutions with event horizons. D-branes
are represented by a specific subclass of these solutions that have
the lowest possible mass per unit volume. They have a universal form
as follows. The metric is an extremal black $p$-brane
solution,\footnote{See Ref.~\cite{Stelle:1996tz} for a thorough
review of supersymmetric solution of this types.}
\begin{equation}
g=f^{-1/2}(-dy_0^2+dy_1^2+dy_2^2 +\cdots+dy_p^2) +f^{1/2}
(dx_{p+1}^2+\cdots +dx_9^2)
\end{equation}
where for $n$ D$p$-branes, located at $\vec x= \vec X_{(i)}$ (i=1,
2, \dots, n), $f$ is a harmonic function on $R^{9-p}$ of the form
\begin{equation}
f=1+Q_p\sum_{i=1}^n \frac{1}{|\vec x - \vec X_{(i)}|^{7-p}}
\end{equation}
with $Q_p$  a quantized dimensionful quantity. This solution has
event horizons at $\vec x = \vec X_{(i)}$, and can be thought of as
an analog of the extremal Reissner-Nordstr\"om black holes that appear
in the four-dimensional Einstein-Maxwell theory.

The dilaton $\phi$ and the Ramond-Ramond tensor field $C^{(p+1)}$
are also fixed in terms of the same harmonic function $f$ via
\begin{eqnarray}
e^{\phi} &=& e^{\phi_0}f^{(3-p)/4} \cr dC^{(p+2)} &=&
e^{-\phi_0}dx_I \wedge dy_0\wedge dy_1\wedge\cdots\wedge dy_p\times
\partial_I\left(\frac{1}{f}\right)   \, .
\end{eqnarray}

\subsection{D is for Dirichlet}

These solutions are called D-branes because strings can end on them;
i.e., they satisfy a Dirichlet boundary condition at $\vec x= \vec
X_{(i)}$ \cite{Polchinski:1995mt}. While we must understand how
D-branes are realized in the full string theory in order to show
this fact, it turns out that there is a more heuristic picture of
why this happens \cite{Lee:1997xh,Yi:1998tx}. Here we will follow
Ref.~\cite{Yi:1998tx} and consider a Nambu-Goto string propagating
in the background of two parallel D-branes, and ask what would
happen if part of that string happened to meet the horizon at $\vec
x= \vec X_{(i)}$.  If we denote the induced metric on the
world-sheet by $h_{\mu\nu}$, the action is
\begin{equation}
S=-\frac{1}{2\pi\alpha'}\int d\sigma^2 \sqrt{-{\rm Det}\, h} \, ,
\end{equation}
up to couplings to the dilaton and to the antisymmetric tensor $B$.
(The latter is absent in a D-brane background, while the dilaton
coupling occurs at higher order in $\alpha'$ and will be
subsequently ignored.)

The spacetime geometry for a pair of parallel D$p$-branes located at
$\vec{x}=\vec X_{(1)}$ and $\vec{x}=\vec{X}_{(2)}$ is completely
determined by the harmonic function
\begin{equation}
f= 1+\frac{Q_p}{|\vec x-\vec X_{(1)}|^{7-p}}
+\frac{Q_p}{|\vec{x}-\vec{X}_{(2)}|^{7-p}}   \, .
\end{equation}
We consider a string segment stretched between such a pair and
denote its embedding into the coordinates $y_\mu$ and $x_I$ by
$Y_\mu(\sigma^s)$ and $X_I(\sigma^s)$, where the $\sigma^s$ are the
two worldvolume coordinates. For the sake of simplicity, we will
choose $\sigma_1=\sigma$ to run from 0 to 1 and adopt a static gauge
where the world-volume time $\sigma_0=\tau$ is identified with that
of spacetime, $y_0$, so that $Y_0(\tau,\sigma)=\tau$. The induced
metric is
\begin{equation}
h= -f^{-1/2}d\tau^2 + f^{-1/2}\partial_s Y^n\partial_t Y^n d\sigma^s
d\sigma^t +f^{1/2}\partial_s X^I\partial_t X^I d\sigma^s d\sigma^t
\, . \label{induced:string}
\end{equation}
Taking its determinant, we find
\begin{equation}
{\rm Det}\,h= -(\partial_\sigma X^I)^2 - f^{-1}(\partial_\sigma
Y^n)^2 +{\rm Det} \left( f^{-1/2}\partial_s Y^n\partial_t Y^n+
f^{1/2}\partial_s X^I\partial_t X^I \right) \, .
\end{equation}
Note that the third term contains two factors of the time
derivatives $
\partial_\tau X$ and $\partial_\tau Y$. This implies that there exists
a static solution
\begin{equation}
\vec{X}=\sigma\vec{L}\, ,\qquad \qquad \partial_s {Y}^n={0} \, ,
\end{equation}
with $\vec L=\vec{X}_{(1)}-\vec X_{(2)}$, that corresponds to a
straight BPS string segment located at a constant $y^n$ coordinate.
The action per unit time for a static configuration is the energy,
so we find the ground state energy to be
\begin{equation}
\frac{1}{2\pi\alpha'}L \equiv \frac{1}{2\pi\alpha'}|\vec L| \, .
\end{equation}
We find that the BPS mass of this stretched string is insensitive to
the gravitational radius of the background.  However, there is a
subtlety here, in that the distance that enters the mass formula is
not the proper distance but rather a coordinate distance in a
preferred coordinate system, widely known as the isotropic
coordinate system.

Consider small fluctuations around this ground state of the
stretched string. Let $\vec{X}=\sigma\vec{L}+
\vec{\epsilon}(\tau,\sigma)$, with $\vec \epsilon$ orthogonal to
$\vec{L}$, and ${Y}^n={\eta}^n(\tau,\sigma)$. To the first
nonvanishing order, the determinant can be expanded as
\begin{eqnarray}
{\rm Det}\,h = &-& L^2 + L^2\,(\partial_\tau \eta^n)^2-f^{-1}
(\partial_\sigma \eta^n)^2 \cr
  &+& f L^2\,(\partial_\tau \epsilon^I)^2-
(\partial_\sigma \epsilon^I)^2 + \cdots \, .
\end{eqnarray}
The ellipsis represents terms that are at least quartic in the small
fluctuations, and $f$ here is to be evaluated along the ground state
of the string, so $f(\vec{x})=f(\sigma\vec{L})$. The Lagrangian is
obtained by taking the square root and expanding in powers of
$\epsilon^I$ and $\eta^n$.

For fluctuations orthogonal to the background D$p$-branes, the
combination $fL^2$ is the effective (inertial) mass density. A
finite energy motion must have a finite integrated value of
$f\,(\partial_\tau \epsilon^I)^2$ and, in addition,
$f({\epsilon}^I)^2$ must integrate to a finite number for any
eigenmode of the Hamiltonian.  With the divergence of $f\sim
(\Delta\sigma)^{p-7}$ near either end of the string (at least for
small enough $p$), this immediately implies that the $\vec\epsilon$
part of the fluctuation must obey Dirichlet boundary conditions. In
contrast, no such condition is imposed on the other fluctuations,
${Y}^n={\eta}^n$, which are parallel to the background D$p$-brane.
The boundary value of $\vec\epsilon$ represents a fluctuation that
would take the endpoint of the string off the D$p$-brane, so the
Dirichlet boundary condition means that the string cannot break away
from the D$p$-brane.  This gives a classical picture that tells us
that D$p$-branes are places where a string can end and become an
open string.  A byproduct of this heuristic observation is that the
coordinates $X^i$ of the isometric coordinate system are the ones
corresponding to the world-sheet fields that must be quantized with
Dirichlet boundary condition.

Finally we must caution the readers to be wary of this picture where
we have effectively ``put the cart before the horse." As is well
known, the curved geometry of the D-branes can be thought of as a
higher-order effect from the viewpoint of the open string.  Here we
have used this curved geometry to argue for the possibility of open
strings ending on the D-branes.  This is one of many phenomena that
must be present if the D-brane story is to be a self-consistent one.
Later in this section we will provide further heuristic reasoning,
based on charge conservation, as to why open strings can end on
D-branes. For this, however, we must first understand what kinds of
fields live on the worldvolumes of D-branes.

\subsection{Low-energy interactions between D-branes}

When we wrote the supergravity solution for many D$p$-branes, we did
not specify what their positions $\vec X_{(i)}$ should be; that was
because these are moduli parameters.  As with ordinary solitons, we
may imagine a low-energy approximation to the dynamics of these
D-branes, i.e., a moduli space approximation that includes the
D-brane positions as massless fields. An important constraint on
such an attempt comes from supersymmetry. The D-brane solution above
preserves precisely half of the spacetime supersymmetry, and thus
must respect 16 supercharges. Recall that the number of propagating
field theory degrees of freedom is essentially independent of
dimension and is fixed solely by the supersymmetry. On the other
hand, each position vector $\vec X_{(i)}$ carries $9-p$ parameters,
so we must include additional bosonic, as well as the fermionic,
degrees of freedom. We must look for an appropriate supermultiplet
into which the moduli parameters can be organized.

Except in two or six dimensions, where a chiral form of
supersymmetry is possible, the smallest BPS supermultiplet in
theories with 16 supersymmtries is unique. Furthermore, this
universal multiplet has exactly $9-p$ scalars in it and generically
has a single gauge field as a superpartner carrying $p-1$ degrees of
freedom. Let us call this the maximal vector multiplet. Thus, the
low-energy effective action of a single D-brane must involve a
single maximal vector multiplet. Its action fits into the
Dirac-Born-Infeld action
\cite{Cederwall:1996pv,Aganagic:1996pe,Cederwall:1996ri,Bergshoeff:1996tu,Aganagic:1996nn},
whose bosonic part has two pieces. The first
term \cite{Dai:1989ua,Leigh:1989jq,Li:1995pq},
\begin{equation}
    -\mu_{p}e^{-\phi}
    \sqrt{-{\rm Det}\,(g_{\mu\nu}+B_{\mu\nu}+2\pi\alpha' F_{\mu\nu})}  \, .
\end{equation}
is a nonlinear
kinetic term that dictates how the worldvolume moves.  Here the
tension of the D$p$-brane is $\mu_p e^{-\phi_0}$, where
\begin{equation}
\mu_p=\frac{2\pi}{(2\pi\sqrt{\alpha'})^{p+1}}
\end{equation}
is determined by the string tension and $\phi_0$, the asymptotic
value of the dilaton, is related to the asymptotic value of the
string coupling constant by $g_s=e^{\phi_0}$.

Given a spacetime metric $G$, the induced metric that enters the
action can be written as
\begin{equation}
g_{\mu\nu}= \sum_{I=0}^9\partial_\mu Z^I\partial_\nu Z^J G_{IJ}
\end{equation}
with the Greek indices running over $0,1,2,\dots,p$, and $Z^I$
embedding the D-brane worldvolume into spacetime. Similarly,
\begin{equation}
   B_{\mu\nu}=
   \sum_{I=0}^9\partial_\mu Z^I\partial_\nu Z^J {\cal B}_{IJ}
\end{equation}
is the pull-back of the NS-NS 2-form tensor ${\cal B}$. (Throughout
this report we will consider only backgrounds with ${\cal B}\equiv
0$.)  The dilaton is given by
\begin{equation}
e^\phi=e^{\phi_0}f^{(3-p)/4} \, .
\end{equation}

The second, topological, term \cite{Green:1996dd,Cheung:1997az},
\begin{equation}
\mu_p \left[\sum_{n=0}^{[(p+1)/2]} C^{(p+1-2n)}\wedge
e^{B+2\pi\alpha'F} \right]_{ (p+1)-{\rm form}}   \, ,
\label{braneTopologicalTerm}
\end{equation}
has no analogue in the usual gauge theory, since it dictates how
worldvolume fields couple to the spacetime Ramond-Ramond fields.
This generalization of minimal coupling has far-reaching
consequences in what follows.  One of its implications is that a
worldvolume configuration with nontrivial Chern-character, (i.e.,
nonzero integrals of expressions like $F^n$) couples minimally to a
lower-rank Ramond-Ramond field and behaves as if it were a D-brane
of lower dimensions.

Let us ask how such D-branes interact with each other in the
low-energy limit.  One way to isolate the long-range interactions
between these objects is to ask how a test D-brane responds to
another D-brane located far from the test D-brane. This is the same
sort of approximation that we adopted for determining the asymptotic
form of the moduli space metric for well-separated monopoles. To
make it a valid approximation, we would typically have to introduce
many coincident D-branes, which would have the effect of multiplying
the charge $Q_p$ by a large integer.  Since this does not change the
overall structure of the interaction, we will drop this step and
pretend that we are studying the interactions of just a pair of
D$p$-branes.

Thus, let us hold one D$p$-brane at a fixed point, $\vec X_{(1)}$,
and ask for the low-energy action of the other. This is simply
achieved by inserting the background generated by the first
D$p$-brane into the worldvolume action of the other. For instance,
the $(p+1)\times (p+1)$ matrix that enters the Born-Infeld term
should be
\begin{eqnarray}
g_{\mu\nu}+2\pi\alpha' F_{\mu\nu} &=&f^{-1/2}_{12}\eta_{\mu\nu}+
f^{1/2}_{12}\sum_{K=p+1}^9\partial_\mu X_{(2)}^K\partial_\nu
X_{(2)}^K +2\pi\alpha' F_{(2)\mu\nu}\nn &=&f_{12}^{-1/2}\left[
\eta_{\mu\nu} + f_{12} \sum_{K=p+1}^9\partial_\mu
X_{(2)}^K\partial_\nu X_{(2)}^K+ 2\pi\alpha' f^{1/2}_{12}
F_{(2)\mu\nu} \right]   \cr &&
\end{eqnarray}
where we have chosen to use the $y^\mu=(y_0, y_1, y_2, \dots,
y_p)$ that appear in the D-brane solution as the worldvolume
coordinates and to encode the position of the second D-brane in
$9-p$ functions $X_{(2)}^K(y)$. The effect of the first D-brane is
encoded in
\begin{equation}
f_{12}=1+Q_p\frac{1}{|\vec X_{(2)}(y) - \vec X_{(1)}|^{7-p}} \, .
\end{equation}
Here we have kept an explicit subscript for $X_{(2)}$ and $F_{(2)}$
to emphasize that these are fields defined on the worldvolume of the
second D-brane. The function $f_{12}$ also enters the action via
other background fields, $\phi$ and $C^{(p+1)}$.  The first term in
the derivative expansion of the Born-Infeld action is
\begin{equation}
-\mu_p e^{-\phi} f_{12}^{-(p+1)/4}= -\mu_p e^{-\phi_0}
\frac{1}{f_{12}}  \, .
\end{equation}
It appears that there is a potential term here from $f_{12}$, but
this interaction is precisely cancelled by the minimal coupling,
from Eq.~(\ref{braneTopologicalTerm}), to the background
$C^{(p+1)}$, so the leading term in the derivative expansion is
actually\footnote{As usual, there is an additive ambiguity in
$C^{(p+1)}$, since only the field strength $dC^{(p+1)}$ is fixed by
the solution. This ambiguity can be resolved by asking that the
leading constant term of the worldvolume action be due entirely to
the constant tension of the brane.}
\begin{equation}
-\mu_p e^{-\phi_0}   \, .
\end{equation}
The next terms in the expansion, with two derivatives, are
\begin{eqnarray}
-\mu_p e^{-\phi_0}  \frac{1}{f_{12}} \,
\left[\frac{1}{2}f_{12}\,\partial_\mu X^K_{(2)}\partial^\mu
X^K_{(2)} +\frac{1}{4}f_{12} (2\pi\alpha')^2
 F_{(2)\mu\nu}F_{(2)}^{\mu\nu}\right] \, .
\end{eqnarray}
These simplify considerably upon the introduction of a scalar field
$\Phi_{(2)}^K=X_{(2)}^K/2\pi\alpha'$ and become
\begin{equation}
-\frac{1}{2\pi e^{\phi_0}}\frac{1}{(2\pi\sqrt{\alpha'})^{p-3}}
\left[\frac12 (\partial \Phi_{(2)}^K)^2+\frac14 F_{(2)}^2\right] \,
.
\end{equation}
The dependence on the background has again disappeared, showing
that, up to two derivative terms, one D-brane does not feel the
presence of the other.

In fact, supersymmetry combined with gauge symmetry is so
restrictive that we cannot write down any low-energy interactions
between the D-branes if we stick only to terms with two or fewer
derivatives.  Only when we include higher-order terms, such as
(velocity)$^4$ or (field strength)$^4$, do we begin to see
long-range interactions between the D-branes.  For example,
expanding the Born-Infeld term up to fourth order in derivatives
gives long-range interactions of the form
\begin{equation}
e^{-\phi_0} (\sqrt{\alpha'})^{7-p}f_{12}\times\left[ F_{(2)}^4
\hbox{ or } (\partial \Phi_{(2)}^K)^4\right]
\label{quarticDerivBraneterms}
\end{equation}
from the position-dependent part of $f_{12}$.

\subsection{Yang-Mills description and open strings}

This latter form for the interaction is, at best, cumbersome to
handle. A remarkable fact about D-branes, however, is that these
higher-derivative interactions can be encoded in a perfectly
sensible two-derivative action by including additional massive
fields. To reclaim the correct long-range interaction, we must take
the somewhat unusual path of quantizing the theory and then
integrating out these additional fields. These auxiliary fields are
charged and, order by order,  generate the correct long-range
effective interaction between the original massless U(1) fields.  Of
course, this is no accident. The additional charged fields have a
natural stringy interpretation as open strings stretched between the
two D$p$-branes.   We will now finally come to the point and discuss
how the worldvolume dynamics at low energy is encoded in a
SYM theory.

For the proposed two-derivative action for $n$ parallel D$p$-branes,
let us start with the sum of the two-derivative terms from the
Born-Infeld actions of the individual D$p$-branes,\footnote{Note
that our
  conventions in this chapter differ somewhat from those of the
  previous chapters.  We use a (-+++) metric, and have rescaled the
  gauge fields by a factor of $g_{YM}=e/\sqrt{2}$.}
\begin{equation}
-\frac{1}{g_{YM}^2}\sum_{i=1}^n \left[\frac14 (F_{(i)})^2
   +\frac12(\partial \Phi_{(i)}^K)^2\right] \, ,
\end{equation}
where
\begin{equation}
g_{YM}^2=2\pi e^{\phi_0}\left(2\pi\sqrt{\alpha'}\right)^{p-3} \, .
\label{gymdefinition}
\end{equation}
This is precisely the bosonic part of the action of a ${\rm U}(1)^n$
gauge theory with maximal supersymmetry in any dimension.  The
proposal is simply to elevate this action to that of the maximally
supersymmetric ${\rm U}(n)$ theory,
\begin{equation}
-\frac{1}{g_{YM}^2}\,\tr\left[\frac14 F^2+\frac12(D \Phi^K)^2
-\frac14\sum_{K,M}[\Phi^K,\Phi^M]^2\right] \, ,
\label{elevatedaction}
\end{equation}
with the $\Phi^K$ being in the adjoint representation
\cite{Witten:1995im}.

The dictionary for recovering individual D-branes is well-known.  If
we go to the Coulomb phase of this non-Abelian theory, with the
Higgs expectation value in diagonal form, then the identification is
\begin{equation}
\Phi^K=\left(\begin{array}{cccc}
\Phi_{(1)}^K & 0 & 0 & \cdots  \\
0& \Phi_{(2)}^K  & 0 & \cdots  \\
0 & 0 & \Phi_{(3)}^K & \cdots \\
\cdots & \cdots & \cdots & \cdots\end{array}\right)
+\hbox{off-diagonal parts}
\end{equation}
and similarly for the gauge field part.  Since
$\Phi_{(i)}^K=X_{(i)}^K/2\pi\alpha'$, the diagonal parts of the
adjoint scalar fields encode the positions of the individual
D-branes. When the eigenvalues of the vev are all distinct, the
fields corresponding to the off-diagonal parts, $A_{ij}$ and
$\Phi_{ij}^K$ with $i\neq j$, are all massive and do not correspond
to moduli of D-branes.  Rather, they behave as massive fields that
are charged with respect to the diagonal ${\rm U}(1)^n$ theory.

\begin{figure}[t]
\begin{center}
\scalebox{1.2}[1.2]{\includegraphics{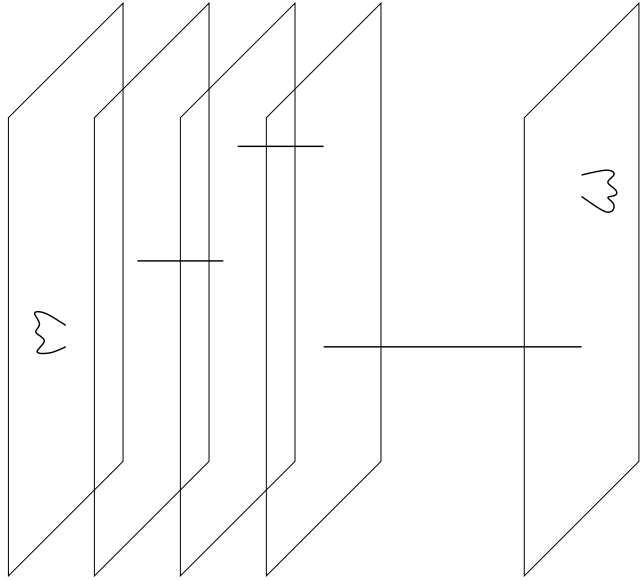}}
\par
\vskip-2.0cm{}
\end{center}
\begin{quote}
\caption{\small Parallel D-branes. The dynamics of D-branes is
described by the elementary excitations of the open strings ending
on them. When the D-branes are separated, the corresponding
Yang-Mills gauge group ${\rm U}(n)$ is broken to ${\rm U}(1)^n$, and
open strings with both ends on the same D-brane give rise to the
unbroken Abelian gauge theory on the D-brane. The open strings
connecting two parallel and separated D-branes produce massive
vector mesons, which correspond to the off-diagonal parts of the
${\rm U}(n)$ gauge fields. 
\label{D} }
\end{quote}
\end{figure}

The origin of the off-diagonal components is also clear, once we know
that D-branes allow open strings to terminate on their
worldvolume. Pictorially, we associate the components $A_{ij}$ and
$\Phi_{ij}^K$ (and their superpartners) with the lowest lying modes of
a supersymmetric open string with ends on the $i$th and $j$th D-branes
(see Fig.\ref{D}).  The mass of such straight stretched strings
should be, as we saw above in the classical approximation, the string
tension times the distance $L_{ij}$ between the two D-branes.  In the
supersymmetric case, this naive classical formula actually gives the
correct energy,
\begin{equation}
  E =\frac{1}{2\pi\alpha'} \, L_{ij}  \, ,
\end{equation}
of the lowest-lying mode (after the GSO projection) of such a
string. The massive particles corresponding to $A_{ij}$ and
$\Phi^I_{ij}$ have masses
\begin{equation}
\sqrt{\sum_I [\Phi^I_{(i)}-\Phi^I_{(j)}]^2}=\frac{1}{2\pi\alpha'}
\sqrt{\sum_I [X^I_{(i)}-X^I_{(j)}]^2}
      =\frac{1}{2\pi\alpha'} L_{ij} \, ,
\end{equation}
thus supporting the claim that they correspond to these lowest-lying
modes.

In the case of two parallel D-branes, corresponding to a ${\rm
U}(2)={\rm U}(1)\times {\rm SU}(2)$ theory, we identify the
traceless part of the $2\times 2$ matrices with the fundamental
representation of SU(2). The normalization is such that the
Yang-Mills coupling here is related to the ($3+1$)-dimensional
electric coupling constant by $e^2=2g_{YM}^2$.  Recall that our
conventions are such that $e$ is the electric charge, in terms of
canonically normalized gauge fields, of the vector meson that
becomes massive when the SU(2) symmetry is broken to U(1). 

The electric charge of the massive vector meson is also consistent
with such a picture, thanks to the coupling of $B_{\mu\nu}$ to
$2\pi\alpha' F_{\mu\nu}$. This coupling in the Dirac-Born-Infeld
action generates an additional source term for ${\cal B}$ such that
\begin{equation}
d*d{\cal B}=\delta_{\rm string} +\delta_{\rm D}\wedge \frac{\partial
{\cal L}_{DBI}}{\partial (2\pi\alpha' F)}
\end{equation}
where ${\cal L}_{DBI}$ is to be understood as a $(p+1)$-form
density. In the absence of magnetic sources for ${\cal B}$, the
left-hand side is an exact eight-form, so the two terms on the
right-hand side must cancel each other when evaluated on any compact
eight-dimensional hypersurface. Whenever $n$ fundamental strings end
on a D$p$-brane, giving a net contribution from the first source
term, the second source term must be there to provide an equal and
opposite contribution.  The latter is precisely the electric charge
on the worldvolume. In other words, the fundamental string flux
which is a gauge charge for ${\cal B}$ is transmuted to an electric
flux on the worldvolume, making the endpoint appear as a point
charge.

Starting from this, the effective interactions between D-branes are
reproduced by integrating out these additional, massive fields.
Because of supersymmetry, only terms with four or more derivatives
survive, with the leading terms reproducing precisely the
four-derivative interaction given previously.  For the simple case
of a pair of D0-branes, the procedure of integrating out the massive
and charged off-diagonal part has been carried out up to two loops,
and has been successfully compared to the prediction from long-range
supergraviton exchange.

The interactions among D-branes are reproduced by a quantum
radiative correction. When the $L_{ij}/2\pi\alpha'$ are finite, the
fields corresponding to the off-diagonal parts of the matrices are
all massive.  The Wilsonian effective action is obtained by
integrating out all these massive fields, thus generating additional
interactions among the diagonal entries. If we lift the above
bosonic action to that of a maximally supersymmetric ${\rm U}(n)$
gauge theory, the leading one-loop terms are (up to a multiplicative
numerical constant)
\begin{equation}
\sim \frac{1}{|{\vec \Phi}_{(i)}- {\vec \Phi}_{(j)}|^{7-p}}
{\times\left[ (F_{(i)}-F_{(j)})^4 \hbox{ or } (\partial
\Phi_{(i)}^K-\partial \Phi_{(j)}^K)^4\right]}  \, .
\end{equation}
This has exactly the right factors of $\alpha'$ and the string
coupling to match with the long-range interaction of
Eq.~(\ref{quarticDerivBraneterms}) that was found by expanding the
Born-Infeld action of one D-brane in the background of the other.
In fact, the coefficient has been found to match
precisely.\footnote{Quite a few computations of this kind have been
performed in recent years.  Some of the more explicit examples can
be found in Refs.~\cite{Becker:1997wh}\cite{Becker:1997xw}, which
considered the case of D0-branes.}

Strictly speaking, these two computations are really justified in
two different regimes. The open string picture is based on the
regime where $\alpha'$ and $L_{ij}$ are taken to zero simultaneously
while keeping $L_{ij}/\alpha'$ finite. The previous (closed string)
picture is valid when we consider larger separations $L_{ij}$ while
keeping the kinetic terms small (in units of $1/\alpha'$) so that
$\alpha' F \ll 1$ and $\alpha' \partial\Phi \ll 1$.  In particular,
this is why an $F^4$ term is absent from the self-energy in the open
string picture while it is present in the Born-Infeld action.  It is
the maximal supersymmetry enjoyed by the D-branes that allows the
naive extrapolation between the two regimes and renders the
comparison here possible.

\section{Yang-Mills solitons on D3-branes}
\label{Dbranesolitons}

Let us concentrate on the case of many D3-branes parallel to each
other, with positions $X^I_{(i)}$ in $R^6$. According to the above
discussion of the low-energy dynamics, the worldvolume dynamics is
then described by a maximally supersymmetric ${\rm U}(n)$ Yang-Mills
theory in a Coulomb phase, with the six adjoint scalars having
diagonal vevs $\langle \Phi^I_{ii}\rangle=X^I_{(i)}/2\pi\alpha'$.
In such a theory there should be magnetic monopoles that appear as
solitons. In this section, we will describe how these solitons are
represented in the D-brane picture, and how their low-energy
dynamics is again described by a lower-dimensional Yang-Mills
theory.

\subsection{Magnetic monopoles as deformations of D3-branes}

We must not forget that the D-brane action also contained
topological terms,
\begin{equation}
\mu_p \; \sum_i \left[\sum_{n=0}^{[(p+1)/2]} C^{(p+1-2n)}\wedge
e^{2\pi\alpha'F_{(i)}} \right]_{ (p+1)-{\rm form}}  \, ,
\end{equation}
that must be similarly elevated to a non-Abelian form.  The leading
term, involving $C^{(p+1)}$, was already incorporated into the above
Yang-Mills form of the action; it was used to cancel the static
force coming from the NS-NS sector via the Born-Infeld term. Once we
have carried out the derivative expansion, the remaining terms from
the topological part of the action can be similarly expanded and
elevated into a Yang-Mills form as
\begin{equation}
   \mu_p \sum_i \left[
 C^{(p-1)}\wedge ({2\pi\alpha'F_{(i)}})+  C^{(p-3)}\wedge \frac12
({2\pi\alpha'F_{(i)}} )^2+\cdots\right] \label{yangmillsTopological}
\end{equation}
where the $i$th term is to be evaluated on, and integrated over, the
$i$th worldvolume.

With this in mind, let us consider an ${\cal N}=4$ ${\rm U}(2)$
theory spontaneously broken to ${\rm U}(1)\times {\rm U}(1)$ by an
adjoint Higgs, rescaled as in Eq.~(\ref{elevatedaction}),
with
\begin{equation}
   \langle\Phi\rangle
   =\frac12\left(\begin{array}{cc} -v &0 \\
    0 & v\end{array}\right)  \, .
\end{equation}
As we discussed above, the corresponding D-brane picture is a pair
of parallel D3-branes, separated from each other by a distance
$2\pi\alpha' v$. Without loss of generality, we may choose the
separation to be along the $x_9$-direction, which means that we
should identify $\Phi_9$ as the adjoint scalar with the above
expectation value.

The BPS monopole of this theory has a very specific profile, which
in the unitary, or string, gauge take the form
\begin{equation}
\Phi= - \frac{\sigma^3}{2} \, \left[v\coth(vr)-\frac{1}{r}\right] \, .
\end{equation}
In terms of the Abelian fields associated with each of the
D3-branes, we have
\begin{equation}
\Phi_{(1)}=-\Phi_{(2)}= - \frac{1}{2} \, \left[v\coth(vr)
-\frac{1}{r}\right]  \, .
\end{equation}
Note that these scalar fields vanish at the origin. On the other
hand, we gave an interpretation of these scalar fields as positions
of the individual D-branes. Visualizing the shape of the two
D3-branes, then, we conclude that the two D3-branes bend themselves
and touch each other along the middle hyperplane, $x_9=0$, precisely
at the center of the monopole core.

In this gauge, the diagonal part of the gauge field satisfies an
Abelian Bianchi identity and must have the profile of a Dirac
monopole,
\begin{equation}
A^3=(\cos\theta -1) d\phi  \, ,
\end{equation}
in the usual $R^3$ spherical coordinates. The magnetic flux
associated with this long-range Abelian gauge field consists of two
diagonal fields,
\begin{equation}
F_{(1)}=-F_{(2)}=-\frac12\sin\theta \, d\theta \, d\phi = -
\frac{{\bf r}\cdot d{\bf r}}{2r^3}  \, ,
\end{equation}
that represent $2\pi$ flux flowing from the first brane and flowing
into the second brane. The apparent singularity at the origin is
smoothed out by the non-Abelian nature of the true gauge field,
whose off-diagonal part,
\begin{equation}
A^1+iA^2=\frac{ivr}{\sqrt2 \sinh (vr)} \left(d\theta +i\sin\theta
d\phi\right)  \, ,
\end{equation}
becomes important near the origin but has an exponentially
suppressed asymptotic behavior.

\begin{figure}[t]
\begin{center}
\scalebox{1}[1]{\includegraphics{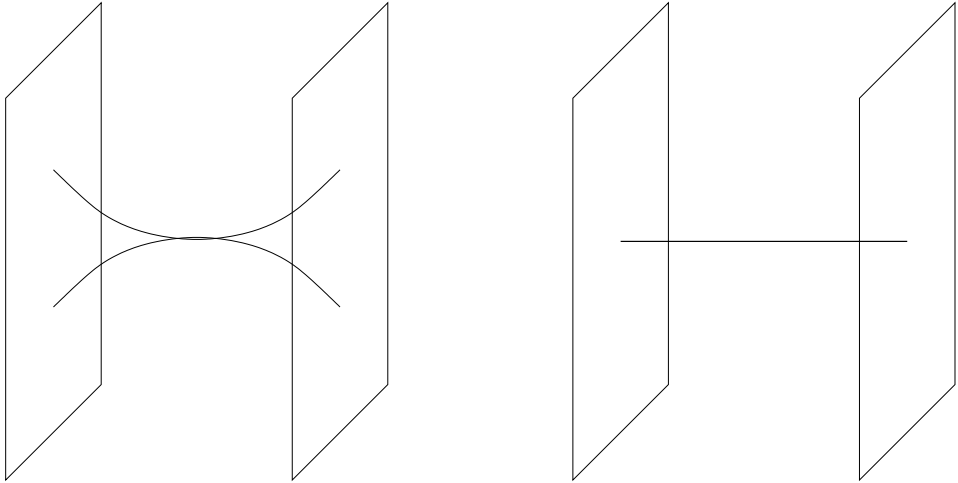}}
\par
\vskip-2.0cm{}
\end{center}
\begin{quote}
\caption{\small Two alternatative pictures of an SU(2) magnetic monopole
(charged vector meson) in terms of D-branes. The figure on the left
shows D3-branes deforming themselves to meet at a point.
Supersymmetry requires that a net magnetic (or electric) charge flow
from one D3-brane to the other, resulting in a BPS magnetic monopole
(a charged vector meson). In terms of conserved charges, this is
equivalent to a segment of D1-string (fundamental string) suspended
between the D3-branes, as shown on the right. 
\label{D3D1}}
\end{quote}
\end{figure}

For an even clearer picture, let us go to the limit where the vev $v$
is very large. The core of the monopole, where the deformation of the
D3-brane worldvolume is most pronounced, is small --- of order $1/v$
--- while the protruding part of the worldvolume becomes elongated
along the $x_9$ direction and is roughly of length $2\pi\alpha' v$.
This looks like a long thin tube, with a pinched middle point,
connecting the two D3-branes, as illustrated in Fig.~\ref{D3D1}.  The
pinching of the tube is related to the fact that we can view the
asymptotic regions as two D3-branes, rather than as a D3- and an
anti-D3-brane.  (Without the pinching, the two parallel objects would
necessarily have opposite orientations.) Thus, we may view the
magnetic monopole as a localized and tubular deformation that connects
two parallel D3-branes.

\subsection{Magnetic monopoles as D1-brane segments}
\label{d1branesegents}

When this configuration is viewed in terms of closed string fields,
Eq.~(\ref{yangmillsTopological}) gives the topological coupling
\begin{equation}
\mu_3
 C^{(2)}\wedge {2\pi\alpha'F} \, ,
\end{equation}
which induces a D1-brane charge, coupled minimally to $C^{(2)}$, on
the tube. Since the flux is quantized in units of $2\pi$, the
D1-brane charge per unit length along the tube is
\begin{equation}
\mu_3 \pi\alpha' \oint_{S^2} F=\mu_3 (2\pi\sqrt{\alpha'})^2=\mu_1 \,
,
\end{equation}
which is exactly the charge per unit length of a D1-brane.  Thus the
tube, if we ignore its girth, looks exactly like a segment of a
D1-brane (or D-string) stretched between two D3-branes.  The length
of the segment is $2\pi\alpha'v$, the same as the distance between
the two D3-branes.

A less precise way of seeing this is to start with the the picture
of two D3-branes connected by a D-string segment.  Because of the
same topological coupling, but seen from the opposite viewpoint, the
gauge fields on the D3-branes see the end points of the D-string
segment as sources of the $\mp 2\pi$ magnetic flux.  By itself, this
does not show the precise structure of the monopole solution, but it
suffices as far as the conserved charge goes
\cite{Strominger:1995ac}.

This crude picture should be no stranger than our earlier assertion
that massive vector mesons are stretched fundamental string segments
between a pair of D3-branes \cite{Callan:1997kz}.  The only
difference here is that in the weak coupling limit the monopole is a
large solitonic object amenable to semiclassical treatment, while
the vector meson is small and must be treated quantum mechanically.
If we go to the opposite extreme of very large Yang-Mills coupling,
monopoles will appear very small while vector mesons are very large,
so there is no fundamental distinction between a fundamental string
segment and a D-string segment.  In ordinary field theories, the
interpolation between the weakly coupled and the strong coupled
regime is dangerous, but for the case at hand, where we are
considering 1/2-BPS objects, the large number of supersymmetries
protects these pictures.

\subsection{1/4-BPS dyons and string webs}

In the context of this symmetric view of monopoles and vector
mesons, the construction of some dyonic states follows naturally.
The trick is to realize that, in addition to the fundamental strings
and D-strings, there are other varieties of (1+1)-dimensional
string-like objects, known as $(q,p)$ strings.  These are tightly
bound states of $q$ fundamental strings and $p$ D1-branes, with $q$
and $p$ required to be coprime integers.  {}From the D1-brane
viewpoint, a $(q,1)$ string is nothing but a D1-brane carrying $q$
units of quantized electric flux. When $q$ and $p$ are coprime
integers, $p$ D1-branes cannot share $q$ quantized electric fluxes
equally among themselves, and must therefore be at the same location
in order to be able to carry such a charge and yet remain
supersymmetric.

Type IIB superstring theory possesses an ${\rm SL}(2,Z)$ duality,
similar to that of ${\cal N}=4$ SU(2) SYM theory, except that
it acts on these string-like objects instead of on the charged
particles. The appearance of these additional strings is again a
consequence of the ${\rm SL}(2,Z)$.  Having a segment of $(q,p)$
string ending on a pair of D3-branes generates $q$ units of vector
meson charge and $p$ units of monopole charge, leading to a simple
$(q,p)$ dyon of the SU(2) SYM theory. Thus, the ${\rm
SL}(2,Z)$ of ${\cal N}=4$ SYM theories is a direct
consequence of the ${\rm SL}(2,Z)$ of type IIB superstring theory.
Perhaps a more accurate way of phrasing this is to say that the
existence of $(q,p)$ dyons in the SYM theory is important
evidence for the ${\rm SL}(2,Z)$ duality of the type IIB theory.

At the same time, it is clear that most of the dyons we have found
cannot be realized in this simple manner. As we have seen, in a
theory with gauge group of rank $\ge 2$, the electric and magnetic
charges of a generic dyon do not correspond to parallel vectors in
the Cartan subalgebra. From the D-brane viewpoint, such a dyon
cannot be made from a single $(q,p)$ string segment connecting a
pair of D3-branes. Instead the desired configuration must involve
strings with ends on more than two D3-branes, which is possible for
rank 2 and higher gauge groups. The simplest case would involve
three types of strings, each with one end on a different D3-brane
and the other at the junction of the three strings.

\begin{figure}[t]
\begin{center}
\scalebox{1.0}[1.0]{\includegraphics{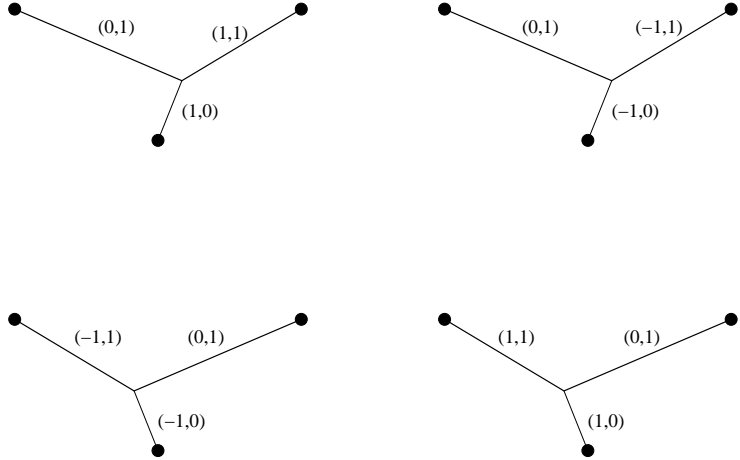}}
\par
\vskip-2.0cm{}
\end{center}
\begin{quote}
\caption{\small The four simplest types of string junction
corresponding to 1/4-BPS dyons. The circles represent the D3-branes
on which the strings end, while the strings are labelled with their
charges. Each of these four types preserves a different 1/4 of the
${\cal N}=4$ supersymmetry.\label{prong} }
\end{quote}
\end{figure}

For instance, a $(1,0)$ string and a $(0,1)$ string can join to
become a $(1,1)$ string \cite{Dasgupta:1997pu}\cite{Sen:1997xi},
with the ends of this ``three-pronged" configuration each on a
different D3-brane \cite{Bergman:1997yw}.  In the SU(3) theory, this
corresponds (in a suitably chosen basis) to a dyon with magnetic
charge corresponding to $\bbeta_1+\bbeta_2$ and electric charge
corresponding to $\pm\bbeta_1$.  Consideration of the energetics
alone shows that the location of the junction point is determined
solely by the positions of the D3-branes.  Each of the three strings
has a definite tension, regardless of its length, so the positions
of the D3-branes define three attractive force vectors acting on the
junction. The balance of forces determines where the junction will
be, as shown in Fig.~\ref{prong}.  For this three-pronged string
configuration, the balance of forces is enough to guarantee its BPS
nature.  Just as in the field theory computation, these dyons would
preserve 1/4 of the ${\cal N}=4$ supersymmetry.\footnote{An
interesting realization of this configuration in a
gravitational setting  is given in Ref.~\cite{Gauntlett:1999xz},
where the D3-branes at the ends of two of the three prongs are replaced by
a gravitational background.  See also Ref.~\cite{deMelloKoch:1999ui}.}

More generally, we can consider a web of $(q,p)$ strings with many
junctions and many external ends ending on D3-branes
\cite{Bergman:1998gs}.  With more than three external lines,
however, the balance of forces is not enough to guarantee the
1/4-BPS property. We saw from the field theory BPS equations that at
most two adjoint Higgs fields can be involved in the formation of
1/4-BPS dyons. Since the adjoint Higgs field encodes the
configuration of the D3-branes and the strings, this translates to
the condition that the string web be planar.  Furthermore, the field
theory BPS equation has only two overall sign choices, one for the
primary BPS equation and another for the secondary BPS equation.
This translates to the condition that the orientation of string
segments be consistent with each other.  Thus, for example, two
$(1,0)$ string segments in different parts of the web should be
directed the same way.

\begin{figure}[t]
\begin{center}
\scalebox{1.0}[1.0]{\includegraphics{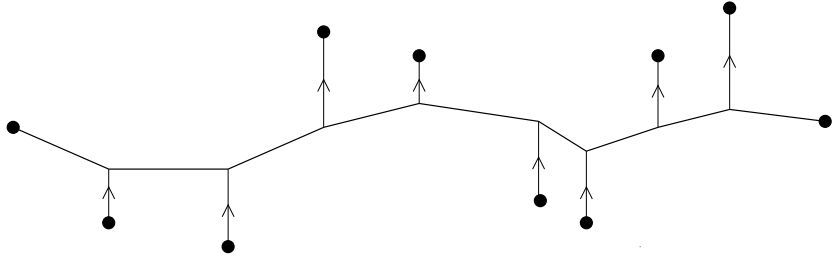}}
\par
\vskip-2.0cm{}
\end{center}
\begin{quote}
\caption{\small A string web corresponding to a 1/4-BPS dyon made
from a sequence of distinct fundamental monopoles. The horizontal
line has a D-string charge while the vertical lines are fundamental
strings. Note that the fundamental strings are all flowing along the
same direction.   \label{bamboo}}
\end{quote}
\end{figure}

Figure~\ref{bamboo} illustrates the string web corresponding to the
dyonic bound states made from of a sequence of distinct monopoles,
as in the previous chapter.  It has one D1-brane connecting two
D3-branes and passing by many nearby D3-branes.  Fundamental strings
shoot out from the latter set of D3-branes to meet the D1-brane. The
fundamental string charges can then be immersed into the worldvolume
of the D1-brane as electric fields.  This pattern is uniquely
determined by the magnetic and electric charges on the D1-string and
by how many fundamental strings come out of each of the D3-branes.
Apart from the balancing of forces at each and every junction, the
BPS condition requires that all fundamental string segments are
directed in the same way: all up or all down.  While the string web
picture is not particularly useful for the counting of states, it
proves to be a handy way of cataloging whether a given dyonic state
exists. In terms of the restrictions found in the index computations
for 1/4-BPS dyons in ${\cal N}=4$ SYM theory, the following correspondence
can be established:

\begin{itemize}

\item
$\pm \tilde a_Aq_A >0$ $\Rightarrow$ unidirectional property of
$(q,p)$ strings in any given web.  The same type of string cannot
appear twice with opposite orientations.

\item
$|q_A| < |\tilde a_A|$ $\Rightarrow$ existence of three-point
junctions. Too much electric charge (or too many fundamental strings)
will pull the junction to the side and destroy it. The resulting
string web configuration is not supersymmetric, and the
corresponding field theory configuration involves two or more
charged particles that repel each other.

\end{itemize}

\section{T-Duality and monopoles as instanton partons }
\label{instantonpartons}

Before proceeding to the Nahm data, let us consider a variation on
the above D-brane/Yang-Mills soliton picture.  Instead of
considering supersymmetric configurations of D3-branes, we will
consider D4-branes. Just as an open D1-brane acts like a monopole in
a D3-brane, a D0-brane can be embedded into a D4-brane and act like
an instanton soliton.  Not only is this phenomenon of interest on
its own but, after T-dualizing the configurations, we will find
important implications for monopole physics.  In this section, we
start with the D0/instanton correspondence, then explain how
T-dualization acts on the classical field theory degrees of freedom,
and finally arrive at the conclusion that monopoles can be
considered as partons of an instanton soliton when the latter is
defined on $R^3\times S^1$.  This will naturally lead us to the ADHM
and ADHMN constructions in the next section.

\subsection{An instanton soliton as an embedded D0-brane}

The line of thought of Sec.~\ref{d1branesegents} can be extended
immediately to the next topological coupling,
\begin{equation}
\mu_p\, C^{(p-3)}\wedge \frac12 \tr ({2\pi\alpha'F} )^2  \, .
\end{equation}
For instance, we can consider a stack of $n$ coincident D4-branes.
The worldvolume theory is a maximally supersymmetric ${\rm U}(n)$
Yang-Mills theory, and this coupling implies that a classical
configuration with
\begin{equation}
\int_{R^4}\tr F\wedge F \neq 0
\end{equation}
generates a D0-brane charge, as seen by spacetime
\cite{Strominger:1995ac}. We are already familiar with such
configurations as the instantons of four-dimensional Euclidean
Yang-Mills theory. In the ($4+1$)-dimensional setting, these
instanton solutions exist as solitons, again solving the familiar
self-dual equation,
\begin{equation}
F=\pm *F    \, .
\end{equation}
The quantization of the instanton charge is such that
\begin{equation}
\frac{1}{8\pi^2}\int_{R^4}\tr F\wedge F =k
\end{equation}
for integral $k$, and the instanton \cite{Belavin:1975fg} soliton
with this charge generates a D0 charge
\begin{equation}
\mu_4 \times \frac12 (2\pi\alpha' )^2\times 8\pi^2 k =\mu_4 \times
\left(2\pi\sqrt{\alpha'}\right)^4\times k =\mu_0\times k
\end{equation}
that represents precisely $k$ units of D0 charge. One major
difference from the mono\-pole case is that this solution does not
need the scalar fields. Since the latter dictate the actual shape of
the D4-branes, it means that D0-branes do not induce any deformation
of the D4-brane worldvolume.  All that happens is that, when the
D0-branes are absorbed by the D4-brane worldvolume, their point-like
charges are converted into self-dual Yang-Mills flux of arbitrary
width.

The vacuum condition on D0-branes in the presence of D4-branes leads
to the familiar ADHM construction of instantons.  As the first step
towards this, we consider $k$ D0-branes embedded inside $n$
D4-branes. From the worldvolume perspective, the configuration is
$k$ instanton solitons of a $(4+1)$-dimensional, maximally
supersymmetric ${\rm U}(n)$ Yang-Mills theory.  As with magnetic
monopoles, the dynamics of such solitons can be described by a
moduli space approximation. Instead of doing so, however, we will
stick to the D-brane interpretation of the instanton solitons and
ask what type of Yang-Mills theory lives on their worldlines.

\subsection{T-duality maps on Yang-Mills theories}\label{T}

Let us consider the Yang-Mills field theory associated with an
infinite number of parallel D$p$-branes separated at equal distances
along the $x^9$ direction.  Furthermore, let us constrain their
motions in such a way that the motion of a single D$p$-brane is
exactly mimicked by all the other D$p$-branes. In other words, we
require the fields labelled by the gauge index pair $(i,j)$ to
behave exactly like those with $(i+k,j+k)$, for any integer $k$.  To
ensure this, it is sufficient to require that
\begin{equation}
(A_\mu)_{i+1,j+1}=(A_\mu)_{ij}
\end{equation}
for all integer pairs $(i,j)$. (As a matter of convenience, we have
partially fixed the gauge so that the constraint can be written in a
particularly simple form. Resuscitating the full gauge symmetry at
the end of the day is straightforward.)  The one exception to this
rule is for $\Phi^{p+1}$, which encodes the positions of the
D$p$-branes along $x^{p+1}$. For this latter, the restriction we
should require is that
\begin{equation}
\Phi^{p+1}_{i+1,j+1}=\Phi^{p+1}_{ij} +\frac{2\pi R}{2\pi\alpha'}
\,\delta_{ij}
\end{equation}
where $2\pi R$ is the distance between successive pairs of
D$p$-branes. The other adjoint Higgs fields, $\Phi^{\tilde K}$ with
$\tilde K=p+2,\dots,9$, obey the same constraint as the gauge field,
\begin{equation}
\Phi^{\tilde K}_{i+1,j+1}=\Phi^{\tilde K}_{ij}  \, .
\end{equation}

This set of constraints is naturally imposed if we view the system
in a slightly different way, that is, by dividing it by a $2\pi R$
shift of along $x^{p+1}$ \cite{Taylor:1996ik}. {}From this
viewpoint, we consider all the D$p$-branes as mirror images of each
other and effectively study a single D$p$-brane sitting at a point
on a circle of radius $R$. What is the mass spectrum of the
elementary particles of this theory? Since we are effectively in a
Coulomb phase of a ${\rm U}(\infty)$ theory broken to ${\rm
U}(1)^\infty$, we expect to find an infinite number of massive
vector mesons. In fact, from the form of the $\Phi_{p+1}$ that is
responsible for the symmetry breaking, we can see that the
off-diagonal fields, such as $\vec A_{i,i+n}$, have masses given by
\begin{equation}
   m_n^2 =  \left(\frac{nR}{\alpha'}\right)^2
\end{equation}
for every integer $n$. In fact, there is exactly one maximal vector
multiplet for each $n$.

We are also familiar with another situation where one gets an
infinite tower of massive fields with such an integer-spaced mass
formula.  This happens when a field theory is compactified on a
circle, say of radius $\tilde R$,
 and then described in terms of a field theory in one fewer dimension.
The squared masses of the so-called Kaluza-Klein tower are then
\begin{equation}
  {\tilde m}_n^2 =  \left(\frac{n}{\tilde R}\right)^2
\end{equation}
for all integer $n$. For now, we note that the two mass formulas
coincide if
\begin{equation}
R\tilde R=\alpha'  \, .
\end{equation}
What we wish to show in the rest of this section is that the above
worldvolume theory of a single D$p$-brane sitting on a circle of
radius $R$ is equivalent to a worldvolume theory of a D$(p+1)$-brane
whose $(p+1)$th direction is wrapping a circle of radius $\tilde
R=\alpha'/R$.  The same kind of statements hold for multiple
D$p$-branes and multiple D$(p+1)$-branes; establishing these
requires no more than adding additional internal indices in what
follows.

To actually prove the above statement, it is convenient to introduce
a new parameter $\sigma$, with period $2\pi \tilde R$, and organize
the matrices $\vec A_{ij}$, $\Phi^{p+1}_{ij}$, and $\Phi^{\tilde
K}_{ij}$ into bilocal quantities
\begin{eqnarray}
A_\nu (y^\mu;\sigma,\sigma')&\equiv&\frac{1}{2\pi\tilde R}
 \sum_{mk} (A_\nu)_{mk} (y^\mu)\,
e^{-im\sigma/\tilde R}e^{ik\sigma'/\tilde R} \nn \Phi^{p+1}
(y^\mu;\sigma,\sigma')&\equiv& \frac{1}{2\pi\tilde R} \sum_{mk}
\Phi^{p+1}_{mk} (y^\mu)\, e^{-im\sigma/\tilde R}e^{ik\sigma'/\tilde
R} \nn \Phi^{\tilde K} (y^\mu;\sigma,\sigma')&\equiv&
\frac{1}{2\pi\tilde R} \sum_{mk} \Phi^{\tilde K}_{mk} (y^\mu)\,
e^{-im\sigma/\tilde R}e^{ik\sigma'/\tilde R}  \, .
\end{eqnarray}
The choice of the Fourier basis is, of course, dictated by the
periodic nature of the allowed configurations.

Imposing the periodicity constraint effectively reduces the number
of degrees of freedom in such a way that we can replace the matrices
by column vectors or, equivalently, reduce these general bilocal
expressions to local ones. It is a matter of straightforward
computation to see that the three types of fields can be written in
the form \cite{Michishita:2001ph}
\begin{eqnarray}
A_\nu (y^\mu;\sigma,\sigma') &=& A_\nu(y^\mu;
\sigma)\,\delta(\sigma-\sigma') \cr\cr \Phi^{p+1}
(y^\mu;\sigma,\sigma') &=& \left[A_{p+1}(y^\mu;
\sigma)+i\frac{\partial}{\partial\sigma}\right]
\delta(\sigma-\sigma')  \cr \cr \Phi^{\tilde K}
(y^\mu;\sigma,\sigma') &=& \Phi^{\tilde K}(y^\mu;
\sigma)\,\delta(\sigma-\sigma')
\end{eqnarray}
where all quantities on the right-hand side are local fields in
terms of $y^\mu$ and $\sigma$.

The derivative operator in $\Phi^{p+1}$ can be understood as
follows. The original matrix quantities have a natural operation
among themselves, namely matrix multiplication. When we replace the
matrices by bilocal quantities, this matrix multiplication carries
over to an integration: if $Z_{kn}=\sum_m X_{km}Y_{mn}$, then their
bilocal versions obey
\begin{equation}
Z(\sigma,\sigma')= \int d\sigma''
X(\sigma,\sigma'')Y(\sigma'',\sigma')  \, .
\end{equation}
Thus, each bilocal quantity is an operator acting on the right, and
the derivative with respect to $\sigma$ should be understood as
such.

The actual SYM theory on a D$p$-brane has three types of
purely bosonic terms in the action,
\begin{eqnarray}
  S_{\rm bos} = \frac{1}{g_{YM}^2}
\int d^{p+1}y\;\left(-\frac{1}{4} \tr F_{\mu\nu}F^{\mu\nu} -
\frac{1}{2}\tr D_\mu \Phi^K D^\mu \Phi^K +\frac{1}{2}\tr \sum_{K<M}
[\Phi^K,\Phi^M]^2\right) \, . \cr
\end{eqnarray}
[Here $g_{YM}$ is again given by Eq.~(\ref{gymdefinition}).] If we
follow the procedure described above, this becomes
\begin{eqnarray}
  S_{\rm bos} &=& \frac{1}{g_{YM}^2} \int d\sigma \, d\sigma' \,
     \delta(\sigma'-\sigma) \int d^{p+1}y \;\delta(\sigma'-\sigma)
\left( -\frac{1}{4}  F_{\mu\nu}F^{\mu\nu} 
\qquad \qquad \qquad
\right.  \cr
 & &
- \frac{1}{2} D_\mu \Phi^{\tilde K} D^\mu \Phi^{\tilde K} -\frac12
[D_\mu, A_{p+1}+i\partial_\sigma] [D^\mu, A_{p+1}+i\partial_\sigma]
\cr
 & & \left.
+\frac{1}{2} \sum_{\tilde K< \tilde M} [\Phi^{\tilde K},\Phi^{\tilde
M}]^2 +\frac{1}{2} \sum_{\tilde K} [A_{p+1}+i\partial_\sigma
,\Phi^{\tilde K}]
 [A_{p+1}+i\partial_\sigma  ,\Phi^{\tilde K}] \right)  \, ,
\end{eqnarray}
where the integrations over $\sigma$ and $\sigma'$ and one of the
delta functions come from taking the trace;

After the integration over $\sigma'$, the two delta functions reduce
to $\delta(0)$.  This infinite factor is an artifact of counting
mirror D-branes as if they were real, and should be replaced by the
inverse volume factor, $1/2\pi\tilde R$. The correct normalization
of this factor can be found by tracing back to the discrete index
notation and dropping precisely one summation. If we now define a
$(p+2)$-dimensional coordinate system $y^{\tilde \mu}= (y^\mu,
\sigma)$, with $y^{p+1}=\sigma$, and regard $A_{p+1}$ as a component
of the gauge field along the new direction, we obtain
\begin{eqnarray}
    S_{\rm bos} =
\frac{1}{\tilde g_{YM}^2}\int d^{p+2}y\;\tr \left(-\frac{1}{4}
F_{\tilde\mu\tilde\nu}F^{\tilde\mu\tilde\nu}
-\frac{1}{2}D_{\tilde\mu} \Phi^{\tilde K} D^{\tilde \mu}
\Phi^{\tilde K} +\frac{1}{2} \sum_{\tilde K < \tilde M}
[\Phi^{\tilde K},\Phi^{\tilde M}]^2\right)  \, .  \cr
\end{eqnarray}
This is precisely the bosonic piece of the ($p+2$)-dimensional
maximally supersymmetric Yang-Mills theory on a circle of radius
$\tilde R$. The modified Yang-Mills coupling
\begin{equation}
\tilde g_{YM}^2 =g_{YM}^2  2\pi\tilde R= \left(2\pi e^{\phi_0}
\frac{\tilde R}{\sqrt{\alpha'}}\right)
 \left(2\pi \sqrt{\alpha'}\right)^{p-2}
\end{equation}
can be interpreted as the correct Yang-Mills coupling on
D$(p+1)$-branes in a string theory with a different string coupling
constant,
\begin{equation}
e^{\phi_0}\rightarrow e^{\tilde \phi_0}= e^{\phi_0} \frac{\tilde
R}{\sqrt{\alpha'}}= e^{\phi_0} \frac{\sqrt{\alpha'}}{R} \, .
\end{equation}
What we discover from this is that there is no real distinction
between a D$p$-brane sitting on a circle of radius $R$ and a
D$(p+1)$-brane wrapping a dual circle of radius $\tilde
R=\alpha'/R$, provided that we tweak the Yang-Mills couplings of the
two sides appropriately \cite{Taylor:1996ik}.

While this is shown here at the level of low-energy dynamics, it has
of course a deeper origin in string theory, which goes by the name
of T-duality. This T-duality transformation also flips a GSO
projection, and actually maps between type IIA theory (which has
only even $p$ D-branes) and type IIB theory (which has only odd $p$
D-branes). This well-known fact will not be of relevance for our
purposes; we refer interested readers to the standard string theory
textbooks.

In the SYM theory, the background geometry is not part of the
configuration space, so we cannot quite consider this operation as a
discrete local symmetry. This is why it is sometimes stated that
there can be no T-duality in local field theories. The T-duality
here can be understood as a sort of
tautology, since as far as the purely field theoretical picture
goes, there is only one sensible description of the setup as a
$(p+2)$-dimensional theory compactified on a circle. Nevertheless,
its ``T-dual picture'', with an infinite number of $(p+1)$
dimensional fields, becomes quite useful once we visualize it in
terms of D$p$-branes.  In the next section, we will see the most
famous example of this, and see how these two different geometrical
pictures allow a simple understanding of the mysterious relationship
between monopoles and instantons.

\subsection{Monopoles are partons of periodic instantons}

A slight modification of the D0-D4 system occurs when we consider
D4-branes compactified on a circle. A D0-brane on $n$ D4-branes is
then a ${\rm U}(n)$ instanton on $R^3\times S^1$.  Let us further
imagine turning on some Wilson lines along $S^1$, thus effectively
breaking the ${\rm U}(n)$ gauge symmetry to a smaller group,
generically to the maximal torus ${\rm U}(1)^n$. Having a Wilson
line can be viewed as having one component of the gauge field,
$A_4$, acquire an expectation value. In the T-dual picture, however,
$A_4$ came from the adjoint scalar field associated with the
D3-branes. The latter encodes where these D3-branes are sitting
along the dual circle $\tilde S^1$. So we can also visualize the
Coulomb phase due to a Wilson lines as a separation of D-branes
along some spatial direction. The only difference is that this
direction is now compact.

Let us ask what happens to the D0-brane upon such a T-duality
mapping. By the nature of T-duality, it has to be converted to a
D1-brane winding along the dual circle, $\tilde S^1$. Along the
$\tilde S^1$, on the other hand, are $n$ D3-branes distributed
according to the value of the Wilson line; these meet the D1-branes
at right angles. In particular, the D1-brane can be split up into
$n$ segments, each connecting adjacent pairs of D3-branes. Note that
each of these D1-brane segments behaves exactly like a BPS monopole
with respect to the relevant pair of D3-branes. We find that, in the
presence of a Wilson line, $n$ such mutually distinct monopoles
compose a single instanton on $R^3\times S^1$
\cite{Lee:1997vp,Lee:1998vu,Lee:1998bb,Kraan:1998kp,Kraan:1998pm,Kraan:1998sn,Garland:bv}.

In fact, this is the simplest way to understand why an instanton of
${\rm U}(n)$ theory has exactly $4n$ moduli parameters; $4n$ equals
$4$ times $n$, and the moduli are really coming from the fact that
the instanton is composed of $n$ monopoles, each of which always
carries four moduli.\footnote{This remarkable fact allowed us a
better understanding of gaugino condensate in ${\cal N}=1$ super
QCD. By compactifying the Euclidean time, the basic instantonic
objects in $R^3\times S^1$ are these monopole solutions which have
exactly the right zero mode structure to contribute to a
superpotential and  a fermion bilinear consisting of gaugino. See
Ref.~\cite{Davies:1999uw,Davies:2000nw,Kim:2004xx} for more detail.}
This also leads to a new type of solution, in which we start with an
instanton on $R^3\times S^1$ and send one or more of the monopoles
away to infinity. We end up with a solitonic configuration carrying
both quantized magnetic charge and fractional instanton charge.

Even more drastically, we could collapse the $S^1$ and thus expand
the $\tilde S^1$ until we have $R^3\times \tilde R^1$ on the dual
side, where we could maintain some number $\le n$ of D3-branes
sitting at finite positions along $\tilde R^1$.  Taking various
limits of sending some of the D1-brane segments to infinity, we end
up with perfectly ordinary BPS monopoles on $R^3$. This gives a
natural map relating the worldvolume theory of a D0-brane on
D4-branes to that of open D1-branes ending on D3-branes, which is
essentially how one obtains the Nahm data from the ADHM
construction. We will come back to this relationship in the next
section, after we have first examined in some detail the relation
between the ADHM construction and the D0-D4 system.

\section{ADHM and ADHMN constructions}
\label{DbraneNahm}

We have seen that a D0-brane absorbed by a stack of D4-branes acts
like an instanton soliton, and have mentioned that the conventional
ADHM data is nothing but the specification of a supersymmetric
configuration of a D0-brane under the influence of D4-branes
\cite{Witten:1995gx,Douglas:1995bn,Douglas:1996uz,Hori:1999me}.
Here we will make this more precise and describe the ADHM data from
this viewpoint. Upon T-dualizing this picture we find a D3-D1
complex in which the D1-brane vacuum configurations become the Nahm
data, leading us to the desired connection between Nahm data and BPS
monopoles.\footnote{A somewhat different approach, in which the D3-D1
configuration is obtained
from the decay of unstable D4-branes, has been
proposed recently in Refs.~\cite{Hashimoto:2005yy,Hashimoto:2005qh}.}

\subsection{ADHM from D0-D4}

In the absence of D4-branes, the low-energy theory on a D0-brane
must be the unique SYM quantum mechanics with
gauge group ${\rm U}(k)$ and 16 supercharges. It has one gauge field
(with only a time component), nine adjoint scalars, and eight
complex fermions, also in the adjoint representation.  The action is
the dimensional reduction of ten-dimensional SYM theory
down to $(0+1)$ dimensions. When D4-branes are present, half of the
supersymmetry is broken by the D4-branes, so the vector multiplet is
smaller than otherwise. The time-like gauge field and the five
adjoint scalars transverse to the D4-branes combine with half of the
fermions into a vector multiplet. Let us denote these five scalars
by $X_i$.

The other parts of the maximal supermultiplet survive, and organize
themselves into an adjoint hypermultiplet that contains four real
adjoint scalars and the other half of the fermions. The
hypermultiplets are what we are interested in, and we will write
their scalars in terms of two complex adjoint scalar fields $\bar
H_1$ and $\bar H_2$. Another modification is that we can now have
open strings connecting the D0-brane and the D4-brane. This induces
hypermultiplets in the fundamental representation of ${\rm U}(k)$.
We will denote the two complex scalar fields of these
hypermultiplets as $Q_\alpha^f$, with $\alpha=1$ or 2 while $f= 1,2,
\dots, n$ labels the flavors.

Since we are dealing with a simple mechanical system, the vacuum
condition demands that all fields take constant values in such a way
that the potential vanishes identically. There are two types of
potential terms. The commutator term,
\begin{equation}
\int dt \; \,{\rm tr}\left(-\frac12\sum_{i<k}[X_i,X_k]^2 +
\sum_{i,\alpha} \left| [X_i, \bar H_\alpha]\right|^2\right) \, ,
\end{equation}
can be set to zero by insisting that $X_k=0$, which amounts to
saying that the D0-branes are stuck to the D4-branes.  The other
part of the potential arises from the so-called D-terms,
\begin{equation}
\frac12\int dt \sum_{b=1}^3 \tr \ {\cal D}_a^2
   = \frac12\int dt \sum_{b=1}^3 \tr \left[
    \sum_{\alpha,\beta}(\tau_a)^{\alpha\beta}
\left([\bar H_\alpha,\bar H^{\dagger}_\beta]+ \sum_f
Q^f_\alpha\otimes Q^{f\dagger }_{\beta}\right) \right]^2 \, ,
\label{DtermforADHM}
\end{equation}
where the $\tau_a$ are the Pauli matrices and the trace and the
tensor product here refer to the (implicit) U($k$) gauge group
indices.

The point of this is that the supersymmetric vacuum condition ${\cal
D}_a=0$ is exactly the same as the ADHM equation
\cite{Christ:1978jy} if we identify the $(n+k)\times k $
quaternionic ADHM matrix with
\begin{equation}
\left(\begin{array}{cc} Q_1+Q_2j \\
\\ \bar H_1+\bar H_2j\end{array}\right)
\end{equation}
where $j$ is one of the three quaternionic imaginary units.
Starting from this picture, one can derive the ADHM prescription for
obtaining the instanton configuration on $R^4$.  Since this goes
well beyond the scope of this review, we will simply refer
interested readers to Ref.~\cite{Hori:1999me}.

\subsection{Nahm data from D1-D3}

Suppose we compactify one spatial direction on the D4-brane. From
the D0-brane viewpoint this means that one real scalar field from
the adjoint hypermultiplet has to become a covariant derivative
along the dual circle; the argument for this exactly parallels that
in Sec.~\ref{T}.  We rewrite the adjoint scalars as four real
fields,
\begin{equation}
\bar H_1 = \frac{1}{\sqrt 2}\left(Y_3+iY_0\right) \, , \qquad \bar
H_2=\frac{1}{\sqrt 2}\left(Y_1-iY_2 \right) \, .
\end{equation}
If we take the $Y_0$-direction to be the one that is compactified,
then $Y_0$ turns into a covariant derivative along the dual circle
parameterized by $s$, while the remaining three $Y_i$ become
functions of $s$; i.e.,
\begin{eqnarray}
   Y_0 &\rightarrow& i\delta(s-s')D_0\equiv i\delta(s-s')\left[iT_0(s)
   +\frac{\partial}{\partial s}\right]   \cr\cr
    Y_i &\rightarrow& \delta(s-s')T_i(s) \, .
\end{eqnarray}
The two gauge indices in the matrices turn into two continuous
variables, $s$ and $s'$, living on the dual circle. We thus find
\begin{equation}
\sum_{\alpha,\beta}(\tau_i)^{\alpha\beta} [\bar  H_\alpha,\bar
H^{\dagger}_\beta]\rightarrow -\delta(s-s')\left(D_0 T_i
  +\frac{i}{2}\epsilon_{ijk}[T_j,T_k] \right)  \, .
\label{firstpartNahm}
\end{equation}
Aside from the delta function and an overall sign, this is precisely
the right-hand side of the Nahm equation,
Eq.~(\ref{fullNahmequation}).

To complete the reconstruction of the Nahm data, we must determine
what the T-duality transformation does to the fundamental
hypermultiplets, $Q^f_\alpha$.  In the dual D3-D1 picture, we would
proceed as in Sec.~\ref{T} to find an effective theory of D3-branes
transverse to a circle. The trick here is to regard the circle as
$R/Z$ and consider an infinite set of mirror image D3-branes that
repeat themselves with $2\pi R$ shifts in the covering space.  In
the D4-D0 picture, with $k$ D0-branes, we start with a ${\rm
U}(k\times \infty)$ theory and, for all fields but $Y_0$, identify
any given $k\times k$ block with the next $k\times k$ block along
the diagonal direction.  For $Y_0$, the diagonal entries are to be
shifted by $R/\alpha'$ when the indices are shifted by $k$.

The question, then, is precisely what kind of (quasi-)periodicity
condition we should impose on the fundamental scalars. Since these
are charged with respect to the ${\rm U}(n)$ on the D4-branes, the
rule for them must reflect the configurations of the D4-branes.  In
the D3-D1 picture, the $n$ D3-branes are spread out and separated
from each other.  From the $(4+1)$-dimensional Yang-Mills theory
viewpoint, this position information is encoded in the Wilson line
of the gauge field along the compactification circle as
\begin{equation}
\left(Pe^{i\oint A}\right)_{\hbox{D4-Branes}}=
\left(\begin{array}{ccccc}
e^{i  s_1/\tilde R} &0&0& \cdots &0 \\
0&e^{i  s_2/\tilde R} &0&\cdots &0  \\
0&0&e^{i s_3/\tilde R} &\cdots &0 \\
\cdots &\cdots &\cdots &\cdots&\cdots\\
0&0&0&\cdots & e^{i  s_n/\tilde R}
\end{array}\right)
\end{equation}
with $s_f$ being the position of the $f$th D3-brane on the dual
circle. (This can be seen by inverting the process we used in
Sec.~\ref{T}.)  We have normalized the $s_f$ so that they are
periodic in $2\pi \tilde R =2\pi\alpha'/R$; roughly speaking, $s$
lives along the original circle in the eigenvalue space of
$A_4/2\pi\alpha'$.

Since the fundamental hypermultiplet couples to the gauge field on
the D4-branes, the only sensible prescription is to parallel
transport the $Q^f_\alpha$ along the $2\pi R$ shift. This means that
we should require
\begin{equation}
(Q^f_\alpha)_{j+k}= e^{ i s_f/\tilde R}(Q^f_\alpha)_j
\end{equation}
for any value of the color ${\rm U}(k)$ index $j$ and the flavor
${\rm U}(n)$ index $f$. The upshot is that on the dual circle we
find
\begin{equation}
Q^f_\alpha\rightarrow
   \sqrt{2\pi \tilde R} \;\delta(s-s_f)\;Q_f(s)  \, .
\end{equation}

The reason that the $Q^f_\alpha$ depend only on $s$, instead on both
$s$ and $s'$, is that they come from a fundamental representation,
which has only one ${\rm U}(k)$ index, rather than from an adjoint
representation, which has two. Furthermore, because of the factor of
$\delta(s-s_f)$, they are really just a set of numbers sitting at
$s=s_f$, rather than functions of $s$. In the bilinear $Q\otimes
Q^\dagger$, there is no summation over a ${\rm U}(k)$ gauge index,
so the two delta functions remain intact. Thus we find the map
\begin{equation}
Q^f_\alpha\otimes Q^{\dagger f}_\beta\rightarrow 2\pi \tilde
R\;\delta(s-s_f)\,\delta(s'-s_f)\; Q^f_\alpha(s_f)\otimes Q^{\dagger
f}_\beta(s_f)
\end{equation}
or, equivalently,
\begin{equation}
Q^f(s)_\alpha\otimes Q^{\dagger f}_\beta(s)\rightarrow 2\pi \tilde
R\;\delta(s-s') \left[\delta(s-s_f)\;
    Q^f_\alpha\otimes Q^{\dagger f}_\beta\right] \, ,
\end{equation}
where we have dropped the arguments of the $Q^f_\alpha$ on the
right-hand side since they are only defined on the D3-brane
positions $s=s_f$.  Combining this with Eqs.~(\ref{DtermforADHM})
and (\ref{firstpartNahm}), we find that the D-term condition of the
ADHM construction transforms into
\begin{equation}
D_0 T_i +\frac{i}{2}\epsilon_{ijk}[T_j,T_k] = \sum_f 2\pi \tilde
R\;\delta(s-s_f)\sum_{\alpha,\beta}
(\tau_i)^{\alpha\beta}Q^f_\alpha\otimes Q^{\dagger f}_\beta \, .
\end{equation}
The jumping data from the $Q^f_\alpha$ encode the spatial
separations between distinct fundamental monopoles, that is, between
two sets of D1-brane segments on opposite sides of a D3-brane.

When considered locally in $s$, this is precisely the Nahm equation
for a configuration of $k$ distinct fundamental monopoles in an
SU($N$) gauge theory that is spontaneously broken by an adjoint
Higgs whose vev has eigenvalues $(\dots,s_{f-1}, s_f,
s_{f+1},\dots)$ \cite{Kapustin:1998pb,Tsimpis:1998zh}.  The normalization of the
coordinates is such that the mass of the vector meson corresponding
to an open string between the $f$th and the $(f+1)$th D3-branes is
$(s_{f+1}-s_f)/2\pi\alpha'$.  As discussed in
Sec.~\ref{NahmBiggroups}, the Nahm construction for this case
requires jumping data, consisting of a pair of complex $k$-vectors
$a_1$ and $a_2$; these correspond to $\sqrt{2\pi \tilde R}\;Q_1$ and
$\sqrt{2\pi \tilde R}\;Q_2$.

Because we started with a D0-D4 system and then T-dualized, with the
$s_f$ being periodic variables associated with Wilson lines, the
global interpretation of this system of equations differs from that
of the usual Nahm data. However, this is easy to fix: We consider
the limit where the radius $\tilde R=\alpha'/R$ of the dual circle
goes to infinity while (some of) the $s_f$ remain finite. In the
process, we will find that at least one D1-brane segment becomes
infinitely long. We then remove some D1-brane segments to infinity.
We must find a boundary condition for this limit that is consistent
with the monopole interpretation of the D1 segments.

For instance, suppose that we keep $k$ D1-brane segments in one
interval, say on $(s_1,s_2)$, while taking the segments in all other
intervals to spatial infinity.  Without loss of generality, we can
take $-s_1=s_2=\pi\alpha' v$, so that this generates $k$ SU(2)
monopoles of mass $4\pi v/e$. We then have a one-dimensional
self-duality equation on a finite open interval, $(-\pi\alpha'
v,\pi\alpha' v)$,
\begin{equation}
D_0 T_i +\frac{i}{2}\epsilon_{ijk}[T_j,T_k] =0
\end{equation}
with  delta-function sources at $s=\pm \pi\alpha' v$.  Furthermore,
since the other D1-brane segments have been removed to spatial
infinity, the $Q^f_\alpha$ at the ends, and therefore the source
terms, must diverge. The proper thing to do would be to first solve
for the Nahm data, and then in the solution take the limit of some
branes going to infinity. On the other hand, since we are dealing
with a local field theory on the D1-branes, there must be some
boundary conditions at $s=\pm \pi\alpha' v$ that effectively emulate
this procedure and give the correct Nahm data without having to go
through this limiting procedure.

As we discussed in Sec.~\ref{nahmsection}, the Nahm equation without
the source terms allows singular behavior at the boundaries.  The
requirement that the divergence in the commutation term balance
against that in $D_0T$ constrains the singularity at $s=-\pi\alpha'
v$ to be a pole of the form
\begin{equation}
T_i(s)= - \frac{t_i}{s + \pi\alpha' v }+\cdots
\end{equation}
with
\begin{equation}
t_i=\frac{i}{2}\epsilon_{ijk}[t_j,t_k]
\end{equation}
and with similar behavior at $s_2=\pi\alpha' v$. The singular
boundary behavior is determined by the choice of the SU(2)
representation $t_a$.  (Regular behavior also may be included in
this discussion by taking the trivial representation, $t_a=0$.)
This matches the known boundary condition for the Nahm data, given
in Sec.~\ref{nahmsection}, if and only if the $k\times k$ matrices
$t_a$ form an irreducible representation of SU(2).

\begin{figure}[t]
\begin{center}
\scalebox{1}[1]{\includegraphics{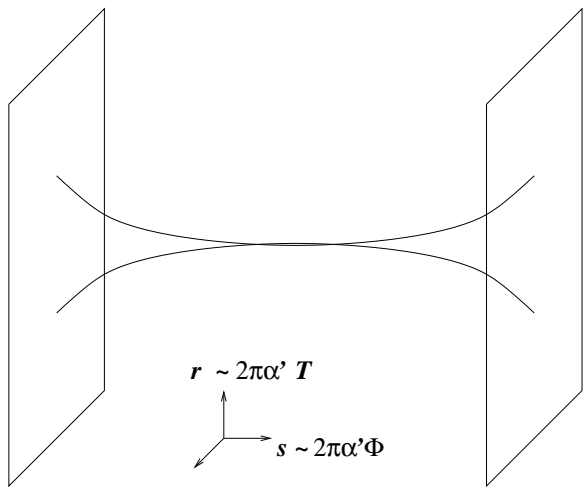}}
\par
\vskip-2.0cm{}
\end{center}
\begin{quote}
\caption{\small Coordinates and scalar fields on the D1-brane and
the D3-brane have a reciprocal relationship. The linear coordinate
$s$ on the D1-brane is encoded in the adjoint Higgs field $\Phi$ on
the D3-brane, while the spatial coordinates $r_a$ on the D3-brane
are encoded in the adjoint Higgs fields $T^a$ on the D1-brane.
\label{D-Nahm} }
\end{quote}
\end{figure}

While it remains quite difficult to show that the irreducible
representation is the only acceptable boundary behavior, there is a
simple physical motivation for this choice. Recall that the D1-brane
picture is itself motivated from the shape of monopole solitons on
D3-branes. As we saw earlier, the monopole solution can be regarded
as a tubular deformation of flat D3-branes, with the tube carrying
D1-brane charge. For large charge or small electric coupling,
however, it is clear that the shape of the tube can be reliably
described by the classical monopole solution where the tube widens
and continuously merges into the D3-branes. The asymptotic form of
the charge $k$ SU(2) monopole solution in the unitary gauge scales
as
\begin{equation}
\Phi_{(2)}-\left(-\frac{v}{2}\right)=\frac{k}{2r}+\cdots  \, .
\end{equation}
On the other hand, near $s_1/2\pi\alpha'= -v/2$ the coordinate
$s/2\pi\alpha'$ on the D1-brane encodes the value of the quantity
$\Phi_{(2)}$, since the latter is really the transverse position
coordinate of the deformed D3-brane (see Fig.~\ref{D-Nahm}). 
In the same spirit, the $T_i$
specify the position of the tube along the D3-brane worldvolume
directions, and are related to the position 3-vector $r_i$ by
$T_iT_i\simeq r^2/(2\pi\alpha')^2I_{k\times k}$. This map tells us
that the monopole solution on the D3-brane would appear as a
configuration of the $T_i$ on the D1-brane such that
\begin{equation}
s-s_1 \simeq \frac{k}{2\sqrt{T_iT_i}}
\end{equation}
or, more precisely,
\begin{equation}
T_iT_i\simeq \left(\frac{k/2}{s-s_1}\right)^2 I_{k\times k} \, ,
\end{equation}
which suggests that the boundary condition must be chosen so that
\begin{equation}
t_it_i\simeq(k/2)^2 I_{k\times k} \, .
\end{equation}
The quadratic Casimir of the $k$-dimensional irreducible
representation is $(k^2-1)/4\simeq (k/2)^2$, so the Nahm data
boundary condition gives us a D1-brane picture consistent with the
D3-brane viewpoint. A reducible representation would be difficult to
reconcile with this in two aspects. The ``size" of $T_iT_i$ would be
smaller since $l^2+(k-l)^2=k^2-2l(k-l) < k^2$ for any positive
$l<k$.  Even apart from the issue of the size, a reducible
representation would make the configuration appear as if there were
two or more unrelated tubes whose sizes were labelled by the size of
the irreducible blocks.

This observation generalizes immediately to the case with arbitrary
numbers of D1-brane segments. If two families of D1-branes, with $k$
and $k'<k$ components, respectively, are on opposite sides of a
D3-brane, the net magnetic charge on the D3-brane is $k-k'$, and the
asymptotics of the BPS monopole is determined by $k-k'$. We
therefore conclude that the boundary condition across the D3-brane
should contain a pole-like divergence on a $(k-k')\times(k-k')$
irreducible block.  This again matches the known Nahm data boundary
condition precisely. A mathematical proof, from string theory, of
whether and why this generically singular boundary condition is the
only consistent choice is still absent as far as we know.  A more
detailed discussion of the relationship between different Nahm data
boundary conditions was given in {Ref.~\cite{Chen:2002vb}}.

\newpage
\centerline{\bf Acknowledgments}

\bigskip

\noindent We are indebted to our collaborators, Dongsu Bak, Xingang Chen, Jerome
Gauntlett, Huidong Guo, Conor Houghton, Chanju Kim,
Nakwoo Kim, Choonkyu Lee, Jaemo Park, Mark Stern, and, especially,
Kimyeong Lee, from whom we have learned much over the years.  We are
also grateful for conversations with Philip Argyres, Oren Bergman,
Frederik Denef, Gary Gibbons, Koji Hashimoto, M\aa ns Henningson, Kentaro Hori, Barak Kol,
Nick Manton, Michael Murray, Paul Sutcliffe, and David Tong.  We would also like to 
thank Jerome Gauntlett and Choonkyu Lee for helpful comments on the
manuscript.

E.W. thanks the Aspen Center for Physics and the Korea Institute for
Advanced Study, and P.Y. thanks the Aspen Center for Physics, Columbia
University, the Fields Institute, and the Institute for Advanced Study, for
their hospitality during the time that this review was written.

This work was supported in part by the Science Research Center Program
of the Korea Science and Engineering Foundation through the Center for
Quantum Spacetime(CQUeST) of Sogang University with grant number
R11-2005-021, and by the U.S. Department of Energy.

\appendix{}

\chapter{ Complex geometry and the geometry of zero modes}

\section{Complex geometry}

While the moduli space always comes with a metric that defines the
affine connection and curvature tensor, the moduli spaces of
supersymmetric solitons are often endowed with additional structures,
such as the K\"ahler and hyper-K\"ahler structures that are required
by the constraints on the dynamics imposed by the supersymmetry. In
this brief appendix we outline some basic concepts and ideas in
complex geometry that are of some relevance for the monopole moduli space.

\subsection{Complex structure and integrability}

A manifold is ``almost complex" if there is a tensor field
$J^m{}_n$
that
rotates any tangent vector by 90 degrees.  Since rotating
by 90 degree twice reverses direction, an invariant way to
state this condition is to say that
\begin{equation}
J^m{}_n J^n{}_k=-\delta^m{}_k \, .
\end{equation}
An example of this is, of course, the complex plane, where the action
of $J$ is induced by multiplication of complex numbers by $i$. In
terms of holomorphic and antiholomorphic vector fields, the
action of $J$ is diagonal:
\begin{eqnarray}
\frac{\partial}{\partial z}&\rightarrow&
-i\frac{\partial}{\partial z}\nn \frac{\partial}{\partial \bar
z}&\rightarrow& +i\frac{\partial}{\partial \bar z} \, .
\label{trivial}
\end{eqnarray}
However, this particular $J$ satisfies many more properties than a
generic ``almost complex structure."

The idea of a complex manifold should be that we can model the
manifold locally by  $C^n$, just as a real manifold is something
that is locally $R^m$. The reason why a manifold with $J^2=-1$, is
called {\it almost} complex is that, without a further
integrability condition on $J$, there is no guarantee that a
holomorphic coordinate system $z^k$ exists and that we can write
$J$ in a simple form as above. For the manifold to be truly
complex, an almost complex structure must satisfy the
integrability condition
\begin{equation}
0=J^m_{\;\;\;n}(\partial_m J^k_{\;\;\;l}-\partial_l J^k_{\;\;\;m})
-J^m_{\;\;\;l}(\partial_m J^k_{\;\;\;n}-\partial_n J^k_{\;\;\;m}) \, .
\end{equation}
The expression on the right-hand-side is known as the Nijenhuis
tensor. If this condition holds, the manifold can be equipped with
holomorphic and antiholomorphic coordinates on which $J$ acts
diagonally, as in the example of the complex plane. Such a
tensor $J$ is called a complex structure.

The Nijenhuis tensor can be regarded as a mapping $N$ of a pair of
vector fields to a third vector field,
\begin{equation}
\frac12N(X,Y)=[JX,JY]-[X,Y]-J[JX,Y]-J[X,JY] \, .
\end{equation}
Note that if there exist a holomorphic coordinate system, as above,
we can write any vector field $X$ in terms of this coordinate basis so
that it can be decomposed into a
$(1,0)$
type vector field and $(0,1)$ type vector field,
\begin{equation}
X=x^k\frac{\partial}{\partial z^k}+ \bar
x^k\frac{\partial}{\partial \bar z^k} \, .
\end{equation}
It is then straightforward to show that the Nijenhuis tensor acting on
any pair of vector fields is zero, which shows that the $N\equiv 0$
condition is necessary for integrability. The only facts we need
to use for this are that $J$ acts linearly and that
\begin{equation}
JX=-ix^k\frac{\partial}{\partial z^k}+ i\bar
x^k\frac{\partial}{\partial \bar z^k} \, .
\end{equation}
More abstractly, the integrability $N\equiv 0$ follows if the
commutator of a pair of vector fields of $(1,0)$ type is again of
$(1,0)$ type and if the corresponding statement holds for vector fields of
$(0,1)$ type. The converse statement is, of course, the difficult
part, for which we refer the reader to the mathematics literature.

\subsection{K\"ahler and hyper-K\"ahler manifolds}
\label{KandHyperKappendix}

A manifold can be equipped with a metric, as in the case of the moduli
space. A complex structure is defined without reference to the metric,
but one may still ask if there should be a compatibility condition
between these two superstructures. One obvious thing to require is
that ``rotation by 90 degree" by $J$ be a symmetry of the metric. That
is, it should leave the metric invariant, so that
\begin{equation}
g(X,Y)=g(JX,JY)
\end{equation}
for any pair of vectors $X$ and $Y$. This is called the
Hermiticity condition.

Furthermore, when a manifold has a metric, it also inherits a
Levi-Civita connection that enables one to parallel transport
tensors. So a natural compatibility condition to ask of any
superstructure on a manifold with metric is that it be
covariantly constant. A complex manifold with a complex structure
$J$ and a Hermitian metric $g$ is called K\"ahler if
\begin{equation}
\nabla J=0  \, .
\end{equation}
Because $J$ is covariantly constant, the curvature tensor must act
trivially on $J$.  This restricts the possible holonomy to be
unitary. In other words, when a $2n$-dimensional manifold is K\"ahler,
one finds a $U(n)$-valued curvature tensor instead of the usual
$SO(2n)$-valued curvature tensor.  The group generated by the
holonomy of the manifold is the structure group of the tangent bundle,
so a fancy way of characterizing a K\"ahler manifold would be to say
that the manifold has a unitary structure group.

Note that for the manifold to be complex
$J$ must already satisfy the integrability condition,
which constrains the gradient of $J$ somewhat. Furthermore, because
the derivatives in the definition of the Nijenhuis tensor are promoted
to covariant derivatives on a manifold with a Hermitian metric and
its affine connection, some terms in $\nabla J$ already vanish
if the manifold is complex. It turns out that the remaining terms
can be grouped into another tensor, which vanishes if and only if
the so-called K\"ahler two-form
\begin{equation}
w_{mn}=-w_{nm}=g_{mk}J^k_{\;\;\;n}
\end{equation}
is closed,
\begin{equation}
dw=0 \, .
\end{equation}
Conversely, if $\nabla J=0$ and the metric is Hermitian, the
Nijenhuis tensor vanishes and the K\"ahler form is closed.
Such a manifold is called K\"ahler.

A manifold is called quaternionic if there are three such complex
structures,\footnote{In the literature there also exists a slightly
different definition of a quaternionic manifold, which refers to a
manifold with three such complex structures and an ${\rm Sp}(2)\times
{\rm Sp}(2k)$ holonomy that allows the three complex structures to
mix among themselves upon parallel transport.}
$(J^{(s)})^m_{\;\;\;n}$, that satisfy the integrability condition and
that obey the algebraic relationship
\begin{equation}
(J^{(s)})^m_{\;\;\;n} (J^{(t)})^n_{\;\;\;k}=
-\delta^{st}\delta^m_{\;\;\;k}+\epsilon^{stu}(J^{(u)})^m_{\;\;\;k}
\end{equation}
at every point.  The idea is, again, that the manifold can be equipped
with a local coordinate system that is modelled after that on the
quaternionic space $H^n = R^{4n}$.

When a quaternionic manifold has a metric, then, we may similarly
require that this metric be Hermitian with respect to all three
complex structures and that, in addition, the three K\"ahler forms
\begin{equation}
w_{mn}^{(s)}= g_{mk}(J^{(s)})^k_{\;\;\;n}
\end{equation}
obey
\begin{equation}
dw^{(s)}=0 \, .
\end{equation}
If a quaternionic structure has these properties, the manifold is
called hyper-K\"ahler. Because there are three covariantly conserved
complex structures, the curvature tensor is even more severely
restricted. A $4n$-dimensional hyper-K\"ahler manifold has a
symplectic structure group; i.e., the structure group of its tangent
bundle is ${\rm Sp}(2n)$.

One unexpected aspect of the hyper-K\"ahler condition is that the
integrability conditions are actually a lot simpler than suggested
above. In fact, the vanishing of the three Nijenhuis tensors is
implied by the three conditions $dw^{(s)}=0$
\cite{Atiyah-Hitchin}. We start by writing any vector field as a sum
of two pieces, each of which is an eigenvector of the almost complex
structure. The $J^2=-1$ condition implies that the only allowed
eigenvalues are $\pm i$; the eigenvectors with these eigenvalues are
of type $(1,0)$ and $(0,1)$ respectively. As already noted, the
integrability condition on a complex structure is equivalent to the
statement that each of these eigensectors is preserved under the
commutator action of vector fields.  Now recall that given an almost
complex structure $J$ and a Hermitian metric $g$ we have
\begin{equation}
w(X,Y)= g(X, JY)
\end{equation}
for any pair of the vector fields $X$ and $Y$. With three complex
structures that form a hyper-K\"ahler structure, we also
have a series of algebraic identities of the form
\begin{equation}
w^{(2)}(X,Y)=g(X,J^{(2)}Y)=g(X,J^{(3)}J^{(1)}
Y)=w^{(3)}(X,J^{(1)}Y)  \, .
\end{equation}
Because of this, a vector field $X$ being of type $(1,0)$ with
respect to $J^{(1)}$ is equivalent to the statement that
\begin{equation}
w^{(2)}(X,Z)=iw^{(3)}(X, Z)
\end{equation}
for any vector field $Z$.

The integrability of the complex structure $J^{(1)}$ follows if,
for any such vector fields $X$ and $Y$, the same relationship
holds for the commutator $[X,Y]$ as well. On the other hand,
$[X,Y]$ is the Lie derivative of $Y$ with respect to $X$, so
\begin{eqnarray}
 w^{(2)}([X,Y],Z) &=&{\cal L}_X (w^{(2)}(Y,Z))
-w^{(2)}(Y,{\cal L}_X Z) - ({\cal L}_X w^{(2)})(Y,Z) \cr &=&
i{\cal L}_X(w^{(3)}(Y,Z)) -iw^{(3)}(Y,{\cal L}_X Z) - ({\cal L}_X
w^{(2)})(Y,Z)  \cr &&
\end{eqnarray}
for any vector field $Z$.  Furthermore, if $dw^{(2)}=0$ we
find
\begin{equation}
{\cal L}_X w^{(2)} = d\langle X, w^{(2)}\rangle =i d\langle X,
w^{(3)}\rangle = i{\cal L}_X w^{(3)}   \, .
\end{equation}
Combining these results gives the identity
\begin{equation}
w^{(2)}([X,Y],Z)=iw^{(3)}([X,Y],Z)  \, ,
\end{equation}
showing that the commutator $[X,Y]$ of a pair of $(1,0)$ type vector
fields is again of $(1,0)$ type.

\subsection{Symplectic and hyper-K\"ahler quotients}

The symplectic quotient should be familiar from
classical mechanics. The phase space of a classical mechanical system is
always a symplectic manifold with the symplectic two-form
\begin{equation}
\Omega=\sum_m dx^m \wedge dp_m  \, ,
\end{equation}
where the $x^m$ are the coordinates and the $p_m$ their conjugate
momenta.  (Recall that a symplectic form is a closed two-form,
$d\Omega=0$, that is nowhere degenerate, and that we call a manifold
symplectic if such a two-form is given.) The phase space is a
particular example of a symplectic manifold, and has the general form
of a cotangent bundle $T^*(X)$ where $X$ is the space spanned by the
configuration space. Together with the Hamiltonian $H(p,x)$, its
symplectic two-form is used to generate the equation of motion.

If one of the coordinates happens to be cyclic,
we can reduce the mechanics problem by removing
the associated degrees of freedom.  This procedure can be
generalized to any symplectic manifold as the ``symplectic quotient."
With the phase space example above, this goes as follows. The
momentum $\nu$ conjugate to a cyclic coordinate $\xi$ is a
constant of motion and can be set equal to a fixed value for any
motion. Recalling that the symplectic form has a term
$d\xi\wedge d\nu$,
we find an invariant way to isolate the conjugate momentum $\nu$
by computing
\begin{equation}
\left\langle \frac{\partial}{\partial \xi}, \Omega\right\rangle
=d\nu
\end{equation}
where the inner product between a vector field $V$ and a
differential $p$-form $\Lambda$ is defined as
\begin{equation}
\left \langle V,\Lambda\right \rangle_{i_1i_2\cdots i_{p-1}}
=\frac{1}{k}\sum_{k=1}^p V^q(-1)^{k-1} \Lambda_{i_1i_2\cdots
i_{k-1}\,q\,i_{k}\cdots i_{p-1}}  \, .
\end{equation}
Setting $\nu$ to a constant value, say $f$, will reduce the
phase space dimension by one, while we actually wish to remove
$\xi$ as well. This second step is achieved by considering a new
phase space
\begin{equation}
\nu^{-1}(f)/G \, ,
\end{equation}
where $G$ is the translational group acting on the phase space as
$\xi\rightarrow\xi+\hbox{constant}$.

The resulting reduced phase space is again symplectic. The reduced
symplectic form is obtained in two steps. First, we pull-back
$\Omega$ to $\nu^{-1}(f)$ by
\begin{equation}
\Omega'=i^*\Omega
\end{equation}
where $i$ is the embedding map of  $\nu^{-1}(f)$ into $T^*(X)$.
Then, we choose any (local) lift map $\sigma$ from $\nu^{-1}(f)/G$
into $\nu^{-1}(f)$ and define
\begin{equation}
\Omega''=\sigma^*\Omega'  \, .
\end{equation}
This two-form on $\nu^{-1}(f)/G$  is closed, since
the pull-back and
exterior derivative always commute,
\begin{equation}
[d,i^*]=[d,\sigma^*] = 0  \, .
\end{equation}
There is a potential ambiguity in this procedure, since $\sigma$ is
not unique. Recall that a lift map is defined by the property that
when followed by the projection $\nu^{-1}(f)\rightarrow
\nu^{-1}(f)/G$, the combined action is an identity map on
$\nu^{-1}/G$.  This leaves a lot of freedom in the choice of $\sigma$.
However, this ambiguity is harmless as long as
$\Omega'(V,\cdot\,)=0$. Thus, $\Omega''$ is naturally a symplectic
form on the quotient manifold.

This quotient procedure generalizes straightforwardly to any
symplectic manifold if we replace the pair $(T^*(X),\Omega)$ by an
arbitrary symplectic manifold $(M,\omega)$ with $d\omega=0$. The role
of $\partial/\partial\xi$ is taken over by any vector field $V$ that
preserves the symplectic form,
\begin{equation}
{\cal L}_V\omega=0 \, .
\end{equation}
The latter condition is essential because
\begin{equation}
0={\cal L}_V\omega=\left\langle V,d\omega\right\rangle
+d\left\langle V,\omega\right\rangle=d\left\langle
V,\omega\right\rangle
\end{equation}
ensures that the moment map is well-defined and also that
\begin{equation}
\omega''=\sigma^*(i^*(\omega))
\end{equation}
is a good symplectic form on the quotient manifold.

When a manifold is hyper-K\"ahler, it is equipped with three
complex structures $J^{(s)}$, a Hermitian metric $g$, and finally
the three K\"ahler two-forms $w^{(s)}$ that are related to the
first two by
\begin{equation}
w^{(s)}(X,Y)=g(X,J^{(s)}Y)
\end{equation}
for any vector fields $X$ and $Y$. The K\"ahler forms
$w^{(s)}$ are nowhere degenerate, since neither $g$ nor  $J^{(s)}$
is degenerate, and are also closed. Therefore, a K\"ahler form
is always a symplectic two-form, and a hyper-K\"ahler manifold is a
symplectic manifold. If there exists a vector field $V$ that
preserves both $w^{(s)}$ and $g$,
\begin{equation}
{\cal L}_V\omega^{(s)}=0,\qquad {\cal L}_Vg=0 \, ,
\end{equation}
one can proceed similarly to perform a symplectic quotient, except
that now there are three ``conjugate momenta", or moment maps,
$\nu^{(s)}$. The quotient manifold,
\begin{equation}
{\cal Q}=\left(\nu_1^{-1}(f_1)\cap
\nu_2^{-1}(f_2)\cap\nu_3^{-1}(f_3)\right) /G \equiv {\cal S}/G \, ,
\end{equation}
is referred to as the hyper-K\"ahler quotient.

This reduced manifold is again hyper-K\"ahler. The triplet of
symplectic forms $\omega^{(s)}$ induce a triplet of symplectic
forms
\begin{equation}
w''^{(s)}=\sigma^*(i^*(w^{(s)}))
\end{equation}
on the reduced manifold, as before,
with $i$ being the embedding map of ${\cal S}$ into $M$ and
$\sigma$  any lift map from the quotient manifold ${\cal Q}$
into ${\cal S}$. This triplet of two-forms are all closed, and
constitute three K\"ahler forms on ${\cal Q}$.

\section{The index bundle and the geometry of zero modes \label{geometry}}

Expanding a massless charged Dirac field in a monopole background, one
encounters fermionic zero modes.  Although the number of zero modes is
invariant under continuous changes of the background, their precise
form depends on the details of the monopole background, and thus the
zero modes can be thought of as forming a bundle over the monopole
moduli space, commonly known as the index bundle \cite{Manton:aa}.  The geometry of
this index bundle encodes a great deal of the information about the
low-energy monopole dynamics, as is shown in Chap.~8 for the case of
pure ${\cal N}=2$ and ${\cal N}=4$ SYM theory, and in
Appendix~B for ${\cal N}=2$ SYM theory  with additional matter
supermultiplets.


The index bundles are equipped with a natural connection
whose holonomies are nothing but the Berry phase associated with
the zero mode equation, namely the time-independent Dirac
equation \cite{Manton:aa}.
To see this clearly, let us concentrate on the part of the action
that couples the Dirac fermion directly to the monopole
background.  For a Dirac fermion in a hypermultiplet this is a term
\begin{equation}
   S_{\rm hyper} = -i\int dx^4\: \left( \bar\Psi\gamma^j D_j \Psi
      +  \bar\Psi b \Psi \right)
   = -i\int dx^4\: \Psi^\dagger\Gamma^a \bar D_a \Psi
\end{equation}
where the $\Gamma_a$ are defined by Eq.~(\ref{chap4eucdirac}), $\bar
D_j = D_j$, and $\bar D_4 = -D_4 = b$.  The fact that the sign in the
$D_4$ term is the opposite of that in the Dirac operator for the
adjoint representation fermion zero modes [as in, e.g.,
Eqs.(\ref{chap4eucdiracoperator}) and (\ref{chap8eucdirac})] is simply
a consequence of the standard conventions for fermions in vector
multiplets and hypermultiplets; the sign can be reversed by
multiplying the fermion field on the left by $\gamma^5$.  Our choice
of signs here is such that the hypermultiplet zero modes are chiral
with respect to $\Gamma_5=\Gamma_1\Gamma_2\Gamma_3\Gamma_4$.

Denoting the zero modes by $\Psi_A$ ($A=1,...,l$), we expand
$\Psi$ as
\begin{equation}
\Psi=\sum_A\psi^A(t)\Psi_A({\bf x},z(t))\, , \qquad
\Psi^\dagger=\sum_A\bar\psi_A(t)[\Psi^A({\bf x},z(t))]^\dagger
\end{equation}
where the $\psi^A$ and their complex conjugates $\bar\psi_A$ are
collective coordinates. The dependence of the zero mode on the
background is summarized by its $z$-dependence. We will normalize
the zero modes so that
\begin{equation}
\int dx^3\;(\Psi^A)^\dagger\Psi_B=\delta^A_B  \, .
\end{equation}
Inserting this back into the action, and integrating over space,
we find
\begin{equation}
  S_{\rm hyper} =
i\int dt \: \left(\bar\psi_A\dot\psi^A + \bar\psi_{A} \dot z^m
{\cal A}^{\;\;\;A}_{m\;\; B}\psi^B  \right)
\label{basicf}
\end{equation}
where
\begin{equation}
{\cal A}^{\;\;\;A}_{m\;\; B}\equiv \int dx^3\: (\Psi^A)^\dagger
\partial_m\Psi_B
\end{equation}
is the holomorphic part of the natural connection.

A natural Hermitian metric
\begin{equation}
h^{\bar AB}=h_{\bar AB}=\delta_{\bar AB}
\end{equation}
exists on the bundle and can be used to raise and lower indices.
One can consistently keep track of the barred (antiholomorphic)
and unbarred (holomorphic) indices with two basic rules:

1) Raising and lowering indices changes barred indices to unbarred
indices and vice versa.

2) Complex conjugation changes barred indices to unbarred indices
and vice versa.

By raising and lowering the indices on $\bar\psi_A$ and $\psi^B$ and
exchanging their order in the above effective Lagrangian, we find
that the natural completion of the connection $\cal A$ is
\begin{equation}
{\cal A}^{\;\;\;\bar B}_{m\;\; \bar A}=- {\cal A}_{m\bar
A}^{\;\;\;\;\;\;\bar B}  \, .
\end{equation}
This can be used to show
\begin{equation}
{\cal A}^{\;\;\;\bar B}_{m\;\; \bar A}= ({\cal A}_{m\bar
B}^{\;\;\;\;\;\;\bar A})^*= ({\cal A}^{\;\;\;B}_{m\;\; A})^*
\end{equation}
where we have used the anti-Hermiticity of ${\cal
A}^{\;\;\;A}_{m\;\; B}$ that follows from its definition. Finally,
${\cal A}^{\;\;\;A}_{m\;\;\bar B} ={\cal A}^{\;\;\;\bar
A}_{m\;\; B} =0$. With this, the tensor defined by
\begin{equation}
I_A^{\;\;\;B}= i\delta_A^{\;\;\;B}, \qquad I_{\bar A}^{\;\;\;\bar
B}=-i \delta_{\bar A}^{\;\;\;\bar B}, \qquad I_A^{\;\;\;\bar
B}=I_{\bar A}^{\;\;\; B} =0
\end{equation}
is covariantly constant. This tensor is nothing but the complex
structure of the index bundle, so the bundle comes with the
unitary structure group ${\rm U}(l)$.

When the hypermultiplet is in a real or a pseudoreal
representation, the structure group gets smaller. To see this,
consider the charge conjugation operation
\begin{equation}
\Psi\rightarrow i\gamma^5 C\bar\Psi^T =i\gamma^0\gamma^5 C\Psi^*
\end{equation}
where the $4\times 4$ real antisymmetric matrix $C$ acts on
spinor indices and satisfies  $C^2=-1$ and $\gamma^\mu C
=-C(\gamma^\mu)^T$. It follows that $\Gamma^a C
=-\,C(\Gamma^a)^T =-C\,(\Gamma^a)^*$. Taking the complex
conjugate of the Dirac equation and using the fact that the zero
modes are antichiral with respect to
$\Gamma^5=-i\gamma^0\gamma^5$, one finds that
\begin{equation}
\Gamma_a \bar D_a^* (C \Psi^*)=0 \, ,
\end{equation}
where $\bar D_a^*$ is the complex conjugation of $\bar D_a$.

An irreducible representation of a Lie algebra is real or
pseudoreal if there exists a constant matrix $R$, acting only on
gauge indices, such that
\begin{equation}
t= R^{-1} t^*R
\end{equation}
where $t$ is any of the anti-Hermitian generators of the gauge group
acting on this representation.
Repeating the complex
conjugation twice, and using Schur's lemma, we find that $RR^*$ and
$R^TR^{-1}$ are both proportional to the identity matrix.
By rescaling $R$ appropriately, then, we can make $R$ to be unitary
and to satisfy
\begin{equation}
RR^*=\pm 1 \, .
\end{equation}
The representation is called real (pseudoreal) when $RR^*=1$
($RR^*=-1$).

When the spinor $\Psi$ is in a real or pseudoreal representation
of the gauge group, we have
\begin{equation}
R^{-1}\bar D_aR =\bar D_a^*  \, ,
\end{equation}
so the transformation
\begin{equation}
\Psi\rightarrow \tilde\Psi \equiv (R\otimes C) \;\Psi^*
\end{equation}
maps a zero mode to another zero mode of the same Dirac field. If
we denote the complete basis of zero modes by $\Psi_A$, as above,
then there must be a matrix ${\cal C}$ such that
\begin{equation}
\tilde\Psi^A= (R\otimes C) (\Psi_A)^* ={\cal C}^{AB}\,\Psi_B \, .
\end{equation}
This leads to the complex conjugate relation
\begin{equation}
(\tilde\Psi_A)^*={\cal C}_{AB}\,(\Psi^B)^*
\end{equation}
with ${\cal C}_{AB}=({\cal C}^{AB})^*$. Taking this conjugation
twice, we find that
\begin{equation}
{\cal C}^2\,\Psi =-RR^*\,\Psi \, .
\end{equation}
(Note that the contraction of any tensor, including $\cal C$, must be
carried out via the bundle metric $h$.)

The most important property of ${\cal C}$ is that it is
covariantly constant. This can be seen by evaluating the
connection in the $\tilde\Psi$ basis and converting it to
the $\Psi$ basis in two different ways.  We find
\begin{equation}
-{\cal A}_{\bar B}^{\;\;\;\bar A}= {\cal C}^{\bar A}_{\;\;\;E}\,
{\cal C}_{\bar B}^{\;\;\;F} {\cal A}^E_{\;\;\;F}
\end{equation}
or, equivalently,
\begin{equation}
{\cal A}\,{\cal C}+{\cal C}{\cal A}^T=0 \, ,
\end{equation}
where indices are contracted using the canonical metric $h$. Thus,
in addition to the metric $h$ and the complex structure $I$, we
find another covariantly constant tensor $\cal C$.
Covariantly constant
tensors always imply reduced structure groups.  Thus:

\begin{itemize}
\item For a Dirac spinor in a complex representation, the index
bundle is complex with the structure group ${\rm U}(l)$.

\item For a Dirac spinor in a pseudoreal representation
$(RR^*=-1,{\cal C}^2=+1)$, the index bundle is real, with the
structure group being at most ${\rm O}(l)$. The tensor $\cal C$
provides a new symmetric bilinear form on the index bundle that is
preserved by ${\rm O}(l)$.

\item For a Dirac spinor in a real representation $(RR^*=+1,{\cal
C}^2=-1)$, the index bundle is symplectic.  The number of zero modes,
$l$, is always even\footnote{One can easily see that $l$ is always
even when ${\cal C}^2= - 1$ by taking the determinant of ${\cal
C}_{AB}\, {\cal C}^{BC}= ({\cal C}^{AB})^* \,{\cal C}^{BC}
= -\delta_A^C$.  The left-hand side gives $|{\rm Det}\,(C^{AB})|^2$,
which is always positive, while the right hand side gives $(-1)^l$
which is positive only when $l$ is even.}, and the structure group is
at most ${\rm Sp}(l)$. The three complex structures that are
associated with this symplectic structure group are $I$, ${\cal C}$,
and $I{\cal C}$.
\end{itemize}

One immediate consequence of this is that the index bundle associated
with any adjoint fermion must be symplectic. Since supersymmetry
forces the index bundle of the adjoint fermion to be identical to the
cotangent bundle of the moduli space manifold, we essentially again
recover the fact that the monopole moduli space is always a
hyper-K\"ahler manifold. The integrability conditions of the three
complex structures were already demonstrated in
Sec.~\ref{moduliSpaceProperties}.

\chapter{Moduli space dynamics with potential in general ${\cal N}=2$ SYM}
\label{nequal2appendix}

In Chap.~\ref{modspacefermionchap} we studied the low-energy
dynamics of monopoles for pure ${\cal N}=2$ and ${\cal N}=4$ SYM theories.
In this appendix we extend this program by considering ${\cal N}=2$ SYM
theory with arbitrary hypermultiplets. In most cases, a hypermultiplet
contributes to the low-energy monopole dynamics via its fermion zero
modes. These zero modes of the matter fermions reside in the
so-called index bundle over the moduli space, and supersymmetry
constrains the geometry of this index bundle in much the same way
that it constrains the geometry of the cotangent bundle where the
adjoint fermion zero modes reside.

In addition, there are instances where the scalar field of a
hypermultiplet can develop a vev without inducing further breaking of
the gauge symmetry. In such cases, the additional vev also affects the
low-energy dynamics. In this appendix, we first derive how the matter
fermion zero modes affect the low-energy dynamics, and then consider
cases where a hypermultiplet also develops a nonzero scalar vev.
We then discuss the resulting modifications to the supersymmetry
algebra and the quantization procedures.  Finally, in
Sec.~\ref{twofromfour}, we show how the results of Chap.~8 for ${\cal
N}=4$ SYM theory can be recovered by viewing this theory as ${\cal N}=2$ SYM
theory with an adjoint representation hypermultiplet.

\section{Monopoles coupled to matter fermions}

\subsection{Pure ${\cal N}=2$ SYM theories revisited}

Let us start by briefly recalling the results of
Sec.~\ref{lowELagInSYM} for pure ${\cal N}=2$ SYM theory, whose Lagrangian we
wrote as
\begin{eqnarray}
\label{susylagab}
 {\cal L} &=& \Tr\Biggl\{-{1\over 2}F_{\mu\nu}F^{\mu\nu}
+  (D_\mu a)^2 +(D_\mu b)^2 + e^2[a,b]^2
\nonumber
\\ && \hskip 1cm + i\bar\chi\gamma^\mu D_\mu\chi -e \bar \chi[b,\chi]
+ie\bar\chi\gamma_5[a,\chi]\Biggr\} \, .
\end{eqnarray}

Given the low-energy ansatz
\begin{eqnarray}
A_a &=&A_a({\bf x},z(t))\nn \chi&=&\delta_q A_a\Gamma^a \,
\zeta \lambda^q(t)
\end{eqnarray}
for the vector multiplet,
the equations of motion imply
\begin{eqnarray}
\label{ansatztwo} A_0&=&\dot z^q\epsilon_q
   -{i\over 4}\phi_{qr}\lambda^q \lambda^r \nn
 a&=&\bar a
    -{i\over 4} \phi_{qr}\lambda^q\lambda^r
\end{eqnarray}
where $\bar a$ is induced by a nonzero vacuum expectation value and
obeys
\begin{equation}\label{cat} D_a \bar a =
  G^q\delta_q A_a
\end{equation}
for an appropriate triholomorphic gauge isometry $G$.  The first
term in $a$ induces a bosonic potential energy, while the interference
between the first and second terms produces a fermionic bilinear term
proportional to $G$.  The final result for the low-energy effective
action for pure ${\cal N}=2$ SYM theory is
\begin{equation}
\label{actiontoo} S={1\over 2}\int dt[\dot z^q \dot z^r g_{qr}+
ig_{qr} \lambda^q D_t \lambda^r - G^q G^r g_{qr} - i\nabla_q G_r
\lambda^q \lambda^r] -{\bf b}\cdot {\bf g}  \, ,
\end{equation}
where
\begin{equation}
 D_t\lambda^q=\dot \lambda^q
+\Gamma^q_{rs}\dot z^r\lambda^s  \, .
\end{equation}

\subsection{Coupling to ${\cal N}=2$ matter fermions\label{matterfermion}}

A massless hypermultiplet with complex scalar fields $H_1$ and $H_2$
and a fermion field $\Psi$ enters the ${\cal N}=2$ SYM Lagrangian
through the additional terms
\begin{eqnarray}
\label{matterlag}
 {\cal L}_{\rm H}&=&\frac12 D_\mu\tilde H^{\dagger i} D^\mu \tilde H_i
      + i \bar\Psi\gamma^\mu D_\mu \Psi
      - e\bar\Psi(b - i\gamma_5 a)\Psi \nn
   &&+ e\tilde H^{\dagger 1}\bar\chi\Psi +e \bar\Psi\chi \tilde H_1
     + ie\tilde H^{\dagger 2}\bar\chi^c\gamma_5\Psi
     + ie\bar\Psi\gamma_5\chi^c \tilde H_2 \nn
   &&-\frac{e^2}2 \tilde H^{\dagger i} (b^2 + a^2) \tilde H_i
     -\frac{e^2}8 [(\tau_i)_\alpha^\beta \tilde H^{\dagger \alpha} t^s
      \tilde H_\beta]^2
\end{eqnarray}
where $\chi^c$ is the charge conjugate of $\chi$. The $t^s$ are
Hermitian generators in the matter representation and $\tau^a$
are the three Pauli matrices, acting as generators of the SU(2)
R-symmetry.

The hypermultiplet zero modes $\Psi^A$ enter the Lagrangian through
the ansatz
\begin{equation}
\label{hyperzeromodeansatz} \Psi=\psi^A(t)\Psi_A  \, ,
\end{equation}
with the Grassmanian complex variables $\psi^A(t)$ playing the role of
collective coordinates. These live in the index bundle and show up
in the low-energy effective Lagrangian coupled to the connection
${\cal A}$ of the index bundle.

In addition, there are interaction terms that are generated because
the excitation of $\Psi$ contributes to the equations of motion of the
bosonic fields. To leading order these excitations determine the
scalar fields in the same hypermultiplet via
\begin{eqnarray}\label{ansatzfour}
   \tilde H_1&=&{2e\over D^2}(\bar\chi\Psi)\nn \tilde H_2&=&{2ie\over
D^2}(\bar\chi^c\gamma^5\Psi)
\end{eqnarray}
with $\Psi$ given by Eq.~(\ref{hyperzeromodeansatz}).  They also
change the expressions for the vector multiplet fields in
Eq.~(\ref{ansatztwo}) to
\begin{eqnarray}
\label{ansatzthree}
   A_0&=&\dot z^q\epsilon_q - {i\over 4} \phi_{qr}\lambda^{q}
    \lambda^r +\frac{ie}{D^2}(\Psi^\dagger t^s\Psi) t^s
 \nn a&=&\bar a - {i\over 4}\phi_{qr}\lambda^{q} \lambda^r
  +\frac{ie}{D^2}(\Psi^\dagger t^s\Psi) t^s  \, .
\end{eqnarray}

These interactions generate terms in the effective action that are
quartic and quadratic in the fermionic variables. The quartic term,
\begin{equation}
{\cal F}_{qr\bar A B}\lambda^q\lambda^r\psi^{\bar A}\psi^B  \, ,
\end{equation}
which couples a pair of $\psi^A$ and a pair of $\lambda^q$ via the
curvature tensor of the index bundle, arises in a manner similar to
the quartic term in the ${\cal N}=4$ SYM case studied in Chap.~8.
Because the index bundle is now more general, the expression for the
curvature tensor in terms of fields is more involved. Following the
procedure of Cederwall et al.~\cite{Cederwall:1995bc}, one can show
that the field strength for the index bundle of $\Psi$ is
\begin{equation}
{\cal F}_{qr\bar AB}=<{\cal D}_q\Psi_{\bar A}|\Pi {\cal D}_r\Psi_B>
-<{\cal D}_r\Psi_{\bar A}|\Pi {\cal D}_q\Psi_B> +e<\Psi_{\bar
A}|\phi_{qr}\Psi_B>   \, ,
\end{equation}
where the zero modes are related
via the completeness relation
\begin{equation}
\label{completeness} |\Psi_A>\delta^{A \bar B}<\Psi_{\bar B}|+\Pi
+{1-\Gamma_5\over 2}=1
\end{equation}
to the projection operator
\begin{equation}
  \Pi=\gamma_5{\Ds}{1\over D_a D_a}{\Ds}{1+\Gamma_5\over 2}\gamma_5
\end{equation}
that projects onto the chiral non-zero
modes.

A less familiar term arises from the interaction between $\bar a$ and
the $\psi^A$ in the Yukawa coupling.  This term has the general form
\begin{equation}
\label{holland} -i\psi^{\bar A}\psi^B T_{\bar AB}
\end{equation}
with $\psi^{\bar A}$ being the complex conjugate of $\psi^{A}$
and
\begin{eqnarray}
T_{\bar AB}= &=&e\int d^3  x \Psi^{\dagger A} \bar a \Psi^B\nn
&\equiv & e\langle\Psi_{\bar A}|\bar a\Psi_B\rangle
\end{eqnarray}
satisfying the consistency
condition\footnote{ To obtain this condition, consider
\begin{equation}
\label{rain}
\partial_q T_{\bar AB}= e\langle{\cal D}_q\Psi_{\bar A}|\bar a\Psi_B\rangle +
e\langle\Psi_{\bar A}|({\cal D}_q\bar a)\Psi_B\rangle +
e\langle\Psi_{\bar A}|\bar a {\cal D}_q\Psi_B\rangle  \, .
\end{equation}
By using Eq.~(\ref{completeness}), the first term in
Eq.~(\ref{rain}) can be rewritten using
\begin{equation}
   \langle{\cal D}_q\Psi_{\bar A}|\bar a\Psi_B\rangle =
 \langle{\cal D}_q\Psi_{\bar A}|\Psi_C\rangle \delta^{C\bar C}
\langle\Psi_{\bar C}|\bar a\Psi_B\rangle + \langle{\cal
D}_q\Psi_{\bar A}|\Pi\bar a\Psi_B\rangle \, .
\end{equation}
The first term is $-{\cal A}_{q\bar A C}\delta^{C\bar C} T_{\bar
CB}$. Using the identity
\begin{equation}
\Ds\gamma_5(e\bar a\Psi_{A} - G^q {\cal D}_q\Psi_A)=0 \, ,
\end{equation}
which can be proven by acting with $G^q {\cal D}_q$ on $\Ds \gamma_5
\Psi_A=0$, we can rewrite the second term as
\begin{equation}
G^r\langle{\cal D}_q\Psi_{\bar A}|\Pi {\cal D}_r\Psi_B\rangle \, .
\end{equation}
The last term in Eq.~(\ref{rain}) can be manipulated in a similar
manner. Putting all this together, we obtain
\begin{equation}
\nabla_q T_{\bar AB}= G^r\{\langle{\cal D}_q\Psi_{\bar A}|\Pi {\cal
D}_r\Psi_B\rangle -\langle{\cal D}_r\Psi_{\bar A}|\Pi
  {\cal D}_q\Psi_B\rangle
  + e\langle\Psi_{\bar A}|\phi_{qr}\Psi_B\rangle\} \, .
\end{equation}
The expression inside the bracket is precisely the curvature of the
index bundle, which gives us the condition in Eq.~(\ref{tues}).}
\begin{equation}
\label{tues} T_{\bar AB;q}= {\cal F}_{qr \bar AB}G^r  \, .
\end{equation}
Because $T$ is anti-Hermitian, in a real basis Eq.~\p{holland}
becomes $-i \psi^M\psi^N T_{MN}/2$ with $T_{MN}=-T_{NM}$.

Combining these terms with the results for the pure ${\cal N}=2$ SYM
theory given in Eq.~(\ref{actiontoo}), and employing a real basis for all
fermions, we obtain \cite{Gauntlett:2000ks},
\begin{eqnarray}
 L &=&{1\over 2} \biggl( g_{qr} \dot{z}^q \dot{ z}^r +
ig_{qr} \lambda^q D_t \lambda^r
 - g^{qr} G_q G_r - i\nabla_q G_r  \lambda^q \lambda^r \nn
&&+i\psi^M{\cal D}_t\psi^M + {1\over 2}{\cal F}_{qr MN}\lambda^q
\lambda^r\psi^M\psi^N - i T_{MN}\psi^M\psi^N\biggr) -{\bf b\cdot g}
\, . \cr && \label{action1}
\end{eqnarray}

\subsection{Massive matter fields \label{massivematter}}

Adding a bare mass term for the hypermultiplet slightly modifies the
above action. The mass term for the fermions is
\begin{equation}
\label{masspert}
  {\cal L}_M = m_R\bar\Psi\Psi - m_I i\bar \Psi\gamma_5\Psi \, .
\end{equation}
There is a global $U(1)$ rotation in ${\cal N}=2$ SYM theory that mixes the
real and the imaginary parts of the fermion mass and also rotates the
two adjoint Higgs fields of the vector multiplet into each other.
Once we have fixed this rotation by our choice of the adjoint
Higgs fields $b$ and $a$, the real and imaginary parts of the fermion
mass are similarly determined.

If we regard this mass term as a small perturbation of the same order of magnitude
as $a$,
the new terms in the low-energy Lagrangian are obtained by
substituting our ansatz, Eq.~(\ref{hyperzeromodeansatz}). Because the
zero modes are chiral with respect to $\Gamma_5$, only the second term
in Eq.~(\ref{masspert}) is nonvanishing, leading to
\begin{equation}
 L_M =  m_I\psi^{\bar A}\psi^B \delta_{\bar A B}
  = {i\over 2}m_I\; \psi^M I_{MN}\,\psi^N \, .
\end{equation}
In the second equality we have converted to a real basis, with $I$
being the natural complex structure on the index bundle. This mass
term is readily incorporated in the supersymmetric quantum mechanics
by adding it to $T_{MN}$ via
\begin{equation}
T\rightarrow T-m_II  \, ,
\end{equation}
since the differential condition on $T$ allows a shift of $T$ by a
covariantly constant piece.

Note that $I$ exists for any index bundle, regardless of the gauge
representation of the hypermultiplet.  The structure group of the
index bundle here is unitary by default. When the matter fermion is
in a real or pseudoreal representation of the gauge group, the
structure group becomes smaller and degenerates to an orthogonal or
symplectic group, but $I$ always remains a part of the index bundle
structure.

\section{Monopoles coupled to a hypermultiplet vev in a
real representation }

For certain hypermultiplets it is possible to turn on a scalar vev
while leaving the U(1) gauge symmetries of the Coulomb phase
intact. More specifically, this can be done when the matter
representation contains a zero-weight vector. In such cases, this
leads to additional potential energy terms in the low-energy monopole
dynamics, in much the same way as the vector multiplet scalars did.
Of particular importance to us is the case of a hypermultiplet in the
adjoint representation, but other cases include symmetric tensors for
${\rm SO}(k)$ and antisymmetric tensors for ${\rm Sp}(2k)$. All three
of these are real representations, and we will restrict ourselves to
this subclass.\footnote{The most general low-energy dynamics with a
hypermultiplet vev in an arbitrary representation has recently been
worked out \cite{Kim:2006tq}.}

A low-energy ansatz that solves the equations of motion to
leading order is obtained by shifting the scalar fields so that
Eq.~(\ref{ansatzfour}) becomes
\begin{eqnarray}
\label{ansatzsix}
\tilde H_1&=&\bar H_1+{2e\over D^2}(\bar\chi\Psi)\nn \tilde
H_2&=&\bar H_2+{2ie\over D^2}(\bar\chi^c\gamma^5\Psi)
\end{eqnarray}
where the $\bar H_i$ solve the covariant Laplace equation in the
monopole background,
\begin{equation}
D^2 \bar H_i=0   \, ,
\label{D2barH}
\end{equation}
and are equal to the corresponding nonzero expectation values at
spatial infinity.  The new terms that arise from this shift can be
either linear or quadratic in the $\bar H_i$. The former generate
terms with fermionic bilinears, while the latter correspond to bosonic
potential energy terms. Since we are assuming that the hypermultiplet
is in a real representation, it is convenient to introduce four real
fields $H_a$ that are obtained from the two complex scalars $\bar H_i$
via
\begin{eqnarray}
    &&\bar H_1 =  H_3
+ i H_0 \nn &&\bar H_2 = - H_1 + i H_2  \, .
\end{eqnarray}

Now note that Eq.~(\ref{D2barH}) implies that if $\zeta$ is an
arbitrary spinor with positive chirality under $\Gamma_5$, then $\Ds
H_a \zeta$ is annihilated by $\Ds=\Gamma^bD_b$.  It follows that we
can expand the former quantity in terms of the zero modes of
$\gamma^5\Psi$ and write
\begin{equation}
  \Ds H_a\zeta =-i\gamma^5\sqrt{2}\sum_A \tilde
   K_a^A\Psi_A   \, .  \label{DH}
\end{equation}
This defines sections $\tilde K_a^A$ over the moduli space.  We will
find it more convenient to re-express these in terms of a real basis,
denoted by $K^M_a$.
Because the index bundle of $\Psi$ is symplectic, there are
three complex structures, $I^{(i)}$, that act naturally on both
$\psi^M$ and $K_a^M$.  As in the pure ${\cal N}=2$ SYM case, these
additional geometric quantities are tightly constrained by
supersymmetry and the zero-mode equations. Like the $J^{(i)}$, the
$I^{(i)}$ are all covariantly constant, and the section $K$ must obey
\begin{eqnarray}
&&\nabla_q K^M_a=(J^{(k)}\nabla)_q(I^{(k)}K)_a^M  \label{holo}
\end{eqnarray}
for $k=1$, 2, or 3.  (No summation on $k$ is implied here.)

A detailed derivation of the low-energy effective action can be
found in Ref.~\cite{Gauntlett:2000ks}.  Here we simply quote the
results. The bosonic potential energy,
\begin{equation}
\frac12\sum_{a=0}^3 K_a^M K_{aM} \, ,
\end{equation}
is reminiscent of the bosonic potential energy
from the adjoint Higgs field.  The fermion
bilinears from the Yukawa terms are
\begin{equation}
   -i\lambda^q \nabla_q
K_{0M}\psi^M +i\sum_{k=1}^3\lambda^q J^{(k)r}_q\nabla_r K_{kM}\psi^M
\,.
\end{equation}
The identity Eq.~\p{holo} allows us to combine these into a single sum,
\begin{equation}
\label{lastequation} - i\sum_{a=0}^3\lambda^q
I^{(a)N}_MK_{aN;q}\psi^M  \, ,
\end{equation}
by defining
$I^{(0)}=I$.

Adding these contributions to those found previously we find the
low-energy effective Lagrangian for the case of one real
hypermultiplet \cite{Gauntlett:2000ks},
\begin{eqnarray}
\label{aaa2}  L &=&{1\over 2} \biggl( g_{qr} \dot{z}^q \dot{ z}^r +
ig_{qr} \lambda^q D_t \lambda^r +i\psi^M{\cal D}_t\psi^M  + {1\over
2}{\cal F}_{qr MN}\lambda^q \lambda^r\psi^M\psi^N  \nn && - g^{qr}
G_q G_r - i\nabla_q G_r  \lambda^q \lambda^r -i
T_{MN}\psi^M\psi^N\nn && -\sum_{a=0}^3 K_a^M K_{aM} - 2i
\sum_{a=0}^3I_{M}^{(a)\,N} K_{aN;q}\lambda^q\psi^M\biggr) \, .
\end{eqnarray}

The action we have written here is appropriate for ${\cal N}=2$ SYM
theory with a single real hypermultiplet, with a vev, that contributes the
tensors $K^M$ and the Grassmann variables $\psi^M$.  For other
hypermultiplets without vevs, we can simply turn off the $K^M$ and
keep the $\psi^M$.  The index bundle for the $\psi^M$ is symplectic,
with structures $I^{(k)}$, only if the hypermultiplet is in a real
representation.  If the representation is pseudoreal or complex the
$I^{(k)}$ are absent, but they are not needed once we drop the
$K^M$.

\section{Symmetries and Superalgebra\label{LSUSY}}

Just as in the case of pure ${\cal N}=2$ SYM theory, the low-energy
effective action for the theory with additional matter hypermultiplets
is invariant under four supersymmetry transformations.  In this
section we will briefly describe these invariances and discuss the
quantization of the theory.  Many of the equations here follow closely
those in Sec.~\ref{n=2q}, but with modifications arising from the presence
of the additional hypermultiplets.

\subsection{Symmetries and Constraints}

The low-energy effective action obtained by integrating
Eq.~(\ref{aaa2}) over time is invariant under the four supersymmetry
transformations
\begin{eqnarray}
\delta z^q &=& -i\ep\lambda^q -i\sum_{k=1}^3\ep_{(k)}
\lambda^r {J_{\;r}^{(k)q}}\nn
   \delta \lambda^q &=&\ep(\dot z^q -G^q)+i\ep\lambda^r\Gamma^q_{rs}\lambda^s
  +\sum_{k=1}^3 \ep_{(k)}\left[- (\dot z^r - G^r){J_{\;r}^{(k)q}}
+i\lambda^r  {J_{\;r}^{(k)t}} \Gamma^q_{ts}\lambda^s\right] \nn
  \delta\psi^M&=&-{{{\cal A}_q}^M}_N\delta z^q\psi^N-
\ep \sum_{a=0}^3I^{(a)M}{}_N K_a^N
   -\sum_{k=1}^3\sum_{a=0}^3 \ep_{(k)} I^{(a)M}{}_N I^{(k)N}{}_L K_a^L
  \cr &&
\end{eqnarray}
where $\epsilon$ and the three $\ep_{(k)}$ are constant Grassmann-odd
parameters. As in Eq.~(\ref{N2transformation}), the
$\Gamma^q_{rs}\lambda^r\lambda^s$ term in the second line vanishes
identically but has been kept for the sake of a symmetrical appearance
of the four possible supersymmetry transformations.

The action is also invariant under the symmetry
transformation
\begin{eqnarray}
\delta z^q &=&k G^q\nn
  \delta\lambda^q &=&k{G^q}_{,r}\lambda^r\nn
\delta\psi^M&=&k {T^M}_N\psi^N-{\cal A}^M_{q N}\delta z^q\psi^N \, ,
\end{eqnarray}
with $k$ a small real number, that is generated by the
triholomorphic Killing vector field $G$.

Demonstrating the invariance of the action under these symmetry
transformations requires certain geometric properties of various
quantities on the moduli space.  In addition to the hyper-K\"ahler
property of the moduli space and the triholomorphic Killing condition
on $G$, we have new constraints on $K$, $I$, and $T$.  These must satisfy
Eqs.~(\ref{tues}) and (\ref{holo}), as well as the differential
constraint
\begin{equation}
  G^q\nabla_q K_{aM}= T_{M}^{\;\;\;N}K_{aN} \label{GK}
\end{equation}
and the algebraic constraints
\begin{eqnarray}
&&K_s^MI^{(k)}_{MN} K_t^N=0 \label{abcd} \\ \cr
&& I^{(k)L}{}_N T_{LM}= I^{(k)L}{}_M T_{LN} \, .
\label{abcde}
\end{eqnarray}
We refer readers to the original literature \cite{Gauntlett:2000ks}
for a complete derivation of these constraints from the supersymmetric
field theory.

\subsection{Quantization \label{n=2algebra}}

As in Sec.~\ref{n=2q}, to
quantize the effective action we first introduce a frame $e^E_q$
and define $\lambda^E=\lambda^q e_q^E$ that commute with all
bosonic variables. The remaining canonical commutation relations are
then given by
\begin{eqnarray}
 [z^q,p_r]&=&i\delta^q_r   \nn
\{\lambda^E,\lambda^F\}&=&\delta^{EF} \nn
\{\psi^M,\psi^N\}&=&\delta^{MN} \, .
\end{eqnarray}
This algebra is realized in terms of
spinors on the moduli space by
letting $\lambda^E=\gamma^F/{\sqrt 2}$, where the $\gamma^F$ are gamma
matrices. The states must also provide a representation of the
Clifford algebra generated by the $\psi^M$. The supercovariant
momentum operator defined by
\begin{equation}
\pi_q=p_q-{i\over 4}\omega_{q EF}[\lambda^E,\lambda^F] -{i\over 2}
{\cal A}_{qMN}\psi^M \psi^N,
\end{equation}
where $\omega_{q\, F}^{\, \, E}$ is the spin connection, then
becomes the covariant derivative acting on spinors twisted in an
appropriate way by $\cal A$. Note that
\begin{eqnarray}
{[}\pi_q,\lambda^r{]}&=&i\Gamma^r_{qs}\lambda^s \nn
{[}\pi_q,\psi^M{]}&=&i{{{\cal A}_q}^M}_N\psi^N \nn
{[}\pi_p,\pi_q{]}&=& -{1\over 2}R_{pqrs}\lambda^r\lambda^s -{1\over
2}{\cal F}_{pq MN}\psi^M\psi^N \, .
\end{eqnarray}

The four supersymmetry charges take the form
\begin{eqnarray}
Q&=&\lambda^q(\pi_q-G_q)-\psi^M \sum_{a=0}^3 [I^{(a)} K_a]_M \nn
Q_{j}&=&\lambda^q {J^{(j)r}_{\,\, q}}(\pi_r-G_r) - \psi^M
\sum_{a=0}^3 [ I^{(a)} I^{(j)}K_{a}]_M \, , \qquad j=1,2,3 \, .
\end{eqnarray}
They again satisfy
\begin{eqnarray}
  \label{algebra}
\{Q,Q\}&=&2({ H}-{ Z})   \nn \{Q_j,Q_k\}&=&2\,\delta_{jk}({ H}-{
Z})  \nn \{Q,Q_j\}&=&0 \, ,
\end{eqnarray}
but with the Hamiltonian ${ H}$ and the central charge ${ Z}$ now
given by
\begin{eqnarray}
{ H}&=& {1\over 2\sqrt{g}}\pi_q \sqrt{g }g^{qr}\pi_r + {1\over 2}G_q
G^q  + {i\over 2}\lambda^q\lambda^r \nabla_q G_r \nn && +{i\over 2}
\psi^M \psi^N T_{MN} -{1\over 4}{\cal F}_{qr
MN}\lambda^q\lambda^r\psi^M \psi^N \nn && +\frac{1}{2}\, K_a^M
K_{aM} + i I_{M}^{(a)\,N} K_{aN;q}\lambda^q\psi^M \nn { Z}&=& G^q
\pi_q -{i\over 2}  \lambda^q\lambda^r(\nabla_q G_r) +{i\over
2}\psi^M \psi^N T_{MN} \, .
\label{hamiltonian2}
\end{eqnarray}
The operator $i{ Z}$ is  the Lie derivative ${\cal L}_G$
acting on a spinor, twisted by $T$.

\section{Recovering ${\cal N}=4$  from ${\cal N}=2$}
\label{twofromfour}

A special case of the ${\cal N}=2$ theory we have considered is when
there is a single massless adjoint hypermultiplet, and no other matter
fields.  The field theory then possesses enhanced supersymmetry and
is, in fact, just ${\cal N}=4$ SYM theory.  This thus
provides a second route to the low-energy action that we obtained in
Sec.~\ref{n=2q}.  Let us examine in more detail how this comes about.

We start with the observation that the fermionic coordinates $\psi^M$
now live in the cotangent bundle of the moduli space, just as the
$\lambda^q$ do. These are then naturally combined into a doublet,
\begin{equation}
\eta^q=\left(\begin{array}{cc}\lambda^q \\
\psi^q\end{array}\right)  \, .
\end{equation}
The curvature ${\cal F}$ on the index bundle for the $\psi$
naturally becomes the curvature tensor of the moduli space,
\begin{equation}
{\cal F}_{pqMN}\rightarrow R_{pqrs}\label{FR}  \, .
\end{equation}

The vevs (if any) of the hypermultiplet scalar fields
lead to bosonic potential energy terms, as well as to terms that are
bilinear in both the vector multiplet and hypermultiplet fermions.
All of these terms contain tensors on the moduli
space. Since the sections $K_a$ are now induced by
adjoint scalar fields, they must become triholomorphic Killing
vectors, on an equal footing with $G$. We therefore introduce the
notation $G_I$ by
\begin{equation}
 K_j\rightarrow G_j\, ,\quad j=1,2,3 \, ;\qquad G\rightarrow G_4
     \, ;\qquad
    K_0\rightarrow G_5   \, .
\end{equation}
As we saw in Chap.~\ref{modspacefermionchap}, these five Killing
vectors can be intermingled by the action of an SO(5) subgroup of
the ${\rm SO}(6)_R$ symmetry of ${\cal N}=4$ SYM theory in such a way as to
preserve the low-energy dynamics.

The antisymmetric tensor $T$ and the $I^{(j)}$ must also be
associated with the moduli space itself. Using the mapping of
Eq.~(\ref{FR}) in Eq.~(\ref{tues}), we find that $T_{ab}$ becomes
proportional to $dG$. More specifically, we have
\begin{equation}
T_{MN}\rightarrow T_{qr}=-\nabla_q G_r  \, ,
\end{equation}
with the antisymmetrization of indices implicit because of the
Killing properties of $G$. Finally, the $I^{(j)}$ must become the
three complex structures, $J^{(j)}$, on the moduli space.

When all of these changes are incorporated, we recover precisely the
low-energy action for ${\cal N}=4$ SYM theory that was given in
Eq.~(\ref{N4lowenergyLag}).

Furthermore, the various conditions on the $K_i$ and on $G$ must be
lifted to conditions on the $G_I$ in such a way that the five $G_I$
are on equal footing.  Since $G=G_4$ has to be a triholomorphic
Killing vector, so should all of the $G_I$,
\begin{equation}
{\cal L}_{G_{\bI}} g=0 \, ;\qquad
    {\cal L}_{G_{\bI}} J^{(j)}=0 \, ,\quad
   j=1,2,3 \label{tri}   \, .
\end{equation}
Furthermore, $T=-dG_4$ implies that the condition of Eq.~(\ref{GK})
becomes $[G_4, G_I]=0$. Since the five $G_I$ must be on equal
footing, we must have
\begin{equation}
[G_I,G_J]={\cal L}_{G_{\bI}} G_\bJ=0 \label{commu}
\end{equation}
for all pairs. Equations~(\ref{holo}) and (\ref{abcde}) imply that
the two-forms $dG_I$ are all of type $(1,1)$ with respect to each of
the three complex structures on the moduli space. That is,
\begin{equation}
[J^{(j)}\nabla]_q[J^{(j)}G_I]_r=\nabla_q G_{Ir}\label{holo'}
\end{equation}
for $j=1$, 2, or 3, and all five values of $I$.

Using the condition that the K\"ahler forms be closed, $dw^{(j)}=0$,
and Eqs.~(\ref{commu}) and (\ref{holo'}), we find that
\begin{equation}
\nabla\left[G_{\bI}^q \,w^{(j)}_{qr}\,
G_{\bJ}^r\right]=0   \, .
\end{equation}
However, the supersymmetry actually requires that the quantity
inside the parenthesis vanish, so
\begin{equation}
G_{\bI}^q\,w^{(j)}_{qr}\, G_{\bJ}^r =0
\end{equation}
for all three values of $j$. This is the condition lifted from
Eq.~(\ref{abcd}).

\end{document}